\documentclass[12pt,a4paper]{article}
\usepackage[dvips]{epsfig}

\begin{document}
\author{\normalsize\bf Yu.A. Markov$\!\,$\thanks{e-mail:markov@icc.ru}
$\,$and M.A. Markova$^*$}
\title{Nonlinear dynamics of soft fermion\\
excitations in hot QCD plasma  II:\\
Soft-quark\,--\,hard-particle scattering\\
and energy losses}
\date{\it Institute for System Dynamics\\
and Control Theory Siberian Branch\\
of Academy of Sciences of Russia,\\
P.O. Box 1233, 664033 Irkutsk, Russia}

\thispagestyle{empty}
\maketitle{}


\def\theequation{\arabic{section}.\arabic{equation}}

\[
{\bf Abstract}
\]
In general line with our first work \cite{markov_NPA_06} within
the framework of semiclassical approximation a general theory for
the scattering processes of soft (anti)quark excitations off hard
thermal particles in hot QCD-medium is thoroughly considered. The
dynamical equations describing evolution for the usual classical color
charge $Q^a(t)$ and Grassmann color charges $\theta^i(t),
\theta^{\dagger i}(t)$ of hard particle taking into account the soft
fermion degree of freedom of the system are suggested. On the basis of
these equations and the Blaizot-Iancu equations iterative
procedure of calculation of effective currents and sources
generating the scattering processes under consideration is defined
and their form up to third order in powers of free soft quark
field, soft gluon one, and initial values of the color charges of hard
particle is explicitly calculated. With use of the generalized
Tsytovich principle a connection between matrix elements of the
scattering processes and the effective currents and sources is
established. In the context of the effective theory suggested 
for soft and hard fermion excitations new mechanisms of energy losses 
of high-energy parton propagating through QCD-medium are considered.


\newpage

\section{Introduction}
\setcounter{equation}{0}

In the second part of our work we proceed with our analysis of
dynamics of fermion excitations in hot QCD-medium at the soft
momentum scale, started in \cite{markov_NPA_06} (to be referred to
as ``Paper I'' throughout this text) in the framework of the
{\it hard thermal loop} effective theory. Here we focus our research on
study of the scattering processes of soft quark plasma
excitations off hard thermal (or external) particle within real
time formalism based on Boltzmann type kinetic equations for
soft quark, antiquark and gluon modes. This problem has been
already discussed earlier in our work \cite{markov_PRD_01} in 
somewhat other technique of calculation than presented here. In
this work we consider this problem in great detail and by
systematic way using the technique of construction of effective
currents and sources generating the scattering processes we are 
interested in.

Our approach is based on the complete system of dynamical equations derived by
Blaizot and Iancu \cite{blaizot} complemented by the Wong equation \cite{wong}
describing a change of the classical color charge
$Q=(Q^a),\,a=1,\ldots,\,N_c^2-1$ of hard color-charged particle and by the
dynamical equations for the Grassmann color charges $\theta=(\theta^i)$
and $\theta^{\dagger}=(\theta^{\dagger i}),\,i=1,\ldots,\,N_c$ for the first
time proposed in the papers \cite{barducii}. The use of the last equations from
above-mentioned ones is the main new ingredient of the effective theory for
nonlinear interaction of hard and soft modes in a hot quark-gluon plasma
developed in the present work. Introduction in consideration of the Grassmann
color charges of hard particle on the equal footing with usual color
charge enables us to enter a so-called color (Grassmann) source of
a spin-1/2 hard particle along with classical color current.
We add this Grassmann source to the right-hand side of the Dirac equation
for soft-quark field just as we add the usual color current of hard particle
to the right-hand side of the Yang-Mills equation \cite{markov_AOP_04}.
This allows us to obtain closed self-consistent description of nonlinear
interaction dynamics of soft and hard excitations both Fermi and Bose statistics
(within the framework of semiclassical approximation).

Unfortunately, Wong's equation and equations for the
Grassmann charges in that form as they have been derived in
original works \cite{wong, barducii} are insufficient for obtaining complete 
and gauge-invariant expressions for matrix elements of the scattering
processes under consideration. The reason of this is that these
equations have been obtained on the assumption that there exists only (regular 
and/or stochastic) gluon field $A_{\mu}^a(x)$ in the system. In this work, we 
suggest a minimal extension of these equations to the case of the presence of
soft (stochastic) quark field $\psi_{\alpha}^i(x)$ in system. These
generalized equations generate new gauge-covariant additional
color currents and sources of the hard test particle, which we add to
the right-hand side of the corresponding field equations. In this case only
we are able to calculate complete and gauge-covariant expressions for 
effective currents and sources generating the scattering processes of soft
quark excitations off hard particle.

We apply the current approach to study of the propagating a high
energy parton (gluon or quark) through the hot QCD-medium and energy
losses associated with this motion. We show that the account of an
existence in the medium of soft excitations obeying Fermi statistics
results in appearance of new channels for energy losses and write
out complete explicit expressions determining the energy losses
to the first orders in interaction with soft stochastic fields of the plasma.
As a special case we have obtained the expression for so-called polarization 
losses caused by large distance `inelastic' collisions under which a type
of initial hard parton changes. This expression supplements well-known
expression for the polarization losses caused by large distance
'elastic' scattering \cite{braaten}.

The paper II is organized as follows. In Section 2, preliminary
comments with regard to derivation of a system of the
Boltzmann equations describing a change of the number
densities for soft quark and soft gluon excitations due to their
scattering off hard thermal partons of medium are given. In
Section 3, the self-consistent system of the nonlinear integral
equations for the gauge potential $A_{\mu}^a(k)$ and quark wave
function $\psi_{\alpha}^i(q)$ taking into account presence in
the system of color current and color Grassmann source of hard test
particle is written out. On the basis of perturbative solutions of
these equations the notions of the effective currents and sources
are introduced. In Section 4, examples of calculation of the simplest
effective currents and sources are given. It is shown that
some of the expressions obtained are either incomplete or
gauge-noncovariant. In Section 5, the more general expressions of dynamical
equations describing a change of the usual color charge $Q^a(t)$
and the Grassmann color charges $\theta^i(t),\,\theta^{\dagger
i}(t)$ of the hard test particle on interaction of the last one 
with both soft gluon and soft quark fluctuation fields of system are
suggested. On the basis of these equations it is determined all
necessary additional color currents and sources allowing to derive
complete and gauge covariant expressions for the effective currents and 
sources at least up to third order in powers of free soft gluon field
$A^{\!(0)a}_{\mu}(k)$, free soft quark field
$\psi^{(0)i}_{\alpha}(q)$, and initial value of the color
charges $Q_{0}^a,\,\theta_{0}^i$ and $\theta_0^{\dagger i}$. In
Sections 6 and 7, and Appendices B, D, and E, the explicit examples
of such calculations are given. In Section 8, due to
the Tsytovich correspondence principle the effective currents and
sources derived in previous sections are used for calculation of
the matrix elements of the scattering processes under
consideration. Section 9 is concerned with analysis of a
structure of simplest scattering probability of plasmino by hard particle. 
Here a problem of generalization of results obtained for the case when hard 
test parton is in a partial polarization state is also considered. 
In Section 10 we deal with definition of semiclassical expressions for energy
losses of energetic color particle propagating through hot
quark-gluon plasma taking into account the scattering processes
off soft-plasma excitations carrying a half-integer spin. In
particular an expression for the polarization losses induced by large
distance collisions with a change of statistics of energetic particle
is given.

In Sections 11 and 12, thoroughly the so-called 'non-diagonal'
contributions to the energy losses are analyzed and conditions for
cancellation of off mass-shell singularities of complete
expressions for the energy losses are written out. In Conclusion some
features of dynamics of soft and hard excitations of Fermi and Bose
statistics considered in this work are briefly discussed. Finally,
in Appendix A we give an explicit form for the action varying which one 
obtains the equation of motion for color charge of hard test 
parton and also the soft gluon and quark field equations.
In Appendix C origin of singularity generated by classical soft-quark and
soft-gluon one-loop corrections (Section 7) to the effective currents and 
sources is analyzed and in Appendix F all formulae of the paper
\cite{markov_PRD_01} necessary for analysis of a structure of the simplest
scattering probability are given.

\section{Preliminaries}
\setcounter{equation}{0}

As was mentioned in Introduction in this paper we consider a change of
the number densities of colorless soft-quark
$n^{\pm\,ij}_{\bf q}=\delta^{ij} n^{\pm}_{\bf q}$, soft-antiquark
$\bar{n}^{\pm\,ij}_{\bf q}=\delta^{ij}\bar{n}^{\pm}_{\bf q}$,
and soft-gluon $N_{\bf k}^{t\!,\,l\,ab} = \delta^{ab} N_{\bf k}^{t\!,\,l}$
excitations as a result of their scattering off hard thermal quarks, antiquarks
and hard transverse gluons. Hereafter, we denote momenta of the soft-quark
fields by $q,\,q^{\prime},\,q_1,\ldots$, momenta of the soft-gauge fields by
$k,\,k^{\prime},\,k_1,\ldots\,$, and momenta of the hard thermal particles
by $p,\,p^{\,\prime},\ldots\,.$

We expect that a time-space evolution of scalar functions
$n_{\bf q}^{(f)},\,f=\pm$ and $N_{\bf k}^{(b)},\,b=t,\,l$ will be described by
a self-consistent system of kinetic equations
\begin{equation}
\frac{\partial n_{\bf q}^{(f)}}{\partial t} +
{\bf v}_{\bf q}^{(f)}\cdot\frac{\partial n_{\bf q}^{(f)}}{\partial {\bf x}} =
- n_{\bf q}^{(f)}\,\Gamma_{\rm d}^{(f)}
[n_{\bf q}^{\pm},N_{\bf k}^{t,\,l},f_{\bf p}^{G\!,\,Q}] +
\Bigl(1 - n_{\bf q}^{(f)}\Bigr)\Gamma_{\rm i}^{(f)}
[n_{\bf q}^{\pm},N_{\bf k}^{t,\,l},f_{\bf p}^{G\!,\,Q}]
\label{eq:2q}
\end{equation}
\[
\equiv
-\,{\rm C}^{(f)}[n_{\bf q}^{\pm},N_{\bf k}^{t,\,l},f_{\bf p}^{G\!,\,Q}],
\]
\begin{equation}
\frac{\partial N_{\bf k}^{(b)}}{\partial t} +
{\bf v}_{\bf k}^{(b)}\cdot\frac{\partial N_{\bf k}^{(b)}}{\partial {\bf x}} =
- N_{\bf k}^{(b)}\,\Gamma_{\rm d}^{(b)}
[n_{\bf q}^{\pm},N_{\bf k}^{t,\,l},f_{\bf p}^{G\!,\,Q}] +
\Bigl(1 + N_{\bf k}^{(b)}\Bigr)\Gamma_{\rm i}^{(b)}
[n_{\bf q}^{\pm},N_{\bf k}^{t,\,l},f_{\bf p}^{G\!,\,Q}]
\label{eq:2w}
\end{equation}
\[
\equiv
-\,{\rm C}^{(b)}[n_{\bf q}^{\pm},N_{\bf k}^{t,\,l},f_{\bf p}^{G\!,\,Q}],
\]
where  ${\bf v}_{\bf q}^{(f)}=\partial\omega_{\bf
q}^{(f)}/\partial{\bf q}$, ${\bf v}_{\bf
k}^{(b)}=\partial\omega_{\bf k}^{(b)}/\partial{\bf k}$ are
the group velocities of soft fermionic and bosonic excitations and
$f_{\bf p}^{G}\equiv f^{G}({\bf p},x)$, $f_{\bf p}^{Q}\equiv
f^{Q}({\bf p},x)$ are distribution functions of hard thermal
gluons and quarks, which in general case obey own kinetic
equations. Here for the sake of brevity we drop a dependence on
soft and hard antiquark occupation numbers $\bar{n}_{\bf q}^{\pm}$
and $f_{\bf p}^{\bar{Q}}$ on the right-hand side of
Eqs.\,(\ref{eq:2q}) and (\ref{eq:2w}). The equation for
$\bar{n}_{\bf q}^{(f)}$ is obtained from (\ref{eq:2q}) by
replacement $n_{\bf q}^{(f)}\rightleftharpoons
(1-\bar{n}_{\bf q}^{(f)})$.

We present, as it usually is, decay and regenerating rates in the form of the
functional expansion in powers of the soft-(anti)quark and soft-gluon number
densities
\begin{equation}
\Gamma_{\rm d}^{(f\!,\,b)}
[n_{\bf q}^{\pm},N_{\bf k}^{t,\,l},f_{\bf p}^{G\!,\,Q}] =
\sum_{n = 1}^{\infty}
\Gamma_{\rm d}^{(f\!,\,b)(2n + 1)}
[n_{\bf q}^{\pm},N_{\bf k}^{t,\,l},f_{\bf p}^{G\!,\,Q}],
\label{eq:2e}
\end{equation}
\[
\Gamma_{\rm i}^{(f\!,\,b)}
[n_{\bf q}^{\pm},N_{\bf k}^{t,\,l},f_{\bf p}^{G\!,\,Q}] =
\sum_{n = 1}^{\infty}
\Gamma_{\rm i}^{(f\!,\,b)(2n + 1)}
[n_{\bf q}^{\pm},N_{\bf k}^{t,\,l},f_{\bf p}^{G\!,\,Q}],
\]
where
$\Gamma_{{\rm d},\,{\rm i}}^{(f\!,\,b)(2n + 1)}
[n_{\bf q}^{\pm},N_{\bf k}^{t,\,l},f_{\bf p}^{G\!,\,Q}]$
collect the contributions of the total $n$th power in
$n_{\bf q}^{\pm},\,\bar{n}_{\bf q}^{\pm}$, and $N_{\bf k}^{t,\,l}$.
As well as in a case of the nonlinear interaction of soft fermion and soft
boson excitations among themselves (Paper I), the general structure of the
expressions for arbitrary $n$ is rather cumbersome and therefore here we
write out only in an explicit form the decay and regenerating rates to
lowest order in the nonlinear interaction ($n=0,\,1$).

\noindent
The fermion decay rate is written in the form
\begin{equation}
\Gamma_{\rm d}^{(f)}[n_{\bf q}^{\pm},\!N_{\bf k}^{t,\,l},f_{\bf p}^{G\!,\,Q}]
=\int\!\frac{d {\bf p}}{(2\pi)^3}\,
\Bigl[\,f_{\bf p}^{G}\Bigl(1-f_{{\bf p}^{\prime}}^{Q}\Bigr) +
f_{\bf p}^{\bar{Q}}\Bigl(1+f_{{\bf p}^{\prime}}^{G}\Bigr)\Bigr]
\label{eq:2r}
\end{equation}
\[
\times
\Biggl\{\sum\limits_{\,b\,=\,t,\,l}
\int\!d{\cal T}_{q\rightarrow {\rm g}}^{(f;\,b)}
\,{\it w}_{\,q\rightarrow {\rm g}}^{(f;\,b)}
({\bf p}|\,{\bf q};{\bf k})\Bigl(1+N_{{\bf k}}^{(b)}\Bigr)
\hspace{4cm}
\]
\[
+\!\!\sum\limits_{\,b_1,\,b_2\,=\,t,\,l}\biggl[\,
\int\!d{\cal T}_{q\rightarrow {\rm g}{\rm g}}^{(f;\,b_1b_2)}
\,{\it w}_{\,q\rightarrow {\rm g}{\rm g}}^{(f;\,b_1b_2)}
({\bf p}|\,{\bf q};{\bf k}_1,{\bf k}_2)\Bigl(1+N_{{\bf k}_1}^{(b_1)}\Bigr)
\Bigl(1+N_{{\bf k}_2}^{(b_2)}\Bigr)
\]
\[
\hspace{0.7cm}
+\int\!d{\cal T}_{q{\rm g}\rightarrow {\rm g}}^{(f b_1;\,b_2)}
\,{\it w}_{\,q{\rm g}\rightarrow {\rm g}}^{(fb_1;\,b_2)}
({\bf p}|\,{\bf q},{\bf k}_1;{\bf k}_2)N_{{\bf k}_1}^{(b_1)}
\Bigl(1+N_{{\bf k}_2}^{(b_2)}\Bigr)\,\biggr]
\]
\[
\hspace{1.2cm}
+\!\!\sum\limits_{\,f_1,\,f_2\,=\,\pm\,}\biggl[\,
\int\!d{\cal T}_{q\rightarrow \bar{q}_1q_2}^{(f;\,f_1f_2)}
\,{\it w}_{\,q\rightarrow \bar{q}_1q_2}^{(f;\,f_1f_2)}
({\bf p}|\,{\bf q};{\bf q}_1,{\bf q}_2)\Bigl(1-\bar{n}_{{\bf q}_1}^{(f_1)}
\Bigr)
\Bigl(1-n_{{\bf q}_2}^{(f_2)}\Bigr)
\hspace{1.1cm}
\]
\[
\hspace{2.2cm}
+
\int\!d{\cal T}_{qq_1\rightarrow q_2}^{(ff_1;\,f_2)}
\,{\it w}_{\,qq_1\rightarrow q_2}^{(ff_1;\,f_2)}
({\bf p}|\,{\bf q},{\bf q}_1;{\bf q}_2)\,n_{{\bf q}_1}^{(f_1)}
\Bigl(1-n_{{\bf q}_2}^{(f_2)}\Bigr)
\,\biggr]\Biggr\}\,+\,\ldots
\]
\[
+\!\sum\limits_{\zeta=Q,\,\bar{Q},\,G}
\int\!\!\frac{d {\bf p}}{(2\pi)^3}
\Bigl[\,f_{\bf p}^{(\zeta)}
\Bigl(\,1+f_{{\bf p}^{\prime\prime}}^{(\zeta)}\Bigr)\Bigr]
\,\Biggl\{
\sum\limits_{\,f_1=\,\pm}
\int\!d{\cal T}_{q\rightarrow q}^{(f;\,f_1)}
\,{\it w}_{\,q\rightarrow q}^{(\zeta)(f;\,f_1)}
({\bf p}|\,{\bf q};{\bf q}_1)\Bigl(1-n_{{\bf q}_1}^{(f_1)}\Bigr)
\]
\[
+\!\sum\limits_{\,b\,=\,t,\,l}\sum\limits_{\,f_1=\,\pm}\biggl[\,
\int\!d{\cal T}_{q\rightarrow {\rm g}q_1}^{(f;\,bf_1)}
\,{\it w}_{\,q\rightarrow {\rm g}q_1}^{(\zeta)(f;\,bf_1)}
({\bf p}|\,{\bf q};{\bf k},{\bf q}_1)
\Bigl(1+N_{{\bf k}}^{(b)}\Bigr)
\Bigl(1-n_{{\bf q}_1}^{(f_1)}\Bigr)
\]
\[
\hspace{0.8cm}
+\int\!d{\cal T}_{q{\rm g}\rightarrow q_1}^{(fb;\,f_1)}
\,{\it w}_{\,q{\rm g}\rightarrow q_1}^{(\zeta)(fb;\,f_1)}
({\bf p}|\,{\bf q},{\bf k};{\bf q}_1)
N_{{\bf k}}^{(b)}\Bigl(1-n_{{\bf q}_1}^{(f_1)}\Bigr)
\]
\[
\hspace{2.6cm}
+\int\!d{\cal T}_{q\bar{q}_1\rightarrow {\rm g}}^{(ff_1;\,b)}
\,{\it w}_{\,q\bar{q}_1\rightarrow {\rm g}}^{(\zeta)(ff_1;\,b)}
({\bf p}|\,{\bf q},{\bf q}_1;{\bf k})
\Bigl(1+N_{{\bf k}}^{(b)}\Bigr)\bar{n}_{{\bf q}_1}^{(f_1)}\,\biggr]
\Biggr\}\,+\,\ldots\,.
\]
The fermion regenerating rate is
\begin{equation}
\Gamma_{\rm i}^{(f)}[n_{\bf q}^{\pm},\!N_{\bf k}^{t,\,l},f_{\bf p}^{G\!,\,Q}]
=\int\!\frac{d {\bf p}}{(2\pi)^3}\,
\Bigl[\,f_{{\bf p}^{\prime}}^{Q}\Bigl(1+f_{\bf p}^{G}\Bigr) +
f_{{\bf p}^{\prime}}^{G}\Bigl(1-f_{\bf p}^{\bar{Q}}\Bigr)\Bigr]
\label{eq:2t}
\end{equation}
\[
\Biggl\{\sum\limits_{\,b\,=\,t,\,l}
\int\!d{\cal T}_{q\rightarrow {\rm g}}^{(f;\,b)}
\,{\it w}_{\,q\rightarrow {\rm g}}^{(f;\,b)}
({\bf p}|\,{\bf q};{\bf k}) N_{{\bf k}}^{(b)}
+\!\!\sum\limits_{\,b_1,\,b_2\,=\,t,\,l}\biggl[\,
\int\!d{\cal T}_{q\rightarrow {\rm g}{\rm g}}^{(f;\,b_1b_2)}
\,{\it w}_{\,q\rightarrow {\rm g}{\rm g}}^{(f;\,b_1b_2)}
({\bf p}|\,{\bf q};{\bf k}_1,{\bf k}_2) N_{{\bf k}_1}^{(b_1)}
N_{{\bf k}_2}^{(b_2)}
\]
\[
\hspace{3.1cm}
+\int\!d{\cal T}_{q{\rm g}\rightarrow {\rm g}}^{(f b_1;\,b_2)}
\,{\it w}_{\,q{\rm g}\rightarrow {\rm g}}^{(fb_1;\,b_2)}
({\bf p}|\,{\bf q},{\bf k}_1;{\bf k}_2)
\Bigl(1+N_{{\bf k}_1}^{(b_1)}\Bigr)N_{{\bf k}_2}^{(b_2)}\,
\biggr]
\]
\[
+\!\!\sum\limits_{\,f_1,\,f_2\,=\,\pm\,}\biggr[\,
\int\!d{\cal T}_{q\rightarrow \bar{q}_1q_2}^{(f;\,f_1f_2)}
\,{\it w}_{\,q\rightarrow \bar{q}_1q_2}^{(f;\,f_1f_2)}
({\bf p}|\,{\bf q};{\bf q}_1,{\bf q}_2)\,\bar{n}_{{\bf q}_1}^{(f_1)}
n_{{\bf q}_2}^{(f_2)}
\]
\[
\hspace{4.1cm}
+
\int\!d{\cal T}_{qq_1\rightarrow q_2}^{(ff_1;\,f_2)}
\,{\it w}_{\,qq_1\rightarrow q_2}^{(ff_1;\,f_2)}
({\bf p}|\,{\bf q},{\bf q}_1;{\bf q}_2)(1-n_{{\bf q}_1}^{(f_1)})
n_{{\bf q}_2}^{(f_2)}\,\biggr]\Biggr\}\,+\,\ldots
\]
\[
+\!\sum\limits_{\zeta=Q,\,\bar{Q},\,G}
\int\!\!\frac{d {\bf p}}{(2\pi)^3}
\Bigl[\,f_{{\bf p}^{\prime\prime}}^{(\zeta)}
\Bigl(\,1+f_{\bf p}^{(\zeta)}\Bigr)\Bigr]
\,\Biggl\{
\sum\limits_{\,f_1=\,\pm}
\int\!d{\cal T}_{q\rightarrow q}^{(f;\,f_1)}
\,{\it w}_{\,q\rightarrow q}^{(\zeta)(f;\,f_1)}
({\bf p}|\,{\bf q};{\bf q}_1)\,n_{{\bf q}_1}^{(f_1)}
\]
\[
+\!\sum\limits_{\,b\,=\,t,\,l}\sum\limits_{\,f_1=\,\pm}\biggl[\,
\int\!d{\cal T}_{q\rightarrow {\rm g}q_1}^{(f;\,bf_1)}
\,{\it w}_{\,q\rightarrow {\rm g}q_1}^{(\zeta)(f;\,bf_1)}
({\bf p}|\,{\bf q};{\bf k},{\bf q}_1)
N_{{\bf k}}^{(b)} n_{{\bf q}_1}^{(f_1)}
\hspace{0.3cm}
\]
\[
\hspace{2.7cm}
+\int\!d{\cal T}_{q{\rm g}\rightarrow q_1}^{(fb;\,f_1)}
\,{\it w}_{\,q{\rm g}\rightarrow q_1}^{(\zeta)(fb;\,f_1)}
({\bf p}|\,{\bf q},{\bf k};{\bf q}_1)
\Bigl(1+N_{{\bf k}}^{(b)}\Bigr)n_{{\bf q}_1}^{(f_1)}
\]
\[
\hspace{4.6cm}
+\int\!d{\cal T}_{q\bar{q}_1\rightarrow {\rm g}}^{(ff_1;\,b)}
\,{\it w}_{\,q\bar{q}_1\rightarrow {\rm g}}^{(\zeta)(ff_1;\,b)}
({\bf p}|\,{\bf q},{\bf q}_1;{\bf k})
N_{{\bf k}}^{(b)}\Bigl(1-\bar{n}_{{\bf q}_1}^{(f_1)}\Bigr)\,\biggr]
\Biggr\}\,+\,\ldots\,.
\]
Here, for brevity we have used a somewhat symbolical denotation.
We have taken out factors
$\left[\,f_{\bf p}^{G}(1-f_{{\bf p}^{\prime}}^{Q}) +
f_{\bf p}^{\bar{Q}}(1+f_{{\bf p}^{\prime}}^{G})\right]$,
$\left[\,f_{\bf p}^{(\zeta)}(\,1+f_{{\bf p}^{\prime\prime}}^{(\zeta)})\right]$
etc. as the general multipliers. However it is necessary to mean
that they differ for each term in braces by values of momenta
${\bf p}^{\prime}$ and ${\bf p}^{\prime\prime}$. Thus, for example, for
terms linear in $N_{{\bf k}}^{(b)}$ and $n_{{\bf q}}^{(f)}$ respectively
it is necessary to mean
${\bf p}^{\prime}\equiv {\bf p}+{\bf q}-{\bf k},\,
{\bf p}^{\prime\prime}\equiv {\bf p}+{\bf q}-{\bf q}_1$. Furthermore for the
terms quadratic in $N_{{\bf k}}^{(b)}$ and $n_{{\bf q}}^{(f)}$
it is necessary to mean
${\bf p}^{\prime}\equiv {\bf p}+{\bf q}-{\bf k}_1-{\bf k}_2$
(or ${\bf p}^{\prime}\equiv {\bf p}+{\bf q}-{\bf q}_1-{\bf q}_2$),
${\bf p}^{\prime\prime}\equiv {\bf p}+{\bf q}-{\bf k}-{\bf q}_1$ and so on.
The function
${\it w}_{\,q\rightarrow {\rm g}}^{(f;\,b)}({\bf p}|\,{\bf q};{\bf k})$
defines a probability of two processes: (1) the process of absorption of
soft-quark excitation with frequency $\omega^{(f)}_{\bf q}$ (and momentum
{\bf q}) by
hard thermal gluon\footnote{In the subsequent discussion hard thermal
particles of plasma undergoing the scattering processes off soft excitations,
will be called also {\it test} particles.} with consequent conversion
of the latter into hard thermal quark and radiation of soft-gluon excitation
with frequency $\omega^{(b)}_{\bf k}$ (and momentum {\bf k}) and (2) the
process of annihilation of soft-quark excitation with hard thermal antiquark
into soft-gluon excitation and hard transverse gluon. In other words, this
probability defines the scattering processes of `inelastic' type
\[
q+G\rightarrow {\rm g}+Q,\qquad
q+\bar{Q}\rightarrow {\rm g}+G,
\]
where $q$, ${\rm g}$ are plasma collective excitations and $G,\,
Q$ and $\bar{Q}$ are excitations with typical momenta of
temperature order $T$ and above. Furthermore, the function ${\it
w}_{\,q\rightarrow q}^{(\zeta)(f;\,f_1)}({\bf p}|\,{\bf q};{\bf
q}_1)$ defines a probability for an `elastic' scattering of
soft-quark excitation off hard thermal quark, antiquark, and
gluon:
\[
q+Q\rightarrow q+Q,\qquad
q+\bar{Q}\rightarrow q+\bar{Q},\qquad
q+G\rightarrow q+G.
\]

The functions ${\it w}_{\,q\rightarrow {\rm g}{\rm
g}}^{(f;\,b_1b_2)} ({\bf p}|\,{\bf q};{\bf k}_1,{\bf k}_2)$, ${\it
w}_{\,q\rightarrow \bar{q}_1q_2}^{(f;\,f_1f_2)} ({\bf p}|\,{\bf
q};{\bf q}_1,{\bf q}_2)$, ${\it w}_{\,q\rightarrow {\rm
g}q_1}^{(\zeta)(f;\,bf_1)} ({\bf p}|\,{\bf q};{\bf k},{\bf q}_1)$,
and so on define probabilities of more complicated scattering
processes connected with an interaction of three soft plasma waves
with a hard test particle. Thus the first two of them are the
scattering probabilities for the processes of type
\[
q+G\rightarrow {\rm g}+{\rm g}+Q,\quad
q+\bar{Q}\rightarrow {\rm g}+{\rm g}+G,
\]
\[
q+G\rightarrow \bar{q}_1+q_2+Q,\quad
q+\bar{Q} \rightarrow \bar{q}_1+q_2+G
\]
correspondingly, and  the third function is the scattering
probability for the processes of type
\[
q+G\rightarrow q_1+g+G,\quad
q+Q\rightarrow q_1+g+Q,\quad
q+\bar{Q}\rightarrow q_1+g+\bar{Q}.
\]
Let us specially emphasize that in the first case the type of the hard 
particle changes and in the second one it doesn't. Such a division of the
probabilities will take place to all orders of nonlinear interaction of soft
and hard modes. This is reflected in particular in our notation of generalized
rates (\ref{eq:2r}) and (\ref{eq:2t}).

The phase-space measures for these processes are
\begin{equation}
\int\!d{\cal T}_{q\rightarrow {\rm g}}^{(f;\,b)}=
\int\!\!\frac{d{\bf k}}{(2\pi)^3}\,
2\pi\,\delta\Bigl(E_{\bf p}+\omega_{\bf q}^{(f)}-
E_{{\bf p}+{\bf q}-{\bf k}}-\omega_{\bf k}^{(b)}\Bigr),
\label{eq:2y}
\end{equation}
\[
\int\!d{\cal T}_{q\rightarrow q}^{(f;\,f_1)}=
\int\!\!\frac{d{\bf q}_1}{(2\pi)^3}\,
2\pi\,\delta\Bigl(E_{\bf p}+\omega_{\bf q}^{(f)}-
E_{{\bf p}+{\bf q}-{\bf q}_1}-\omega_{{\bf q}_1}^{(f_1)}\Bigr),
\]
\[
\int\!d{\cal T}_{q\rightarrow {\rm g}{\rm g}}^{(f;\,b_1b_2)}=
\int\!\!\frac{d{\bf k}_1}{(2\pi)^3}\,\frac{d{\bf k}_2}{(2\pi)^3}\,
2\pi\,\delta\Bigl(E_{\bf p}+\omega_{\bf q}^{(f)}-
E_{{\bf p}+{\bf q}-{\bf k}_1-{\bf k}_2}-
\omega_{{\bf k}_1}^{(b_1)}-\omega_{{\bf k}_2}^{(b_2)}\Bigr),
\]
\[
\int\!d{\cal T}_{q\rightarrow \bar{q}q}^{(f;\,f_1f_2)}=
\int\!\!\frac{d{\bf q}_1}{(2\pi)^3}\,\frac{d{\bf q}_2}{(2\pi)^3}\,
2\pi\,\delta\Bigl(E_{\bf p}+\omega_{\bf q}^{(f)}-
E_{{\bf p}+{\bf q}-{\bf q}_1-{\bf q}_2}
-\omega_{{\bf q}_1}^{(f_1)}-\omega_{{\bf q}_2}^{(f_2)}\Bigr),
\]
\[
\int\!d{\cal T}_{q\rightarrow {\rm g}q}^{(f;\,bf_1)}=
\int\!\!\frac{d{\bf k}}{(2\pi)^3}\,\frac{d{\bf q}_1}{(2\pi)^3}\,
2\pi\,\delta\Bigl(E_{\bf p}+\omega_{\bf q}^{(f)}-
E_{{\bf p}+{\bf q}-{\bf k}-{\bf q}_1}-
\omega_{\bf k}^{(b)}-\omega_{{\bf q}_1}^{(f_1)}\Bigr),
\hspace{0.6cm}
\]
etc. Here $E_{\bf p}\equiv|{\bf p}|$ for massless hard gluons and
(anti)quarks. The $\delta$-functions in Eqs.\,(\ref{eq:2y}) represent the
energy conservation for the scattering processes under consideration.

We particularly note that for generalized decay and regenerated rates
(\ref{eq:2r}) and (\ref{eq:2t}) it was assumed that the scattering
probabilities satisfy symmetry relations over permutation of incoming and
outgoing soft quark and gluon momenta. For the simplest
probabilities, e.g., we have
\begin{equation}
{\it w}_{\,q\rightarrow {\rm g}}^{(f;\,b)}({\bf p}|\,{\bf q};{\bf k})=
{\it w}_{\,{\rm g}\rightarrow q}^{(b;\,f)}({\bf p}|\,{\bf k};{\bf q}),
\quad
{\it w}_{\,q\rightarrow q}^{(\zeta)(f;\,f_1)}({\bf p}|\,{\bf q};{\bf q}_1)=
{\it w}_{\,q\rightarrow q}^{(\zeta)(f_1;\,f)}({\bf p}|\,{\bf q}_1;{\bf q}).
\label{eq:2u}
\end{equation}
These relations are a consequence of more general relations for exact
scattering probabilities depending on initial and final values of momenta
of hard particles, namely,
\[
{\it w}_{\,q\rightarrow {\rm g}}^{(f;\,b)}
({\bf p},\,{\bf p}^{\prime}|\,{\bf q};{\bf k})=
{\it w}_{\,{\rm g}\rightarrow q}^{(b;\,f)}
({\bf p}^{\prime}\!,\,{\bf p}|\,{\bf k};{\bf q}),\quad
{\it w}_{\,q\rightarrow q}^{(\zeta)(f;\,f_1)}
({\bf p},{\bf p}^{\prime\prime}|\,{\bf q};{\bf q}_1)=
{\it w}_{\,q\rightarrow q}^{(\zeta)(f_1;\,f)}
({\bf p}^{\prime\prime}\!,{\bf p}|\,{\bf q}_1;{\bf q}).
\]
They present detailed balancing principle in scattering
processes and in this sense they are exact.  The scattering
probabilities in Eqs.\,(\ref{eq:2r}), (\ref{eq:2t}) are
obtained by integrating initial probabilities over ${\bf
p}^{\prime}$ or ${\bf p}^{\prime\prime}$ with regard to momentum
conservation law:
\[
{\it w}_{\,q\rightarrow {\rm g}}^{(f;\,b)}({\bf p}|\,{\bf q};{\bf k})
2\pi\,\delta(E_{\bf p}+\omega_{\bf q}^{(f)}-
E_{{\bf p}+{\bf q}-{\bf k}}-\omega_{\bf k}^{(b)})
\]
\[
=\!\int\!{\it w}_{\,q\rightarrow {\rm g}}^{(f;\,b)}
({\bf p},\,{\bf p}^{\prime}|\,{\bf q};{\bf k})
2\pi\,\delta(E_{\bf p}+\omega_{\bf q}^{(f)}-
E_{{\bf p}^{\prime}}-\omega_{\bf k}^{(b)})
(2\pi)^3
\delta({\bf p}+{\bf q}-{\bf p}^{\prime}-{\bf k})
\,\frac{d{\bf p}^{\prime}}{(2\pi)^3}\,,
\]
\[
{\it w}_{\,q\rightarrow q}^{(\zeta)(f;\,f_1)}({\bf p}|\,{\bf q};{\bf q}_1)
2\pi\,\delta(E_{\bf p}+\omega_{\bf q}^{(f)}-
E_{{\bf p}+{\bf q}-{\bf q}_1}-\omega_{{\bf q}_1}^{(f_1)})
\]
\[
=\!\int\!{\it w}_{\,q\rightarrow q}^{(\zeta)(f;\,f_1)}
({\bf p},\,{\bf p}^{\prime\prime}|\,{\bf q};{\bf q}_1)
2\pi\,\delta(E_{\bf p}+\omega_{\bf q}^{(f)}-
E_{{\bf p}^{\prime\prime}}-\omega_{\bf q}^{(f_1)})
(2\pi)^3
\delta({\bf p}+{\bf q}-{\bf p}^{\prime\prime}-{\bf q}_1)
\,\frac{d{\bf p}^{\prime\prime}}{(2\pi)^3}\,.
\]
The scattering probabilities obtained satisfy relations
(\ref{eq:2u}) in the limit of interest to us, i.e., when we neglect
by (quantum) recoil of hard particles. In the general case
expressions (\ref{eq:2u}) are replaced by more complicated ones (see, e.g.,
Ref.\,\cite{markov_AOP_04}, Section 10).

For the boson sector of plasma excitations the generalized rates
$\Gamma_{\rm d}^{(b)}$ and $\Gamma_{\rm i}^{(b)}$ to lowest order in the
nonlinear interactions of soft and hard modes have a similar structure.
For this reason, their explicit form is not given here.

Taking into account that in semiclassical approximation we have
$|{\bf p}|\gg |{\bf k}|,\,|{\bf q}|,\,|{\bf q}_1|,\dots\,$, the energy
conservation laws can be represented in the form of the following
resonance conditions:
\begin{equation}
\omega_{\bf q}^{(f)}-\omega_{\bf k}^{(b)}-
{\bf v}\cdot({\bf q}-{\bf k})=0\,,
\quad
\omega_{\bf q}^{(f)}-\omega_{{\bf q}_1}^{(f_1)}-
{\bf v}\cdot({\bf q}-{\bf q}_1)=0,
\label{eq:2i}
\end{equation}
\[
\omega_{\bf q}^{(f)}-\omega_{{\bf k}_1}^{(b_1)}-\omega_{{\bf k}_2}^{(b_2)}-
{\bf v}\cdot({\bf q}-{\bf k}_1-{\bf k}_2)=0,
\]
\[
\omega_{\bf q}^{(f)}-\omega_{{\bf q}_1}^{(f_1)}-\omega_{{\bf q}_2}^{(f_2)}
-{\bf v}\cdot({\bf q}-{\bf q}_1-{\bf q}_2)=0,
\]
\[
\omega_{\bf q}^{(f)}-\omega_{{\bf k}}^{(b)}-\omega_{{\bf q}_1}^{(f_1)}-
{\bf v}\cdot({\bf q}-{\bf k}-{\bf q}_1)=0,
\hspace{0.2cm}
\]
and so on. Here, ${\bf v}={\bf p}/|{\bf p}|$ is a velocity of the hard test
particle.

In subsequent discussion for the sake of simplicity of the problem we
suppose that a characteristic time for nonlinear relaxation of the soft
oscillations is much less than a time of relaxation of the distribution of
hard partons $f_{\bf p}^G,\,f_{\bf p}^Q$ and $f_{\bf p}^{\bar{Q}}$.
In other words under the conditions when the intensity of soft plasma
excitations are sufficiently small and they cannot essentially change such
`crude' equilibrium parameters of plasma as particle density,
temperature, and thermal energy, we can neglect by a space-time
change of the distribution functions of hard partons assuming that these
functions are specified and describe the global (baryonfree) equilibrium
state of hot non-Abelian plasma
\[
f_{\bf p}^G = 2\,\frac{1}{{\rm e}^{E_{\bf p}/T}-1},\quad
f_{\bf p}^{Q(\bar{Q})} = 2\,\frac{1}{{\rm e}^{E_{\bf p}/T}+1}.
\]
Here the coefficient 2 takes into account that hard gluon and hard
(anti)quark have two helicity states. 

In Section 8 for determining an explicit form of the required probabilities 
we need in some specific approximation of collision
terms. We use the fact that the occupation numbers $N^{(b)}_{{\bf
p}}$ are more large than one, i.e., $1+N_{{\bf p}}^{(b)}\simeq
N_{{\bf p}}^{(b)}$. Furthermore, we present the integration measure
as
\[
\int\!\frac{d{\bf p}}{(2\pi)^3} =
\int\!\frac{|{\bf p}|^2\,d|{\bf p}|}{2\pi^2}
\int\!\frac{d\Omega_{\bf v}}{4\pi}\,,
\]
where the solid integral is over the directions of unit vector ${\bf v}$.
Setting $f_{{\bf p}^{\prime}}^G\simeq f_{\bf p}^G$,
$f_{{\bf p}^{\prime}}^{Q(\bar{Q})}\simeq f_{\bf p}^{Q(\bar{Q})}$ and also
$1\pm f_{\bf p}^{G\!,\,Q(\bar{Q})}\simeq
1\pm f_{{\bf p}^{\prime}}^{G\!,\,Q(\bar{Q})}\simeq 1$ by virtue of
$f_{\bf p}^{G\!,\,Q(\bar{Q})},\;f_{{\bf p}^{\prime}}^{G\!,\,Q(\bar{Q})}\ll1$,
in the limit of vanishing fermion intensity
$n_{\bf q}^{(f)}\rightarrow 0$ we have the following expression for collision
term ${\rm C}^{(f)}[n_{\bf q}^{\pm},N_{\bf k}^{t,\,l},f_{\bf p}^{G\!,\,Q}]$:
\[
{\rm C}^{(f)}[n_{\bf q}^{\pm},N_{\bf k}^{t,\,l},f_{\bf p}^{G\!,\,Q}]
\simeq
-\!\int\!\frac{|{\bf p}|^2\,d|{\bf p}|}{2\pi^2}\,
\Bigl[\,f_{\bf p}^{Q}+f_{\bf p}^{G}\Bigr]
\int\!\frac{d\Omega_{\bf v}}{4\pi}\,
\Biggl\{\sum\limits_{\,b\,=\,t,\,l}
\int\!d{\cal T}_{q\rightarrow {\rm g}}^{(f;\,b)}
\,{\it w}_{\,q\rightarrow {\rm g}}^{(f;\,b)}
({\bf p}|\,{\bf q};{\bf k}) N_{{\bf k}}^{(b)}
\]
\begin{equation}
+\!\!\sum\limits_{\,b_1,\,b_2\,=\,t,\,l}
\int\!d{\cal T}_{q\rightarrow {\rm g}{\rm g}}^{(f;\,b_1b_2)}
\,{\it w}_{\,q\rightarrow {\rm g}{\rm g}}^{(f;\,b_1b_2)}
({\bf p}|\,{\bf q};{\bf k}_1,{\bf k}_2) N_{{\bf k}_1}^{(b_1)}
N_{{\bf k}_2}^{(b_2)}
\label{eq:2o}
\end{equation}
\[
\hspace{1.2cm}
+\!\!\sum\limits_{\,f_1,\,f_2\,=\,\pm}
\int\!d{\cal T}_{q\rightarrow \bar{q}_1q_2}^{(f;\,f_1f_2)}
\,{\it w}_{\,q\rightarrow \bar{q}_1q_2}^{(f;\,f_1f_2)}
({\bf p}|\,{\bf q};{\bf q}_1,{\bf q}_2)\bar{n}_{{\bf q}_1}^{(f_1)}
n_{{\bf q}_2}^{(f_2)}+\,\ldots\Biggr\}
\]
\[
-\!\sum\limits_{\zeta=Q,\,\bar{Q},\,G\,}
\int\!\frac{|{\bf p}|^2\,d|{\bf p}|}{2\pi^2}\,
f_{\bf p}^{(\zeta)}\!
\int\!\frac{d\Omega_{\bf v}}{4\pi}\,
\Biggl\{\,
\!\sum\limits_{\,f_1\,=\,\pm}
\int\!\frac{d\Omega_{\bf v}}{4\pi}
\int\!d{\cal T}_{q\rightarrow q}^{(f;\,f_1)}
\,{\it w}_{\,q\rightarrow q}^{(\zeta)(f;\,f_1)}
({\bf p}|\,{\bf q};{\bf q}_1)n_{{\bf q}_1}^{(f_1)}
\]
\[
+\!\sum\limits_{\,b\,=\,t,\,l}\sum\limits_{\,f_1=\,\pm}
\int\!d{\cal T}_{q\rightarrow {\rm g}q_1}^{(f;\,bf_1)}
\,{\it w}_{\,q\rightarrow {\rm g}q_1}^{(\zeta)(f;\,bf_1)}
({\bf p}|\,{\bf q};{\bf k},{\bf q}_1)
N_{{\bf k}}^{(b)} n_{{\bf q}_1}^{(f_1)} + \,\ldots\Biggr\}_{.}
\hspace{0.3cm}
\]
The kinetic equation (\ref{eq:2q}) with collision term in the form
of (\ref{eq:2o}) defines a change of the soft-quark number density
$n_{\bf q}^{(f)}$ caused by so-called {\it spontaneous} scattering processes
of soft-quark excitations off hard thermal partons.

\section{\bf Soft-field equations}
\setcounter{equation}{0}

We consider ${\rm SU}(N_c)$ gauge theory with $n_f$ flavors of massless quarks.
The color indices for the adjoint representation $a,b, \ldots$ run from 1 to
$N_c^2-1$ while ones for the fundamental representation $i,j, \ldots$ run
from 1 to $N_c$. The Greek indices $\alpha, \beta, \ldots$ for the spinor
representation run from 1 to 4.

In this section we discuss the equations of motion for soft-gluon and
soft-quark plasma excitations, which will play a main role in our subsequent
discussion. We have already written out these equations in Paper I
(Eqs.\,(I.3.1), (I.3.2)). Here they should be correspondingly extended to
take into account the presence of a color current caused by hard test particle
passing through the hot QCD plasma.

One expects the world lines of the hard modes to obey classical trajectories in
the manner of Wong \cite{wong} since their coupling to the soft modes is weak
at a very high temperature. Considering this circumstance, we
add the color current of color point charge
\begin{equation}
j_Q^{a\mu}(x)=gv^{\mu}Q^a(t){\delta}^{(3)}({\bf x}-{\bf v}t)
\label{eq:3q}
\end{equation}
to the right-hand side of the Yang-Mills field equation (I.3.1).
Here, $Q^a=Q^a(t)$ is a color classical charge satisfying the Wong equation
\begin{equation}
\frac{dQ^a(t)}{dt} + igv^{\mu}A^b_{\mu}(t,{\bf v}t)(T^b)^{ac}Q^c(t)=0,\quad\;
\left.Q^a_0=Q^a(t)\right|_{\,t=0},
\label{eq:3w}
\end{equation}
where $v^{\mu}=(1,{\bf v}),\,(T^a)^{bc}\equiv -if^{abc}$ and $t$
is a coordinate time. The gauge potential $A^b_{\mu}(t,{\bf x})$
in Eq.\,(\ref{eq:3w}) is determined on a straight-line trajectory of a
hard parton, i.e., at ${\bf x}={\bf v}t$. The explicit form of
a solution of Eq.\,(\ref{eq:3w}) in the momentum representation is
given in Ref.\,\cite{markov_AOP_04}. Thus we lead to the following
integral equation for gauge potential $A_{\mu}^a(k)$ instead of
Eq.\,(I.3.1)
\[
\,^{\ast}{\cal D}^{-1}_{\mu\nu}(k) A^{a\nu}(k) =
- j^{A(2)a}_{\mu}(A,A)(k) - j^{A(3)a}_{\mu}(A,A,A)(k) -
j^{\Psi(0,2)a}_{\mu}(\bar{\psi},\psi)(k)
\]
\begin{equation}
-j^{\Psi(1,2)a}_{\mu}(A,\bar{\psi},\psi)(k)
-j^{(0)a}_{Q\mu}(k)-j^{(1)a}_{Q\mu}(A)(k)-j^{(2)a}_{Q\mu}(A,A)(k)
\label{eq:3e}
\end{equation}
\[
\equiv
-j^{a}_{\mu}[A,\bar{\psi},\psi,Q_0](k)-j^{(0)\,a}_{Q\mu}(k).
\]
On the right-hand side in the expansion of the induced currents
$j^A,\,j^{\psi}$
and current of hard color-charged parton we keep the terms up to the third order
in interacting fields and initial value of color charge $Q_0^a$. The explicit
form of induced currents $j^{A(2)a}_{\mu},\,j^{\Psi(0,2)a}_{\mu}$ and
$j^{\Psi(1,2)a}_{\mu}$ in the hard thermal loop (HTL) approximation is defined
by Eq.\,(I.3.3). The expansion terms of hard parton current have the
following structure:
\begin{equation}
j_Q^{(0)a\mu}(k)=\frac{\,g}{(2\pi)^3}\,v^{\mu}Q^a\delta(v\cdot k),
\label{eq:3r}
\end{equation}
\[
j_Q^{(1)a\mu}(A)(k)=
\frac{\,g^2}{(2\pi)^3}\,v^{\mu}\!
\int\!\!\frac{1}{(v\cdot k_1)}\,(v\cdot A^{a_1}(k_1))
\,\delta(v\cdot (k-k_1))dk_1\,(T^{a_1})^{ab}Q_0^b,
\]
\[
j_Q^{(2)a\mu}(A,A)(k)=
\frac{\,g^3}{(2\pi)^3}\,v^{\mu}\!
\int\!\!\frac{1}{(v\cdot (k_1+k_2))(v\cdot k_2)}\,(v\cdot A^{a_1}(k_1))
(v\cdot A^{a_2}(k_2))
\]
\[
\times\,
\delta(v\cdot (k-k_1-k_2))dk_1dk_2\,(T^{a_1}T^{a_2})^{ab}Q_0^b.
\]

Now we consider the Dirac field equations (I.3.2). To take into account 
an existence of current of a test color particle in system it is necessary to
put a certain additional terms into the right-hand side of soft-quark field
equations. For determination of an explicit form of these terms we introduce
a Grassmann color charge $\theta^i=\theta^i(t)$ (and conjugate color charge
$\theta^{\dagger\, i}=\theta^{\dagger\, i}(t)$) of hard particle satisfying
the following fundamental equation proposed in \cite{barducii}
\begin{equation}
\frac{d\theta^i(t)}{dt} + igv^{\mu}A^a_{\mu}(t,{\bf v}t)(t^a)^{ij}
\theta^j(t)=0, \quad\;
\left.\theta^i_0=\theta^i(t)\right|_{\,t=0}
\label{eq:3t}
\end{equation}
and correspondingly
\begin{equation}
\hspace{0.6cm}
\frac{d\theta^{\dagger\, i}(t)}{dt} - igv^{\mu}A^a_{\mu}(t,{\bf v}t)
\theta^{\dagger j}(t)(t^a)^{ji}=0, \quad
\left.\theta^{\dagger\, i}_0=\theta^{\dagger\, i}(t)\right|_{\,t=0}.
\label{eq:3y}
\end{equation}
The Grassmann color charge $\theta^i$ is associated with the usual
color charge $Q^a$ by relation 
\begin{equation}
Q^a(t)=\theta^{\dagger\, i}(t)(t^a)^{ij}\theta^{j}(t).
\label{eq:3u}
\end{equation}
By analogy with classical color current (\ref{eq:3q}) we write the following 
expression for a Grassmann color `current' (further called a Grassmann
color source) of hard test particle
\begin{equation}
\eta_{\theta\alpha}^i(x)=
g\theta^i(t)\chi_{\alpha}{\delta}^{(3)}({\bf x}-{\bf v}t)
\label{eq:3i}
\end{equation}
and correspondingly its conjugation
\begin{equation}
\bar{\eta}_{\theta\alpha}^i(x)=
g\theta^{\dagger\, i}(t)\bar{\chi}_{\alpha}
{\delta}^{(3)}({\bf x}-{\bf v}t).
\label{eq:3o}
\end{equation}
Here, $\chi_{\alpha}$ is a spinor independent of time $t$. The
physical sense of this spinor will be discussed in the following
sections.

Taking into account (\ref{eq:3i}) and (\ref{eq:3o}), we lead to the nonlinear
integral equations for soft-quark interacting fields $\psi^i_{\alpha}$ and
$\bar{\psi}^i_{\alpha}$ instead of (I.3.2)
\[
\,^{\ast}\!S^{-1}_{\alpha\beta}(q)\psi^i_{\beta}(q) =
-\,\eta^{(1,1)\,i}_{\alpha}(A,\psi)(q) -
\eta^{(2,1)\,i}_{\alpha}(A,A,\psi)(q)
\]
\begin{equation}
-\,\eta^{(0)i}_{\theta\,\alpha}(q)
-\,\eta^{(1)i}_{\theta\,\alpha}(A)(q)
-\,\eta^{(2)i}_{\theta\,\alpha}(A,A)(q)
\equiv
-\,\eta^{i}_{\alpha}[A,\psi,\theta_0](q)
-\,\eta^{(0)\,i}_{\theta\,\alpha}(q),
\label{eq:3p}
\end{equation}
\vspace{0.1cm}
\[
\bar{\psi}^i_{\beta}(-q)\,^{\ast}\!S^{-1}_{\beta\alpha}(-q) =
\bar{\eta}^{(1,1)\,i}_{\alpha}(A^{\ast},\bar{\psi})(-q) +
\bar{\eta}^{(2,1)\,i}_{\alpha}(A^{\ast},A^{\ast},\bar{\psi})(-q)
\]
\[
+\,\bar{\eta}^{(0)i}_{\theta\,\alpha}(-q) +
\bar{\eta}^{(1)i}_{\theta\,\alpha}(A^{\ast})(-q) +
\bar{\eta}^{(2)i}_{\theta\,\alpha}(A^{\ast},A^{\ast})(-q)
\equiv
\bar{\eta}^{i}_{\alpha}[A^{\ast},\bar{\psi},\theta_0^{\dagger\,}](-q) +
\bar{\eta}^{(0)i}_{\theta\,\alpha}(-q),
\]
where on the right-hand side in the expansion of induced sources
$\eta,\,\bar{\eta}$ and the Grassmann test particle sources
$\eta_{\theta},\,\bar{\eta}_{\theta}$, we keep the terms up to the third
order in interacting fields and initial values of the Grassmann color charges
$\theta_0^i,\,\theta_0^{\dagger\, i}$. Equations (I.3.4) and (I.3.5) define an
explicit form of induced sources
$\eta^{(1,1)\,i}_{\alpha},\,\eta^{(2,1)\,i}_{\alpha},\ldots$ in the HTL
approximation. By virtue of identical structure of evolution
equations (\ref{eq:3w}) and (\ref{eq:3t}) by trivial replacements it is easy
to define the expansion terms from (\ref{eq:3r}) for the Grassmann source
$\eta_{\theta\alpha}^i$ in the momentum representation
\begin{equation}
\eta^{(0)\,i}_{\theta\,\alpha}(q)
=\frac{\,g}{(2\pi)^3}\,\theta_0^i\chi_{\alpha}\delta(v\cdot q),
\label{eq:3a}
\end{equation}
\[
\eta^{(1)\,i}_{\theta\,\alpha}(A)(q)=
\frac{\,g^2}{(2\pi)^3}\,\chi_{\alpha}\!
\int\!\!\frac{1}{(v\cdot q_1)}\,(v\cdot A^{a_1}(q_1))
\,\delta(v\cdot (q-q_1))\,dq_1\,(t^{a_1})^{ij}\theta_0^j\,,
\]
\[
\eta^{(2)\,i}_{\theta\,\alpha}(A,A)(q)=
\frac{\,g^3}{(2\pi)^3}\,\chi_{\alpha}\!
\int\!\!\frac{1}{(v\cdot (q_1+q_2))(v\cdot q_2)}\,(v\cdot A^{a_1}(q_1))
(v\cdot A^{a_2}(q_2))
\]
\[
\times\,
\delta(v\cdot (q-q_1-q_2))\,dq_1dq_2\,(t^{a_1}t^{a_2})^{ij}\theta_0^j\,.
\]
The expressions for conjugate terms
$\bar{\eta}^{(s)\,i}_{\theta\alpha},\,s=0,1,2$ can be obtained from
(\ref{eq:3a}) by rule:
$\bar{\eta}^{\,i}_{\theta\alpha}(-q)
=\eta^{\dagger\, i}_{\theta\beta}(q)\gamma^0_{\beta\alpha}$.

Now we rewrite equations (\ref{eq:3e}) and (\ref{eq:3p}) in the following
form:
\[
A_{\mu}^{a}(k) = A_{\mu}^{(0)a}(k)
-\!\,^{\ast}{\cal D}_{\mu\nu}(k) j^{(0)a\nu}_Q(k)
-\!\,^{\ast}{\cal D}_{\mu\nu}(k)
j^{\,a\nu}[A,\bar{\psi},\psi,Q_0](k),
\]
\begin{equation}
\psi^{i}_{\alpha}(q) = \psi^{(0)i}_{\alpha}(q)
-\!\,^{\ast}\!S_{\alpha\beta}(q)\,
\eta^{(0)i}_{\theta\,\beta}(q)
-\!\,^{\ast}\!S_{\alpha\beta}(q)\,
\eta^{\,i}_{\beta}[A,\psi,\theta_0](q),
\hspace{1.2cm}
\label{eq:3s}
\end{equation}
\[
\hspace{0.05cm}
\bar{\psi}^{\,i}_{\alpha}(-q) = \bar{\psi}^{(0)i}_{\alpha}(-q)
+\bar{\eta}^{(0)i}_{\theta\,\beta}(-q)
\,^{\ast}\!S_{\beta\alpha}(-q)
+\bar{\eta}^{\,i}_{\beta}[A^{\ast},\bar{\psi},\theta_0^{\dagger\,}\,]
\,^{\ast}\!S_{\beta\alpha}(-q),
\]
where $A_{\mu}^{(0)a}(k),\,\psi^{(0)i}_{\alpha}(q)$ and
$\bar{\psi}^{\,i}_{\alpha}(-q)$ are free-field solutions. The second terms on
the right-hand sides of these equations represent fields induced 
by a test `bare' particle moving in the medium. In the weak-field limit the 
system of the nonlinear integral equations can be perturbatively solved by the
approximation scheme method. Formally, a solution of system (\ref{eq:3s}) can
be presented in the following form:
\[
A_{\mu}^{a}(k) = A_{\mu}^{(0)a}(k)
-^{\ast}{\cal D}_{\mu\nu}(k) j^{(0)a\nu}_Q(k)
-^{\ast}{\cal D}_{\mu\nu}(k)
\tilde{j}^{\,a\nu}
[A^{(0)},\bar{\psi}^{(0)},\psi^{(0)},Q_0,\theta_0^{\dagger\,},\theta_0\,](k),
\]
\begin{equation}
\psi^{i}_{\alpha}(q) = \psi^{(0)i}_{\alpha}(q)
-\,^{\ast}\!S_{\alpha\beta}(q)\,
\eta^{(0)i}_{\theta\,\beta}(q)
-\,^{\ast}\!S_{\alpha\beta}(q)\,
\tilde{\eta}^{\,i}_{\beta}[A^{(0)},\psi^{(0)},Q_0,\theta_0](q),
\hspace{0.9cm}
\label{eq:3d}
\end{equation}
\[
\bar{\psi}^{\,i}_{\alpha}(-q) = \bar{\psi}^{(0)i}_{\alpha}(-q)
+\tilde{\bar{\eta}}^{(0)i}_{\theta\,\beta}(-q)
\,^{\ast}\!S_{\beta\alpha}(-q)
+\tilde{\bar{\eta}}^{\,i}_{\beta}
[A^{(0)\ast},\bar{\psi}^{(0)},Q_0,\theta_0^{\dagger\,}\,]
\,^{\ast}\!S_{\beta\alpha}(-q).
\hspace{0.7cm}
\]
Here the functions $\tilde{j}^{\,a\nu},\,\tilde{\eta}^{\,i}_{\beta}$ and
$\tilde{\bar{\eta}}^{\,i}_{\beta}$ represent some new (effective) currents
and sources being functionals of free fields and initial values of color
charges. Our main purpose is calculation of an explicit form of these
effective currents and sources.

Unfortunately, as was earlier shown \cite{markov_NPA_06} a direct employment of 
the approximation scheme method for determination of desired effective currents
and sources is very complicated already on the second step of iteration and
as a consequence it is ineffective. Here we use a simpler approach
to calculation
of effective currents and sources suggested in our previous works 
\cite{markov_AOP_02, markov_AOP_04, markov_AOP_05, markov_NPA_06}.
It is based on the fact that by virtue of the structure of the right-hand
sides of Eqs.\,(\ref{eq:3s}) and (\ref{eq:3d}) the following system of
equations should be carried out identically
\begin{equation}
j^{a}_{\mu}[A,\bar{\psi},\psi,Q_0](k)=
\tilde{j}^{a}_{\mu}
[A^{(0)},\bar{\psi}^{(0)},\psi^{(0)},Q_0,\theta_0^{\dagger\,},\theta_0\,](k),
\label{eq:3f}
\end{equation}
\begin{equation}
\eta^{\,i}_{\beta}[A,\psi,\theta_0](q)=
\tilde{\eta}^{\,i}_{\beta}[A^{(0)},\psi^{(0)},Q_0,\theta_0](q),
\hspace{2cm}
\label{eq:3g}
\end{equation}
\begin{equation}
\bar{\eta}^{\,i}_{\beta}[A^{\ast},\bar{\psi},\theta_0^{\dagger\,}\,](-q)=
\tilde{\bar{\eta}}^{\,i}_{\beta}
[A^{(0)\ast},\bar{\psi}^{(0)},Q_0,\theta_0^{\dagger\,}\,](-q),
\hspace{0.9cm}
\label{eq:3h}
\end{equation}
under the condition that on the left-hand sides of
Eqs.\,(\ref{eq:3f})\,--\,(\ref{eq:3h}) the interacting fields
$A^a_{\mu},\,\psi^{\,i}_{\alpha}$ and $\bar{\psi}^{\,i}_{\alpha}$ are defined by
expressions (\ref{eq:3d}). For deriving an explicit form of the effective 
currents and sources we functionally differentiate the right- and left-hand 
sides of equations (\ref{eq:3f})\,--\,(\ref{eq:3h}) with respect to free fields
$A^{(0)a}_{\mu},\,\psi^{(0)i}_{\alpha},\,\bar{\psi}^{(0)i}_{\alpha}$ and
initial color charges $Q_0^a,\,\theta_0^{\,i}$, and $\theta^{\dagger\, i}_0$
considering Eqs.\,(I.3.3), (\ref{eq:3r}), (I.3.4), (I.3.5), (\ref{eq:3a}) and
so on for differentiation on the left-hand sides, and set
$A^{(0)a}_{\mu}=\psi^{(0)i}_{\alpha}=\bar{\psi}^{(0)i}_{\alpha}
=Q_0^a=\theta_0^{\dagger\,i}=\theta_0^{\, i}=0$ at the end of calculations.
In Sections 4, 6, and 7 we will give some examples.

\section{\bf Effective currents and sources of lower order}
\setcounter{equation}{0}

\begin{figure}[hbtp]
\begin{center}
\includegraphics*[scale=0.4]{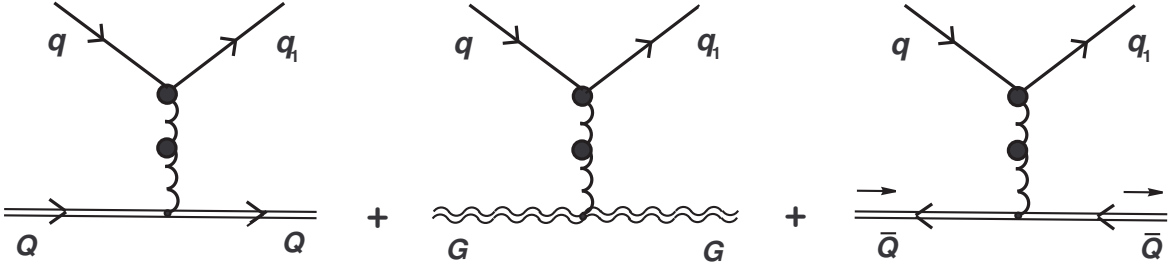}
\end{center}
\caption{\small The process of elastic scattering of soft fermion
excitation off the hard test particles. The blob stands for the HTL resummation
and the double line denotes hard test particles ($G$ is a hard gluon,
$Q\,(\bar{Q})$ is a hard quark (antiquark)).}
\label{fig1}
\end{figure}

The first simplest effective source arises from second derivative of
the source $\eta_{\alpha}^i[A,\psi,\theta_0](q)$ with respect to free
soft-quark field $\psi^{(0)}$ and initial usual color charge $Q_0$.
Taking into account Eq.\,(I.3.4) and the derivative
\begin{equation}
\left.\frac{\delta A_{\mu}^a(k)}
{\delta Q_0^b}\,\right|_{\,0}
=-\frac{\,g}{(2\pi)^3}\,^{\ast}{\cal D}_{\mu\nu}(k)v^{\nu}\delta^{ab}
\delta(v\cdot k),
\label{eq:4q}
\end{equation}
we find
\begin{equation}
\left.\frac{\delta^2\eta_{\alpha}^i[A,\psi,\theta_0](q)}
{\delta\psi^{(0)i_1}_{\alpha_1}(q_1)\,\delta Q_0^a}\,
\right|_{\,0}
=
\left.\frac{\delta^2\tilde{\eta}_{\alpha}^i
[A^{(0)},\psi^{(0)},Q_0,\theta_0](q)}
{\delta\psi^{(0)i_1}_{\alpha_1}(q_1)\,\delta Q_0^a}\,
\right|_{\,0}
\label{eq:4w}
\end{equation}
\[
=
-\frac{\,g^2}{(2\pi)^3}\,(t^a)^{ii_1}
\,^{\ast}\Gamma^{(Q)\mu}_{\alpha\alpha_1}(q-q_1;q_1,-q)
\,^{\ast}{\cal D}_{\mu\nu}(q-q_1)v^{\nu}
\delta(v\cdot (q-q_1)).
\]
Here and henceforth the symbol $``\,|_{\,0}\,"$ means that derivatives are
taken for values
$A^{(0)}=\psi^{(0)}=\bar{\psi}^{(0)}=Q_0=\theta_0^{\dagger}=\theta_0=0$.
Thus in linear approximation in free quark field $\psi^{(0)}$ and color
charge $Q_0$, we have
\begin{equation}
\tilde{\eta}^{(1)i}_{\alpha}(\psi^{(0)},Q_0)(q) =
\frac{\,g^2}{(2\pi)^3}\,(t^a)^{ii_1}Q_0^a\int\!
K_{\alpha\alpha_1}^{(Q)}(\chi,\bar{\chi}|\,q,-q_1)
\psi^{(0)i_1}_{\alpha_1}(q_1)\,\delta(v\cdot (q-q_1))dq_1,
\label{eq:4e}
\end{equation}
where the coefficient function in the integrand is
\begin{equation}
K_{\alpha\alpha_1}^{(Q)}(\chi,\bar{\chi}|\,q,-q_1)\equiv
-\,^{\ast}\Gamma^{(Q)\mu}_{\alpha\alpha_1}(q-q_1;q_1,-q)
\,^{\ast}{\cal D}_{\mu\nu}(q-q_1)v^{\nu}.
\label{eq:4r}
\end{equation}
The effective source (\ref{eq:4e}) within the framework of semiclassical
approximation generates the simplest elastic scattering process of
soft-quark excitation off the hard test particle, i.e., the scattering
process occurring without change of statistics of soft and hard
excitations (Fig.\,\ref{fig1}).

\begin{figure}[hbtp]
\begin{center}
\includegraphics[width=0.81\textwidth]{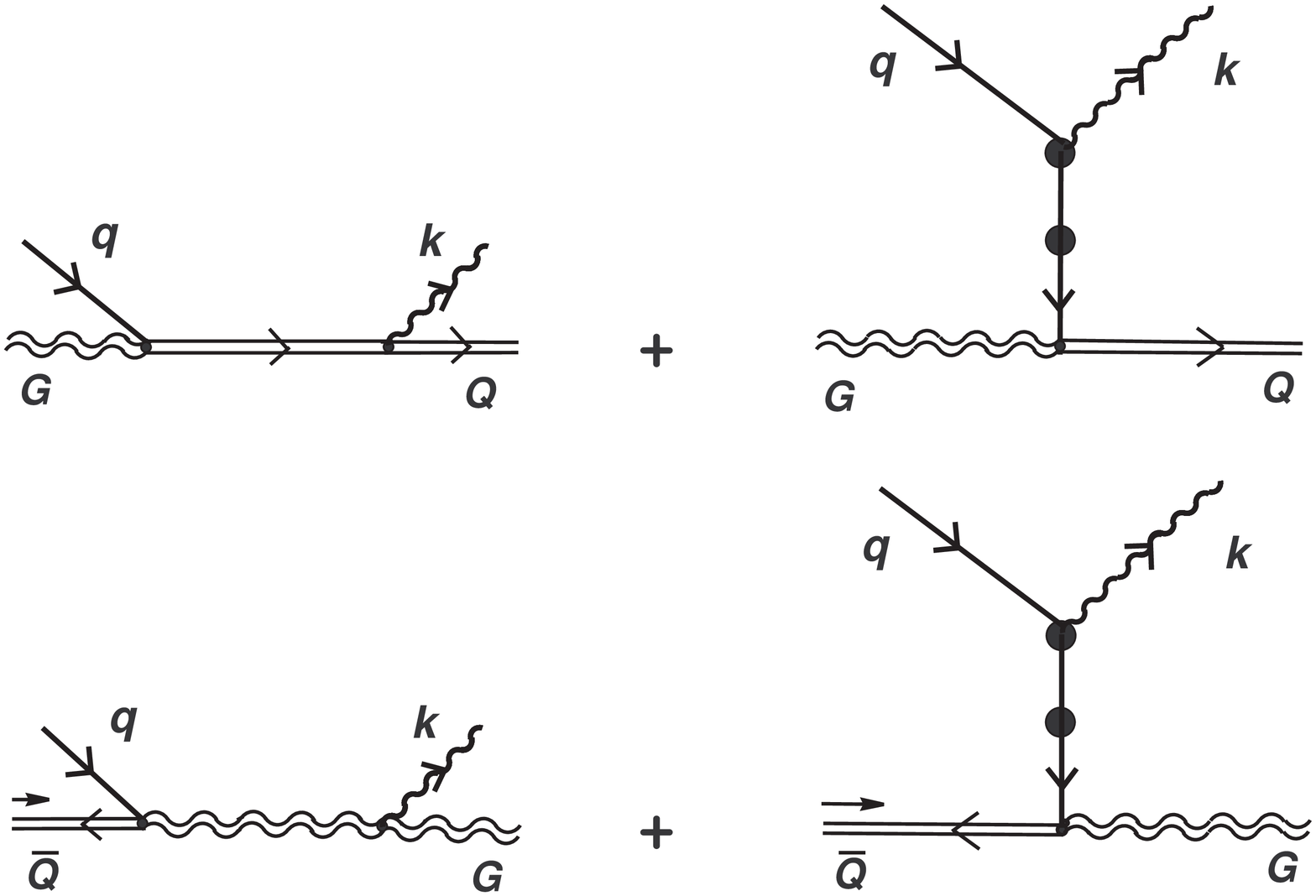}
\end{center}
\caption{\small The lowest order scattering process of soft fermion
excitations off the hard test particles with a change of statistics
of hard and soft excitations.}
\label{fig2}
\end{figure}

Furthermore, we calculate the second derivative of the source
$\eta_{\alpha}^i[A,\psi,\theta_0](q)$ with respect to $A^{(0)}$
and $\theta_0$. Taking into account Eq.\,(\ref{eq:3a}), (I.3.4)
and derivative
\begin{equation}
\left.\frac{\delta\psi_{\alpha}^i(q)}
{\delta\theta_0^j}\,\right|_{\,0}
=-\frac{\,g}{(2\pi)^3}\,^{\ast}S_{\alpha\beta}(q)\chi_{\beta}\,\delta^{ij}
\delta(v\cdot q)
\label{eq:4t}
\end{equation}
after simple calculations, we get
\[
\left.\frac{\delta^2\eta_{\alpha}^i[A,\psi,\theta_0](q)}
{\delta A^{\!(0)a}_{\mu}(k)\,\delta\theta_0^j}\,
\right|_{\,0}
=
\left.\frac{\delta^2\tilde{\eta}_{\alpha}^i
[A^{(0)},\psi^{(0)},Q_0,\theta_0](q)}
{\delta A^{\!(0)a}_{\mu}(k)\,\delta\theta_0^j}\,
\right|_{\,0}
\]
\[
=
\frac{\,g^2}{(2\pi)^3}\,(t^a)^{ij}
K_{\alpha}^{(Q)\mu}({\bf v},\chi|\,k,-q)
\delta(v\cdot (k-q)),
\]
where the coefficient function $K_{\alpha}^{(Q)\mu}$ is defined by the 
expression
\begin{equation}
K_{\alpha}^{(Q)\mu}({\bf v},\chi|\,k,-q)\equiv
\frac{v^{\mu}\chi_{\alpha}}{v\cdot q}\,-
\,^{\ast}\Gamma^{(Q)\mu}_{\alpha\beta}(k;q-k,-q)
\,^{\ast}\!S_{\beta\beta^{\prime}}(q-k)\chi_{\beta^{\prime}}.
\label{eq:4y}
\end{equation}
Thus we have found one more effective source additional to (\ref{eq:4e})
linear in free gauge field $A^{(0)}$ and the Grassmann color
charge $\theta_0$
\begin{equation}
\tilde{\eta}^{(1)i}_{\alpha}(A^{(0)}\!,\theta_0)(q) =
\frac{\,g^2}{(2\pi)^3}\,(t^a)^{ij}\,\theta_0^j\!\int\!
K_{\alpha}^{(Q)\mu}({\bf v},\chi|\,k,-q)
A^{(0)a}_{\mu}(k)\,\delta(v\cdot (k-q))dk.
\label{eq:4u}
\end{equation}
This effective source generates now somewhat more complicated
scattering process of soft-quark excitation off hard thermal
particles bringing into change of statistics of hard and soft
modes. The diagrammatic interpretation of two terms in expression
(\ref{eq:4y}) is presented in Fig.\,\ref{fig2}, where in first
line a `direct' channel of scattering is depicted and in the second one
`annihilation' channel is drawn\footnote{In the semiclassical
approximation, the first term in coefficient function
(\ref{eq:4y}) also contains processes, where the soft gluon is
emitted prior to soft-quark absorption. Hereafter for the sake of
brevity the diagrams of these scattering processes will be sequentially
omitted.}.
For convenience of the further references we write out also an explicit form
of effective source Dirac conjugate to (\ref{eq:4u})
\begin{equation}
\tilde{\bar{\eta}}^{(1)i}_{\alpha}(A^{\ast(0)}\!,\theta_0^{\dagger})(-q) =
\frac{\,g^2}{(2\pi)^3}\,\,\theta_0^{\dagger j}(t^a)^{ji}\!\int\!
\bar{K}_{\alpha}^{(Q)\mu}({\bf v},\bar{\chi}|\,k,-q)
A^{\ast(0)a}_{\mu}(k)\,\delta(v\cdot (k-q))dk.
\label{eq:4i}
\end{equation}
Here the coefficient function $\bar{K}_{\alpha}^{(Q)\mu}$ equals
\begin{equation}
\bar{K}_{\alpha}^{(Q)\mu}({\bf v},\bar{\chi}|\,k,-q)=
\frac{v^{\mu}\bar{\chi}_{\alpha}}{v\cdot q}\,+
\bar{\chi}_{\beta^{\prime}}
\,^{\ast}\!S_{\beta^{\prime}\beta}(-q+k)
\,^{\ast}\Gamma^{(Q)\mu}_{\beta\alpha}(-k;-q+k,q).
\label{eq:4o}
\end{equation}

It should be mentioned that in calculating derivatives with respect to
the Grassmann charges $\theta_0^i$ and $\theta_0^{\dagger i}$ we must
consider variables $Q_0^a,\,\theta_0^i$ and $\theta_0^{\dagger i}$ as
completely independent, in spite of the fact that there exists relation
(\ref{eq:3u}). Otherwise in calculating effective currents and sources we
obtain the coefficient functions containing terms of different order
in the coupling constant.

Now we consider derivatives of the current
$j^{\,a\nu}[A,\bar{\psi},\psi,Q_0](k)$. To lowest order in the coupling
here there exist two nontrivial derivatives linear in free soft-quark fields
$\bar{\psi}^{(0)},\,\psi^{(0)}$ and the Grassmann charges $\theta_0^i$,
$\theta_0^{\dagger i}$. The first of them has a form
\begin{equation}
\left.\frac{\delta^{2}\!j^{\,a\mu}[A,\bar{\psi},\psi,Q_0](k)}
{\delta\bar{\psi}^{(0)\,i}_{\alpha}(-q)\,\delta\theta_0^j}\,
\right|_{\,0}
=
\left.\frac{\delta^2 \tilde{j}^{\,a\mu}
[A^{(0)},\bar{\psi}^{(0)},\psi^{(0)},Q_0,\theta_0^{\dagger},\theta_0\,](k)}
{\delta\bar{\psi}^{(0)\,i}_{\alpha}(-q)\,\delta\theta_0^j}\,
\right|_{\,0}
\label{eq:4p}
\end{equation}
\[
=\frac{\,g^2}{(2\pi)^3}\,(t^a)^{ij}\,
\,^{\ast}\Gamma^{(G)\mu}_{\alpha\beta}(k;q,-k-q)
\,^{\ast}\!S_{\beta\beta^{\prime}}(k+q)\chi_{\beta^{\prime}}
\,\delta(v\cdot(k+q)).
\]
In deriving the expression we have used Eqs.\,(I.3.3) and
(\ref{eq:4t}). Another variation has a similar structure
\begin{equation}
\left.\frac{\delta^{2}\!j^{\,a\mu}[A,\bar{\psi},\psi,Q_0](k)}
{\delta\psi^{(0)j}_{\alpha}(q)\,\delta\theta_0^{\dagger \,i}}
\right|_{\,0}
=
\left.\frac{\delta^2 \tilde{j}^{\,a\mu}
[A^{(0)},\bar{\psi}^{(0)},\psi^{(0)},Q_0,\theta_0^{\dagger},\theta_0\,](k)}
{\delta\psi^{(0)j}_{\alpha}(q)\,\delta\theta_0^{\dagger \,i}}\,
\right|_{\,0}
\label{eq:4a}
\end{equation}
\[
=\frac{\,g^2}{(2\pi)^3}\,(t^a)^{ij}\,
\bar{\chi}_{\beta^{\prime}}
\,^{\ast}S_{\beta^{\prime}\beta}(k-q)
\,^{\ast}\Gamma^{(G)\mu}_{\beta\alpha}(k;-k+q,-q)
\,\delta(v\cdot(k-q)).
\]

Let us recall that we consider the quark wave functions
$\psi^{(0)}_{\alpha}$ and $\bar{\psi}^{(0)}_{\alpha}$ as Grassmann variables
similarly to color charges $\theta_0^i$ and $\theta_0^{\dagger\,i}$, i.e.,
they obey anticommutation relations. Besides we require in addition fulfilment
of rules
\[
\{\bar{\psi}^{(0)},\theta_0\}=\{\psi^{(0)},\theta_0^{\dagger}\}=
\{\bar{\psi}^{(0)},\theta_0^{\dagger}\}=\{\psi^{(0)},\theta_0\}=0,
\]
where $\{\,,\}$ denotes anticommutator\footnote{In conjunction
operation of product of two (and more) anticommutating functions it
is necessary to follow the rule
\[
(ab)^{\dagger}=b^{\dagger} a^{\dagger},
\]
i.e., $a^{\dagger}$ and $b^{\dagger}$ are rearranged without change of a sign
\cite{berezin}.}. Consideration of these relations is important for
obtaining correct signs ahead of the terms in the coefficient functions of
the effective currents and sources.

It is easy to show that the coefficient functions in the effective
sources (\ref{eq:4e}) and (\ref{eq:4u}) are gauge invariant if
they are determined on mass-shell of soft plasma modes. In the
first case this implies independence of the function (\ref{eq:4r})
from a choice of gauge for soft-gluon propagator. Let us consider in
more detail the second case. We convolve function (\ref{eq:4y})
with longitudinal projector $\bar{u}_{\mu}(k)=k^2u_{\mu} -
k_{\mu}(k\cdot u)$ in a covariant gauge. Making use the effective
Ward identity
\begin{equation}
\,^{\ast}\Gamma^{(Q)\mu}(k;q-k,-q)k_{\mu} =
\,^{\ast}\!S^{-1}(q-k) -\!\,^{\ast}\!S^{-1}(q),
\label{eq:4s}
\end{equation}
we find
\begin{equation}
\bar{u}_{\mu}(k)
K_{\alpha}^{(Q)\mu}({\bf v},\chi|\,k,-q)=
k^2\biggl\{
\frac{\chi_{\alpha}}{v\cdot q}\,-
\,^{\ast}\Gamma^{(Q)0}_{\alpha\beta}(k;q-k,-q)
\,^{\ast}S_{\beta\beta^{\prime}}(q-k)\chi_{\beta^{\prime}}
\!\biggr\}
\label{eq:4d}
\end{equation}
\[
\hspace{3cm}
-\,k^0\Bigl\{\chi_{\alpha}-
\Bigl(\,^{\ast}\!S^{-1}_{\alpha\beta}(q-k) -
\!\,^{\ast}\!S^{-1}_{\alpha\beta}(q)\Bigr)
\,^{\ast}\!S_{\beta\beta^{\prime}}(q-k)
\chi_{\beta^{\prime}}\biggr\}.
\]
On mass-shell of soft fermion excitations the equation
$\,^{\ast}\!S^{-1}(q)=0$ is fulfilled and therefore, the second term on
the right-hand side of Eq.\,(\ref{eq:4d}) vanishes. Furthermore, we
consider a convolution of coefficient function (\ref{eq:4y}) with
longitudinal projector in the temporal gauge
$\tilde{u}_{\mu}(k)=k^2(u_{\mu}(k\cdot u)-k_{\mu})/(k\cdot u)$.
The reasonings similar to previous ones result the convolution
$\tilde{u}_{\mu}(k)K_{\alpha}^{(Q)\mu}|_{\rm on-shell}$ in the
expression, which exactly equals the first term on the right-hand
side of Eq.\,(\ref{eq:4d}). By this means we have shown that at least in
the class of temporal and covariant gauges coefficient
function (\ref{eq:4y}) is gauge invariant.

If we repeat the same reasonings for coefficient function
generated by derivatives (\ref{eq:4p}) (or (\ref{eq:4a})) 
using the Ward identity for HTL-resummed vertex
$\,^{\ast}\Gamma^{(G)\mu}(k;q,-k-q)$, then we see that the function 
is not gauge invariant. The analysis of this fact suggests that for
restoration of gauge symmetry of effective currents it is
necessary to add an additional current to the right-hand side of the
Yang-Mills equation (\ref{eq:3e}). This current is directly
connected with an existence of spin-half hard partons with the
Grassmann color charge. The next section is concerned with
consideration of this problem in more details.

\section{\bf Additional color currents and sources}
\setcounter{equation}{0}

Let us define an additional current, which should be added to the
right-hand side of the Yang-Mills equation in order to restore a gauge
invariance of the coefficient function determined by derivative (\ref{eq:4p})
(and also (\ref{eq:4a})). At first we consider this problem 
in the coordinate
representation. It is clear that this current is to be real, gauge-covariant,
vanishing in the absence of the hard test particle with the Grassmann charge
(i.e., for $\theta_0^{\dagger}=\theta_0=0$). Besides, the current
as far as possible should not generate additional terms, which not needed for
restoration of gauge invariance of coefficient functions in the effective 
currents. Let us write out usual color current (\ref{eq:3q}) taking into 
account the relation (\ref{eq:3u})
\[
j_{Q\mu}^{\,a}[A](x)=gv_{\mu}\,
\theta^{\dagger i}(t)(t^a)^{ij}\theta^{j}(t)\,
{\delta}^{(3)}({\bf x}-{\bf v}t)
\]
and transform this current as follows. We perform the replacements
\[
\theta^{\dagger i}(t)\rightarrow \theta^{\dagger i}(t)\;\,
\Bigl(\equiv\theta^{\dagger\,i^{\prime}}_0U^{i^{\prime}i}(t_0,t)\Bigr),
\]
\[
\theta^{j}(t)\rightarrow ig\!\int\limits_{t_0}^{t}\!
U^{jj^{\,\prime}}(t,\tau)
\Bigl(\bar{\chi}_{\alpha}\psi^{j^{\,\prime}}_{\alpha}
(\tau,{\bf v}\tau)\Bigr)\,d\tau,
\]
where $U(t,\tau)={\rm T}\exp\{-ig\!\int_{\tau}^t (v\cdot
A^a(\tau^{\,\prime},{\bf v}\tau^{\,\prime\,}))\,t^ad\tau^{\,\prime}\}$ is
the evolution operator in the fundamental representation.
Furthermore, in the same way we replace in the initial current
\[
\theta^{\dagger i}(t)\rightarrow
-\,ig\!\int\limits_{t_0}^{t}
\Bigl(\bar{\psi}^{i^{\,\prime}}_{\alpha}
(\tau,{\bf v}\tau)\chi_{\alpha}\Bigr)
U^{i^{\prime}i}(\tau,t)d\tau,
\]
\[
\theta^{j}(t)\rightarrow \theta^{j}(t)\;\,
\Bigl(\equiv U^{jj^{\,\prime}}(t,t_0)\theta_0^{j^{\,\prime}}\Bigr)
\]
and combine together two in such a manner the expressions obtained. As a
result we obtain new color current in the following form:
\begin{equation}
j_{\theta\mu}^{\,a}[A,\bar{\psi},\psi](x)=ig^2v_{\mu}
\!\int\limits_{t_0}^{t}
\Bigl(\bar{\psi}^{i^{\,\prime}}_{\alpha}
(\tau,{\bf v}\tau)\chi_{\alpha}\Bigr)
U^{i^{\prime}i}(\tau,t)d\tau\,
(t^a)^{ij}\theta^{j}(t)\,
{\delta}^{(3)}({\bf x}-{\bf v}t)
\label{eq:5q}
\end{equation}
\[
\hspace{2.3cm}
-\,ig^2v_{\mu}\,
\theta^{\dagger i}(t)(t^a)^{ij}
\!\int\limits_{t_0}^{t}\!
U^{jj^{\,\prime}}(t,\tau)
\Bigl(\bar{\chi}_{\alpha}\psi^{j^{\,\prime}}_{\alpha}
(\tau,{\bf v}\tau)\Bigr)d\tau
{\delta}^{(3)}({\bf x}-{\bf v}t).
\]
In order to distinguish this current\footnote{Note that by using
differentiation rule of the link operator $U(t,\tau)$, we can represent also
current (\ref{eq:5q}) in more compact form
\[
j_{\theta\mu}^{\,a}[A,\bar{\psi},\psi](x)=g\,
{\delta}^{(3)}({\bf x}-{\bf v}t)
\]
\[
\times
\frac{\delta}{\delta A^{a\mu}(t,{\bf v}t)}\,
\left\{\theta_0^{\dagger i}\!
\!\int\limits_{t_0}^{t}\!
U^{ij}(t_0,\tau)
\Bigl(\bar{\chi}_{\alpha}\psi^{j}_{\alpha}
(\tau,{\bf v}\tau)\Bigr)d\tau
-
\!\int\limits_{t_0}^{t}
\Bigl(\bar{\psi}^{i}_{\alpha}
(\tau,{\bf v}\tau)\chi_{\alpha}\Bigr)
U^{ij}(\tau,t_0)d\tau\,\theta_0^j
\right\}.
\]} from (\ref{eq:3q}) we use notation $j_{\theta}$ instead of
$j_Q$. The current (\ref{eq:5q})
satisfies all properties listed above. Unlike color current (\ref{eq:3q}) and
color sources (\ref{eq:3i}), (\ref{eq:3o}) expression (\ref{eq:5q}) explicitly
(linearly) depends on interacting soft-quark fields $\bar{\psi}$ and $\psi$.
Now we turn to the momentum representation
\begin{equation}
j_{\theta\mu}^{\,a}[A,\bar{\psi},\psi](k)=
\int\!{\rm e}^{ik\cdot x}
j_{\theta\mu}^{\,a}[A,\bar{\psi},\psi](x)
\,\frac{dt}{2\pi}\,\frac{d{\bf x}}{(2\pi)^3}
\label{eq:5w}
\end{equation}
and define the first term in expansion of $j_{\theta\mu}^a$ in powers of
interacting fields $A,\,\psi,\,\bar{\psi}$, and initial values of the
Grassmann color charges $\theta_0^i$, $\theta_0^{\dagger\,i}$. For this
purpose in (\ref{eq:5q}) we set
\[
\theta^j(t)\rightarrow \theta_0^j,\quad
U^{i^{\prime}i}(\tau,t)\rightarrow\delta^{i^{\prime}i},\quad
\Bigl(\bar{\psi}^{i^{\,\prime}}_{\alpha}
(\tau,{\bf v}\tau)\chi_{\alpha}\Bigr)=
\int\!{\rm e}^{-i(v\cdot q)\tau}
\Bigl(\bar{\psi}^{i^{\,\prime}}_{\alpha}(-q)\chi_{\alpha}\Bigr)dq
\]
and so on. Discarding all terms containing initial time $t_0$, we
obtain after simple calculations
\begin{equation}
j_{\theta\mu}^{(1)\,a}(\bar{\psi},\psi)(k) =
\frac{\,g^2}{(2\pi)^3}\,v_{\mu}\!\int\!\frac{1}{(v\cdot q)}\,
\Bigl(\bar{\psi}^{i}_{\alpha}(-q)\chi_{\alpha}\Bigr)
(t^a)^{ij}\theta_0^j\,\delta(v\cdot(k+q))dq
\label{eq:5ww}
\end{equation}
\[
\hspace{2.6cm}
+\,\frac{\,g^2}{(2\pi)^3}\,v_{\mu}\,
\theta^{\dagger\,i}_0(t^a)^{ij}\!\int\!\frac{1}{(v\cdot q)}
\Bigl(\bar{\chi}_{\alpha}\psi^{j}_{\alpha}(q)\Bigr)
\,\delta(v\cdot(k-q))dq.
\]

We add additional current (\ref{eq:5w}), (\ref{eq:5q}) to the initial one
$j_{\mu}^{\,a}[A,\bar{\psi},\psi,Q_0](k)$ on the right-hand side of field
equation (\ref{eq:3e}). Taking into account the explicit expression
$j_{\theta\mu}^{(1)\,a}$ written above, we obtain instead of (\ref{eq:4p})
\[
\left.\frac{\delta^{2}\!\Bigl(j^{\,a\mu}[A,\bar{\psi},\psi,Q_0](k)
+j^{\,a\mu}_{\theta}[A,\bar{\psi},\psi](k)\Bigr)}
{\delta\bar{\psi}^{(0)i}_{\alpha}(-q)\delta \theta_0^j}\,
\right|_{\,0}
=
\left.\frac{\delta^2 \tilde{j}^{\,a\mu}
[A^{(0)},\bar{\psi}^{(0)},\psi^{(0)},Q_0,\theta_0^{\dagger},\theta_0\,](k)}
{\delta\bar{\psi}^{(0)i}_{\alpha}(-q)\delta \theta_0^j}\,
\right|_{\,0}
\]
\[
=
-\,\frac{\,g^2}{(2\pi)^3}\,(t^a)^{ij}
K_{\alpha}^{(G)\mu}({\bf v},\chi|\,k,q)
\,\delta(v\cdot (k+q)),
\]
where the coefficient function $K_{\alpha}^{(G)\mu}$ is defined by
\begin{equation}
K_{\alpha}^{(G)\mu}({\bf v},\chi|\,k,q)\equiv
\frac{v^{\mu}\chi_{\alpha}}{v\cdot q}\,-
\,^{\ast}\Gamma^{(G)\mu}_{\alpha\beta}(k;q,-k-q)
\,^{\ast}S_{\beta\beta^{\prime}}(k+q)\chi_{\beta^{\prime}},
\label{eq:5e}
\end{equation}
and correspondingly instead of (\ref{eq:4a}), we have
\[
\left.\frac{\delta^{2}\!\Bigl(j^{\,a\mu}[A,\bar{\psi},\psi,Q_0](k)
+j^{\,a\mu}_{\theta}[A,\bar{\psi},\psi](k)\Bigr)}
{\delta\psi^{(0)i}_{\alpha}(q)\,\delta \theta_0^{\dagger j}}
\right|_{\,0}
=
\left.\frac{\delta^2 \tilde{j}^{\,a\mu}
[A^{(0)},\bar{\psi}^{(0)},\psi^{(0)},Q_0,\theta_0,\theta_0^{\dagger}\,](k)}
{\delta\psi^{(0)i}_{\alpha}(q)\,\delta \theta_0^{\dagger j}}\,
\right|_{\,0}
\]
\[
=
\frac{\,g^2}{(2\pi)^3}\,(t^a)^{ji}
\bar{K}_{\alpha}^{(G)\mu}({\bf v},\bar{\chi}|\,k,-q)
\,\delta(v\cdot (k-q)),
\]
where
\begin{equation}
\bar{K}_{\alpha}^{(G)\mu}({\bf v},\bar{\chi}|\,k,-q)\equiv
\frac{v^{\mu}\bar{\chi}_{\alpha}}{v\cdot q}\,+
\bar{\chi}_{\beta^{\prime}}
\,^{\ast}S_{\beta^{\prime}\beta}(k-q)
\,^{\ast}\Gamma^{(G)\mu}_{\beta\alpha}(k;-k+q,-q).
\label{eq:5r}
\end{equation}
Now we can write out a total expression for the effective current linear in 
free soft-quark fields $\bar{\psi}^{(0)},\,\psi^{(0)}$ and the Grassmann 
charges $\theta_0^{\dagger i},\,\theta_0^i$:
\begin{equation}
\tilde{j}^{(1)a}_{\theta\mu}(\bar{\psi}^{(0)},\psi^{(0)})(k)
= \,\tilde{j}_{\mu}^{\,\dagger\,ai}(\psi^{(0)})(-k)\,
\theta_0^i\,+\,
\theta_0^{\dagger\,i}\,\tilde{j}_{\mu}^{\,ai}(\psi^{(0)})(k),
\label{eq:5t}
\end{equation}
where
\begin{equation}
\tilde{j}_{\mu}^{\,ai}(\psi^{(0)})(k)\equiv
\frac{\,g^2}{(2\pi)^3}\,\,(t^a)^{ij}\!\!\int\!
\bar{K}_{\alpha,\,\mu}^{(G)}({\bf v},\bar{\chi}|\,k,-q)
\psi^{(0)j}_{\alpha}(q)\,\delta(v\cdot (k-q))dq,
\label{eq:5tt}
\end{equation}
\[
\tilde{j}_{\mu}^{\,\dagger\,ai}(\psi^{(0)})(-k)\equiv
\frac{\,g^2}{(2\pi)^3}\int\!\bar{\psi}^{(0)j}_{\alpha}(-q)
(t^a)^{ji}K_{\alpha,\,\mu}^{(G)}({\bf v},\chi|\,k,q)
\,\delta(v\cdot (k+q))dq.
\]

This effective current generates the scattering processes that
are inverse to the scattering processes depicted in
Fig.\,\ref{fig2}. By a direct calculation it is easy to verify that
this current satisfies the condition of reality:
$\tilde{j}^{\,\ast}(k)=\tilde{j}(-k)$, and coefficient functions
(\ref{eq:5e}) and (\ref{eq:5r}) in contracting with a longitudinal
projector are gauge invariant in the sense as it was discussed in the previous
section.

Further, we  consider a derivation of current (\ref{eq:5q})
proceed from the corresponding generalization of evolution
equations (\ref{eq:3t}) and (\ref{eq:3y}). As was already mentioned in
Introduction these equations was obtained in Ref.\,\cite{barducii} under
the assumption that in medium there is only mean (and/or stochastic) gauge
field $A_{\mu}^a(x)$ in which the classical color-charged particles
move. Here the following question arises: How are the equations 
modified if in addition there exist stochastic soft-quark fields
$\bar{\psi}_{\alpha}^i(x)$,
$\psi_{\alpha}^i(x)$ in the medium? We assume that a minimal extension of
equations (\ref{eq:3t}) and (\ref{eq:3y}) to the soft-fermion degree
of freedom of system with retention of gauge symmetry has the following
form:
\begin{equation}
\frac{d\vartheta^i(t)}{dt} + igv^{\mu}A^a_{\mu}(t,{\bf v}t)(t^a)^{ij}
\vartheta^j(t)
+ig\Bigl(\bar{\chi}_{\alpha}\psi_{\alpha}^i(t,{\bf v}t)\!\Bigr)=0,
\quad\;\;
\left.\theta^i_0=\vartheta^i(t)\right|_{\,t=t_0}
\label{eq:5y}
\end{equation}
\[
\frac{d\vartheta^{\dagger i}(t)}{dt} - igv^{\mu}A^a_{\mu}(t,{\bf v}t)
\vartheta^{\dagger j}(t)(t^a)^{ji}
-ig\Bigl(\bar{\psi}_{\alpha}^i(t,{\bf v}t){\chi}_{\alpha}\Bigr)=0,
\quad
\left.\theta^{\dagger i}_0=\vartheta^{\dagger i}(t)\right|_{\,t=t_0}.
\hspace{0.3cm}
\]
The general solution of these equations is
\begin{equation}
\vartheta^{i}(t) = U^{ij}(t,t_0)\theta_0^{j}
-ig\!\int\limits_{t_0}^{t}\!
U^{ij}(t,\tau)
\Bigl(\bar{\chi}_{\alpha}\psi^{j}_{\alpha}(\tau,{\bf v}\tau)\Bigr)d\tau,
\label{eq:5u}
\end{equation}
\[
\hspace{0.3cm}
\vartheta^{\dagger i}(t) = \theta^{\dagger j}_0U^{ji}(t_0,t)
+ig\!\int\limits_{t_0}^{t}
\Bigl(\bar{\psi}^{j}_{\alpha}
(\tau,{\bf v}\tau)\chi_{\alpha}\Bigr)
U^{ji}(\tau,t)d\tau.
\]
Here we have introduced new symbols $\vartheta^i$ and $\vartheta^{\dagger i}$
for the Grassmann charges having kept the old symbols $\theta^i$
and $\theta^{\dagger i}$ for solutions of homogeneous equations
(\ref{eq:3t}), (\ref{eq:3y}).

If we now substitute these solutions into (\ref{eq:3u}) and take into account
the identity
\begin{equation}
U(\tau,t)t^aU(t,\tau)=\tilde{U}^{ab}(t,\tau)t^b,
\label{eq:5uu}
\end{equation}
where $\tilde{U}(t,\tau)={\rm T}\exp\{-ig\!\int_{\tau}^t
(v\cdot A^a(\tau^{\,\prime},{\bf v}\tau^{\,\prime\,}))\,T^a d\tau^{\,\prime}\}$ 
is the evolution operator in the adjoint representation, then we obtain the
following expression\footnote{The function (\ref{eq:5i}) is formally
a solution of the equation
\begin{equation}
\frac{d{\cal Q}^a(t)}{dt} + igv^{\mu}A^b_{\mu}(t,{\bf v}t)(T^b)^{ac}
{\cal Q}^c(t)
+ig\Bigl[\,\vartheta^{\dagger j}(t)(t^a)^{ji}
\Bigl(\bar{\chi}_{\alpha}\psi_{\alpha}^i(t,{\bf v}t)\!\Bigr)-
\Bigl(\bar{\psi}_{\alpha}^i(t,{\bf v}t){\chi}_{\alpha}\Bigr)
(t^a)^{ij}\vartheta^j(t)\Bigr]=0,
\label{eq:5o}
\end{equation}
with initial condition $\left.Q^a_0={\cal Q}^a(t)\right|_{\,t=0}$. This
equation follows from system (\ref{eq:5y}). The functions
$\vartheta^{\dagger j}(t)$ and $\vartheta^j(t)$ here, are defined by
Eq.\,(\ref{eq:5u}).} for the color charge ${\cal Q}^a(t)$:
\begin{equation}
{\cal Q}^a(t)=\tilde{U}^{ab}(t,t_0)\,Q_0^b+ig\left\{\,
\int\limits_{t_0}^{t}
\Bigl(\bar{\psi}^{j^{\,\prime}}_{\alpha}
(\tau,{\bf v}\tau)\chi_{\alpha}\Bigr)
U^{j^{\prime}\!j}(\tau,t)\,d\tau\,(t^a)^{ji}
\Bigl[\,U^{ii^{\prime}}(t,t_0)\,\theta_0^{i^{\prime}}\Bigr]
\right.
\label{eq:5i}
\end{equation}
\[
\hspace{4.5cm}
\left.
-\,\Bigl[\,\theta_0^{\dagger i^{\prime}}U^{i^{\prime}i}(t_0,t)\Bigr]
(t^a)^{ij}\int\limits_{t_0}^{t}
U^{jj^{\,\prime}}(t,\tau)
\Bigl(\bar{\chi}_{\alpha}\psi^{j^{\,\prime}}_{\alpha}(\tau,{\bf v}\tau)\Bigr)
d\tau\!\right\}
\]
\[
\hspace{2.5cm}
+\,g^2\!
\int\limits_{t_0}^{t}\!\int\limits_{t_0}^{t}\!
\Bigl(\bar{\psi}^{i^{\prime}}_{\alpha}
(\tau,{\bf v}\tau)\chi_{\alpha}\Bigr)
U^{i^{\prime}i}(\tau,t)\,(t^a)^{ij}
U^{jj^{\,\prime}}(t,\tau^{\,\prime\,})
\Bigl(\bar{\chi}_{\alpha^{\prime}}
\psi^{j^{\,\prime}}_{\alpha^{\prime}}
(\tau^{\,\prime},{\bf v}\tau^{\,\prime\,})\Bigr)d\tau d\tau^{\,\prime},
\]
where we also have introduced a new symbol ${\cal Q}^a$ for usual
color charge, having kept the old symbol $Q^a$ for the solution of
the Wong equation (\ref{eq:3w}). If we now insert (\ref{eq:5i})
into expression for current (\ref{eq:3q}), then the first term
on the right-hand side of Eq.\,(\ref{eq:5i}) determines the usual
`classical' color current of hard parton. The expression in braces
in Eq.\,(\ref{eq:5i}) determines current (\ref{eq:5q}), which we
have introduced above for recovering gauge invariance of
scattering amplitude. The physical meaning of the last term in
(\ref{eq:5i}) is less clear, as well as the physical meaning of
the last terms on the right-hand sides of solutions
(\ref{eq:5u}) in inserting the last into Grassmann color sources
(\ref{eq:3i}) and (\ref{eq:3o}). Unfortunately, these terms
generate additional contributions redefining the effective currents
and sources. For example, in color source (\ref{eq:3i}) there
exists a term linear in interacting soft-quark field $\psi$
\[
\eta_{\theta\alpha}^i=\eta_{\theta\alpha}^{(0)i}
+\frac{\,g^2}{(2\pi)^3}\,\frac{\chi_{\alpha}}{(v\cdot q)}
\int\!\Bigl(\bar{\chi}_{\beta}\,\psi^{i}_{\beta}(q^{\,\prime\,})\Bigr)
\delta(v\cdot(q-q^{\,\prime\,}))dq^{\,\prime}\,+\,\ldots\,.
\]
This in particular leads to the fact that it is necessary to redefine 
the left-hand side of the Dirac field equation (\ref{eq:3p}) to the following
form
\[
\,^{\ast}\!S^{-1}_{\alpha\beta}(q)\psi^{i}_{\beta}(q)\rightarrow
\int\!dq^{\,\prime}\biggl\{\!
\,^{\ast}\!S^{-1}_{\alpha\beta}(q^{\,\prime\,})\,\delta(q-q^{\,\prime\,})
+\frac{\,g^2}{(2\pi)^3}\,\frac{\chi_{\alpha}\bar{\chi}_{\beta}}
{(v\cdot q)}\,\delta(v\cdot(q-q^{\,\prime\,}))\biggr\}\,
\psi^{i}_{\beta}(q^{\,\prime\,}).
\]
There are additional contributions to all HTL-induced sources:
$\eta^{(1,1)}(A,\psi)$, $\eta^{(2,1)}(A,A,\psi)$ and so on. The
last term on the right-hand side of Eq.\,(\ref{eq:5i}) generates
new contributions to the HTL-induced currents
$j^{\psi(0,2)}(\bar{\psi},\psi)$,
$j^{\psi(1,2)}(A,\bar{\psi},\psi),\ldots\,.$ By virtue of this fact we
can say nothing about physical sense of these new contributions, 
during all work we  simply ignore them. However the account of
the functions generating these contributions is very important. With the 
help of these functions we can construct new
gauge-covariant currents and sources, which generate new terms in
scattering amplitudes. These terms have already quite concrete
physical sense. This will be the subject of discussion just
below. Here we would like to make one more additional remark. As
it is well known evolution equation (\ref{eq:3w}) admits an
integral of motion
\[
Q^a(t)Q^a(t)=Q_0^aQ_0^a=const,
\]
and equations (\ref{eq:3t}), (\ref{eq:3y}) admit, correspondingly
\[
\theta^{\dagger i}(t)\theta^i(t)=\theta_0^{\dagger i}\theta_0^i=const.
\]
The new evolution equations (\ref{eq:5y}) and (\ref{eq:5o}) allows no
such integrals of motion. In other words, in the presence of soft-quark field  
fluctuations in the medium, `the length' of color vector of classical
particle is not conserved any more.

\begin{figure}[hbtp]
\begin{center}
\includegraphics[width=0.95\textwidth]{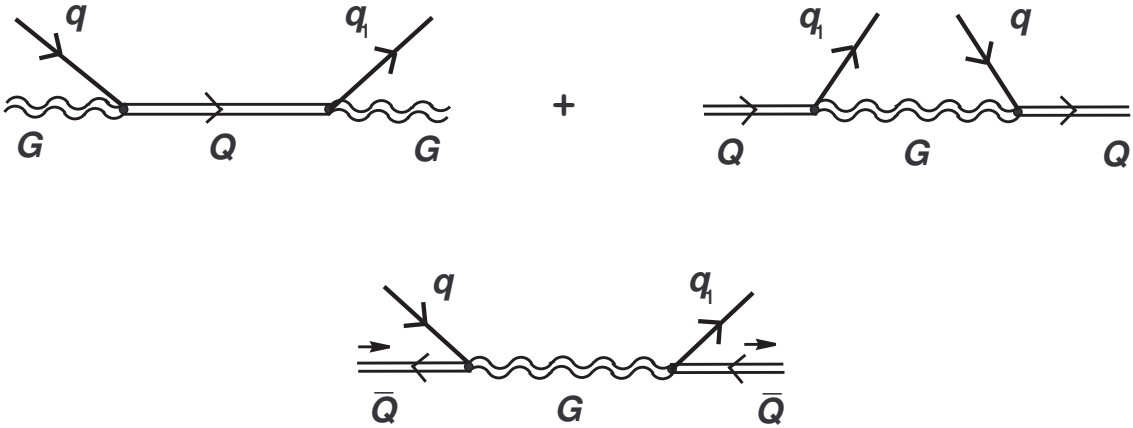}
\end{center}
\caption{\small The eikonal contributions to the elastic scattering process
of soft fermion excitation off the hard test particle (Fig.\,\ref{fig1}).}
\label{fig3}
\end{figure}

Now we introduce the following notations:
\begin{equation}
\Omega^i(t)\equiv
-ig\!\int\limits_{t_0}^{t}\!
U^{ij}(t,\tau)
\Bigl(\bar{\chi}_{\alpha}\psi^{j}_{\alpha}(\tau,{\bf v}\tau)\Bigr)d\tau,
\quad
\Omega^{\dagger i}(t) =
ig\!\int\limits_{t_0}^{t}
\Bigl(\bar{\psi}^{j}_{\alpha}
(\tau,{\bf v}\tau)\chi_{\alpha}\Bigr)
U^{ji}(\tau,t)d\tau,
\label{eq:5p}
\end{equation}
\[
\Xi^a(t)\equiv
g^2\!
\int\limits_{t_0}^{t}\!\int\limits_{t_0}^{t}\!
\Bigl(\bar{\psi}^{i^{\prime}}_{\alpha}
(\tau,{\bf v}\tau)\chi_{\alpha}\Bigr)
U^{i^{\prime}i}(\tau,t)\,(t^a)^{ij}
U^{jj^{\prime}}(t,\tau^{\,\prime})
\Bigl(\bar{\chi}_{\alpha^{\prime}}
\psi^{j^{\prime}}_{\alpha^{\prime}}
(\tau^{\,\prime},{\bf v}\tau^{\,\prime\,})\Bigr)d\tau d\tau^{\,\prime}
\]
\[
=\frac{i}{g}\,\frac{\delta}{\delta(v_{\mu}A^{a\mu}(t,{\bf v}t))}\,
\Bigl(\Omega^{\dagger i}(t)\Omega^i(t)\Bigr).
\]
In the last equality we have used the differentiation rule of the evolution
operator
\[
\frac{\delta U^{ij}(t,\tau)}{\delta(v_{\mu}A^{a\mu}
(t^{\prime\,},{\bf v}t^{\,\prime\,}))}=-\,ig\,
U^{ii^{\prime}}(t,t^{\,\prime})(t^a)^{i^{\prime}\!j^{\prime}}
U^{j^{\prime}\!j}(t^{\,\prime},\tau).
\]
Let us define a new additional source setting by the definition
\begin{equation}
\eta_{Q\alpha}^i(x)\equiv i\alpha\,\chi_{\alpha}Q^a(t)\!
\left(\frac{\delta\,\Omega^i(t)}{\delta(v_{\mu}A^{a\mu}(t,{\bf v}t))}\right)
\!{\delta}^{(3)}({\bf x}-{\bf v}t)
\label{eq:5a}
\end{equation}
\[
=-i\alpha\,g^2\chi_{\alpha}Q^a(t)(t^a)^{ij}
\!\int\limits_{t_0}^{t}\!\!
U^{jj^{\prime}}(t,\tau)
\Bigl(\bar{\chi}_{\beta}\psi^{j^{\prime}}_{\beta}(\tau,{\bf v}\tau)\Bigr)
d\tau\,{\delta}^{(3)}({\bf x}-{\bf v}t)
\]
\[
\equiv
\alpha\,g\chi_{\alpha}Q^a(t)(t^a)^{ij}\Omega^j(t)\,
{\delta}^{(3)}({\bf x}-{\bf v}t).
\]
Here $\alpha$ is a constant, which should be defined from some physical
reasons. Some of them will be discussed later. The multiplier `$i$' is
introduced for convenience subsequently. Under the gauge transformation
non-Abelian charge is transformed by the rule
\[
Q^at^a\rightarrow SQ^at^aS^{-1},
\]
therefore, as it is not difficult to see from explicit expression
(\ref{eq:5a}), this source is transformed properly:
$\eta_{Q\alpha}\rightarrow S\eta_{Q\alpha}$. We turn to
the momentum representation. From (\ref{eq:5a}) it follows the
next nontrivial derivative of the second order with respect to
$\psi^{(0)}$ and $Q_0^a$
\[
\left.\frac{\delta^2\eta_{Q\alpha}^i[A,\psi,\theta_0](q)}
{\delta\psi^{(0)\,i_1}_{\alpha_1}(q_1)\delta Q_0^a}\,
\right|_{\,0}=\alpha\,
\frac{\,g^2}{(2\pi)^3}\,\,(t^a)^{ii_1}\,
\frac{\chi_{\alpha}\bar{\chi}_{\alpha_1}}{(v\cdot q_1)}
\,\delta(v\cdot (q-q_1)).
\]
If we now add this expression to (\ref{eq:4w}), then we obtain total
expression for the effective source $\tilde{\eta}_{\alpha}^{(1)i}$ linear in
free soft-quark field $\psi^{(0)}$ and color charge $Q_0$, instead of
(\ref{eq:4e}), where now as the coefficient function
$K_{\alpha\alpha_1}^{(Q)}(\chi,\bar{\chi}|\,q,-q_1)$
it is necessary to understand the following expression
\begin{equation}
K_{\alpha\alpha_1}^{(Q)}(\chi,\bar{\chi}|\,q,-q_1)=
\alpha\,\frac{\chi_{\alpha}\bar{\chi}_{\alpha_1}}{v\cdot q_1}
-\,^{\ast}\Gamma^{(Q)\mu}_{\alpha\alpha_1}(q-q_1;q_1,-q)
\,^{\ast}{\cal D}_{\mu\nu}(q-q_1)v^{\nu}.
\label{eq:5s}
\end{equation}
The diagrammatic interpretation of new `eikonal' term on the right-hand side
of Eq.\,(\ref{eq:5s}) is presented in Fig.\,\ref{fig3}. These graphs should be
added to Fig.\,\ref{fig1}. Note that the fundamental difference between 
coefficient function (\ref{eq:5s}) and coefficient functions
(\ref{eq:4y}) and (\ref{eq:5e}) consists in the fact that two terms in
(\ref{eq:5s}) are not associated to one another by the
requirement of gauge invariance of the scattering amplitude that,
for example, unambiguously would fix the constant $\alpha$.

For convenience of the further references we write out an explicit
form of the effective source $\tilde{\bar{\eta}}$ Dirac conjugate to
(\ref{eq:4e})
\begin{equation}
\tilde{\bar{\eta}}^{(1)i}_{\alpha}(\bar{\psi}^{(0)}\!,Q_0)(-q)=
\frac{\,g^2}{(2\pi)^3}\,(t^a)^{i_1i}Q_0^a\int\!
\bar{K}_{\alpha_1\alpha}^{(Q)}(\chi,\bar{\chi}|\,q,-q_1)
\bar{\psi}^{(0)i_1}_{\alpha_1}(-q_1)\,\delta(v\cdot (q-q_1))dq_1,
\label{eq:5d}
\end{equation}
where
\begin{equation}
\bar{K}_{\alpha_1\alpha}^{(Q)}(\chi,\bar{\chi}|\,q,-q_1)\equiv
\alpha^{\ast}\,\frac{\chi_{\alpha_1}\bar{\chi}_{\alpha}}{v\cdot q_1}
\,-\,^{\ast}\Gamma^{(Q)\mu}_{\alpha_1\alpha}(-q+q_1;-q_1,q)
\,^{\ast}{\cal D}_{\mu\nu}^{\dagger}(q-q_1)v^{\nu}.
\label{eq:5f}
\end{equation}

From functionals (\ref{eq:5p}) one can form two some more additional sources:
\begin{equation}
\eta_{\Xi\,\alpha}^i(x)=\beta g\,\chi_{\alpha}(t^a)^{ij}\,\theta^j(t)\,
\Xi^a(t)\,{\delta}^{(3)}({\bf x}-{\bf v}t)
\hspace{1.5cm}
\label{eq:5g}
\end{equation}
and
\begin{equation}
\hspace{2cm}
\eta_{\Omega\,\alpha}^i(x)=\beta_{1}\,g\,\chi_{\alpha}
(t^a)^{ii_1}\,\Omega^{i_1}(t)
\Bigl[\,\Omega^{\dagger j_1}(t)(t^a)^{j_1j}\,\theta^j(t)\Bigr]\,
{\delta}^{(3)}({\bf x}-{\bf v}t),
\label{eq:5h}
\end{equation}
where $\beta,\,\beta_1$ are some constants. The sources (\ref{eq:5g}),
(\ref{eq:5h}) generate contributions to the scattering processes of 
higher order in the coupling (see the next sections). For example, the first
nontrivial derivative of source (\ref{eq:5g}) in the momentum
representation
\begin{equation}
\left.\frac{\delta^3\eta_{\Xi\,\alpha}^i[A,\psi,\theta_0](q)}
{\delta\psi^{(0)i_2}_{\alpha_2}(q_2)
\delta\bar{\psi}^{(0)i_1}_{\alpha_1}(-q_1)
\delta\theta_0^j }\,
\right|_{\,0}=
\beta\,\frac{\,g^3}{(2\pi)^3}\,(t^a)^{ij}(t^a)^{i_1i_2}\,
\frac{\chi_{\alpha}\chi_{\alpha_1}\bar{\chi}_{\alpha_2}}
{(v\cdot q_1)(v\cdot q_2)}\,\,
\delta(v\cdot (q+q_1-q_2))
\label{eq:5j}
\end{equation}
defines eikonal contribution to nonlinear interaction process of three
soft-quark excitations with the hard parton (Eqs.\,(\ref{eq:6t}), (\ref{eq:6y})
and Fig.\,\ref{fig6}). The source (\ref{eq:5h}) generates a similar
contribution with replacements in Eq.\,(\ref{eq:5j}):
$\beta\rightarrow\beta_1$ and
$(t^a)^{ij}(t^a)^{i_1i_2}\rightarrow(t^a)^{ii_2}(t^a)^{i_1j}$. By virtue of
a structure of the functionals $\Omega^i(t)$, $\Omega^{\dagger i}(t)$ and
$\Xi^a(t)$, sources (\ref{eq:5g}), (\ref{eq:5h}) are transformed by
covariant way under the gauge transformation.

Finally, we can define another color current supplementing (\ref{eq:5q})
setting by definition
\begin{equation}
j_{\Xi\,\mu}^{\,a}[A,\bar{\psi},\psi,Q_0](x)=
i\sigma\,v_{\mu}Q^b(t)\!\left(
\frac{\delta\,\Xi^a(t)}{\delta(v_{\nu}A^{b\nu}(t,{\bf v}t))}\right)\!
{\delta}^{(3)}({\bf x}-{\bf v}t)
\label{eq:5k}
\end{equation}
\[
=\sigma\,v_{\mu\,} g^3\!
\int\limits_{t_0}^{t}\!\int\limits_{t_0}^{t}\!
\Bigl(\bar{\psi}^{i^{\prime}}_{\alpha}
(\tau,{\bf v}\tau)\chi_{\alpha}\Bigr)
U^{i^{\prime}\!i}(\tau,t)\,\{t^a,t^b\}^{ij}
U^{jj^{\,\prime}}(t,\tau^{\,\prime\,})
\Bigl(\bar{\chi}_{\alpha^{\prime}}
\psi^{j^{\,\prime}}_{\alpha^{\prime}}
(\tau^{\,\prime},{\bf v}\tau^{\,\prime\,})\Bigr)d\tau d\tau^{\,\prime}
\]
\[
\times\,Q^b(t)
\,{\delta}^{(3)}({\bf x}-{\bf v}t),
\]
where $\sigma$ is a new constant. The additional color current (\ref{eq:5k})
similar to additional sources (\ref{eq:5g}), (\ref{eq:5h}) generates
contributions only to higher orders nonlinear interaction of soft
and hard modes. The first nontrivial variation of current (\ref{eq:5k})
(in the momentum representation) defines eikonal contribution to nonlinear
interaction process of two soft-quark and one soft-gluon excitations with
the hard parton (Eqs.\,(\ref{eq:6q}), (\ref{eq:6w}) and Fig.\,\ref{fig4})
\begin{equation}
\left.\frac{\delta^3j_{\Xi\,\mu}^{\,a}[A,\bar{\psi},\psi,Q_0](k)}
{\delta\psi^{(0)j}_{\beta}(q_1)
\delta\bar{\psi}^{(0)i}_{\alpha}(-q)
\delta Q_0^b}\,
\right|_{\,0}=
\sigma\frac{\,g^3}{(2\pi)^3}\,v_{\mu}\,\{t^a,t^b\}^{ij}\,
\frac{\chi_{\alpha}\bar{\chi}_{\beta}}
{(v\cdot q)(v\cdot q_1)}\,\,
\delta(v\cdot (k+q-q_1)).
\label{eq:5l}
\end{equation}

The currents and sources written out in this section and Section 3
enables us to calculate complete expressions for scattering
amplitudes of soft QGP modes off hard thermal particles, at least
up to the third order in free soft-field amplitudes and initial
values of color charges of hard partons. In two latter sections
this will be proved by explicit calculation of effective currents
and sources generating processes of nonlinear interaction of three
plasma waves with the test particle and soft-loop corrections to
the scattering processes considered in Section 4.
Unfortunately, we cannot prove whether it will be necessary to
introduce more complicated in structure additional currents and
sources in calculating the total scattering amplitudes of higher
order in interaction. The direct computations here become very
cumbersome and therefore for the proof of closure (or non-closure)
of the theory it is necessary to use methods and approaches that
are not related to an expansion in a series in powers of
soft-field amplitudes and initial values of color charges.

In Appendix A we give an explicit form of the action ${\cal S}$ varying
which we get the equation for color charge evolution of the hard test parton
and the Yang-Mills and Dirac equations for soft gluon and quark excitations.
The action suggested generates all additional sources introduced in present
section.

\section{\bf Third-order effective currents and sources}
\setcounter{equation}{0}

\begin{figure}[hbtp]
\begin{center}
\includegraphics[width=1\textwidth]{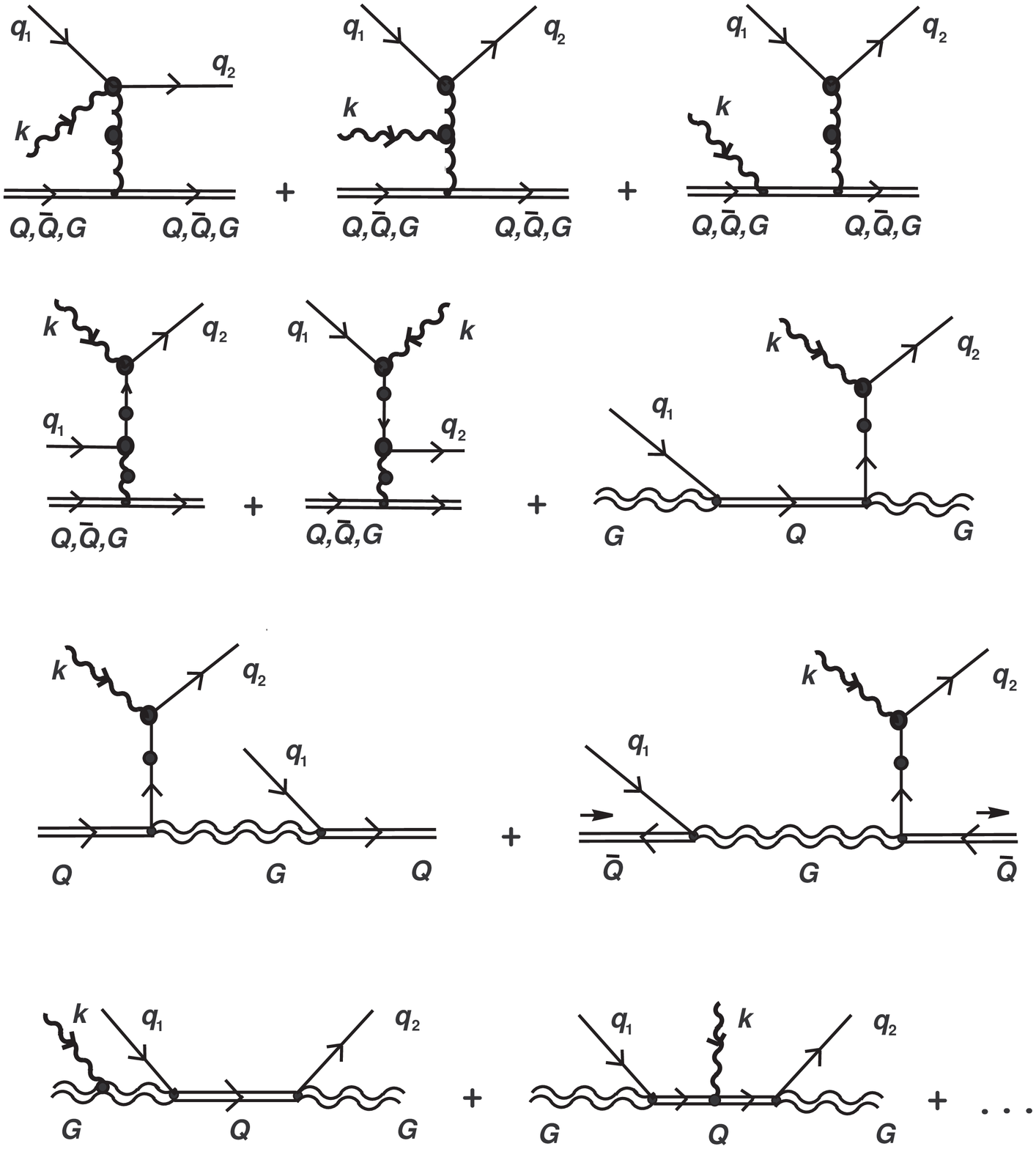}
\end{center}
\caption{\small The processes of absorption of soft-gluon and soft-quark
excitations by hard parton, accompanied by soft-quark radiation.}
\label{fig4}
\end{figure}

In this section examples of calculation of effective currents
and sources next-to-leading order in the coupling $g$ will be given. As a 
first step we consider the third order functional derivative of relation
(\ref{eq:3f}) with respect to $\psi^{(0)}$, $\bar{\psi}^{(0)}$ and $Q_0$.
By current $j_{\mu}^a$ on the left-hand side it is necessary to realize 
a sum of the initial current and additional currents (\ref{eq:5q}) and
(\ref{eq:5k}), i.e., 
$$j^{a}_{\mu}[A,\bar{\psi},\psi,Q_0,\theta^{\dagger}_0,
\theta_0](k)\equiv j^{a}_{\mu}[A,\bar{\psi},\psi,Q_0](k) +
j^{a}_{\theta\mu}[A,\bar{\psi},\psi,\theta_0](k)+
j^{a}_{\Xi\mu}[A,\bar{\psi},\psi,Q_0](k).
$$
Here, we have
\[
\left.\frac{\delta^{3}\!j^{a}_{\mu}[A,\bar{\psi},\psi,Q_0,\theta^{\dagger}_0,
\theta_0](k)}
{\delta\psi^{(0)j}_{\beta}(q_2) 
\delta\bar{\psi}^{(0)i}_{\alpha}(-q_1)
\delta Q_0^b}\,
\right|_{\,0}
=
\left.\frac{\delta^3 \tilde{j}^{a}_{\mu}
[A^{(0)},\bar{\psi}^{(0)},\psi^{(0)},Q_0,\theta_0^{\dagger},\theta_0](k)}
{\delta\psi^{(0)j}_{\beta}(q_2)
\delta\bar{\psi}^{(0)i}_{\alpha}(-q_1)
\delta Q_0^b}\,
\right|_{\,0}
\]
\[
=\!\int\left\{
\frac{\delta^2 j_{\mu}^{A(2)a}(k)}
{\delta A^{a_1^{\prime}\mu_1^{\prime}}(k_1^{\,\prime})
\delta A^{a_2^{\prime}\mu_2^{\prime}}(k_2^{\,\prime})}\,
\frac{\delta A^{a_2^{\prime}\mu_2^{\prime}}(k_2^{\,\prime})}
{\delta Q_0^b}\,
\frac{\delta^2A^{a_1^{\prime}\mu_1^{\prime}}(k_1^{\,\prime})}
{\delta \psi^{(0)j}_{\beta}(q_2)\delta\bar{\psi}^{(0)i}_{\alpha}(-q_1)}\,
\,dk_1^{\,\prime}dk_2^{\,\prime}
\right.
\]
\[
+\,\frac{\delta^3 j_{\mu}^{\Psi(1,2)a}(k)}
{\delta A^{a_1^{\prime}\mu_1^{\prime}}(k_1^{\,\prime})
\delta \psi^{j_1^{\prime}}_{\beta_1^{\prime}}(q_2^{\,\prime})
\delta\bar{\psi}^{i_1^{\prime}}_{\alpha_1^{\prime}}(-q_1^{\,\prime})}\,
\frac{\delta A^{a_1^{\prime}\mu_1^{\prime}}(k_1^{\,\prime})}
{\delta Q_0^b}\,
\frac{\delta \psi^{j_1^{\prime}}_{\beta_1^{\prime}}(q_2^{\,\prime})}
{\delta \psi^{(0)j}_{\beta}(q_2)}\,
\frac{\delta\bar{\psi}^{i_1^{\prime}}_{\alpha_1^{\prime}}(-q_1^{\,\prime})}
{\delta\bar{\psi}^{(0)\,i}_{\alpha}(-q_1)}\,\,
dk_1^{\,\prime}dq_1^{\,\prime}dq_2^{\,\prime}
\]
\[
+\,\frac{\delta^2 j_{Q\mu}^{(1)a}(k)}
{\delta A^{a_1^{\prime}\mu_1^{\prime}}(k_1^{\,\prime})\delta Q_0^b}
\frac{\delta^2A^{a_1^{\prime}\mu_1^{\prime}}(k_1^{\,\prime})}
{\delta \psi^{(0)j}_{\beta}(q_2)\delta\bar{\psi}^{(0)i}_{\alpha}(-q_1)}\,
\,dk_1^{\,\prime}
\hspace{5cm}
\]
\[
+\,\frac{\delta^2 j_{\mu}^{\Psi(0,2)a}(k)}
{\delta \psi^{j_1^{\prime}}_{\beta_1^{\prime}}(q_2^{\,\prime})
\delta\bar{\psi}^{i_1^{\prime}}_{\alpha_1^{\prime}}(-q_1^{\,\prime})}\,
\frac{\delta^2\psi^{j_1^{\prime}}_{\beta_1^{\prime}}(q_2^{\,\prime})}
{\delta \psi^{(0)j}_{\beta}(q_2)\delta Q_0^b}\,
\frac{\delta\bar{\psi}^{i_1^{\prime}}_{\alpha_1^{\prime}}(-q_1^{\,\prime})}
{\delta\bar{\psi}^{(0)i}_{\alpha}(-q_1)}
\,\,dq_1^{\,\prime}dq_2^{\,\prime}
\hspace{2.8cm}
\]
\[
+\,\frac{\delta^2 j_{\mu}^{\Psi(0,2)a}(k)}
{\delta \psi^{j_1^{\prime}}_{\beta_1^{\prime}}(q_2^{\,\prime})
\delta\bar{\psi}^{i_1^{\prime}}_{\alpha_1^{\prime}}(-q_1^{\,\prime})}\,
\frac{\delta \psi^{j_1^{\prime}}_{\beta_1^{\prime}}(q_2^{\,\prime})}
{\delta \psi^{(0)j}_{\beta}(q_2)}\,
\frac{\delta^2\bar{\psi}^{i_1^{\prime}}_{\alpha_1^{\prime}}(-q_1^{\,\prime})}
{\delta \delta\bar{\psi}^{(0)i}_{\alpha}(-q_1)\delta Q_0^b}
\,\,dq_1^{\,\prime}dq_2^{\,\prime}
\hspace{2.6cm}
\]
\[
\left.\left.
+\,\frac{\delta^3 j_{\Xi\mu}^{\,a}(k)}
{\delta \psi^{j_1^{\prime}}_{\beta_1^{\prime}}(q_2^{\,\prime})
\delta\bar{\psi}^{i_1^{\prime}}_{\alpha_1^{\prime}}(-q_1^{\,\prime})
\delta Q_0^b}\,
\frac{\delta \psi^{j_1^{\prime}}_{\beta_1^{\prime}}(q_2^{\,\prime})}
{\delta \psi^{(0)j}_{\beta}(q_2)}\,
\frac{\delta\bar{\psi}^{i_1^{\prime}}_{\alpha_1^{\prime}}(-q_1^{\,\prime})}
{\delta\bar{\psi}^{(0)i}_{\alpha}(-q_1)}\,\,
dq_1^{\,\prime}dq_2^{\,\prime}
\right\}\right|_{\,0},
\hspace{2.1cm}
\]
where on the right-hand side we keep the terms different from zero only.
Taking into account Eqs.\,(I.3.3), (\ref{eq:4q}), (\ref{eq:3r}), (\ref{eq:4w}),
(\ref{eq:5s}) and (\ref{eq:5l}) from the last expression we find the effective
current in the following form
\begin{equation}
\tilde{j}^{(2)a}_{\mu}(\bar{\psi}^{(0)},\psi^{(0)},Q_0)(k)=
\frac{\,g^3}{(2\pi)^3}\int\!\!
K^{(G)ab,\,ij}_{\mu,\,\alpha\beta}
({\bf v},\chi,\bar{\chi}|\,k;q_1,-q_2)
\,\bar{\psi}^{(0)i}_{\alpha}(-q_1)\psi^{(0)j}_{\beta}(q_2)
\label{eq:6q}
\end{equation}
\[
\hspace{2cm}
\times\,\delta(v\cdot(k+q_1-q_2))dq_1dq_2\,Q_0^b\,,
\]
where the coefficient function $K^{(G)ab,\,ij}_{\mu,\,\alpha\beta}$ is
\[
K^{(G)ab,\,ij}_{\mu,\,\alpha\beta}({\bf v},\chi,\bar{\chi}|\,k;q_1,-q_2)\equiv
-\,\delta{\Gamma}^{(G)ab,\,ij}_{\mu\nu,\,\alpha\beta}(k,-k-q_1+q_2;q_1,-q_2)
\,^\ast{\cal D}^{\nu\nu^{\prime}}\!(k+q_1-q_2)
v_{\nu^{\prime}}
\]
\begin{equation}
\hspace{1.9cm}
+\,[t^{a},t^{b}]^{ij}
K_{\mu\nu}({\bf v},{\bf v}|\,k,q_1-q_2)
\,^{\ast}{\cal D}^{\nu\nu^{\prime}}(-q_1+q_2)
\,^{\ast}\Gamma^{(G)}_{\nu^{\prime},\,\alpha\beta}(-q_1+q_2;q_1,-q_2)
\label{eq:6w}
\end{equation}
\[
-\,(t^{a}t^{b})^{ij}
\,^{\ast}\Gamma^{(G)}_{\mu,\,\alpha\gamma}(k;q_1,-k-q_1)
\,^{\ast}\!S_{\gamma\gamma^{\prime}}(k+q_1)
K_{\gamma^{\prime}\beta}^{(Q)}(\chi,\bar{\chi}|\,k+q_1,-q_2)
\hspace{0.45cm}
\]
\[
+\,(t^{b}t^{a})^{ij}
\bar{K}_{\alpha\gamma}^{(Q)}(\chi,\bar{\chi}|\,k-q_2,q_1)
\,^{\ast}\!S_{\gamma\gamma^{\prime}}(k-q_2)
\,^{\ast}\Gamma^{(G)}_{\mu,\,\gamma^{\prime}\beta}(k;-q_2,-k+q_2)
\hspace{0.3cm}
\]
\[
+\,\sigma v_{\mu}\,\{t^a,t^b\}^{ij}\,
\frac{\chi_{\alpha}\bar{\chi}_{\beta}}
{(v\cdot q_1)(v\cdot q_2)}\,.
\hspace{1cm}
\]
Here the function
\[
K_{\mu\nu}({\bf v},{\bf v}|\,k,q_1-q_2)\equiv
-\,\frac{v_{\mu}v_{\nu}}{v\cdot (q_1-q_2)} +
\!\,^\ast\Gamma_{\mu\nu\lambda}(k,q_1-q_2,-k-q_1+q_2)
\,^\ast\!{\cal D}^{\lambda\lambda^{\prime}}\!(k+q_1-q_2)
v_{\lambda^{\prime}}
\]
was introduced in Ref.\,\cite{markov_AOP_04}. It defines (on mass-shell
of soft modes) the amplitude for elastic scattering of
soft-gluon excitations off hard particle. The coefficient
functions
$K_{\gamma^{\prime}\beta}^{(Q)}(\chi,\bar{\chi}|\,k+q_1,-q_2)$ and
$\bar{K}_{\alpha\gamma}^{(Q)}(\chi,\bar{\chi}|\,k-q_2,q_1)$ in the
third and fourth lines are defined by equations (\ref{eq:5s}) and
(\ref{eq:5f}) correspondingly. The last term on the right-hand
side of Eq.\,(\ref{eq:6w}) is produced by additional current
(\ref{eq:5k}), and two the last but one terms contain the eikonal
contributions induced by additional source (\ref{eq:5a}) and its
conjugation. The effective current (\ref{eq:6q}) generates two processes of 
interaction of three soft plasma excitations with hard test particle: 
(i) the scattering process of soft-gluon excitation off hard parton with 
consequent soft-quark pair creation and (ii) the scattering process of 
soft-gluon and soft-quark excitations with consequent soft-quark radiation.
Diagrammatic interpretation of different terms in
coefficient function (\ref{eq:6w}) is shown in
Fig.\,\ref{fig4}. Here the scattering process of soft-gluon and 
soft-quark excitations only with consequent soft-quark radiation
is presented.

\begin{figure}[hbtp]
\begin{center}
\includegraphics[width=0.95\textwidth]{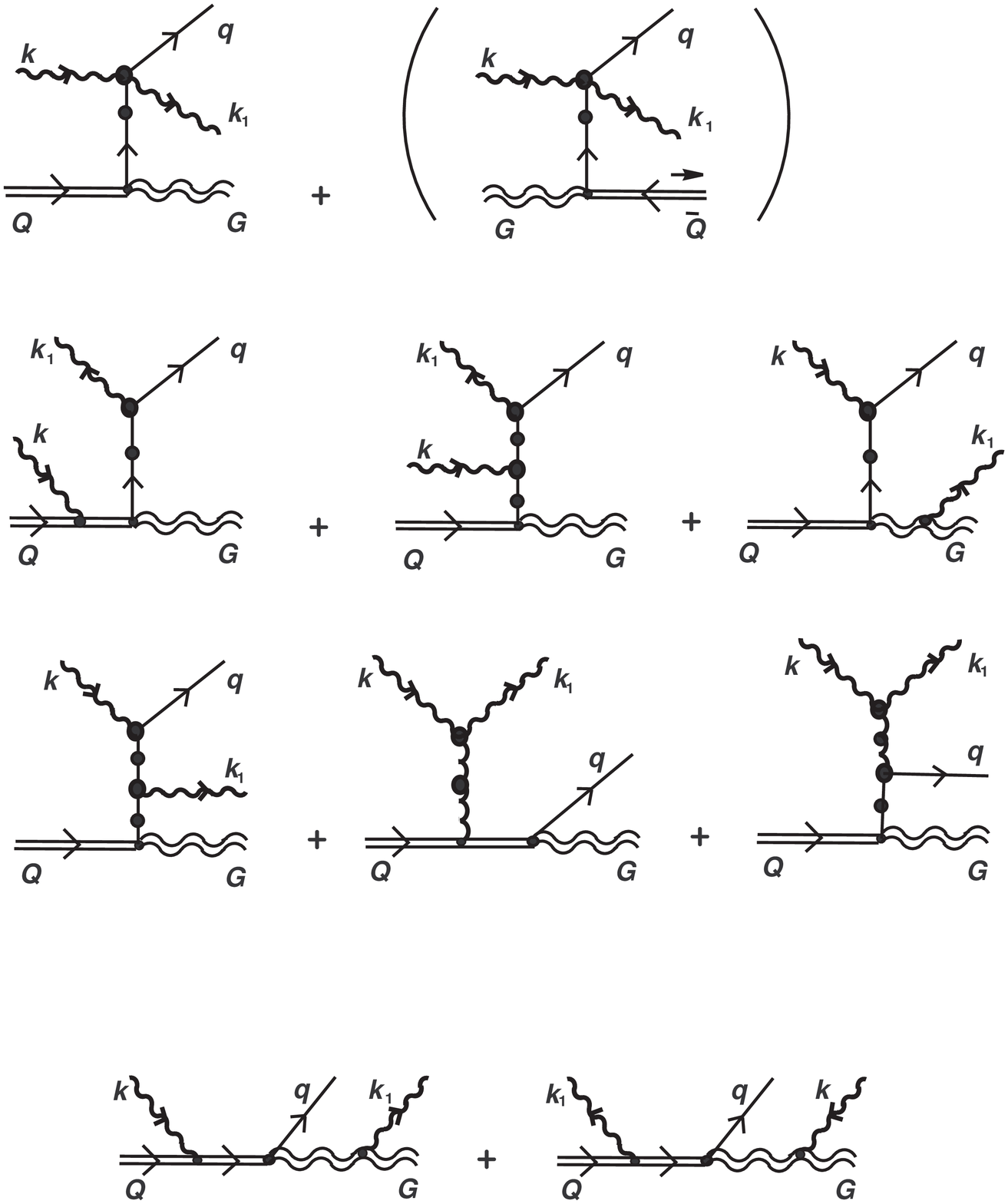}
\end{center}
\caption{\small The scattering process of soft-gluon excitation by
hard parton followed by soft-gluon and soft-quark radiation.}
\label{fig5}
\end{figure}

To define an effective current generating the scattering process with
participation of two soft-gluon excitations and one soft-quark
excitation, it should be considered a functional derivative of the
relation (\ref{eq:3f}) with respect to
$\theta_0^{\dagger}$, $A^{(0)}$, $\psi^{(0)}$ and $\theta_0$,
$A^{\ast(0)}$, $\bar{\psi}^{(0)}$ with the over-all current
$j^{a}_{\mu}[A,\bar{\psi},\psi,Q_0,\theta^{\dagger}_0,
\theta_0](k)$ on the left-hand side. Omitting the details of
calculations, we result at once in the final expression
(cp. (\ref{eq:5t}))
\begin{equation}
\tilde{j}^{(2)a}_{\mu}
(A^{(0)},\bar{\psi}^{(0)},\psi^{(0)},\theta_0^{\dagger},\theta)(k)
= \,\tilde{j}_{\mu}^{\,\dagger\,ai}(A^{(0)},\psi^{(0)})(-k)\,
\theta_0^i\,+\,
\theta_0^{\dagger i}\,\tilde{j}_{\mu}^{\,ai}(A^{(0)},\psi^{(0)})(k),
\label{eq:6e}
\end{equation}
where
\begin{equation}
\tilde{j}_{\mu}^{\,ai}(A^{(0)},\psi^{(0)})(k)\equiv
\frac{\,g^3}{(2\pi)^3}\!\int\!\!
K^{(G)aa_1,\,ij}_{\mu\mu_1,\,\alpha}
({\bf v},{\bf v},\bar{\chi}|\,k;-k_1,-q)
A^{(0)a_1\mu_1}(k_1)\psi^{(0)j}_{\alpha}(q)
\label{eq:6ee}
\end{equation}
\[
\times\,
\delta(v\cdot(k-k_1-q))\,dqdk_1,
\]
\[
\tilde{j}_{\mu}^{\,\dagger\,ai}(A^{(0)},\psi^{(0)})(-k)\equiv
\frac{\,g^3}{(2\pi)^3}\int\!\!
\bar{K}^{(G)aa_1,\,ji}_{\mu\mu_1,\,\alpha}
({\bf v},{\bf v},\chi|\,k;k_1,q)
A^{\ast(0)a_1\mu_1}(k_1)\bar{\psi}^{(0)j}_{\alpha}(-q)
\]
\[
\times\,
\delta(v\cdot(k+k_1+q))\,dqdk_1.
\]
Here the coefficient function $K^{(G)ab,\,ij}_{\mu\mu_1,\,\alpha}$ is defined
by the expression
\[
K^{(G)aa_1,\,ij}_{\mu\mu_1,\,\alpha}
({\bf v},{\bf v},\bar{\chi}|\,k;-k_1,-q)\equiv
-\,\bar{\chi}_{\beta^{\prime}}
\,^{\ast}\!S_{\beta^{\prime}\beta}(k-k_1-q)\,
\delta{\Gamma}^{(G)aa_1,\,ij}_{\mu\mu_1,\,\beta\alpha}(k,-k_1;-k+k_1+q,-q)
\]
\begin{equation}
\hspace{2.1cm}
-\,[t^{a},t^{a_1}]^{ij}
\!\,^\ast\Gamma_{\mu\nu\mu_1}(k,-k+k_1,-k_1)
\,^{\ast}{\cal D}^{\nu\nu^{\prime}}(k-k_1)
\bar{K}_{\nu^{\prime}\!,\,\alpha}^{(G)}({\bf v},\bar{\chi}|\,k-k_1,-q)
\label{eq:6r}
\end{equation}
\[
\hspace{0.2cm}
-\,(t^{a}t^{a_1})^{ij}
\bar{K}_{\mu,\,\beta^{\prime}}^{(G)}({\bf v},\bar{\chi}|\,k,-q-k_1)
\,^{\ast}\!S_{\beta^{\prime}\beta}(q+k_1)
\,^{\ast}\Gamma^{(Q)}_{\mu_1,\,\beta\alpha}(k_1;q,-q-k_1)
\]
\[
-\,(t^{a_1}t^{a})^{ij}
\bar{K}_{\mu_1\!,\,\beta^{\prime}}^{(Q)}({\bf v},\bar{\chi}|\,-k_1,k-q)
\,^{\ast}\!S_{\beta^{\prime}\beta}(k-q)
\,^{\ast}\Gamma^{(G)}_{\mu,\,\beta\alpha}(k;q-k,-q)
\]
\[
+\,(t^{a}t^{a_1})^{ij}\,
\frac{v_{\mu}v_{\mu_1}\bar{\chi}_{\alpha}}
{(v\cdot q)(v\cdot k)}\,\,
-\,(t^{a_1}t^{a})^{ij}\,
\frac{v_{\mu}v_{\mu_1}\bar{\chi}_{\alpha}}{(v\cdot q)(v\cdot k_1)}\,\,.
\hspace{0.3cm}
\]
The second, fourth and fifth terms contain eikonal contributions
generated by additional current (\ref{eq:5q}). Diagrammatic
interpretation of different terms in the coefficient function
(\ref{eq:6r}) are presented in Fig.\ref{fig5}.
These diagrams should be supplemented with graphs describing
interaction of soft modes with hard test antiquark $\bar{Q}$. The
example of such diagrams is shown on the first line in
parentheses.

\begin{figure}[hbtp]
\begin{center}
\includegraphics[width=0.95\textwidth]{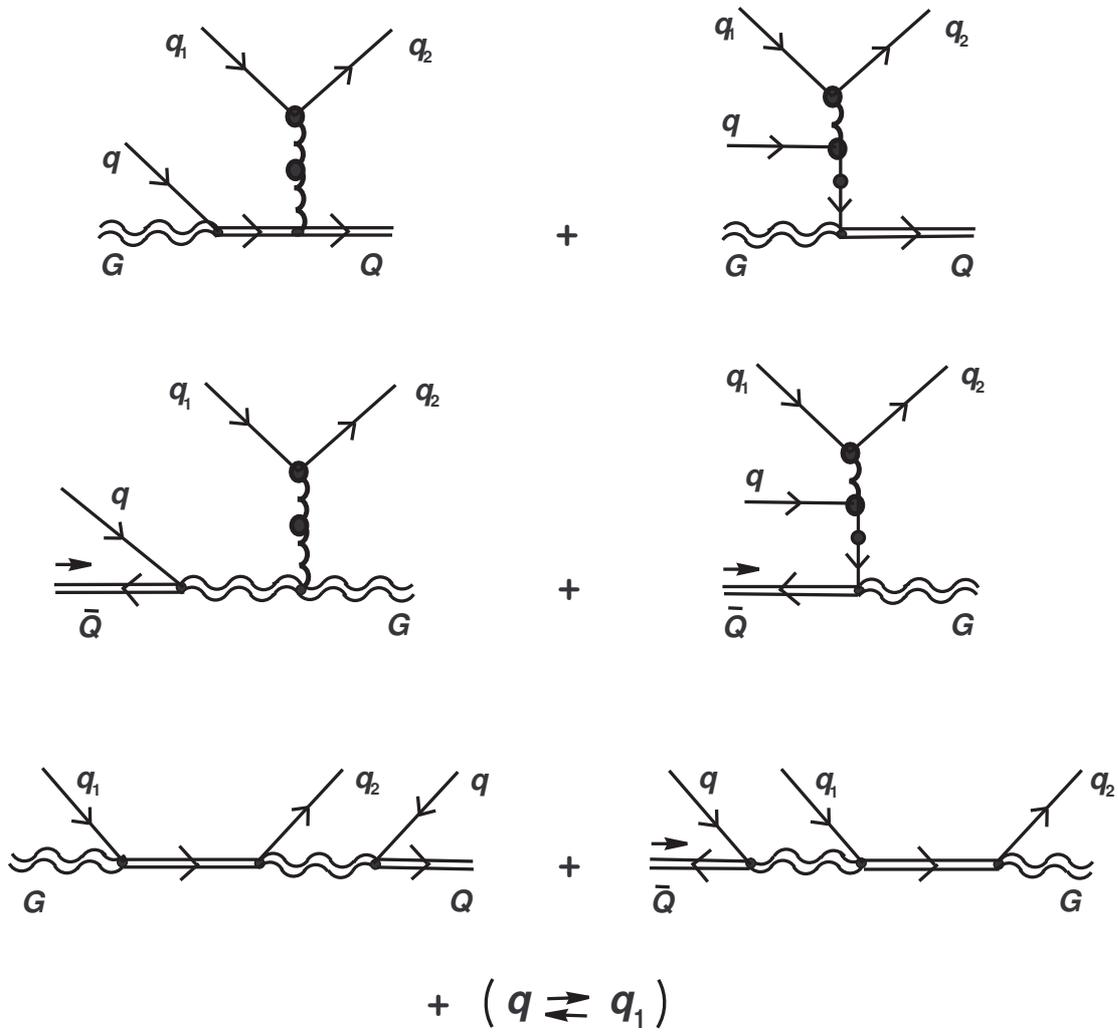}
\end{center}
\caption{\small The processes of interaction of three soft-quark
elementary excitations with hard thermal parton.}
\label{fig6}
\end{figure}

Now we consider a third order functional derivatives of relation
(\ref{eq:3g}). By the source on the left-hand side it is necessary
to mean a sum of initial source (the right-hand side of
Eq.\,(\ref{eq:3p})) and additional sources (\ref{eq:5a}),
(\ref{eq:5g}), and (\ref{eq:5h}):
\[
\eta^{\,i}_{\alpha}[A,\bar{\psi},\psi,Q_0,\theta_0](q)\equiv
\eta^{\,i}_{\alpha}[A,\psi,Q_0](q) +
\eta^{\,i}_{Q\alpha}[A,\psi,\theta_0](q) 
\]
\[
+\,\eta^{\,i}_{\Xi\alpha}[A,\bar{\psi},\psi,\theta_0](q) +
\eta^{\,i}_{\Omega\alpha}[A,\bar{\psi},\psi,\theta_0](q).
\]
Differentiation of (\ref{eq:3g}) with respect to $\psi^{(0)}$,
$\bar{\psi}^{(0)}$ and $\theta_0^j$ yields
\[
\left.\frac{\delta^3\eta^i_{\alpha}[A,\bar{\psi},\psi,Q_0,\theta_0](q)}
{\delta \psi^{(0)i_2}_{\alpha_2}(q_2)\,
\delta \bar{\psi}^{(0)i_1}_{\alpha_1}(-q_1)\,
\delta \theta_0^j}\,
\right|_{\,0}=
\left.\frac{\delta^3\tilde{\eta}^i_{\alpha}
[A^{(0)},\bar{\psi}^{(0)},\psi^{(0)},Q_0,\theta_0](q)}
{\delta \psi^{(0)i_2}_{\alpha_2}(q_2)\,
\delta \bar{\psi}^{(0)i_1}_{\alpha_1}(-q_1)
\delta \theta_0^j}\,
\right|_{\,0}
\]
\[
\hspace{2.5cm}
=\!\int\Biggl\{
\frac{\delta^2 \eta_{\alpha}^{(1,1)i}(q)}
{\delta A^{a_1^{\prime}\mu_1^{\prime}}(k_1^{\,\prime})
\delta \psi^{i_1^{\prime}}_{\alpha_1^{\prime}}(q_1^{\,\prime})}\,
\frac{\delta \psi^{i_1^{\prime}}_{\alpha_1^{\prime}}(q_1^{\,\prime})}
{\delta \theta_0^j}\,
\frac{\delta^2A^{a_1^{\prime}\mu_1^{\prime}}(k_1^{\,\prime})}
{\delta \psi^{(0)i_2}_{\alpha_2}(q_2)
\delta\bar{\psi}^{(0)i_1}_{\alpha_1}(-q_1)}\,
\,dk_1^{\,\prime}dq_1^{\,\prime}
\hspace{3cm}
\]
\[
\hspace{0.5cm}
-\,\frac{\delta^2\eta_{\alpha}^{(1,1)i}(q)}
{\delta A^{a_1^{\prime}\mu_1^{\prime}}(k_1^{\,\prime})
\delta \psi^{i_1^{\prime}}_{\alpha_1^{\prime}}(q_1^{\,\prime})}\,
\frac{\delta^2A^{a_1^{\prime}\mu_1^{\prime}}(k_1^{\,\prime})}
{\delta \theta_0^j\,
\delta\bar{\psi}^{(0)i_1}_{\alpha_1}(-q_1)}\,
\frac{\delta \psi^{i_1^{\prime}}_{\alpha_1^{\prime}}(q_1^{\,\prime})}
{\delta \psi^{(0)i_2}_{\alpha_2}(q_2)}\,
\,\,dk_1^{\,\prime}dq_1^{\,\prime}
\]
\[
+\,\frac{\delta^2\eta_{\theta\alpha}^{(1)i}(q)}
{\delta \theta_0^j\,\delta A^{a_1^{\prime}\mu_1^{\prime}}(k_1^{\,\prime})}\,
\frac{\delta^2A^{a_1^{\prime}\mu_1^{\prime}}(k_1^{\,\prime})}
{\delta \psi^{(0)i_2}_{\alpha_2}(q_2)
\delta\bar{\psi}^{(0)i_1}_{\alpha_1}(-q_1)}\,\,dk_1^{\,\prime}
\hspace{1.8cm}
\]
\[
\hspace{1.1cm}
+\,
\frac{\delta^2\Bigl(\eta_{\Xi\alpha}^{i}(q)+\eta^{\,i}_{\Omega\alpha}(q)\Bigr)}
{\delta\theta_0^j\,
\delta\psi^{i_2^{\prime}}_{\alpha_2^{\prime}}(q_2^{\,\prime})
\delta\bar{\psi}^{i_1^{\prime}}_{\alpha_1^{\prime}}(-q_1^{\,\prime})}\,
\frac{\psi^{i_2^{\prime}}_{\alpha_2^{\prime}}(q_2^{\,\prime})}
{\delta \psi^{(0)i_2}_{\alpha_2}(q_2)}\,
\frac{\bar{\psi}^{i_1^{\prime}}_{\alpha_1^{\prime}}(-q_1^{\,\prime})}
{\delta \bar{\psi}^{(0)i_1}_{\alpha_1}(-q_1)}\,\,
dq_1^{\,\prime}dq_2^{\,\prime}
\Biggl\}\Bigg|_{\,0\,.}
\]
We kept again only the terms different from zero on the right-hand side.
Taking into account Eqs.\,(I.3.4), (I.3.3), (\ref{eq:5e}), (\ref{eq:3a})
and (\ref{eq:5j}), we easily derive an explicit form
of effective source generating the process of nonlinear interaction of
three soft-quark excitations with the hard test particle
\begin{equation}
\tilde{\eta}^{(2)i}_{\alpha}
(\bar{\psi}^{(0)},\psi^{(0)},\theta_0)(q)=
\frac{\,g^3}{(2\pi)^3}\int\!\!
K^{(Q)ii_1i_2j}_{\alpha\alpha_1\alpha_2}(\chi,\chi,\bar{\chi}|\,q,q_1;-q_2)\,
\bar{\psi}^{(0)i_1}_{\alpha_1}(-q_1)\psi^{(0)i_2}_{\alpha_2}(q_2)
\label{eq:6t}
\end{equation}
\[
\hspace{2cm}\times\,
\delta(v\cdot(q+q_1-q_2))\,dq_1dq_2\,\theta_0^j\,,
\]
where the coefficient function $K^{(Q)ii_1i_2j}_{\alpha\alpha_1\alpha_2}$ is
\begin{equation}
K^{(Q)ii_1i_2j}_{\alpha\alpha_1\alpha_2}(\chi,\chi,\bar{\chi}|\,q,q_1;-q_2)
\label{eq:6y}
\end{equation}
\[
\equiv
(t^a)^{ii_2}(t^a)^{i_1j}
\,^{\ast}\Gamma^{(Q)\mu}_{\alpha\alpha_2}(q-q_2;q_2,-q)
\,^{\ast}{\cal D}_{\mu\nu}(q-q_2)
K^{(G)\nu}_{\alpha_1}({\bf v},\chi|\,q-q_2,q_1)
\]
\[
\hspace{1.8cm}
-\,(t^a)^{ij}(t^a)^{i_1i_2}
K^{(Q)\mu}_{\alpha}({\bf v},\chi|\,-q_1+q_2,-q)
\,^{\ast}{\cal D}_{\mu\nu}(-q_1+q_2)
\,^{\ast}\Gamma^{(G)\nu}_{\alpha_1\alpha_2}(-q_1+q_2;q_1,-q_2)
\]
\[
+\,\beta
\,(t^{a})^{ij}(t^{a})^{i_1i_2}\,
\frac{\chi_{\alpha}\chi_{\alpha_1}\bar{\chi}_{\alpha_2}}
{(v\cdot q_1)(v\cdot q_2)}\,
+\,\beta_1
\,(t^{a})^{ii_2}(t^{a})^{i_1j}\,
\frac{\chi_{\alpha}\chi_{\alpha_1}\bar{\chi}_{\alpha_2}}
{(v\cdot q_1)(v\cdot q_2)}\,.
\]
Diagrammatic interpretation of different terms in coefficient function
(\ref{eq:6y}) is presented in Fig.\,\ref{fig6}.

Furthermore, we can define just one more third-order effective source
$\tilde{\eta}^{(2)i}_{\alpha}$ taking variation of relation
(\ref{eq:3g}) with respect to $A^{(0)}$, $\psi^{(0)}$ and $Q_0$.
Omitting calculation details, we give at once final result:
\begin{equation}
\tilde{\eta}^{(2)i}_{\alpha}
(A^{(0)},\psi^{(0)},Q_0)(q)=
\frac{\,g^3}{(2\pi)^3}\int\!\!
K^{(Q)ab,\,ii_1}_{\mu,\,\alpha\alpha_1}({\bf v},\bar{\chi},\chi|\,q;-q_1,-k)\,
A^{(0)a\mu}(k)\psi^{(0)i_1}_{\alpha_1}(q_1)
\label{eq:6u}
\end{equation}
\[
\hspace{2cm}\times\,
\delta(v\cdot(q-q_1-k))\,dq_1dk\,Q_0^b\,,
\]
where the coefficient function $K^{(Q)ab,\,ii_1}_{\mu,\,\alpha\alpha_1}$ is
defined by the following expression:
\[
K^{(Q)ab,\,ii_1}_{\mu,\,\alpha\alpha_1}({\bf v},\bar{\chi},\chi|\,q;-q_1,-k)
\equiv 
-\,\delta{\Gamma}^{(Q)ba,\,ii_1}_{\nu\mu,\,\alpha\alpha_1}
(q-q_1-k,k;q_1,-q)
\,^\ast{\cal D}^{\nu\nu^{\prime}}\!(q-q_1-k)
v_{\nu^{\prime}}
\]
\begin{equation}
\hspace{2.2cm}
+\,[t^{b},t^{a}]^{ii_1}
\,^{\ast}\Gamma^{(Q)}_{\nu^{\prime}\!,\,\alpha\alpha_1}(q-q_1;q_1,-q)
\,^{\ast}{\cal D}^{\nu^{\prime}\nu}(q-q_1)\,
K_{\nu\mu}({\bf v},{\bf v}|\,q-q_1,-k)
\label{eq:6i}
\end{equation}
\[
\hspace{0.45cm}
+\,(t^{b}t^{a})^{ii_1}
K_{\alpha\beta}^{(Q)}(\chi,\bar{\chi}|\,q,-q_1-k)
\,^{\ast}\!S_{\beta\beta^{\prime}}(k+q_1)
\,^{\ast}\Gamma^{(Q)}_{\mu,\,\beta^{\prime}\alpha_1}(k;q_1,-q_1-k)
\]
\[
-\,(t^{a}t^{b})^{ii_1}
\,^{\ast}\Gamma^{(Q)}_{\mu,\,\alpha\beta}(k;q-k,-q)
\,^{\ast}\!S_{\beta\beta^{\prime}}(q-k)\,
K_{\beta^{\prime}\alpha_1}^{(Q)}(\chi,\bar{\chi}|\,q-k,-q_1)
\hspace{0.1cm}
\]
\[
-\,\alpha \,
(t^bt^a)^{ij}\,
\frac{v_{\mu}\chi_{\alpha}\bar{\chi}_{\alpha_1}}
{(v\cdot q)(v\cdot k)}\,+\,\alpha\,
(t^at^b)^{ij}\,
\frac{v_{\mu}\chi_{\alpha}\bar{\chi}_{\alpha_1}}
{(v\cdot q_1)(v\cdot k)}\,.
\]
The effective source (\ref{eq:6u}) generates the scattering processes,
which are reverse to the scattering processes depicted in
Fig.\,\ref{fig4}.

Finally, we give an explicit expression for the effective source
$\tilde{\eta}^{(2)i}_{\alpha}$ arising in calculating the third
order derivative of relation (\ref{eq:3g}) with respect to
$A^{(0)}(k_1)$, $A^{(0)}(k_2)$ and $\theta_0$. The calculations
similar to previous ones result in the following expression for
$\tilde{\eta}^{(2)i}_{\alpha}$:
\[
\tilde{\eta}^{(2)i}_{\alpha}
(A^{(0)},A^{(0)},\theta_0)(q)=
\frac{1}{2!}\,
\frac{\,g^3}{(2\pi)^3}\int\!\!
K^{(Q)a_1a_2,\,ij}_{\mu_1\mu_2,\,\alpha}
({\bf v},{\bf v},\chi|\,q;-k_1,-k_2)\,
A^{(0)a_1\mu_1}(k_1)A^{(0)a_2\mu_2}(k_2)
\]
\begin{equation}
\hspace{2.5cm}\times\,
\delta(v\cdot(q-k_1-k_2))\,dk_1dk_2\,\theta_0^j\,.
\label{eq:6o}
\end{equation}
The explicit form of the coefficient function
$K^{(Q)a_1a_2,\,ij}_{\mu_1\mu_2,\,\alpha}$ and also diagrammatic 
interpretation of different terms are given in Appendix B. The effective source 
(\ref{eq:6o}) closes a set of the effective currents and sources determining 
nonlinear interaction processes of three soft elementary excitations  
with hard test particle in the linear approximation in initial values of the
usual and Grassmann color charges.

\section{\bf Soft-loop corrections}
\setcounter{equation}{0}

The effective currents and sources written out in previous
sections define the scattering processes of soft modes by hard
thermal particles in tree approximation. However, as we have shown
in the case of purely soft-gluon excitations \cite{markov_AOP_04}, there
exists an infinite number of effective currents, in which coefficient
functions define so-called `classical' soft-gluon one-loop
corrections to the tree scattering amplitudes. Although these
effective currents are suppressed by powers of the coupling constant in
comparison with tree-level effective currents, their accounting in
some cases is rather important. Thus, for example, in our work
\cite{markov_AOP_05} it was shown that soft one-loop corrections to soft-gluon
bremsstrahlung generate `off-diagonal' contributions 
to the radiation energy loss of fast parton connected with the
coherent double gluon exchanges. These contributions within the
frameworks of a light-cone path integral approach were first considered
by Zakharov \cite{zakharov} to ensure unitarity. The presence
of soft-quark excitations in the medium results in a new feature:
appearing `classical' soft-quark loops. As well as in the case of
purely soft-gluon excitations \cite{markov_AOP_04} effective currents
and sources containing soft-quark loops arise in calculating
derivatives of higher order with respect to color charges $Q_0$,
$\theta_0^{\dagger}$ and $\theta_0$ of the basic relations
(\ref{eq:3f})\,--\,(\ref{eq:3h}). Below we will give some examples of
concrete calculations and diagrammatic interpretation of the
results.

\begin{figure}[hbtp]
\begin{center}
\includegraphics[width=0.95\textwidth]{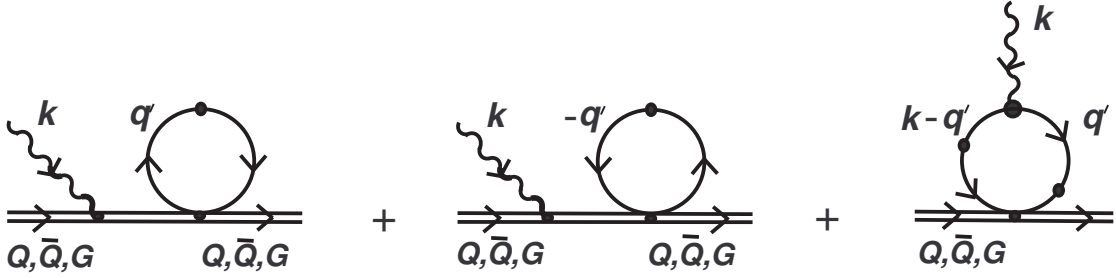}
\end{center}
\caption{\small The one-loop corrections to the initial current
$j_Q^{(0)a\mu}(k)=g/(2\pi)^3\,v^{\mu}Q^a_0\delta(v\cdot k)$.}
\label{fig7}
\end{figure}

The first nontrivial example of this type arises in calculating the
derivative of relation (\ref{eq:3f}) with respect to the Grassmann
charges $\theta_0^{\dagger}$ and $\theta_0$ (with the over-all current on the
left-hand side)
\[
\left.\frac{\delta^{2}\!j^{a}_{\mu}[A,\bar{\psi},\psi,Q_0,\theta^{\dagger}_0,
\theta_0](k)}
{\delta \theta_0^{\dagger\,i}\, \delta \theta_0^{j}}\,
\right|_{\,0}
=
\left.\frac{\delta^2 \tilde{j}^{a}_{\mu}
[A^{(0)},\bar{\psi}^{(0)},\psi^{(0)},Q_0,\theta_0^{\dagger},\theta_0,\,](k)}
{\delta \theta_0^{\dagger\,i}\, \delta \theta_0^{j}}\,
\right|_{\,0}
\]
\[
=
\left.\left(\,\frac{\delta^{2}j^{(1)a}_{\theta\mu}(\bar{\psi},\psi)(k)}
{\delta \theta_0^{\dagger\,i}\, \delta \theta_0^{j}}
+\,\frac{\delta^2 j^{\Psi(0,2)a}_{\mu}(\bar{\psi},\psi)(k)}
{\delta \theta_0^{\dagger\,i}\, \delta \theta_0^{j}}\,
\right)\right|_{\,0}\,,
\]
where in the last line we keep the terms different from zero only.
Taking into account Eqs.\,(\ref{eq:5ww}), (\ref{eq:4t}) and (I.3.3),
it is not difficult to show that the right-hand side of this equation can be
resulted in the following form (for the sake of simplicity here we suppress of
the summation over spinor indices):
\begin{equation}
\frac{\,g^3}{(2\pi)^3}\,\,(t^a)^{ij}\!
\int\Biggl\{\frac{v_{\mu}}{(v\cdot q^{\,\prime\,})}\,
\Bigl[\,\Bigl(\bar{\chi}\,^{\ast}\!S(q^{\,\prime\,})\chi\Bigr)
-\Bigl(\bar{\chi}\,^{\ast}\!S(-q^{\,\prime\,})\chi\Bigr)\Bigr]
\hspace{2.5cm}
\label{eq:7q}
\end{equation}
\[
\hspace{4.3cm}
+\,\Bigl(
\bar{\chi}\,^{\ast}\!S(k-q^{\,\prime\,})
\,^{\ast}\Gamma^{(G)}_{\mu}(k;-k+q^{\,\prime},-q^{\,\prime\,})
\,^{\ast}\!S(q^{\,\prime\,})\chi\Bigr)\!\Biggr\}
\,\delta(v\cdot q^{\,\prime\,})\,dq^{\,\prime}
\,\delta(v\cdot k).
\]
From this point on we will designate virtual momenta by letter with prime.
The diagrammatic interpretation of different terms is presented in
Fig.\,\ref{fig7}. The region of integration in loops is restricted by
the Cherenkov cone $v\cdot q^{\,\prime}=0$. Moreover, by virtue of
condition $v\cdot k=0$ (the last $\delta$-function in
Eq.\,(\ref{eq:7q})) the derivative calculated is not equal to zero
only for off mass-shell plasma excitations. We note especially
that it is impossible to obtain Eq.\,(\ref{eq:7q}) literally from
the right-hand side of Eq.\,(\ref{eq:5t}) by simple replacements
\begin{equation}
\bar{\psi}^{(0)i}_{\alpha}(-q)\rightarrow
\frac{\,g}{(2\pi)^3}\,\,\theta_0^{\dagger\,i}\,\bar{\chi}_{\beta}
\,^{\ast}\!S_{\beta\alpha}(-q)
\,\delta(v\cdot q),
\label{eq:7w}
\end{equation}
\[
\psi^{(0)j}_{\alpha}(q)\rightarrow
-\frac{\,g}{(2\pi)^3}\,
\,^{\ast}\!S_{\alpha\beta}(q)\chi_{\beta}
\,\delta(v\cdot q)\,\theta_0^{j}\,,
\]
since in this case we obtain superfluous term with vertex
$\,^{\ast}\Gamma^{(G)}_{\mu}(k;-k+q^{\,\prime},-q^{\,\prime})$ in comparison 
with (\ref{eq:7q}). Finally, the integrand in (\ref{eq:7q}) contains a 
singularity $\delta(v\cdot q^{\,\prime})/(v\cdot q^{\,\prime})$ generated by 
the eikonal term. Brief analysis of this singularity
is given in Appendix C. We only point to the fact that an origin  of this
singularity is the assumption of linearity of a hard parton
trajectory used by us during all this work. Note that an expression similar to
(\ref{eq:7q}) exists in purely gluonic case \cite{markov_AOP_04}. Here,
instead of the HTL-resummed quark propagators and vertex
$^{\ast}\Gamma_{\mu}^{(G)}$
we have correspondingly the gluon propagators and triple gluon vertex
$^{\ast}\Gamma_{3{\rm g}}$. However, this expression within the framework of
HTL-approximation vanishes by virtue of color factor: $f^{abc} Q_0^b Q_0^c=0$.

\begin{figure}[hbtp]
\begin{center}
\includegraphics*[scale=0.46]{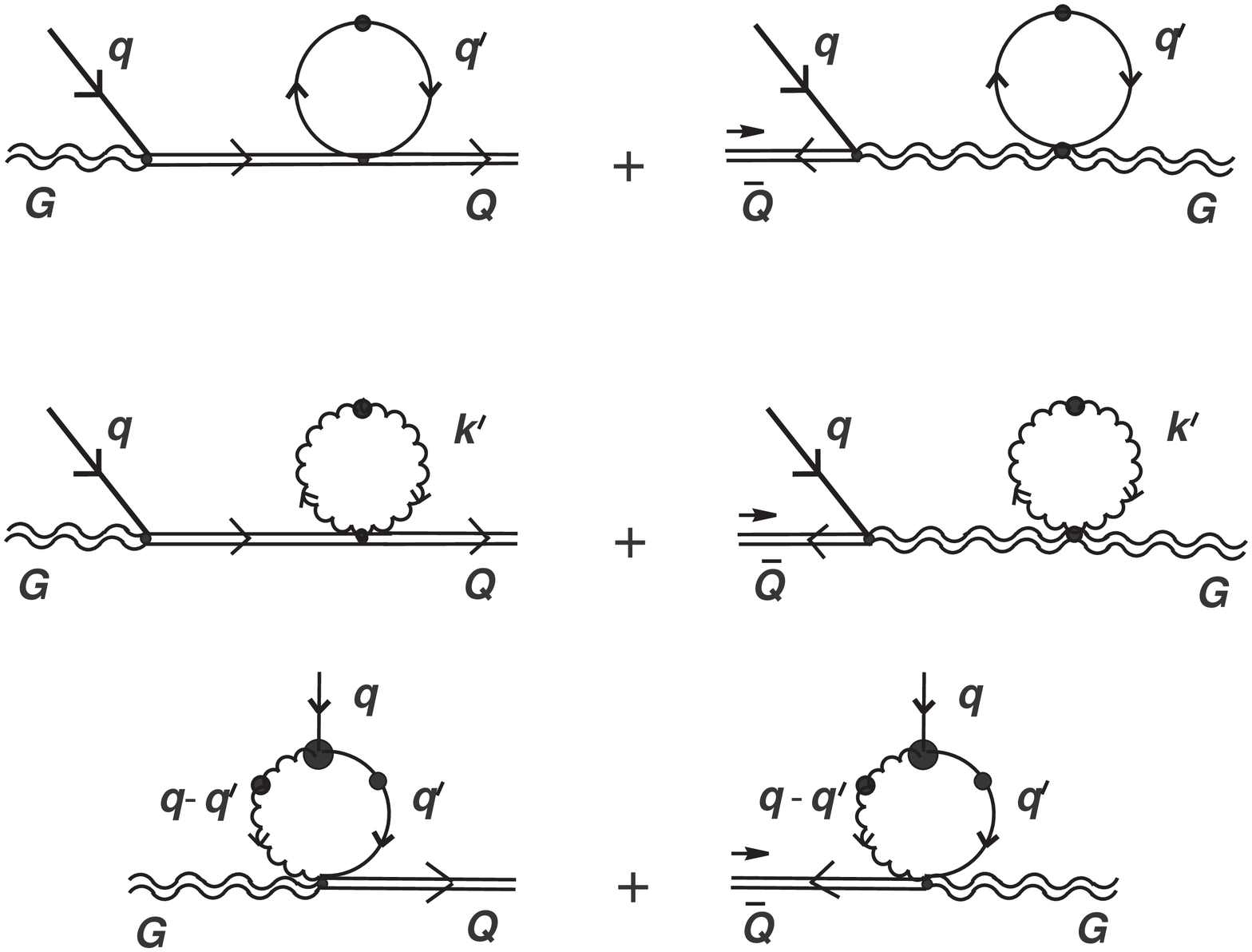}
\end{center}
\caption{\small The one-loop corrections to the initial source
$\eta^{(0)i}_{\theta\,\alpha}(q)
=g/(2\pi)^3\,\theta_0^i\chi_{\alpha}\delta(v\cdot q)$.}
\label{fig8}
\end{figure}

Another nontrivial example of this kind arises in calculating the
derivative of relation (\ref{eq:3g}) with respect to the usual $Q_0$ and
Grassmann $\theta_0$ charges with the total source on the left-hand side
\[
\left.\frac{\delta^2\eta^i_{\alpha}[A,\bar{\psi},\psi,Q_0,\theta_0](q)}
{\delta Q_0^a\,\delta\theta_0^j}\,
\right|_{\,0}=
\left.\frac{\delta^2\tilde{\eta}^i_{\alpha}
[A^{(0)},\bar{\psi}^{(0)},\psi^{(0)},Q_0,\theta_0](q)}
{\delta Q_0^a\,\delta\theta_0^j}\,
\right|_{\,0}
\]
\[
=\!\left.\left(
\frac{\delta^2\eta^{(1)i}_{Q\alpha}(\psi)(q)}
{\delta Q_0^a\,\delta\theta_0^j}\,+\,
\frac{\delta^2\eta^{(1)i}_{\theta\alpha}(A)(q)}
{\delta Q_0^a\,\delta\theta_0^j}\,+\,
\frac{\delta^2\eta^{(1,1)i}_{\alpha}(A,\psi)(q)}
{\delta Q_0^a\,\delta\theta_0^j}
\right)\!\right|_{\,0}.
\]
Taking into account Eqs.\,(\ref{eq:5a}), (\ref{eq:3a}), (\ref{eq:4q}), and
(\ref{eq:4t}) we lead the right-hand side of the last equation to the
following form:
\begin{equation}
-\,\frac{\,g^3}{(2\pi)^3}\,\,(t^a)^{ij}\Biggl\{
\alpha\!\int\!\frac{\chi_{\alpha}}{(v\cdot q^{\,\prime\,})}\,
\Bigl(\bar{\chi}_{\beta}\,^{\ast}\!S_{\beta\beta^{\prime}}
(q^{\,\prime\,})\chi_{\beta^{\prime}}\Bigr)
\,\delta(v\cdot q^{\,\prime\,})\,dq^{\,\prime}
\hspace{2.7cm}
\label{eq:7e}
\end{equation}
\[
+\int\!\frac{\chi_{\alpha}}{(v\cdot k^{\,\prime\,})}\,
\Bigl(v^{\mu}\,^\ast{\cal D}_{\mu\mu^{\prime}}\!(k^{\,\prime\,})
v^{\mu^{\prime}}\Bigr)
\,\delta(v\cdot k^{\,\prime\,})\,dk^{\,\prime}
\]
\[
\hspace{2.8cm}
-\int\Bigl[\,^{\ast}\Gamma^{(Q)\mu}_{\alpha\beta}
(q-q^{\,\prime};q^{\,\prime},-q)
\,^{\ast}\!S_{\beta\beta^{\prime}}(q^{\,\prime\,})\chi_{\beta^{\prime}}\Bigr]
\,^\ast{\cal D}_{\mu\mu^{\prime}}(q-q^{\,\prime\,})
v^{\mu^{\prime}}
\,\delta(v\cdot q^{\,\prime\,})\,dq^{\,\prime}
\!\Biggr\}\,
\delta(v\cdot q).
\]
Diagrammatic interpretation of the different terms in braces is presented in
figure \ref{fig8}.
The expression (\ref{eq:7e}) also cannot be obtained by simply replacements
(\ref{eq:7w}) and
\[
A_{\mu}^{(0)a}(k)\rightarrow
-\,\frac{\,g}{(2\pi)^3}\,
\,^\ast{\cal D}_{\mu\nu}(k)v^{\nu}\delta(v\cdot k)\,Q_0^a
\]
from a sum of effective sources (\ref{eq:4e}) and (\ref{eq:4u}). For such a
replacement a superfluous contribution with the HTL-induced vertex
$\,^{\ast}\Gamma^{(Q)\mu}_{\alpha\beta}(q-q^{\,\prime};q^{\,\prime},-q)$
arises.

\begin{figure}[hbtp]
\begin{center}
\includegraphics*[scale=0.4]{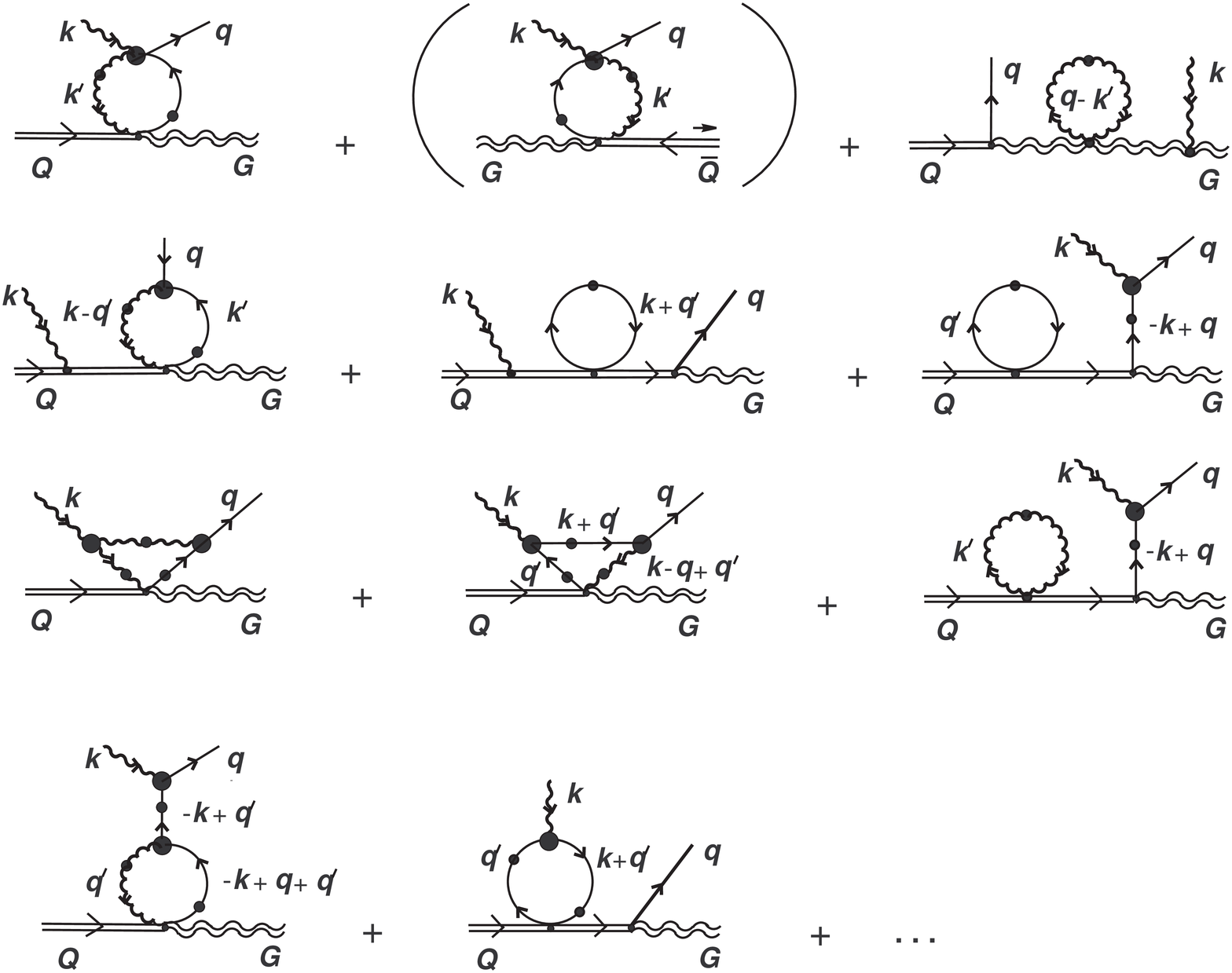}
\end{center}
\caption{\small The one-loop corrections to the scattering process
generated by the effective current (\ref{eq:5t}).}
\label{fig9}
\end{figure}

Physically, more thoughtful examples arise in calculating higher
order derivatives of relations (\ref{eq:3f})\,--\,(\ref{eq:3h}) that contain 
at least one soft free field. We consider at first third order
derivative of relation (\ref{eq:3f}) with respect to the usual
and Grassmann charges $Q_0$, $\theta_0^{\dagger}$ and soft-quark
field $\psi^{(0)}$. Somewhat cumbersome calculations result in
\begin{equation}
\left.\frac{\delta^{3}\!j^{a}_{\mu}[A,\bar{\psi},\psi,Q_0,\theta^{\dagger}_0,
\theta_0](k)}
{\delta Q_0^b\,\delta\theta_0^{\dagger\,i}
\,\delta\psi^{(0)j}_{\alpha}(q)}\,
\right|_{\,0}
=
\left.\frac{\delta^3 \tilde{j}^{a}_{\mu}
[A^{(0)},\bar{\psi}^{(0)},\psi^{(0)},Q_0,\theta_0^{\dagger},\theta_0](k)}
{\delta Q_0^b\,\delta\theta_0^{\dagger\,i}
\,\delta\psi^{(0)j}_{\alpha}(q)}\,
\right|_{\,0}
\label{eq:7r}
\end{equation}
\[
=\frac{\,g^4}{(2\pi)^6}\,
\Biggl\{\int\!\Bigl[\,\bar{\chi}_{\beta^{\prime}}\,^{\ast}
\!S_{\beta^{\prime}\beta}(k-k^{\,\prime}-q)
\delta{\Gamma}^{(G)ab,\,ij}_{\mu\nu,\,\beta\alpha}
(k,-k^{\,\prime};-k+k^{\,\prime}+q,-q)
\,^\ast{\cal D}^{\nu\nu^{\prime}}\!(k^{\,\prime\,})
v_{\nu^{\prime}}\Bigr]
\delta(v\cdot k^{\,\prime\,})dk^{\,\prime}
\]
\[
\hspace{0.3cm}
-\,[t^{a},t^{b}]^{ij}\!\int\!
K_{\mu\nu}({\bf v},{\bf v}|\,k,-q+k^{\,\prime\,})
\,^{\ast}{\cal D}^{\nu\nu^{\prime}}\!(q-k^{\,\prime\,})
\bar{K}^{(G)}_{\nu^{\prime}\!,\,\alpha}
({\bf v},\bar{\chi}|\,-k^{\,\prime}+q,-q)
\,\delta(v\cdot k^{\,\prime\,})dk^{\,\prime}
\]
\[
\hspace{0.1cm}
+\,(t^{a}t^{b})^{ij}\Biggl[\,\int\!
\bar{K}^{(G)}_{\mu,\,\beta^{\prime}}
({\bf v},\bar{\chi}|\,k,-k-q^{\,\prime\,})
\,^{\ast}\!S_{\beta^{\prime}\beta}(k+q^{\,\prime\,})
K_{\beta\alpha}^{(Q)}(\chi,\bar{\chi}|\,k+q^{\,\prime},-q)
\,\delta(v\cdot q^{\,\prime\,})dq^{\,\prime}
\]
\[
\hspace{2cm}
+\,\frac{v_{\mu}\,\bar{\chi}_{\alpha}}{(v\cdot q)(v\cdot k)}\int
\Bigl(v^{\nu}\,^\ast{\cal D}_{\nu\nu^{\prime}}(k^{\,\prime\,})
v^{\nu^{\prime}}\Bigr)
\,\delta(v\cdot k^{\,\prime\,})\,dk^{\,\prime}\Biggr]
\]
\[
+\,(t^{b}t^{a})^{ij}\Biggl[\,\int
\biggl\{
\bar{K}^{(Q)\nu}_{\beta^{\prime}}
({\bf v},\bar{\chi}|\,q-k-q^{\,\prime},-q+k)
\,^\ast{\cal D}_{\nu\nu^{\prime}}(-q+k+q^{\,\prime\,})
v^{\nu^{\prime}}
\hspace{2.1cm}
\]
\[
\hspace{2cm}
-\,\alpha^{\ast}\,
\frac{\bar{\chi}_{\beta^{\prime}}}{(v\cdot q^{\,\prime\,})}\,
\Bigl(\bar{\chi}_{\gamma}\,^{\ast}\!S_{\gamma\gamma^{\prime}}
(-q^{\,\prime\,})\chi_{\gamma^{\prime}}\Bigr)
\biggr\}
\,\delta(v\cdot q^{\,\prime\,})\,dq^{\,\prime}\Biggr]
\,^{\ast}\!S_{\beta^{\prime}\beta}(k-q)
\,^{\ast}\Gamma^{(G)}_{\mu,\,\beta\alpha}(k;q-k,-q)
\]
\[
+\,\sigma \,\{t^a,t^b\}^{ij}\,
\frac{v_{\mu}\,\chi_{\alpha}}
{(v\cdot q)(v\cdot (k-q))}\,
\int\Bigl(\bar{\chi}_{\beta}\,^{\ast}\!S_{\beta\beta^{\prime}}
(-q^{\,\prime\,})\chi_{\beta^{\prime}}\Bigr)
\,\delta(v\cdot q^{\,\prime\,})\,dq^{\,\prime}\Biggr\}
\,\delta(v\cdot (k-q)).
\hspace{1cm}
\]
Diagrammatic interpretation of the different terms on the right-hand side
of this expression is presented in Fig.\,\ref{fig9}. The dots here are
referred to graphs describing interaction with hard gluon in
initial state. The example of a graph of that kind is presented in the first
line in parentheses.

Just one more interesting example arises in differentiating equality
(\ref{eq:3f}) with respect to the Grassmann charges $\theta_0^{\dagger}$,
$\theta_0$ and free soft-gluon field $A^{(0)}$. The explicit form of derivative
$\delta^{3}\!j^{a}_{\mu}[A,\bar{\psi},\psi,Q_0,\theta^{\dagger}_0,\theta_0](k)/
\delta\theta_0^{\dagger i}\,\delta\theta_0^{j}\,
\delta A^{(0)a_1\mu_1}(k_1)|_{\,0}$ is given in Appendix D. The effective
current defined by this derivative generates the soft one-loop corrections to
the nonlinear Landau damping process studied early in
\cite{JOP_G_2000, markov_AOP_04}.
The diagrams with soft-quark loop (Fig.\,\ref{fig14}) in Appendix D should be
added to those of Fig.\,4 in Ref.\,\cite{markov_AOP_04}, which include 
soft-gluon loop.

\begin{figure}[hbtp]
\begin{center}
\includegraphics[width=0.95\textwidth]{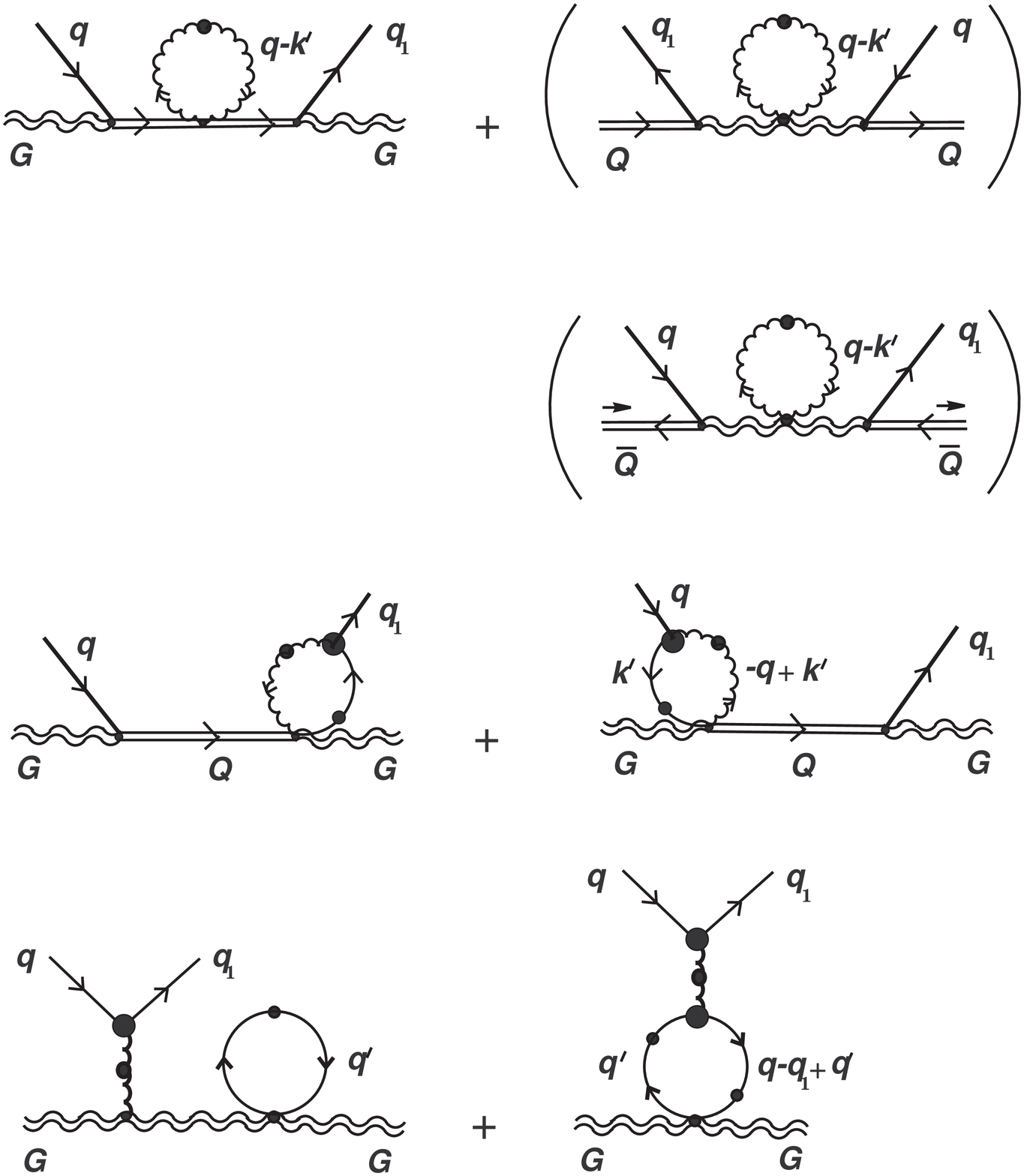}
\end{center}
\caption{\small One-loop corrections to the scattering process
generated by the effective source (\ref{eq:4e}) with coefficient
function (\ref{eq:5s}) (Figs.\,\ref{fig1} and \ref{fig3}).}
\label{fig10}
\end{figure}

Now we  consider differentiation of relation (\ref{eq:3g}). The calculation
of third order derivative with respect to $\theta_0^{\dagger},\,\theta_0$
and $\psi^{(0)}$  leads to the following expression
\begin{equation}
\left.\frac{\delta^3\eta^i_{\alpha}[A,\bar{\psi},\psi,Q_0,\theta_0](q)}
{\delta\theta_0^{\dagger j_1}\,\delta\theta_0^j
\,\delta\psi_{\alpha_1}^{(0)i_1}(q_1)}\,
\right|_{\,0}=
\left.\frac{\delta^3\tilde{\eta}^i_{\alpha}
[A^{(0)},\bar{\psi}^{(0)},\psi^{(0)},Q_0,\theta_0^{\dagger},\theta_0](q)}
{\delta\theta_0^{\dagger j_1}\,\delta\theta_0^j
\,\delta\psi_{\alpha_1}^{(0)i_1}(q_1)}\,
\right|_{\,0}
\label{eq:7t}
\end{equation}
\[
=\!-\frac{\,g^4}{(2\pi)^6}\,\Biggl\{
(t^a)^{ij}(t^a)^{j_1i_1}\!\!
\int\!\!
K^{(Q)\nu}_{\alpha}({\bf v},\chi |\,q-k^{\,\prime},-q)
\!\,^{\ast}{\cal D}_{\nu\nu^{\prime}}(q-k^{\,\prime\,})
\bar{K}^{(G)\nu^{\,\prime}}_{\alpha_1}
({\bf v},\bar{\chi}|\,q-k^{\,\prime},-q_1)
\delta(v\cdot k^{\,\prime\,})dk^{\,\prime}
\]
\[
+\,(t^a)^{ii_1}(t^a)^{j_1j}
\,^{\ast}\Gamma^{(Q)\nu}_{\alpha\alpha_1}(q-q_1;q_1,-q)
\,^{\ast}{\cal D}_{\nu\nu^{\,\prime}}(q-q_1)
\]
\[
\times\!\int\!
\Biggl(\,\frac{v^{\nu^{\,\prime}}}{(v\cdot q^{\,\prime\,})}\,
\Bigl(\Bigl[\,\bar{\chi}\,\,^{\ast}\!S(q^{\,\prime\,})\,\chi\,\Bigr]
-\Bigl[\,\bar{\chi}\,^{\ast}\!S(-q^{\,\prime\,})\,\chi\,\Bigr]\Bigr)
\]
\[
+\Bigl[\,
\bar{\chi}\,^{\ast}\!S(-q^{\,\prime\,})
\,^{\ast}\Gamma^{(G)\nu^{\,\prime}}(q-q_1;q^{\,\prime},-q+q_1-q^{\,\prime\,})
\,^{\ast}\!S(q-q_1+q^{\,\prime\,})\chi\,\Bigr]\Biggr)
\,\delta(v\cdot q^{\,\prime\,})\,dq^{\,\prime}
\Biggr\}
\,\delta(v\cdot(q-q_1))\,.
\]
Diagrammatic interpretation of the different terms on the right-hand side
of this expression is presented in Fig.\,\ref{fig10}. These graphs must be
supplemented with graphs describing interaction with the hard test antiquark
$\bar{Q}$. However, soft-loop corrections (\ref{eq:7t}) do not exhaust all
corrections to the elastic scattering process of soft-quark excitation off hard
parton drawn on Figs.\,\ref{fig1} and \ref{fig3}. The remaining corrections are
determined by variation of relation (\ref{eq:3g}) with respect to usual
color charges $Q_0^a$, $Q_0^b$ and free soft-quark field $\psi^{(0)}$:
$\delta^3 \eta^i_{\alpha}[A,\bar{\psi},\psi,Q_0,\theta_0](q)/
\delta Q_0^a\,\delta Q_0^b
\,\delta\psi_{\alpha_1}^{(0)i_1}(q_1)|_{\,0}\,$. The explicit form and
diagrammatic interpretation of the variation are given in Appendix E.

\begin{figure}[hbtp]
\begin{center}
\includegraphics[width=0.95\textwidth]{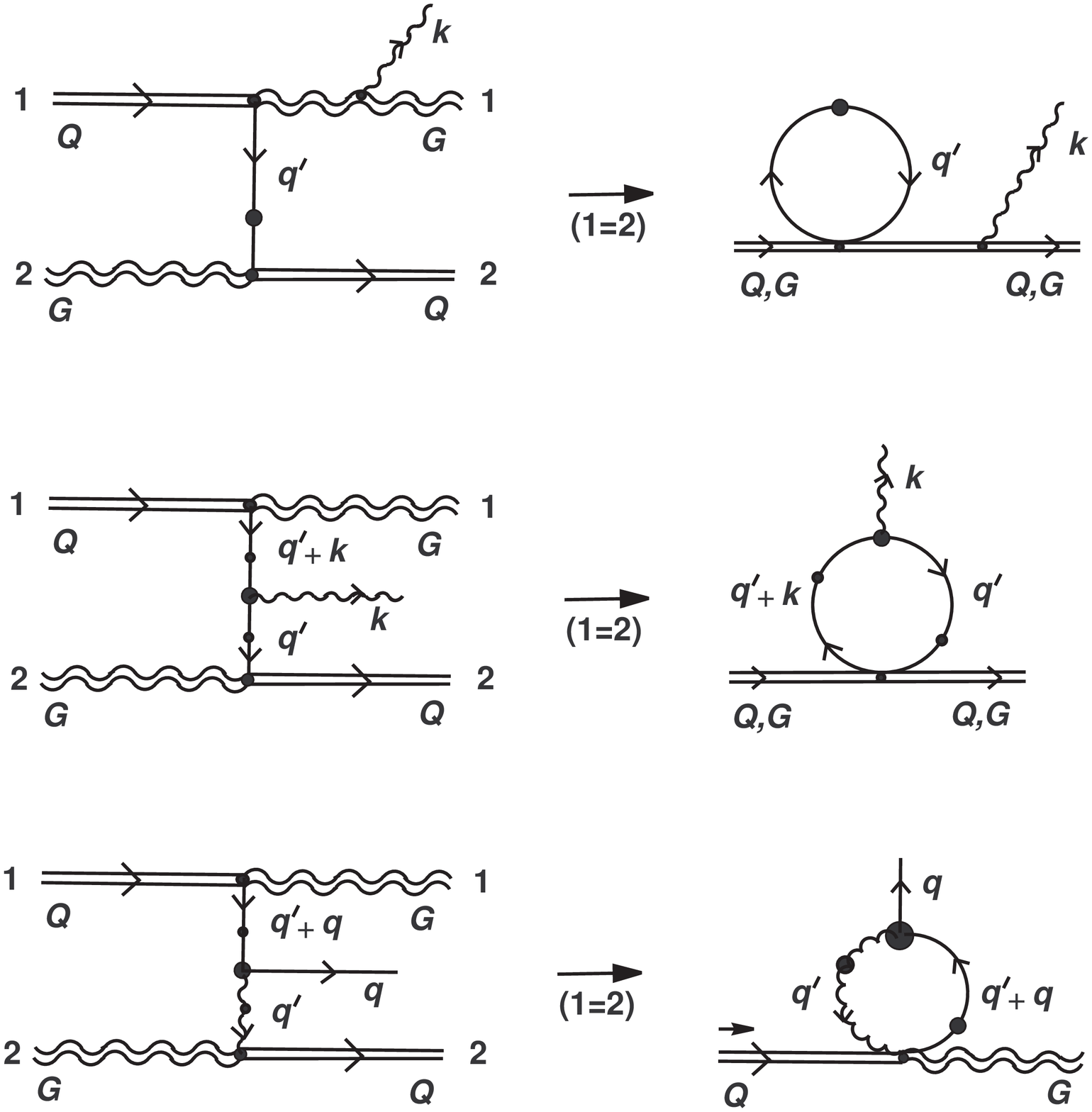}
\end{center}
\caption{\small Graphic illustration of examples for obtaining the
effective currents and sources of the present work including
soft one-loop corrections from the effective currents and sources for the 
processes of soft gluon and soft quark bremsstrahlung \cite{paper_III}.}
\label{fig11}
\end{figure}

In conclusion of this section we note run ahead that in our next
paper \cite{paper_III} we point to another independent way of
deriving the effective currents and sources of (\ref{eq:7q}),
(\ref{eq:7e}) type (and also the other more complicated
expressions obtained in this section) based on the expressions for
the effective currents and sources generating soft-quark
bremsstrahlung in the tree approximation. As known for production
of bremsstrahlung it is necessary that at least two hard
color-charged particles have been involved in the interaction process. The
essence of this approach reduces to a simple identifying in a
final expression for the effective currents and sources (generating
bremsstrahlung) of these two hard particles.
Under this identification soft-quark and soft-gluon propagators
describing the interaction process of two particles among
themselves are effectively closed into loops (with soft virtual
momentum) attached to straight line of hard parton. From the
geometric point of view it can be presented as the imposition
of the straight line of the first hard parton on the straight line
of the second hard parton. Note that there exist two ways for
deriving soft-quark loop resulting in different directions of
circuit of the fermion loop. Fig.\,\ref{fig11} gives graphic interpretation
of the procedure described above.

It should be particularly emphasized that these loop corrections
are not quantum ones. The notion of loops in our consideration has
somewhat conventional character. By this we simply mean the effect of
self-interaction of color classical partons induced by the
surrounding medium in which they move and with which they
interact. This process of interaction with medium 
is more evidently seen from Fig.\,\ref{fig11}, where on the right-hand side
hard test parton ${\bf 1}$ radiates virtual oscillation absorbed by hard
thermal particle ${\bf 2}$ and on the left-hand side the same virtual
oscillation is radiated and absorbed by {\it the same} hard test
parton. The coefficient functions (\ref{eq:7q}), (\ref{eq:7e}),
(\ref{eq:7r}) so on and the effective currents
(sources) connected with them can be interpretated as ``dressing'' initial 
bare current (source) of the test color particle generated by interaction of 
this current (sources) with hot bath.

\section{\bf Scattering probabilities}
\setcounter{equation}{0}

Making use of the explicit form for the effective currents and source
obtained in previous sections, one can define scattering
probabilities of soft quark and soft gluon excitations off hard
test particle. For this purpose according to the {\it Tsytovich
correspondence principle} \cite{markov_AOP_04, markov_AOP_05} it is necessary
to substitute the effective current $\tilde{j}^{a}_{\mu}
[A^{(0)},\bar{\psi}^{(0)},\psi^{(0)},Q_0,\theta_0^{\dagger},\theta_0](k)$
and effective source
$\tilde{\eta}^i_{\alpha}[A^{(0)},\bar{\psi}^{(0)},\psi^{(0)},Q_0,\theta_0](q)$
into expressions (I.8.1) and (I.8.2) correspondingly and define
the emitted radiant power of soft plasma excitations. However,
it needs to be preliminary carried out a little generalization of
expressions (I.8.1), (I.8.2) taking into account a specific of the
problem under consideration.

At first we perform an averaging of the right-hand sides of
Eqs.\,(I.8.1), (I.8.2) over initial value of the usual color charge
$Q_0$ by adding an integration\footnote{As it will be shown below there is no
necessity to enter an averaging over initial values of the Grassmann charges
$\theta_0^{\dagger i}$ and $\theta_0^i$. These charges will always
appear in final expressions in the form of combination
$\theta_0^{\dagger i}\theta_0^i \equiv C_{\theta} = const$.} over the colors
with measure
\[
dQ_0=\prod_{a=1}^{d_A}\!dQ^a_0\,\delta(Q^a_0Q^a_0-C_2^{(\zeta)}),\quad
d_A = N_c^2 - 1,\quad \zeta=G,\,Q,\,\bar{Q}
\]
with the second Casimir $C_2^{(G)} = C_A \,(=N_c)$ for hard gluons
and correspondingly $C_2^{(Q,\,\bar{Q})} = C_F
\,(=(N_c^2-1)/2N_c)$ for hard (anti)quarks normalized such that
$\int\!dQ_0=1$, and thus
\begin{equation}
\int\!dQ_0\,Q^a_0Q^b_0 = \frac{C_A}{d_A}\,\delta^{ab}.
\label{eq:8q}
\end{equation}
Besides, an averaging over distributions of hard particles in thermal
equilibrium should be added. This statistic factor in a certain way
depends on power of Grassmann color charges in expansions of
the effective current and source in powers of soft free fields
and initial value of color charges. If the term in the expansion
contains `not compensated' Grassmann charge $\theta_0$ (or
$\theta_0^{\dagger}$), then it is necessary to introduce an
averaging over hard particle distributions in the following form:
\[
\int\!\frac{d {\bf p}}{(2\pi)^3}\,
\Bigl[\,f_{\bf p}^{Q}+f_{\bf p}^{G}\Bigr]
\qquad\left({\rm or}\;\,
\int\!\frac{d {\bf p}}{(2\pi)^3}\,
\Bigl[\,f_{\bf p}^{\bar{Q}}+f_{\bf p}^{G}\Bigr]
\right).
\]
Otherwise, when every charge $\theta_0$ is compensated by
$\theta_0^{\dagger}$, it is necessary to use an average in the form
\[
\sum\limits_{\zeta=Q,\,\bar{Q},\,G}
\int\!\frac{d {\bf p}}{(2\pi)^3}\,f_{\bf p}^{(\zeta)}\,.
\]
In the former case the effective currents and sources generate the scattering
processes, which change the type of the hard test parton and correspondingly
in the latter case the type of the hard test parton is not varying.
This feature was discussed in Section 2 after Eq.\,(\ref{eq:2t}). In the
remainder of the paper we restrict our consideration only to linear terms
in expansion of the effective current and source, i.e., we set
\begin{equation}
\tilde{j}_{\mu}^{\Psi a}({\bf v},\chi;Q_0,\theta_0|\,k)\simeq
\tilde{j}_{\mu}^{\,\Psi ab}({\bf v},\chi|\,k)Q_0^b +
\Bigl[\,\tilde{j}_{\mu}^{\,\dagger \Psi aj}
({\bf v},\bar{\chi}|\,-\!k)\theta_0^j+
\theta_0^{\dagger j}\,\tilde{j}_{\mu}^{\,\Psi aj}({\bf v},\bar{\chi}|\,k)
\Bigr]\,,
\label{eq:8w}
\end{equation}
\[
\tilde{\eta}^{\,i}_{\alpha}({\bf v},\chi;Q_0,\theta_0|\,q)\simeq
\tilde{\eta}^{\,ib}_{\alpha}({\bf v},\chi|\,q)Q_0^b +
\tilde{\eta}^{\,ij}_{\alpha}({\bf v},\chi|\,q)\theta_0^j.
\hspace{0.3cm}
\]
In the notations $\tilde{j}_{\mu}^{\Psi a}({\bf v},\chi,Q_0,\theta_0|\,k),\,
\tilde{\eta}^{\,i}_{\alpha}({\bf v},\chi,Q_0,\theta_0|\,q),\ldots$ we take
into account that for the globally equilibrium system the effective currents
and sources depend on hard momentum ${\bf p}$ through velocity
${\bf v}={\bf p}/|{\bf p}|$ and also spin state of hard
parton\footnote{Strictly speaking the effective currents and sources are
also implicitly dependent on energy $E$ of hard test particle through spinor
$\chi$ (see Eq.\,(C.2)). It is this dependence that is meant in notation of
the scattering probabilities as functions of hard momentum ${\bf p}$ in
Eqs.\,(\ref{eq:2r}) and (\ref{eq:2t}).} described by $\chi$. Dependence on
soft fields is implicitly implied.

Taking into account above-mentioned we use the following expressions for
emitted powers ${\cal I}_{\,\cal B}$ and ${\cal I}_{\cal F}$ instead
of (I.8.1), (I.8.2):
\begin{equation}
{\cal I}_{\,\cal B}=
\pi\!\lim\limits_{\tau\rightarrow\infty}
\frac{(2\pi)^4}{\tau}
\left(\int\!\frac{|{\bf p}|^2\,d|{\bf p}|}{2\pi^2}\,
\Bigl[\,f_{\bf p}^{Q}+f_{\bf p}^{G}\Bigr]\right)\!
(\theta_0^{\dagger j^{\prime}}\theta_0^{j})
\int\!dQ_0\!\int\!\frac{d\Omega_{\bf v}}{4\pi}\,
\!\int\!dk\,k^0 {\rm sign}(k^0)
\label{eq:8e}
\end{equation}
\[
\times\biggl\{Q^{\mu\mu^{\prime}}\!(k)\Bigl[\,
\Bigl\langle\tilde{j}_{\mu}^{\,\dagger\Psi aj}({\bf v},\bar{\chi}|\,k)
\tilde{j}_{\mu^{\prime}}^{\,\Psi aj^{\prime}}({\bf v},\bar{\chi}|\,k)
\Bigr\rangle +
\Bigl\langle\tilde{j}_{\mu^{\prime}}^{\,\dagger\Psi aj}
({\bf v},\bar{\chi}|\,-k)
\tilde{j}_{\mu}^{\,\Psi aj^{\prime}}({\bf v},\bar{\chi}|\,-k)
\Bigr\rangle\Bigr]
\delta({\rm Re}\,^{\ast}\!\Delta^{\!-1\,l}(k))
\]
\[
\hspace{0,2cm}
+\,P^{\mu\mu^{\prime}}\!(k)
\Bigl[\,
\Bigl\langle\tilde{j}_{\mu}^{\,\dagger\Psi aj}({\bf v},\bar{\chi}|\,k)
\tilde{j}_{\mu^{\prime}}^{\,\Psi aj^{\prime}}({\bf v},\bar{\chi}|\,k)
\Bigr\rangle +
\Bigl\langle\tilde{j}_{\mu^{\prime}}^{\,\dagger\Psi aj}
({\bf v},\bar{\chi}|\,-k)
\tilde{j}_{\mu}^{\,\Psi aj^{\prime}}({\bf v},\bar{\chi}|\,-k)
\Bigr\rangle\Bigr]
\delta({\rm Re}\,^{\ast}\!\Delta^{\!-1\,t}(k))\!\biggr\}
\vspace{0.5cm}
\]
\[
-\pi\!\lim\limits_{\tau\rightarrow\infty}
\frac{(2\pi)^4}{\tau}
\left(\sum\limits_{\,\zeta=Q,\,\bar{Q},\,G}
\int\!\frac{|{\bf p}|^2\,d|{\bf p}|}{2\pi^2}\,f_{\bf p}^{(\zeta)}\!\right)\!
\int\!dQ_0 \,Q_0^bQ_0^{b^{\prime}}
\!\int\!\frac{d\Omega_{\bf v}}{4\pi}\,
\]
\[
\times\!\int\!dk\,k^0 {\rm sign}(k^0)
\,\biggl\{Q^{\mu\mu^{\prime}}\!(k)\,
\Bigl\langle\tilde{j}_{\mu}^{\,\ast\Psi ab}({\bf v},\chi|\,k)
\tilde{j}_{\mu^{\prime}}^{\,\Psi ab^{\prime}}({\bf v},\chi|\,k)
\Bigr\rangle\,\delta({\rm Re}\,^{\ast}\!\Delta^{\!-1\,l}(k))
\]
\[
\hspace{3.4cm}
+\,P^{\mu\mu^{\prime}}\!(k)\,
\Bigl\langle\tilde{j}_{\mu}^{\,\ast\Psi ab}({\bf v},\chi|\,k)
\tilde{j}_{\mu^{\prime}}^{\,\Psi ab^{\prime}}({\bf v},\chi|\,k)
\Bigr\rangle\,\delta({\rm Re}\,^{\ast}\!\Delta^{\!-1\,t}(k))
\biggr\},
\]
and correspondingly
\begin{equation}
{\cal I}_{\cal F}=
-\pi\!\lim\limits_{\tau\rightarrow\infty}
\frac{(2\pi)^4}{\tau}
\left(\int\!\frac{|{\bf p}|^2\,d|{\bf p}|}{2\pi^2}\,
\Bigl[\,f_{\bf p}^{Q}+f_{\bf p}^{G}\Bigr]\right)\!
(\theta_0^{\dagger j}\theta_0^{j^{\prime}})
\int\!dQ_0\!\int\!\frac{d\Omega_{\bf v}}{4\pi}\,
\label{eq:8r}
\end{equation}
\[
\times\!\int\!dq\,q^0 {\rm sign}(q^0)
\,\biggl\{(h_{+}(\hat{\bf q}))_{\alpha\alpha^{\prime}}\,
\Bigl\langle\tilde{\bar{\eta}}^{\,ij}_{\alpha}({\bf v},\bar{\chi}|\,-q)
\tilde{\eta}^{\,ij^{\prime}}_{\alpha^{\prime}}({\bf v},\chi|\,q)
\Bigr\rangle\,\delta({\rm Re}\,^{\ast}\!\Delta^{\!-1}_{+}(q))
\]
\[
\hspace{3.2cm}
+\,(h_{-}(\hat{\bf q}))_{\alpha\alpha^{\prime}}\,
\Bigl\langle\tilde{\bar{\eta}}^{\,ij}_{\alpha}({\bf v},\bar{\chi}|\,-q)
\tilde{\eta}^{\,ij^{\prime}}_{\alpha^{\prime}}({\bf v},\chi|\,q)
\Bigr\rangle\,\delta({\rm Re}\,^{\ast}\!\Delta^{\!-1}_{-}(q))
\biggr\}
\vspace{0.5cm}
\]
\[
-\pi\!\lim\limits_{\tau\rightarrow\infty}
\frac{(2\pi)^4}{\tau}
\left(\sum\limits_{\,\zeta=Q,\,\bar{Q},\,G}
\int\!\frac{|{\bf p}|^2\,d|{\bf p}|}{2\pi^2}\,
f_{\bf p}^{(\zeta)}\!\right)\!
\int\!dQ_0 \,Q_0^bQ_0^{b^{\prime}}
\!\int\!\frac{d\Omega_{\bf v}}{4\pi}\,
\]
\[
\times\!\int\!dq\,q^0 {\rm sign}(q^0)
\,\biggl\{(h_{+}(\hat{\bf q}))_{\alpha\alpha^{\prime}}\,
\Bigl\langle\tilde{\bar{\eta}}^{\,ib}_{\alpha}
({\bf v},\bar{\chi}|\,-q)
\tilde{\eta}^{\,ib^{\prime}}_{{\alpha}^{\prime}}({\bf v},\chi|\,q)
\Bigr\rangle\,\delta({\rm Re}\,^{\ast}\!\Delta^{\!-1}_{+}(q))
\]
\[
\hspace{3.3cm}
+\,(h_{-}(\hat{\bf q}))_{\alpha\alpha^{\prime}}\,
\Bigl\langle\tilde{\bar{\eta}}^{\,ib}_{\alpha}({\bf v},\bar{\chi}|\,-q)
\tilde{\eta}^{\,ib^{\prime}}_{{\alpha}^{\prime}}({\bf v},\chi|\,q)
\Bigr\rangle\,\delta({\rm Re}\,^{\ast}\!\Delta^{\!-1}_{-}(q))
\biggr\}.
\]
In Eq.\,(\ref{eq:8e}) we keep only a contribution from the current
$\tilde{j}^{\Psi a}_{\mu}$ including soft-quark fields and the Grassmann
color charges. The contribution from the current $\tilde{j}^{Aa}_{\mu}$ was
considered in Ref.\,\cite{markov_AOP_04}. Besides, unlike (I.8.1)
and (I.8.2) we remove an integration over volume of the system, since in
the definition of phase-space measures (\ref{eq:2y}) the momentum conservation
laws are explicitly
considered. According to the corresponding principle for determining
the scattering probabilities ${\it w}_{\,q\rightarrow {\rm g}}^{(f;\,b)}
({\bf p}|\,{\bf q};{\bf k}),\, {\it w}_{\,q\rightarrow q}^{(\zeta)(f;\,f_1)}
({\bf p}|\,{\bf q};{\bf q}_1)$ etc. the expression obtained from (\ref{eq:8r})
should be compared with expression determining a change of energy of soft
fermionic plasma excitations generated by the spontaneous processes of
soft-quark and soft-gluon emission only
\begin{equation}
\left( \frac{{\rm d}{\cal E}}{{\rm d} t} \right)^{\!{\rm sp}} =
\sum\limits_{f=\pm}\frac{\rm d}{{\rm d}t}\,\Biggl\{
\int\!\frac{d{\bf q}}{(2 \pi)^3} \;
\omega^{(f)}_{\bf q} \,n^{(f)}_{\bf q}
+
\int\!\frac{d{\bf q}}{(2 \pi)^3} \;
\omega^{(f)}_{\bf q}\,\bar{n}^{(f)}_{\bf q}
\!\Biggr\}
\label{eq:8t}
\end{equation}
\[
=
\sum\limits_{f\,=\,\pm\,}\sum\limits_{\,b\,=t,\,l}
\int\!\frac{|{\bf p}|^2\,d|{\bf p}|}{2\pi^2}\,
\Bigl[\,f_{\bf p}^{Q}+f_{\bf p}^{G}\Bigr]
\int\!\frac{d\Omega_{\bf v}}{4\pi}
\int\!\frac{d{\bf q}}{(2\pi)^3}
\!\!\int d{\cal T}_{q\rightarrow {\rm g}}^{(f;\,b)}
\omega^{(f)}_{\bf q}
\]
\[
\times\,
\Bigl\{{\it w}_{\,q\rightarrow {\rm g}}^{(f;\,b)}
({\bf p}|\,{\bf q};{\bf k}) +
{\it w}_{\,\bar{q}\rightarrow {\rm g}}^{(f;\,b)}
({\bf p}|\,{\bf q};{\bf k})\Bigr\}N_{\bf k}^{(b)}
\vspace{0.5cm}
\]
\[
+
\sum\limits_{f\,=\,\pm\,}\sum\limits_{\,b\,=t,\,l}
\Biggl(\sum\limits_{\,\zeta=Q,\,\bar{Q},\,G\,}
\int\!\frac{|{\bf p}|^2\,d|{\bf p}|}{2\pi^2}\,
f_{\bf p}^{(\zeta)}\Biggr)
\int\!\frac{d\Omega_{\bf v}}{4\pi}
\int\!\frac{d{\bf q}}{(2\pi)^3}
\!\!\int d{\cal T}_{q\rightarrow q}^{(f;\,f_1)}
\omega^{(f)}_{\bf q}
\]
\[
\times\,
\Bigl\{{\it w}_{\,q\rightarrow q}^{(\zeta)(f;\,f_1)}
({\bf p}|\,{\bf q};{\bf q}_1) n_{{\bf q}_1}^{(f_1)} +
{\it w}_{\,\bar{q}\rightarrow \bar{q}}^{(\zeta)(f;\,f_1)}
({\bf p}|\,{\bf q};{\bf q}_1)\Bigl(1-\bar{n}_{{\bf q}_1}^{(f_1)}
\Bigr)\Bigr\} + \,\ldots\,.
\]
In deriving the right-hand side we have taken into account an equality of
integration measures for the processes with participation of soft-quark and
soft-antiquark modes
\[
\int d{\cal T}_{q\rightarrow {\rm g}}^{(f;\,b)}=
\int d{\cal T}_{\bar{q}\rightarrow {\rm g}}^{(f;\,b)},
\quad
\int d{\cal T}_{q\rightarrow q}^{(f;\,f_1)}=
\int d{\cal T}_{\bar{q}\rightarrow \bar{q}}^{(f;\,f_1)}
\]
and also the fact that in conditions of global equilibrium and zero quark
chemical potential an equality
$f_{\bf p}^{Q}=f_{\bf p}^{\bar{Q}}$ takes place. What is more
the kinetic equation (\ref{eq:2q}) with collision term in the limit of a small
intensity $n_{\bf q}^{(f)}\rightarrow 0$ (Eq.\,(\ref{eq:2o})) was used. The
dots designates contributions of higher order scattering processes.

With all required formulas at hand now one can define the
simplest scattering probabilities ${\it w}_{\,q\rightarrow {\rm
g}}^{(f;\,b)},\, {\it w}_{\,q\rightarrow q}^{(\zeta)(f;\,f_1)}$ etc. To be
specific, we consider the scattering processes with participation of normal
soft-quark and transverse soft-gluon modes, i.e., we set $f=f_1=+$
and $b=t$. As the first step we single out on the right-hand side of
(\ref{eq:8r}) the contribution of normal mode of soft fermion
excitations. For this purpose we set
\begin{equation}
\delta({\rm Re}\,^{\ast}\!\Delta^{\!-1}_{\pm}(q))=
{\rm Z}_{\pm}({\bf q})\,
\delta (q^0 - \omega_{\bf q}^{\pm})
+\,{\rm Z}_{\mp}({\bf q})\,
\delta (q^0 + \omega_{\bf q}^{\mp})
\label{eq:8y}
\end{equation}
and expand the spinor projectors $h_{\pm}(\hat{\bf q})$ in terms of
simultaneous eigenspinors of chirality and helicity
\begin{equation}
(h_{+}({\hat{\bf q}}))_{\alpha\alpha^{\prime}}=
\sum\limits_{\lambda=\pm}u_{\alpha}(\hat{\bf q},\lambda)
\bar{u}_{\alpha^{\prime}}(\hat{\bf q},\lambda),
\quad
(h_{-}({\hat{\bf q}}))_{\alpha\alpha^{\prime}}=
\sum\limits_{\lambda=\pm}v_{\alpha}(\hat{\bf q},\lambda)
\bar{v}_{\alpha^{\prime}}(\hat{\bf q},\lambda).
\label{eq:8u}
\end{equation}

Substituting (\ref{eq:8y}), (\ref{eq:8u}) into (\ref{eq:8r}) and
integrating with respect to $dq_0$, we define contribution from the normal
soft-quark modes to the emitted power
\begin{equation}
{\cal I}_{\cal F}=
-\pi\!\lim\limits_{\tau\rightarrow\infty}
\frac{(2\pi)^4}{\tau}
\left(\int\!\frac{|{\bf p}|^{\,2}\,d|{\bf p}|}{2\pi^2}\,
\Bigl[\,f_{\bf p}^{Q}+f_{\bf p}^{G}\Bigr]\right)\!
(\theta_0^{\dagger j}\theta_0^{j^{\prime}})
\int\!dQ_0\!\int\!\frac{d\Omega_{\bf v}}{4\pi}\,
\label{eq:8i}
\end{equation}
\[
\times\sum\limits_{\lambda\,=\,\pm}
\int\!\!d{\bf q}\,\,\omega_{\bf q}^{+}{\rm Z}_{+}({\bf q})
\,\biggl\{\Bigl\langle
\bigl(\tilde{\bar{\eta}}^{\,ij}_{\alpha}({\bf v},\bar{\chi}|\,-q)
u_{\alpha}(\hat{\bf q},\lambda)\bigr)
\bigl(\bar{u}_{\alpha^{\prime}}(\hat{\bf q},\lambda)
\tilde{\eta}^{\,ij^{\prime}}_{\alpha^{\prime}}({\bf v},\chi|\,q)\bigr)
\Bigr\rangle_{q^0\,=\,\omega_{\bf q}^{+}}
\]
\[
\hspace{4.1cm}
+\,\Bigl\langle
\bigl(\tilde{\bar{\eta}}^{\,ij}_{\alpha}({\bf v},\bar{\chi}|\,-q)
v_{\alpha}(\hat{\bf q},\lambda)\bigr)
\bigl(\bar{v}_{\alpha^{\prime}}(\hat{\bf q},\lambda)
\tilde{\eta}^{\,ij^{\prime}}_{\alpha^{\prime}}({\bf v},\chi|\,q)\bigr)
\Bigr\rangle_{q^0=-\omega_{\bf q}^{+}}\biggr\}
\vspace{0.5cm}
\]
\[
-\,\pi\!\lim\limits_{\tau\rightarrow\infty}
\frac{(2\pi)^4}{\tau}
\left(\sum\limits_{\,\zeta=Q,\,\bar{Q},\,G}
\int\!\frac{|{\bf p}|^{\,2}\,d|{\bf p}|}{2\pi^2}\,
f_{\bf p}^{(\zeta)}\!\right)\!
\int\!dQ_0 \,Q_0^bQ_0^{b^{\prime}}
\!\int\!\frac{d\Omega_{\bf v}}{4\pi}\,
\]
\[
\times\sum\limits_{\lambda\,=\,\pm}
\int\!\!d{\bf q}\,\,\omega_{\bf q}^{+}{\rm Z}_{+}({\bf q})
\,\biggl\{\Bigl\langle
\bigl(\tilde{\bar{\eta}}^{\,ib}_{\alpha}({\bf v},\bar{\chi}|\,-q)
u_{\alpha}(\hat{\bf q},\lambda)\bigr)
\bigl(\bar{u}_{\alpha^{\prime}}(\hat{\bf q},\lambda)
\tilde{\eta}^{\,ib^{\prime}}_{\alpha^{\prime}}({\bf v},\chi|\,q)\bigr)
\Bigr\rangle_{q^0\,=\,\omega_{\bf q}^{+}}
\]
\[
\hspace{4.2cm}
+\,\Bigl\langle
\bigl(\tilde{\bar{\eta}}^{\,ib}_{\alpha}({\bf v},\bar{\chi}|\,-q)
v_{\alpha}(\hat{\bf q},\lambda)\bigr)
\bigl(\bar{v}_{\alpha^{\prime}}(\hat{\bf q},\lambda)
\tilde{\eta}^{\,ib^{\prime}}_{\alpha^{\prime}}({\bf v},\chi|\,q)\bigr)
\Bigr\rangle_{q^0=-\omega_{\bf q}^{+}}\biggr\}.
\]

The first contribution on the right-hand side of the last equation enables
us to define the scattering probabilities
${\it w}_{\,q\rightarrow {\rm g}}^{(+;\,t)} ({\bf p}|\,{\bf q};{\bf k})$ and
${\it w}_{\bar{q}\rightarrow {\rm g}}^{(+;\,t)}
({\bf p}|\,{\bf q};{\bf k})$. Let us perform the following replacements
\[
\tilde{\eta}^{\,ij^{\prime}}_{\alpha^{\prime}}({\bf v},\chi|\,q)
\rightarrow
\tilde{\eta}^{\,(1)ij^{\prime}}_{\alpha^{\prime}}(A^{(0)})(q),
\quad
\tilde{\bar{\eta}}^{\,ij}_{\alpha}({\bf v},\bar{\chi}|\,-q)
\rightarrow
\tilde{\bar{\eta}}^{\,(1)ij}_{\alpha}(A^{\ast(0)})(-q),
\]
where the effective sources
$\tilde{\eta}^{\,(1)ij^{\prime}}_{\alpha^{\prime}}$,
$\tilde{\bar{\eta}}^{\,(1)ij}_{\alpha}$ are defined by Eqs.\,(\ref{eq:4u}),
(\ref{eq:4y}) and (\ref{eq:4i}), (\ref{eq:4o}). In this case the contribution
under discussion to the emitted power (\ref{eq:8i}) will be equal
\begin{equation}
-\,\frac{1}{2\pi}\,g^4\!\lim\limits_{\tau\rightarrow\infty}
\frac{1}{\tau}
\left(\int\!\frac{|{\bf p}|^{\,2}\,d|{\bf p}|}{2\pi^2}\,
\Bigl[\,f_{\bf p}^{Q}+f_{\bf p}^{G}\Bigr]\right)\!
\left(\theta_0^{\dagger j}
(t^bt^{b^{\prime}})^{jj^{\prime}}\!\theta_0^{j^{\prime}}\right)
\int\!dQ_0\!\int\!\frac{d\Omega_{\bf v}}{4\pi}\,
\label{eq:8o}
\end{equation}
\[
\times\sum\limits_{\lambda\,=\,\pm}
\int\!\!d{\bf q}\,\,\omega_{\bf q}^{+}
\Biggl(\frac{{\rm Z}_{+}({\bf q})}{2}\Biggr)
\biggl\{\Bigl[\,
\Bigl(\bar{K}_{\alpha}^{(Q)\mu}({\bf v},\bar{\chi}|\,k,-q)
u_{\alpha}(\hat{\bf q},\lambda)\Bigr)
\Bigl(K_{\alpha^{\prime}}^{(Q)\mu^{\prime}}({\bf v},\chi|\,k^{\,\prime},-q)
\bar{u}_{\alpha^{\prime}}(\hat{\bf q},\lambda)\Bigr)
\]
\[
\times\,
\delta(v\cdot(q-k))\delta(v\cdot(q-k^{\,\prime\,}))
\Bigr]_{q^0\,=\,\omega_{\bf q}^{+}}\,+
\]
\[
\Bigl[
\Bigl(\bar{K}_{\alpha}^{(Q)\mu}({\bf v},\bar{\chi}|\,k,-q)
v_{\alpha}(\hat{\bf q},\lambda)\Bigr)
\Bigl(K_{\alpha^{\prime}}^{(Q)\mu^{\prime}}({\bf v},\chi|\,k^{\,\prime},-q)
\bar{v}_{\alpha^{\prime}}(\hat{\bf q},\lambda)\Bigr)
\,\delta(v\cdot(q-k))\delta(v\cdot(q-k^{\,\prime\,}))
\Bigr]_{q^0=-\omega_{\bf q}^{+}}\!\biggr\}
\]
\[
\times
\left\langle A^{\ast(0)b}_{\mu}(k)A^{(0)b^{\prime}}_{\mu^{\prime}}
(k^{\,\prime\,})\right\rangle\,dkdk^{\,\prime}.
\]
The integral in $dQ$ is equal to unit by virtue of the normalization. The
correlation function of random soft-gluon field in conditions of stationary
and homogeneous state of the quark-gluon plasma can be written in the form
\begin{equation}
\left\langle A^{\ast(0)b}_{\mu}(k)A^{(0)b^{\prime}}_{\mu^{\prime}}
(k^{\,\prime\,})\right\rangle
\simeq\delta^{bb^{\prime}}\!I_{\mu\mu^{\prime}}(k)\,\delta(k-k^{\,\prime\,}),
\label{eq:8p}
\end{equation}
where in the spectral density $I_{\mu\mu^{\prime}}(k)$ we keep only
`transverse' part: $P_{\mu\mu^{\prime}}(k)I_k^t$.
In the frame, where $u^{\mu}=(1,0,0,0)$ the transverse projector
$P_{\mu\mu^{\prime}}(k)$ reduces to three-dimensional transverse projector
$P^{ii^{\prime}}(\hat{\bf k}),\,\hat{\bf k}={\bf k}/|{\bf k}|$.
For the transverse mode we introduce polarization vectors
${\rm e}^i(\hat{\bf k},\xi),\,\xi=1,2$ possessing the properties
\[
{\bf k}\cdot{\bf e}(\hat{\bf k},\xi)=0,\quad
{\bf e}^{\ast}(\hat{\bf k},\xi)\cdot
{\bf e}(\hat{\bf k},\xi^{\prime})=
\delta_{\xi{\xi}^{\prime}}.
\]
The three-dimensional transverse projector is associated with the polarization
vectors by relation
\[
P^{ii^{\prime}}(\hat{\bf k})=(\delta^{ii^{\prime}}
-\hat{k}^i\hat{k}^{i^{\prime}})=
\sum_{\xi = 1,2}{\rm e}^{\ast\,i}(\hat{\bf k},\xi)\,
{\rm e}^{i^{\prime}}(\hat{\bf k},\xi).
\]
In what follows we take the  spectral function $I_k^t$
in the form of the quasiparticle approximation.
Turning into the number density of transverse soft gluons
$N_{\bf k}^t\,(=-\,(2\pi)^3\,2\omega_{{\bf k}}^t {\rm Z}_t^{-1}({\bf k})
I_{{\bf k}}^t)$, we finally obtain a substitution rule for the spectral
density $I_{\mu\mu^{\prime}}(k)$:
\begin{equation}
I_{\mu\mu^{\prime}}(k)\rightarrow
\frac{1}{(2\pi)^3}
\left(\frac{{\rm Z}_t({\bf k})}{2\omega_{{\bf k}}^t}\right)\!
\Bigl[\,N_{\bf k}^t\,\delta(k^0-\omega_{{\bf k}}^t)
+N_{-{\bf k}}^t\delta(k^0+\omega_{{\bf k}}^t)\Bigr]
\label{eq:8a}
\end{equation}
\[
\times
\sum_{\xi = 1,2}{\rm e}^{\ast\,i}(\hat{\bf k},\xi)\,
{\rm e}^{i^{\prime}}(\hat{\bf k},\xi).
\]
To take into account weak non-homogeneity and slow evolution of the
medium in time, it is sufficient to replace (within the accuracy accepted)
the equilibrium number density $N_{\bf k}^t$ by
off-equilibrium one in the Wigner form slowly depending on
$x=(t,{\bf x})$.

For the `color' factor with the Grassmann charges considering
$\delta^{\,bb^{\prime}}$ in (\ref{eq:8p}), we have
\[
\theta_0^{\dagger j}
(t^bt^{b})^{jj^{\prime}}\theta_0^{j^{\prime}}=
C_F\theta_0^{\dagger j}\theta_0^{j}\equiv C_F\,C_{\theta}.
\]
The exact value (more precisely, the {\it values})  of the constant
$C_{\theta}$ will be defined in the next section. Here we only need to know
that $C_{\theta}<0$. Substituting (\ref{eq:8p}), (\ref{eq:8a}) into
Eq.\,(\ref{eq:8o}), performing integration with respect to $dk^{\prime}$ and
$dk_0$, and making use of the relation
\[
\left[\,\delta(v\cdot(k-q))\right]^2=
\frac{\tau}{2\pi}\,\delta(v\cdot(k-q)),
\]
we finally obtain an expression for the desired emitted power instead of
(\ref{eq:8o})
\begin{equation}
-\,g^4\,C_F\,C_{\theta}
\left(\int\!\frac{|{\bf p}|^{\,2}\,d|{\bf p}|}{2\pi^2}\,
\Bigl[\,f_{\bf p}^{Q}+f_{\bf p}^{G}\Bigr]\right)\!
\int\!\frac{d\Omega_{\bf v}}{4\pi}\!
\int\!\frac{d{\bf q}}{(2\pi)^3}\,\frac{d{\bf k}}{(2\pi)^3}\,\,
\omega_{\bf q}^{+}N_{\bf k}^{\,t}
\Biggl(\frac{{\rm Z}_{+}({\bf q})}{2}\Biggr)
\Biggl(\frac{{\rm Z}_t({\bf k})}{2\omega_{{\bf k}}^t}\Biggr)
\label{eq:8s}
\end{equation}
\[
\times\!\!
\sum\limits_{\lambda\,=\,\pm}
\sum\limits_{\,\xi\,=\,1,\,2}
\biggl\{\!\Bigl[\,
{\rm e}^{\ast\,i}(\hat{\bf k},\xi)
\bar{K}_{\alpha}^{(Q)i}({\bf v},\bar{\chi}|\,k,-q)
u_{\alpha}(\hat{\bf q},\lambda)\Bigr]
\Bigl[\,
\bar{u}_{\alpha^{\prime}}(\hat{\bf q},\lambda)
K_{\alpha^{\prime}}^{(Q)i^{\prime}}({\bf v},\chi|\,k,-q)
{\rm e}^{i^{\prime}}(\hat{\bf k},\xi)\Bigr]
\hspace{1cm}
\]
\[
+\,\Bigl[\,
{\rm e}^{\ast\,i}(-\hat{\bf k},\xi)
\bar{K}_{\alpha}^{(Q)i}({\bf v},\bar{\chi}|\,-k,q)
v_{\alpha}(-\hat{\bf q},\lambda)\Bigr]
\!\Bigl[\,\bar{v}_{\alpha^{\prime}}(-\hat{\bf q},\lambda)
K_{\alpha^{\prime}}^{(Q)i^{\prime}}({\bf v},\chi|\,-k,q)
{\rm e}^{i^{\prime}}(-\hat{\bf k},\xi)\Bigr]\biggr\}_{\rm on-shell}
\]
\[
\times\,2\pi
\delta(\omega_{\bf q}^{+}-\omega_{\bf k}^{t}-{\bf v}\cdot({\bf q}-{\bf k})).
\]
In deriving this equation we have omitted the contribution
containing the delta-function $\delta(\omega_{\bf
q}^{+}+\omega_{\bf k}^{t}-{\bf v}\cdot({\bf q}+{\bf k}))$. It
defines the processes of simultaneous emission (absorption) of
soft-quark and soft-gluon excitations by hard particle. For the
second term in braces we have made replacements of variables:
${\bf q}\rightarrow -{\bf q},\, {\bf k}\rightarrow -{\bf k}\;
(\omega_{\bf q}^{+}\rightarrow \omega_{\bf q}^{+},\, 
\omega_{\bf k}^{t}\rightarrow  \omega_{\bf k}^{t}).$ Now we introduce the
following matrix elements for soft-quark\,--\,hard-particle and
soft-antiquark\,--\,\,hard-particle `inelastic' scattering,
respectively
\begin{equation}
{\cal M}_{\lambda\xi}^{(+\!,\,t)}({\bf v},\chi|\,{\bf q};{\bf k})\equiv
g^2\left(C_F|C_{\theta}|\right)^{1/2}
\Biggl(\frac{{\rm Z}_{+}({\bf q})}{2}\Biggr)^{\!1/2}\!
\Biggl(\frac{{\rm Z}_t({\bf k})}{2\omega_{{\bf k}}^t}\Biggr)^{\!1/2}
\label{eq:8d}
\end{equation}
\[
\times
\Bigr[\,\bar{u}_{\alpha}(\hat{\bf q},\lambda)
K_{\alpha}^{(Q)i}({\bf v},\chi|\,k,-q)
{\rm e}^{i}(\hat{\bf k},\xi)\Bigr]_{{\rm on-shell}\,,}
\vspace{0.5cm}
\]
\begin{equation}
\bar{\cal M}_{\lambda\xi}^{(+\!,\,t)}({\bf v},\chi|\,{\bf q};{\bf k})\equiv
g^2\left(C_F|C_{\theta}|\right)^{1/2}
\Biggl(\frac{{\rm Z}_{+}({\bf q})}{2}\Biggr)^{\!1/2}\!
\Biggl(\frac{{\rm Z}_t({\bf k})}{2\omega_{{\bf k}}^t}\Biggr)^{\!1/2}
\label{eq:8f}
\end{equation}
\[
\hspace{0.6cm}
\times
\Bigr[\,\bar{v}_{\alpha}(-\hat{\bf q},\lambda)
K_{\alpha}^{(Q)i}({\bf v},\chi|\,-k,q)
{\rm e}^{i}(-\hat{\bf k},\xi)
\Bigr]_{{\rm on-shell}\,}.
\]
Confronting expression obtained (\ref{eq:8s}) with the corresponding terms
in (\ref{eq:8t}), one identifies the desired probabilities
${\it w}_{\,q\rightarrow {\rm g}}^{(+;\,t)}$ and
${\it w}_{\,\bar{q}\rightarrow {\rm g}}^{(+;\,t)}$:
\begin{equation}
{\it w}_{\,q\rightarrow {\rm g}}^{(+,\,t)}
({\bf p}|\,{\bf q};{\bf k})=
\sum\limits_{\lambda\,=\,\pm\,}\sum\limits_{\,\xi\,=\,1,\,2}
\left|\,{\cal M}_{\lambda\xi}^{(+\!,\,t)}({\bf v},\chi|\,{\bf q};{\bf k})
\right|^{\,2},
\label{eq:8g}
\end{equation}
\[
{\it w}_{\,\bar{q}\rightarrow {\rm g}}^{(+,\,t)}
({\bf p}|\,{\bf q};{\bf k})=
\sum\limits_{\lambda\,=\,\pm\,}\sum\limits_{\,\xi\,=\,1,\,2}
\left|\,\bar{\cal M}_{\lambda\xi}^{(+\!,\,t)}({\bf v},\chi|\,{\bf q};{\bf k})
\right|^{\,2}.
\]
The probabilities of the scattering processes with (anti)plasmino and plasmon
are obtained from (\ref{eq:8d})\,--\,(\ref{eq:8g}) with the help of corresponding
replacements such as
\begin{equation}
\left(\!\frac{{\rm Z}_{+}({\bf q})}{2}\!\right)^{\!\!1/2}\!
u_{\alpha}(\hat{\bf q},\lambda)
\rightarrow
\left(\!\frac{{\rm Z}_{-}({\bf q})}{2}\!\right)^{\!\!1/2}\!
v_{\alpha}(\hat{\bf q},\lambda),
\label{eq:8gg}
\end{equation}
\[
\hspace{0.6cm}
\left(\frac{{\rm Z}_t({\bf k})}{2\omega_{{\bf k}}^t}\right)^{\!1/2}\!
{\rm e}^{i}(\hat{\bf k},\xi)
\rightarrow
\left(\frac{{\rm Z}_l({\bf k})}{2\omega_{\bf k}^l}\right)^{\!1/2}\!
\Biggl(\frac{\bar{u}^{\mu}(k)}{\sqrt{\bar{u}^2(k)}}\Biggr),
\]
etc. and proper choice of mass-shell conditions on the right-hand side of
Eqs.\,(\ref{eq:8d}), (\ref{eq:8f}).

Now we turn to deriving the probabilities for the elastic scattering of
soft-quark and soft-antiquark modes off the hard test particle. As in the
previous case we restrict our consideration only to the normal quark
excitations. In the second term on the right-hand side of (\ref{eq:8i})
we perform the following replacements:
\[
\tilde{\bar{\eta}}^{\,i\,b}_{\alpha}({\bf v},\bar{\chi}|\,-q)
\rightarrow
\tilde{\bar{\eta}}^{\,(1)i\,b}_{\alpha}(\bar{\psi}^{(0)})(-q),
\quad
\tilde{\eta}^{\,i\,b^{\prime}}_{\alpha^{\prime}}({\bf v},\chi|\,q)
\rightarrow
\tilde{\eta}^{\,(1)\,i\,b^{\prime}}_{\alpha^{\prime}}(\psi^{(0)})(q),
\]
where the effective sources $\tilde{\bar{\eta}}^{\,(1)i\,b}_{\alpha}$,
$\tilde{\eta}^{\,(1)i\,b^{\prime}}_{\alpha^{\prime}}$ are defined by
Eqs.\,(\ref{eq:5d}), (\ref{eq:5f}) and (\ref{eq:4e}), (\ref{eq:5s})
correspondingly. As a result of this kind replacement the second term on
the right-hand side of Eq.\,(\ref{eq:8i}) is written in the following form:
\begin{equation}
\frac{1}{2\pi}\,g^4\!\lim\limits_{\tau\rightarrow\infty}
\frac{1}{\tau}\,
\Biggl(\sum\limits_{\,\zeta=Q,\,\bar{Q},\,G}
\int\!\frac{|{\bf p}|^{\,2}\,d|{\bf p}|}{2\pi^2}\,
f_{\bf p}^{(\zeta)}\!\Biggr)\!
\left(t^bt^{b^{\prime}}\right)^{jj^{\prime}}\!\!
\!\int\!dQ_0\,Q_0^bQ_0^{b^{\prime}}\!\int\!\frac{d\Omega_{\bf v}}{4\pi}\,
\label{eq:8h}
\end{equation}
\[
\times\sum\limits_{\lambda\,=\,\pm}
\int\!\!d{\bf q}\,\,\omega_{\bf q}^{+}
\Biggl(\frac{{\rm Z}_{+}({\bf q})}{2}\Biggr)
\biggl\{\Bigl[\,
\Bigl(\bar{K}_{\alpha^{\prime}\alpha}^{(Q)}(\bar{\chi},\chi|\,q,-q_1)
u_{\alpha}(\hat{\bf q},\lambda)\Bigr)
\Bigl(\bar{u}_{\beta}(\hat{\bf q},\lambda)
K_{\beta\beta^{\prime}}^{(Q)}(\chi,\bar{\chi}|\,q,-q^{\,\prime\,}_1)\Bigr)
\]
\[
\hspace{0.5cm}
\times\,
\delta(v\cdot(q-q_1))\delta(v\cdot(q-q_1^{\,\prime\,}))
\Bigr]_{q^0\,=\,\omega_{\bf q}^{+}}\,+
\]
\[
\Bigl[
\Bigl(\bar{K}_{\alpha^{\prime}\alpha}^{(Q)}(\bar{\chi},\chi|\,q,-q_1)
v_{\alpha}(\hat{\bf q},\lambda)\Bigr)\!
\Bigl(\bar{v}_{\beta}(\hat{\bf q},\lambda)
K_{\beta\beta^{\prime}}^{(Q)\mu^{\prime}}
(\chi,\bar{\chi}|\,q,-q^{\,\prime\,}_1)\Bigr)
\delta(v\cdot(q-q_1))\delta(v\cdot(q-q_1^{\,\prime\,}))
\Bigr]_{q^0=-\omega_{\bf q}^{+}}\!\biggr\}
\]
\[
\times
\left\langle \bar{\psi}^{(0)j}_{\alpha^{\prime}}(-q_1)
\psi^{(0)j^{\prime}}_{\beta^{\prime}}\!(q_1^{\,\prime\,})
\right\rangle\,dq_1dq_1^{\,\prime}\,.
\]
Under the conditions of stationary and homogeneous state of QGP the
correlation function for random soft-quark field can be presented
as
\begin{equation}
\left\langle \bar{\psi}^{(0)j}_{\alpha^{\prime}}(-q_1)
\psi^{(0)j^{\prime}}_{\beta^{\prime}}\!(q_1^{\,\prime\,})
\right\rangle=
\delta^{jj^{\prime}}\Upsilon_{\beta^{\prime}\alpha^{\prime}}
(q_1^{\,\prime\,})\,\delta(q_1^{\,\prime}-q_1),
\label{eq:8j}
\end{equation}
where in the spectral density $\Upsilon_{\beta^{\prime}\alpha^{\prime}}
(q_1^{\,\prime\,})$ we keep only the normal quark part in the form of the
quasiparticle approximation
\begin{equation}
\Upsilon_{\beta^{\prime}\alpha^{\prime}}
(q_1^{\,\prime\,})\rightarrow\frac{1}{(2\pi)^3}
\Biggl(\frac{{\rm Z}_{+}({\bf q}_1^{\,\prime\,})}{2}\Biggr)\!
\sum\limits_{\lambda_1=\pm}\Bigl[\,
u_{\beta^{\prime}}(\hat{\bf q}_1^{\,\prime},\lambda_1)
\bar{u}_{\alpha^{\prime}}(\hat{\bf q}_1^{\,\prime},\lambda_1)
\,n_{{\bf q}_1^{\,\prime}}^{+}\,
\delta(q_1^{\,\prime \,0}-\omega_{{\bf q}_1^{\,\prime}}^{+})
\hspace{1cm}
\label{eq:8k}
\end{equation}
\[
\hspace{5.3cm}
+\,v_{\beta^{\prime}}(\hat{\bf q}_1^{\,\prime},\lambda_1)
\bar{v}_{\alpha^{\prime}}(\hat{\bf q}_1^{\,\prime},\lambda_1)
\,(1-\bar{n}_{-{\bf q}_1^{\,\prime}}^{+})\,
\delta(q_1^{\,\prime \,0}+\omega_{{\bf q}_1^{\,\prime}}^{+})
\Bigr].
\]
Within the accepted accuracy we also can replace the equilibrium
number densities
$n_{{\bf q}_1^{\prime}}^{+},\,\bar{n}_{-{\bf q}_1^{\prime}}^{+} $
by off-equilibrium ones slowly depending on $x$.

Furthermore, making use of the formula for color averaging (\ref{eq:8q}),
triviality in a color space of the correlation function
(Eq.\,(\ref{eq:8j})), and the equality ${\rm tr}(t^at^a)=d_AT_F$, where
$T_F$ is index of the fundamental representation, we find that
the color factor in (\ref{eq:8h}) equals
$T_F\,C_2^{(\zeta)},\,\zeta=Q,\,\bar{Q},\,G$. Taking into account the
above-mentioned and performing integration with respect to
$dq_1^{\,\prime}dq_1^0$, we define final expression for the desired emitted
power instead of (\ref{eq:8h})
\[
g^4\,T_F
\Biggl(
\sum\limits_{\,\zeta=Q,\,\bar{Q},\,G}\!\!C_2^{(\zeta)}\!
\int\!\frac{|{\bf p}|^{2}\,d|{\bf p}|}{2\pi^2}\,f_{\bf p}^{(\zeta)}\!
\Biggr)\!
\int\!\frac{d\Omega_{\bf v}}{4\pi}
\int\!\frac{d{\bf q}}{(2\pi)^3}
\int\!\frac{d{\bf q}_1}{(2\pi)^3}\,\,
\omega_{\bf q}^{+}
\Biggl(\frac{{\rm Z}_{+}({\bf q})}{2}\Biggr)
\Biggl(\frac{{\rm Z}_{+}({\bf q}_1)}{2}\Biggr)
\]
\begin{equation}
\times
\sum\limits_{\lambda\,=\,\pm}
\sum\limits_{\lambda_1\,=\,\pm}
\biggl\{n_{{\bf q}_1}^{+}\,
\Bigl|\,\bar{u}_{\alpha}(\hat{\bf q},\lambda)
K_{\alpha\alpha_1}^{(Q)}(\chi,\bar{\chi}|\,q,-q_1)
u_{\alpha_1}(\hat{\bf q}_1,\lambda_1)\Bigr|^{\,\,2}_{\,{\rm on-shell}}
\hspace{0.35cm}
\label{eq:8l}
\end{equation}
\[
\hspace{1cm}
+\,
(1-\bar{n}_{{\bf q}_1}^{+})
\Bigl|\,\bar{v}_{\alpha}(-\hat{\bf q},\lambda)
K_{\alpha\alpha_1}^{(Q)}(\chi,\bar{\chi}|\,-q,q_1)
v_{\alpha_1}(-\hat{\bf q}_1,\lambda_1)\Bigr|^{\,\,2}_{\,{\rm on-shell}}
\biggr\}
\]
\[
\times
\,2\pi
\delta(\omega_{\bf q}^{+}-\omega_{{\bf q}_1}^{+}
-{\bf v}\cdot({\bf q}-{\bf q}_1)).
\]
In deriving this expression we have dropped the contribution containing
delta-function
$\delta(\omega_{\bf q}^{+}+\omega_{{\bf q}_1}^{+}
-{\bf v}\cdot({\bf q}+{\bf q}_1))$ and for the second term in braces we have
made the replacement of integration variables:
${\bf q}\rightarrow -{\bf q}$ and ${\bf q}_1\rightarrow -{\bf q}_1$. We
introduce the following matrix elements for soft-quark\,--\,hard-particle and
soft-antiquark\,--\,\,hard-particle `elastic' scatterings, respectively
\begin{equation}
{\cal M}_{\lambda\lambda_1}^{(\zeta)(+,\,+)}
(\chi,\bar{\chi}|\,{\bf q};{\bf q}_1)
\equiv
g^2\left(T_F\,C_2^{(\zeta)}\right)^{1/2}
\Biggl(\frac{{\rm Z}_{+}({\bf q})}{2}\Biggr)^{\!1/2}\!
\Biggl(\frac{{\rm Z}_{+}({\bf q}_1)}{2}\Biggr)^{\!1/2}\!
\label{eq:8z}
\end{equation}
\[
\times\,
\Bigl[\,\bar{u}_{\alpha}(\hat{\bf q},\lambda)
K_{\alpha\alpha_1}^{(Q)}(\chi,\bar{\chi}|\,q,-q_1)
u_{\alpha_1}(\hat{\bf q}_1,\lambda_1)
\Bigr]_{\,{\rm on-shell}\,},
\vspace{0.5cm}
\]
\[
\bar{\cal M}_{\lambda\lambda_1}^{(\zeta)(+,\,+)}
(\chi,\bar{\chi}|\,{\bf q};{\bf q}_1)
\equiv
g^2\left(T_F\,C_2^{(\zeta)}\right)^{1/2}
\Biggl(\frac{{\rm Z}_{+}({\bf q})}{2}\Biggr)^{\!1/2}\!
\Biggl(\frac{{\rm Z}_{+}({\bf q}_1)}{2}\Biggr)^{\!1/2}\!
\]
\[
\times\,
\Bigl[\,\bar{v}_{\alpha}(-\hat{\bf q},\lambda)
K_{\alpha\alpha_1}^{(Q)}(\chi,\bar{\chi}|\,-q,q_1)
v_{\alpha_1}(-\hat{\bf q}_1,\lambda_1)\Bigr]_{\,{\rm on-shell}\,}.
\]
Comparing the expression obtained (\ref{eq:8l}) with the corresponding terms
in (\ref{eq:8t}), we identify the desired probabilities
\begin{equation}
{\it w}_{\,q\rightarrow q}^{(\zeta)(+,\,+)}
({\bf p}|\,{\bf q};{\bf q}_1)=
\sum\limits_{\lambda\,=\,\pm\,}\sum\limits_{\,\lambda_1\,=\,\pm}
\left|\,{\cal M}_{\lambda\lambda_1}^{(\zeta)(+,\,+)}
(\chi,\bar{\chi}|\,{\bf q};{\bf q}_1)
\right|^{\,2},
\label{eq:8x}
\end{equation}
\[
{\it w}_{\,\bar{q}\rightarrow \bar{q}}^{(\zeta)(+,\,+)}
({\bf p}|\,{\bf q};{\bf q}_1)=
\sum\limits_{\lambda\,=\,\pm\,}\sum\limits_{\,\lambda_1\,=\,\pm}
\left|\,\bar{\cal M}_{\lambda\lambda_1}^{(\zeta)(+,\,+)}
(\chi,\bar{\chi}|\,{\bf q};{\bf q}_1)
\right|^{\,2}.
\]
These probabilities depend on type of the hard test particle (through the
Casimirs
$C_2^{(\zeta)}$) on which the scattering of soft-quark modes takes place.
The probabilities of the scattering processes with participation of plasmino
and antiplasmino can be derived from (\ref{eq:8x}), (\ref{eq:8z}) with the help
of the corresponding replacements of the quark wave functions and mass-shell
conditions.

We demand that scattering probabilities (\ref{eq:8x}) satisfy balance
relations (\ref{eq:2u}) for the direct and reverse scattering processes.
A straightforward calculation shows that these relations take place only
under fulfilment of the following conditions:
\[
\alpha^{\ast}=\alpha,\quad
\,^{\ast}\Gamma^{(Q)\mu}(q-q_1;-q,q_1)=
\,^{\ast}\Gamma^{(Q)\mu}(q-q_1;q_1,-q).
\]
The first condition implies that the constant $\alpha$ in definition of
additional source (\ref{eq:5a}) is real. The second condition holds only
when the linear Landau damping of soft-quark on-shell excitations is absent
in the medium.

\section{\bf Structure of scattering probability
${\it w}_{\,q\rightarrow {\rm g}}^{(f;\,b)}$.
Determination of constant $C_{\theta}$}
\setcounter{equation}{0}

In this section we consider in more details a structure of
the scattering probability
${\it w}_{\,q\rightarrow {\rm g}}^{(f;\,b)}({\bf v};\,{\bf q},{\bf k})$
obtained in the previous section and define also an explicit value of the
constant $C_{\theta}\,(\equiv\theta_0^{\dagger i}\theta_0^i)$. Here we make 
use some results of our early work \cite{markov_PRD_01}. For the sake of
convenience of further references all required formulae from this work are
given in Appendix F.

First of all we recall that in the paper \cite{markov_PRD_01} the scattering
probability of plasmino off hard parton with transition to plasmon
${\it w}_{\,q\rightarrow {\rm g}}^{(-;\,l)}({\bf v}|\,{\bf q};{\bf k})$
was defined in a different way by a direct calculation of the following
expression:
\[
{\rm Im}\,{\rm Sp}\!
\left[\,^{\ast}\tilde{\Gamma}^{(Q)a_1a_2,\,ij}_{\mu_1\mu_2}
(k_1,k_2;q_1,-q)\,\delta^{a_1a_2}
\bar{u}^{\mu_1}(k_1)\bar{u}^{\mu_2}(k_2)\,
h_{-}(\hat{\bf q})\right]_{k_1=-k_2=k,\,\,q_1=q}\,,
\]
where `${\rm Sp}$' denotes the Dirac trace and $\bar{u}^{\mu}(k)$ is
the longitudinal projector in the covariant gauge (Eq.\,(C.6)). The
effective amplitude
$\,^{\ast}\tilde{\Gamma}^{(Q)a_1a_2,\,ij}_{\mu_1\mu_2}$ was
defined in Paper I (Eq.\,(I.4.9)). The amplitude determines the
elastic scattering process of soft-quark excitation on soft-gluon
excitation without momentum-energy exchange with hard thermal
partons. Within the framework of the approach \cite{markov_PRD_01}
for description of the scattering process under consideration
there was no necessity to introduce the Grassmann color
charges $\theta_0^{\dagger i}$ and $\theta_0^i$ of hard parton
(and consequently, the constant $C_{\theta}$). Comparing, for
example, the kinetic equations for the plasmino number density $n_{\bf
q}^{-}$ obtained in these two approaches, one can define unknown constant
$C_{\theta}$. However, preliminary we recast the scattering
probability ${\it w}_{\,q\rightarrow {\rm g}}^{(-;\,l)}$ in the
form suggested in Ref.\,\cite{markov_PRD_01}. This makes it possible
in particular correctness of unusual at first sight
structure of the scattering kernel $Q({\bf q},{\bf k})$ given by
Eqs.\,(F.9)\,--\,(F.11) to be independently proved.

At the beginning we write out an explicit form of the matrix elements
defining the scattering processes of plasmino and antiplasmino off
hard parton with the subsequent transition in plasmon. For this
purpose we perform replacements (\ref{eq:8gg}) in Eqs.\,(\ref{eq:8d}),
(\ref{eq:8f}) and relevant replacements of the mass-shell
conditions. Here it is more convenient to use the temporal gauge.
This means that in last replacement in Eq.\,(\ref{eq:8gg}) instead of
the projector $\bar{u}^{\mu}(k)$ it is necessary to use the projector
$\tilde{u}^{\mu}(k)\equiv k^2(u^{\mu}(k\cdot u)-k^{\mu})/(k\cdot u)$. In
the rest system $u^{\mu}=(1,0,0,0)$, we get
\begin{equation}
{\cal M}_{\lambda}^{(-\!,\,l)}({\bf v},\chi|\,{\bf q};{\bf k})\equiv
g^2\left(C_F|C_{\theta}|\right)^{1/2}
\Biggl(\frac{{\rm Z}_{-}({\bf q})}{2}\Biggr)^{\!1/2}\!
\Biggl(\frac{{\rm Z}_l({\bf k})}{2\omega_{{\bf k}}^l}\Biggr)^{\!1/2}\!
\Biggl(\frac{k^2}{{\bf k}^2(\omega_{{\bf k}}^l)^2}\Biggr)^{\!1/2}
\label{eq:9q}
\end{equation}
\[
\times
\Bigr[\,\bar{v}_{\alpha}(\hat{\bf q},\lambda)
\left(K_{\alpha}^{(Q)i}({\bf v},\chi|\,k,-q)
k^{i}\right)\Bigr]_{{\rm on-shell}\,,}
\]
\begin{equation}
\bar{\cal M}_{\lambda}^{(-\!,\,l)}({\bf v},\chi|\,{\bf q};{\bf k})\equiv
g^2\left(C_F|C_{\theta}|\right)^{1/2}
\Biggl(\frac{{\rm Z}_{-}({\bf q})}{2}\Biggr)^{\!1/2}\!
\Biggl(\frac{{\rm Z}_l({\bf k})}{2\omega_{{\bf k}}^l}\Biggr)^{\!1/2}\!
\Biggl(\frac{k^2}{{\bf k}^2(\omega_{{\bf k}}^l)^2}\Biggr)^{\!1/2}
\label{eq:9w}
\end{equation}
\[
\hspace{0.7cm}
\times
\Bigr[\,\bar{u}_{\alpha}(-\hat{\bf q},\lambda)
\left(K_{\alpha}^{(Q)i}({\bf v},\chi|\,-k,q)(-k^i)\right)
\Bigr]_{{\rm on-shell}\,.}
\]
Let us consider at first matrix element (\ref{eq:9q}). Making use of
the definition of the coefficient function $K_{\alpha}^{(Q)\mu}$,
Eq.\,(\ref{eq:4y}), we write out the function
$\bar{v}_{\alpha}(\hat{\bf q},\lambda)(K_{\alpha}^{(Q)i}k^{i})$
in an explicit form (for the sake of brevity we suppress spinor indices)
\begin{equation}
\frac{({\bf v}\cdot {\bf k})}{v\cdot q}\,
(\bar{v}(\hat{\bf q},\lambda)\chi)
-\bar{v}(\hat{\bf q},\lambda)
[\,^{\ast}\Gamma^{(Q)i}(k;q-k,-q)k^i\,]
\,^{\ast}\!S(q-k)\chi=
\frac{({\bf v}\cdot {\bf k})}{v\cdot q}\,
(\bar{v}(\hat{\bf q},\lambda)\chi)
\label{eq:9e}
\end{equation}
\[
+\,\bar{v}(\hat{\bf q},\lambda)\Bigl[
\,h_{-}(\hat{\bf l})
(\!\,^{\ast}\!{\it \Gamma}_{+}^{\,i}k^i)
+h_{+}(\hat{\bf l})
(\!\,^{\ast}\!{\it \Gamma}_{-}^{\,i}k^i)
-2h_{-}(\hat{\bf q})\,{\bf l}^2 \vert {\bf q}\vert
(\!\,^{\ast}\!{\it \Gamma}_{\perp}^{\,i}k^i)
\Bigr]
\Bigl[\,h_{+}(\hat{\bf l}) \,^{\ast}\!\Delta_{+}(l) +
h_{-}(\hat{\bf l}) \,^{\ast}\!\Delta_{-}(l)
\Bigr]\chi\,,
\]
where $l\equiv q-k$. On the right-hand side of the last equation we have used
expansion (F.3) for the vertex $\!\,^{\ast}\Gamma^{(Q)i}$, the properties
(F.2), and the representation
of the soft-quark propagator $\,^{\ast}\!S(l)$ in the form of an expansion
in terms of the
spinor projectors $h_{\pm}(\hat{\bf q})$. Now we multiply together two
expressions in square brackets taking into account the property of nilpotency
$h_{\pm}(\hat{\bf l})h_{\pm}(\hat{\bf l})=0$ and the fact that the wave
function $\bar{v}(\hat{\bf q},\lambda)$ satisfies the equation
$\bar{v}(\hat{\bf q},\lambda)h_{-}(\hat{\bf q})=0$. As a result we obtain
the second term on the right-hand side of Eq.\,(\ref{eq:9e}) in more simple
and symmetric form:
\[
\left[\,\bar{v}(\hat{\bf q},\lambda)h_{-}(\hat{\bf l})
h_{+}(\hat{\bf l})\chi\right]
(\!\,^{\ast}\!{\it \Gamma}_{+}^{\,i}k^i)\,^{\ast}\!\Delta_{+}(l)
+
\left[\,\bar{v}(\hat{\bf q},\lambda)h_{+}(\hat{\bf l})
h_{-}(\hat{\bf l})\chi\right]
(\!\,^{\ast}\!{\it \Gamma}_{-}^{\,i}k^i)\,^{\ast}\!\Delta_{-}(l).
\]
Let us emphasize that the simplification is a direct consequence
of a choice of the expansion (F.3) for convolution
$\!\,^{\ast}\Gamma^{(Q)i}k^i$. The first term on the right-hand
side of Eq.\,(\ref{eq:9e}) can be also written down in a symmetric
form with respect to the matrixes $h_{\pm}(\hat{\bf l})$ if we use an identity
\begin{equation}
1=h_{-}(\hat{\bf l})h_{+}(\hat{\bf l})+
h_{+}(\hat{\bf l})h_{-}(\hat{\bf l}).
\label{eq:9r}
\end{equation}

Furthermore, we collect similar terms and square of the absolute value of
the expression (\ref{eq:9e}). Summing over polarization states of soft-quark
excitations, we obtain an initial expression for the subsequent analysis
\[
\Biggl(\,\sum\limits_{\lambda=\pm}\!
|\,\bar{v}(\hat{\bf q},\lambda)h_{-}(\hat{\bf l})h_{+}(\hat{\bf l})\chi|^{\,2}
\Biggr)\!
\left|\,{\cal M}_{+}^{(-,\,l)}({\bf q},{\bf k})\right|^{\,2}+
\Biggl(\,\sum\limits_{\lambda=\pm}\!
|\,\bar{v}(\hat{\bf q},\lambda)h_{+}(\hat{\bf l})h_{-}(\hat{\bf l})\chi|^{\,2}
\Biggr)\!
\left|\,{\cal M}_{-}^{(-,\,l)}({\bf q},{\bf k})\right|^{\,2}
\]
\begin{equation}
+\,2\sum\limits_{\lambda\,=\,\pm}{\rm Re}\,
\Bigl\{[\bar{v}(\hat{\bf q},\lambda)h_{-}(\hat{\bf l})h_{+}(\hat{\bf l})\chi]
[\bar{v}(\hat{\bf q},\lambda)h_{+}(\hat{\bf l})h_{-}(\hat{\bf l})\chi]^{\ast}
{\cal M}_{+}^{(-,\,l)}({\bf q},{\bf k})
{\cal M}_{-}^{\ast\,(-,\,l)}({\bf q},{\bf k})
\Bigr\}_{\,,}
\label{eq:9t}
\end{equation}
where
\begin{equation}
{\cal M}_{\pm}^{(-,\,l)}({\bf q},{\bf k})\equiv
\frac{{\bf v}\cdot {\bf k}}{v\cdot q}
\,\,+\,^{\ast}\!\Delta_{\pm}(l)\!
\left(\,^{\ast}\!{\it \Gamma}_{\pm}^i(k;l,-q)k^i\right).
\label{eq:9y}
\end{equation}

Let us define an explicit form of coefficients of the scalar amplitudes
${\cal M}_{\pm}^{(-,\,l)}({\bf q},{\bf k})$. At first we consider the
coefficient of $|{\cal M}_{+}^{(-,\,l)}({\bf q},{\bf k})|^2$. Taking into
account identity
(\ref{eq:9r}) and definition of density matrix for a fully unpolarized state of
the hard test (anti)quark (Eq.\,(C.2)), we have the following chain of
equalities:
\[
\sum\limits_{\lambda\,=\,\pm}\!
|\,\bar{v}(\hat{\bf q},\lambda)h_{-}(\hat{\bf l})h_{+}(\hat{\bf l})\chi|^{\,2}
=
-\bar{\chi}h_{-}(\hat{\bf l})h_{+}(\hat{\bf l})
h_{-}(\hat{\bf q})h_{-}(\hat{\bf l})h_{+}(\hat{\bf l})\chi
+
\bar{\chi}h_{-}(\hat{\bf q})h_{-}(\hat{\bf l})h_{+}(\hat{\bf l})\chi
\]
\begin{equation}
=-\frac{1}{2E}\left\{{\rm Sp}\!\left[\,
h_{-}(\hat{\bf l})h_{+}(\hat{\bf l})
h_{-}(\hat{\bf q})h_{-}(\hat{\bf l})h_{+}(\hat{\bf l})\varrho({\bf v})\right]
-{\rm Sp}\!\left[\,
h_{-}(\hat{\bf q})h_{-}(\hat{\bf l})h_{+}(\hat{\bf l})\varrho({\bf v})\right]
\right\}
\label{eq:9u}
\end{equation}
\[
=-\frac{1}{2E}\left\{\frac{1}{2}\,
({\bf v}\cdot(\hat{\bf l}\times(\hat{\bf q}\times\hat{\bf l}))
-\frac{1}{2}\,
\rho_{+}({\bf v};\hat{\bf q},\hat{\bf l})\right\},
\]
where the function $\rho_{+}({\bf v};\hat{\bf q},\hat{\bf l})$ is
defined by Eq.\,(F.10). The coefficient of $|{\cal
M}_{-}^{(-,\,l)}({\bf q},{\bf k})|^2$ in (\ref{eq:9t}) is
defined from the coefficient above by formal
replacement $\hat{\bf l}\rightarrow -\hat{\bf l}$. This reduces to
a simple replacement $\rho_{+}\rightarrow\rho_{-}$ on the
rightmost expression in Eq.\,(\ref{eq:9u}). Finally, it is not
difficult to see that calculation of the coefficient in the
interference term of Eq.\,(\ref{eq:9t}) is reduced to calculation
of the first trace on the second line of Eq.\,(\ref{eq:9u}) and
therefore it is equal to
\[
-\,\frac{1}{2}\,({\bf v}\cdot(\hat{\bf l}\times(\hat{\bf q}\times\hat{\bf l}))
\,\Bigl(\equiv-\frac{1}{2}\,\sigma({\bf v};\hat{\bf q},\hat{\bf l})\Bigr).
\]
Taking into account the above-mentioned and collecting similar terms at
the function $\sigma({\bf v};\hat{\bf q},\hat{\bf l})$, we define instead of
(\ref{eq:9t})
\begin{equation}
\frac{1}{2E}\left\{
\frac{1}{2}\,\rho_{+}({\bf v};\hat{\bf q},\hat{\bf l})
\left|\,{\cal M}_{+}^{(-,\,l)}({\bf q},{\bf k})\right|^{\,2}+
\frac{1}{2}\,\rho_{-}({\bf v};\hat{\bf q},\hat{\bf l})
\left|\,{\cal M}_{-}^{(-,\,l)}({\bf q},{\bf k})\right|^{\,2}
\right.
\label{eq:9i}
\end{equation}
\[
\left.
-\frac{1}{2}\,\sigma({\bf v};\hat{\bf q},\hat{\bf l})
\left|\,{\cal M}_{+}^{(-,\,l)}({\bf q},{\bf k})-
{\cal M}_{-}^{(-,\,l)}({\bf q},{\bf k})\right|^{\,2}\right\}.
\]
A structure of this expression differs from a structure of
the integrand of the scattering kernel (F.9) only by the last
term with the coefficient $\sigma({\bf v};\hat{\bf q},\hat{\bf
l})$. However, this term vanishes (see below) when we average
over the directions of the velocity ${\bf v}$ of the hard test parton.
Thus we have proved by another way a correctness of a structure of the
scattering kernel $Q({\bf q},{\bf k})$ obtained in
\cite{markov_PRD_01} by a direct calculation of the imaginary part of the
effective amplitude for the plasmino-plasmon elastic scattering.

Now we turn our attention to matrix element (\ref{eq:9w})
defining the scattering process with participation of
antiplasmino. By using a property of the coefficient function
$K^{(Q)\mu}$:
\[
K^{(Q)\mu}({\bf v},\chi|\,-k,q)=\gamma^0
\Bigl[\,\bar{K}^{(Q)\mu}({\bf v},\bar{\chi}|\,-k,q)
\Bigr]^{\dagger}
\]
and definition of the (Dirac) conjugate coefficient function
$\bar{K}^{(Q)\mu}$, Eq.\,(\ref{eq:4o}), we find that the modulus
squared of expression in braces in matrix element
(\ref{eq:9w}) can be written in the following form:
\[
\Biggr|\frac{({\bf v}\cdot {\bf k})}{v\cdot q}\,
(\bar{\chi}\,u(-\hat{\bf q},\lambda))
-\bar{\chi}\,^{\ast}\!S(q-k)
[\,^{\ast}\Gamma^{(Q)i}(k;q-k,-q)k^i]
u(-\hat{\bf q},\lambda)\Biggr|^{\,2}.
\]
The advantage of this representation consists in the fact that both the
propagator and vertex here have the same signs at momenta as on the
left-hand side of Eq.\,(\ref{eq:9e}). This enables us at once to perform
calculations fully similar to previous ones and show that
finally we lead to the same expression (\ref{eq:9i}). Thus the
scattering probabilities of plasmino and antiplasmino off the hard
test particle are equal among themselves when the hard test particle is
in fully unpolarized state, i.e.,
\begin{equation}
{\it w}_{\,q\rightarrow {\rm g}}^{(-;\,l)}
({\bf p}|\,{\bf q};{\bf k})\!\!\Bigm|_{\,{\rm unpol.}}\,=\,
{\it w}_{\,\bar{q}\rightarrow {\rm g}}^{(-;\,l)}
({\bf p}|\,{\bf q};{\bf k})\!\!\Bigm|_{\,{\rm unpol.}}.
\label{eq:9o}
\end{equation}
It is evident that this conclusion is valid and for the scattering processes
with participation of soft normal quark and soft transverse gluon modes.

Let us consider now a question of determining the constant $C_{\theta}$.
For this purpose we rewrite kinetic equation (\ref{eq:2q}) for $f=-$ in the
following form:
\[
\frac{\partial n_{\bf q}^{-}}{\partial t} +
{\bf v}_{\bf q}^{-}\cdot\frac{\partial n_{\bf q}^{-}}{\partial {\bf x}} =
\Gamma_{\rm i}^{(-)}[n_{\bf q}^{\pm},N_{\bf k}^{t,\,l},f_{\bf p}^{G\!,\,Q}]
- n_{\bf q}^{-}\,\Bigl\{
\Gamma_{\rm i}^{(-)}[n_{\bf q}^{\pm},N_{\bf k}^{t,\,l},f_{\bf p}^{G\!,\,Q}]
+
\,\Gamma_{\rm d}^{(-)}[n_{\bf q}^{\pm},N_{\bf k}^{t,\,l},f_{\bf p}^{G\!,\,Q}]
\Bigr\}.
\]
The first and second terms on the right-hand side define the spontaneous and
induced  scattering processes correspondingly. Here we are interested in
the second term. This term is precisely one, which necessary for comparing 
with the right-hand side of the kinetic equation (F.7). Within the framework 
of approximations listed at the end of Section 2 this term equals
\begin{equation}
-\,2n_{\bf q}^{-}\!\left\{
\int\!\frac{|{\bf p}|\,d|{\bf p}|}{4\pi^2}\,
\Bigl[\,f_{\bf p}^{Q}+f_{\bf p}^{G}\Bigr]
\int\!\frac{d\Omega_{\bf v}}{4\pi}
\int\!d{\cal T}_{q\rightarrow {\rm g}}^{(-;\,l)}
\,{\it w}_{\,q\rightarrow {\rm g}}^{(-;\,l)}
({\bf v}|\,{\bf q};{\bf k}) N_{{\bf k}}^{l}\,+\,\ldots\,
\right\},
\label{eq:9p}
\end{equation}
where the dots denote the contributions of higher order in the coupling and
the contribution containing soft gluon transverse mode. We have somewhat 
redefined the scattering probability, having explicitly separated dependence 
on $E=|{\bf p}|$, setting by definition
\[
{\it w}_{\,q\rightarrow {\rm g}}^{(-;\,l)}
({\bf p}|\,{\bf q};{\bf k})=
\frac{1}{2E}\,{\it w}_{\,q\rightarrow {\rm g}}^{(-;\,l)}
({\bf v}|\,{\bf q};{\bf k}).
\]
For the global equilibrium plasma statistical factor is equal to
\[
\int\!\frac{|{\bf p}|\,d|{\bf p}|}{4\pi^2}\,
\Bigl[\,f_{\bf p}^{Q}+f_{\bf p}^{G}\Bigr]=\frac{T^2}{8}\,.
\]
Furthermore, the scattering probability ${\it w}_{\,q\rightarrow
{\rm g}}^{(-;\,l)}$ up to kinematic factors equals expression
obtained (\ref{eq:9i}). From the definitions of the scalar
amplitudes ${\cal M}_{\pm}^{(-,\,l)}({\bf q},{\bf k})$ it follows
that the difference ${\cal M}_{+}^{(-,\,l)}({\bf q},{\bf k})- {\cal
M}_{-}^{(-,\,l)}({\bf q},{\bf k})$ is independent of the velocity
${\bf v}$ of hard parton. Therefore substituting (\ref{eq:9i})
into (\ref{eq:9p}) and considering definition of the measure of
integration $\int\!d{\cal T}_{q\rightarrow {\rm g}}^{(-;\,l)}$
(Eq.\,(\ref{eq:2y}) with approximation (\ref{eq:2i})) we find that the last
term in (\ref{eq:9i}) gives contribution proportional to the
integral
\[
\int\!\frac{d\Omega_{\bf v}}{4\pi}\,
({\bf v}\cdot(\hat{\bf l}\times(\hat{\bf q}\times\hat{\bf l}))
\,\delta(\omega_{\bf q}^{-}-\omega_{\bf k}^{l}-
{\bf v}\cdot({\bf q}-{\bf k}))\,
\sim ({\bf l}\cdot(\hat{\bf l}\times(\hat{\bf q}\times\hat{\bf l}))=0.
\]
Considering the above-mentioned and comparing the term written out explicitly
in (\ref{eq:9p}) with the term on the right-hand side of the kinetic
equation (F.7), we find that
\[
C_{\theta}=-\,C_F.
\]

From the other hand we have kinetic equation (F.8) determining a change of
the plasmon number density $N_{\bf k}^l$. It contains on the right-hand side
the same scattering kernel $Q({\bf q},{\bf k})$ as the first equation (F.7)
with different common color multiplier. For determining a connection of
this equation with kinetic equation (\ref{eq:2w}), we write the last one for
$b=l$ rearranging the terms on the right-hand side:
\[
\frac{\partial N_{\bf k}^{l}}{\partial t} +
{\bf v}_{\bf k}^{l}\cdot\frac{\partial N_{\bf k}^{l}}{\partial {\bf x}} =
\Gamma_{\rm i}^{(l)}[n_{\bf q}^{\pm},N_{\bf k}^{t,\,l},f_{\bf p}^{G\!,\,Q}]
+ N_{\bf k}^{l}\,\Bigl\{
\Gamma_{\rm i}^{(l)}[n_{\bf q}^{\pm},N_{\bf k}^{t,\,l},f_{\bf p}^{G\!,\,Q}]
-
\,\Gamma_{\rm d}^{(l)}[n_{\bf q}^{\pm},N_{\bf k}^{t,\,l},f_{\bf p}^{G\!,\,Q}]
\Bigr\}.
\]
Here we are also interested only in the last term since it is necessary for
comparing with the right-hand side of kinetic equation (F.8). Within the
framework of the approximation used for Eq.\,(\ref{eq:9p}) this term is equal
to
\begin{equation}
2N_{\bf k}^{l}\!\left\{
\int\!\frac{|{\bf p}|\,d|{\bf p}|}{4\pi^2}\,
\Bigl[\,f_{\bf p}^{Q}+f_{\bf p}^{G}\Bigr]
\int\!\frac{d\Omega_{\bf v}}{4\pi}
\int\!d{\cal T}_{{\rm g}\rightarrow q}^{(l;\,-)}
\,{\it w}_{\,{\rm g}\rightarrow q}^{(l;\,-)}
({\bf v}|\,{\bf k};{\bf q})\,n_{\bf q}^{-} \,+\,\ldots\,
\right\},
\label{eq:9pp}
\end{equation}
where
\[
\int\!d{\cal T}_{{\rm g}\rightarrow q}^{(l;\,-)}\equiv
\int\!\!\frac{d{\bf q}}{(2\pi)^3}\,
2\pi\,\delta(\omega_{\bf k}^{l}-\omega_{\bf q}^{-}-
{\bf v}\cdot({\bf k}-{\bf q})).
\]
As the scattering probability ${\it w}_{\,{\rm g}\rightarrow q}^{(l;\,-)}
({\bf v}|\,{\bf k};{\bf q})$ here we mean the same expression (\ref{eq:9i}).
Carrying out reasonings completely similar previous ones and comparing
the term written out in (\ref{eq:9pp}) with the term on the right-hand side of
Eq.\,(F.8), we find in this case
\[
C_{\theta}=-\,n_fT_F.
\]
Thus we have a simple rule: {\it the constant $C_{\theta}$ entering into
definition of various probabilities in soft-quark decay and
regenerating rates {\rm (\ref{eq:2r}), (\ref{eq:2t})} should be
considered equal to $(-C_F)$, and in the probabilities of the
soft-gluon decay and regenerating rates in kinetic equation
{\rm (\ref{eq:2w})} it should be set equal to $(-n_fT_F)$.} The last case
is rather obvious since the factor $n_f$ takes into account the
number of possible with respect to flavour channels of interaction
of the soft-gluon excitations with soft-quark ones.

One can somewhat extend the previous results if we consider the
scattering of soft-quark excitations off hard particle taking
place in a partial polarization state. In this case instead of
polarization matrix (C.2) we should use more general expression
\cite{berestetski}
\begin{equation}
\varrho=\varrho({\bf v},\vec{\zeta}\,)=\frac{1}{2}\,(v\cdot\gamma)\!
\left[\,1+\gamma^5(\pm\,\zeta_{\parallel}
+\vec{\zeta}_{\perp}\cdot\vec{\gamma}_{\perp})\right],
\label{eq:9a}
\end{equation}
where $\gamma^5=-i\gamma^0\gamma^1\gamma^2\gamma^3$. The sign
$(+)$ concerns to quark and the sign $(-)$ belongs to antiquark.
The vector $\vec{\zeta}$ is a double average value of a spin
vector in the frame of rest of the test particle. $\zeta_{\parallel}$
is the vector component parallel to (for $\zeta_{\parallel}>0$) or
antiparallel (for $\zeta_{\parallel}<0$) to the momentum of particle;
$\vec{\zeta}_{\perp} =\vec{\zeta}-{\bf v}\,({\bf
v}\cdot\vec{\zeta}\,)$. Now we return to the problem of
calculation of the coefficients in Eq.\,(\ref{eq:9t}). The simple
calculations show that basic expression (\ref{eq:9i}) remains
invariable here, additional interference term only appears
\begin{equation}
\mp\,\zeta_{\parallel}\,\frac{({\bf v}\cdot{\bf n})}{|{\bf q}||{\bf l}|}
\,\,{\rm Im}\!
\left\{{\cal M}_{+}^{(-,\,l)}({\bf q},{\bf k})
{\cal M}_{-}^{\ast(-,\,l)}({\bf q},{\bf k})\right\}.
\label{eq:9s}
\end{equation}
The expression obtained (\ref{eq:9s}) suggests that if the hard test particle
is in a partial polarization state, then the scattering probability of
plasmino off hard gluon with transformation to plasmon and hard quark
(the process $qG\rightarrow{\rm g}Q$) in general case is not equal to
the scattering probability of annihilation of plasmino with hard
antiquark into plasmon and hard gluon (the process
$q\bar{Q}\rightarrow{\rm g}G$). Nevertheless, it is easy to see taking into
account a definition of the functions ${\cal M}_{\pm}^{(-,\,l)}$
in integrating expression (\ref{eq:9s}) over
\[
\int\!\frac{d\Omega_{\bf v}}{4\pi}
\,\delta(\omega_{\bf q}^{-}-\omega_{\bf k}^{l}-
{\bf v}\cdot({\bf q}-{\bf k})),
\]
this additional contribution vanishes. For this reason in the semiclassical
approximation in definition of generalized decay and
regenerating rates (\ref{eq:2r}), (\ref{eq:2t}) we did not make
distinction between the scattering processes such as mentioned above
and thus we have collected the statistical factors
$f_{\bf p}^{G} \Bigl(1-f_{{\bf p}^{\prime}}^{Q}\Bigr)$ and
$f_{\bf p}^{\bar{Q}} \Bigl(1+f_{{\bf p}^{\prime}}^{G}\Bigr)$ together.

If we consider the scattering process with participation of antiplasmino
$(\bar{q})$, then we result in the fact that the scattering probability
${\it w}_{\,\bar{q}\rightarrow {\rm g}}^{(-;\,l)}$ will be proportional
not to a sum of two expressions (\ref{eq:9i}) and (\ref{eq:9s}) but to their
difference. Therefore instead of equality (\ref{eq:9o}) we have more
nontrivial statements
\[
{\it w}_{\,q\bar{Q}\rightarrow {\rm g}G}^{(-;\,l)}
({\bf v}|\,{\bf q};{\bf k})\,=\,
{\it w}_{\,\bar{q}Q\rightarrow {\rm g}G}^{(-;\,l)}
({\bf v}|\,{\bf q};{\bf k}),
\]
\[
{\it w}_{\,qG\rightarrow {\rm g}Q}^{(-;\,l)}
({\bf v}|\,{\bf q};{\bf k})\,=\,
{\it w}_{\,\bar{q}G\rightarrow {\rm g}\bar{Q}}^{(-;\,l)}
({\bf v}|\,{\bf q};{\bf k}).
\]

\section{\bf Energy losses of energetic parton}
\setcounter{equation}{0}

In this section we give general formulae defining energy losses of 
high-energy parton (quark or gluon) traversing the hot QCD medium induced by 
scattering off soft-quark excitations. These formulae supplement expression 
for energy loss generated by the effective current 
$\tilde{j}_{Q\mu}^{Aa}[A^{(0)}](k)$
(Eq.\,(7.5) in Ref.\,\cite{markov_AOP_04}) and thus enables us to obtain 
complete (within a framework of semiclassical approximation) expressions for 
the energy losses of the energetic parton.

As a basic formula for parton energy losses per unit length generated by the
effective current $\tilde{j}_{\mu}^{\Psi a}({\bf v},\chi;Q_0,\theta_0|\,k)$ and
effective source $\tilde{\eta}^{\,i}_{\alpha}({\bf v},\chi;Q_0,\theta_0|\,q)$
we accept the following expression
\[
-\frac{dE}{dx}\,=
\left(\!-\frac{dE}{dx}\,\right)_{\!{\cal B}}+\,
\left(\!-\frac{dE}{dx}\,\right)_{\!{\cal F}}\,,
\]
where
\begin{equation}
\left(\!-\frac{dE}{dx}\,\right)_{\!{\cal B}}\equiv
-\frac{1}{\vert{\bf v}\vert}
\lim\limits_{\tau\rightarrow\infty}
\frac{(2\pi)^4}{\tau}
\int\!dQ_0\!\int\!k^0dk^0d{\bf k}
\label{eq:10q}
\end{equation}
\[
\biggl\{{\rm Im}(^{\ast}{\!\Delta}^t(k))\!\!
\sum\limits_{\xi=1,\,2}\!
\Bigl\langle\left\vert\,\tilde{\bf j}_{\mu}^{\Psi a}
({\bf v},\chi;Q_0,\theta_0|\,k)
\!\cdot\!{\bf e}(\hat{\bf k},\xi)\right\vert^{\,2}\Bigr\rangle\,+\,
\frac{k^2}{k_0^2}\,{\rm Im}(^{\ast}{\!\Delta}^l(k))
\Bigl\langle\left\vert\,\tilde{\bf j}_{\mu}^{\Psi a}
({\bf v},\chi;Q_0,\theta_0|\,k)
\!\cdot\!\hat{\bf k}\right\vert^{\,2}\Bigr\rangle
\!\biggr\},
\]
and
\begin{equation}
\left(\!-\frac{dE}{dx}\,\right)_{\!{\cal F}}\equiv
\frac{1}{\vert{\bf v}\vert}
\lim\limits_{\tau\rightarrow\infty}
\frac{(2\pi)^4}{\tau}
\sum\limits_{\lambda=\pm}
\int\!dQ_0\!\int\!q^0dq^0d{\bf q}
\label{eq:10w}
\end{equation}
\[
\times
\biggl\{{\rm Im}(^{\ast}{\!\Delta}_{+}(q))\,
\langle\vert\,\bar{u}(\hat{\bf q},\lambda)
\tilde{\eta}^{\,i}({\bf v},\chi;Q_0,\theta_0|\,q)
\vert^{\,2}\rangle\,+\,
{\rm Im}(^{\ast}{\!\Delta}_{-}(q))\,
\langle\vert\,\bar{v}(\hat{\bf q},\lambda)
\tilde{\eta}^{\,i}({\bf v},\chi;Q_0,\theta_0|\,q)
\vert^{\,2}\rangle
\!\biggr\}.
\]
The right-hand side of Eq.\,(\ref{eq:10q}) has been written in the temporal 
gauge.

\begin{figure}[hbtp]
\begin{center}
\includegraphics*[scale=0.5]{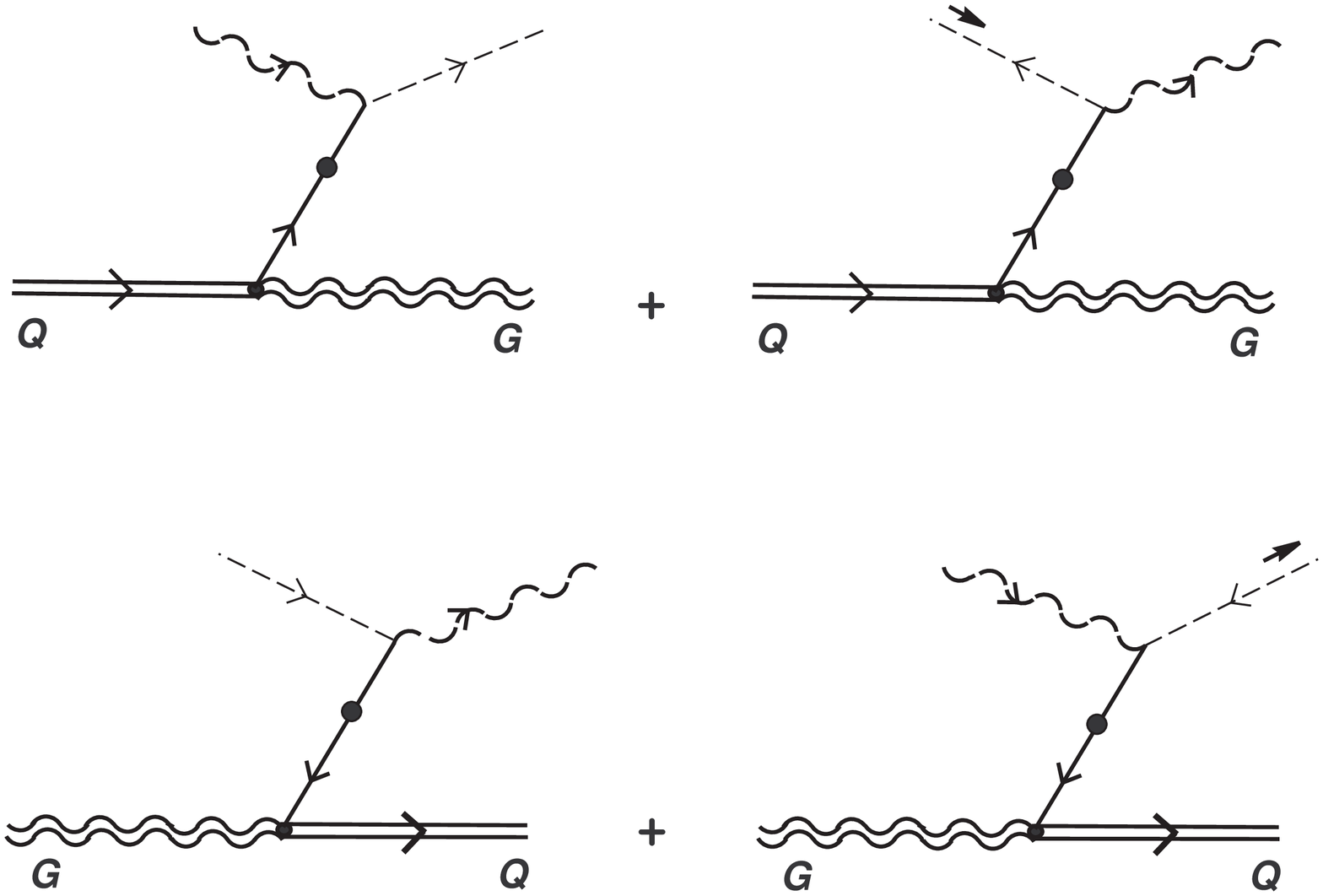}
\end{center}
\caption{\small The energy losses induced by the long-distance collision 
processes wherein a change of a type of the energetic parton takes place. The 
dotted lines denote thermal partons absorbing virtual soft-quark excitation.}
\label{fig12}
\end{figure}

First of all we write the expression for the energy loss associated
with the initial Grassmann color source
$\eta^{(0)\,i}_{\theta\,\alpha}(q)
=g/(2\pi)^3\theta_0^i\,\chi_{\alpha}\delta(v\cdot q)$.
Substituting this source into Eq.\,(\ref{eq:10w}) and taking into
account 
\[
\sum\limits_{\lambda\,=\,\pm}\!
|\,\bar{u}(\hat{\bf q},\lambda)\chi|^{\,2}
=\,\frac{1}{2E}\,{\rm Sp}\left[\,
h_{+}(\hat{\bf q})\varrho({\bf v})\right]
=\frac{1}{2E}\,(1-{\bf v}\cdot\hat{\bf q})\,,
\]
\[
\sum\limits_{\lambda\,=\,\pm}\!
|\,\bar{v}(\hat{\bf q},\lambda)\chi|^{\,2}
=\,\frac{1}{2E}\,{\rm Sp}\left[\,
h_{-}(\hat{\bf q})\varrho({\bf v})\right]
=\frac{1}{2E}\,(1+{\bf v}\cdot\hat{\bf q})\,,
\]
we obtain the energy loss to a zeroth-order in the soft fields
\begin{equation}
\left(\!-\frac{dE^{(0)}}{dx}\right)_{\!{\cal F}}
=\frac{1}{2E}\,\frac{1}{\vert{\bf v}\vert}\,
\Biggl(\frac{C_{\theta}\alpha_s}{2{\pi}^2}\Biggr)\!
\int\!q^0dq^0d{\bf q}\,
{\rm Im}\Bigl[\,{\rm Sp}(\varrho({\bf v})\,^{\ast}\!S(q))\Bigr]
\delta(v\cdot q) 
\label{eq:10e}
\end{equation}
\[
=\frac{1}{2E}\,\frac{1}{\vert{\bf v}\vert}\,
\Biggl(\frac{C_{\theta}\alpha_s}{2{\pi}^2}\Biggr)\!
\int\!q^0dq^0d{\bf q}
\,\Bigl\{(1-{\bf v}\cdot\hat{\bf q})\,
{\rm Im}(^{\ast}{\!\Delta}_{+}(q))\,+\,
(1+{\bf v}\cdot\hat{\bf q})\,
{\rm Im}(^{\ast}{\!\Delta}_{-}(q))
\Bigr\}\,\delta(v\cdot q),
\]
where $\alpha_s = g^2/4\pi$. The equation (\ref{eq:10e}) defines 
so-called the polarization losses of energetic parton related to large 
distance collisions. It supplements the known expression for the polarization 
losses \cite{braaten} induced by the `elastic' scattering off hard thermal 
particles through the exchange of soft virtual gluon. The graphic interpretation of
`inelastic' polarization losses (\ref{eq:10e}) is presented in 
Fig.\,\ref{fig12}. The energy losses (\ref{eq:10e}) decrease with the parton 
energy $E$ as $1/E$. Such suppression results in the fact that polarization
losses (\ref{eq:10e}) are negligible in comparison with usual ones 
\cite{braaten} for asymptotically large parton energy. Nevertheless we can
hope that contribution (\ref{eq:10e}) is important for intermediate values 
of $E$.  

The polarization losses give a leading contribution in the coupling constant, 
provided that the QGP is in thermal equilibrium. However, we can
expect that for rather high level of intensity of plasma 
excitations\footnote{The level of
intensity of plasma excitations is defined by values of the soft-quark and 
soft-gluon occupation numbers.} contributions to energy losses associated with 
the following terms in expansions of the effective current
$\tilde{j}^{\Psi a}_{\mu}$ and source $\tilde{\eta}_{\alpha}^{\,i}$ become
comparable with the polarization losses. Therefore these contributions
to an overall balance of energy losses should be taken into account.
Below we write out general expressions for the energy losses to the
next-to-leading order.

In present and forthcoming sections we restrict our consideration to the 
analysis of expression (\ref{eq:10w}). The analysis of expression 
(\ref{eq:10q}) will be given in Section 12. Let us take the effective sources 
$\tilde{\eta}_{\alpha}^{\,ib}$ and $\tilde{\eta}_{\alpha}^{\,ij}$ on the 
right-hand side of the last equation in (\ref{eq:8w}) up to the second order
approximation in powers of the soft free fields $A^{(0)}$, $\bar{\psi}^{(0)}$ 
and $\psi^{(0)}$:
\begin{equation}
\tilde{\eta}^{\,ib}_{\alpha}({\bf v},\chi|\,q)\,\simeq\,
\tilde{\eta}^{(1)\,ib}_{\alpha}(\psi^{(0)})(q)+
\tilde{\eta}^{(2)ib}_{\alpha}(A^{(0)},\psi^{(0)})(q),
\hspace{4.8cm}
\label{eq:10r}
\end{equation}
\[
\tilde{\eta}^{\,ij}_{\alpha}({\bf v},\chi|\,q)\simeq\,
\delta^{ij}\eta^{(0)}_{\theta\,\alpha}(q)+
\tilde{\eta}^{(1)\,ij}_{\alpha}(A^{(0)})(q)+
\Bigl[\,\tilde{\eta}^{(2)ij}_{\alpha}(\bar{\psi}^{(0)},\psi^{(0)})(q)+
\tilde{\eta}^{(2)ij}_{\alpha}(A^{(0)},A^{(0)})(q)\Bigr],
\]
\[
\eta^{(0)}_{\theta\,\alpha}(q)\equiv
\frac{\,g}{(2\pi)^3}\,\chi_{\alpha}\delta(v\cdot q).
\hspace{10cm}
\]
Here, in the first line the effective sources
$\tilde{\eta}^{(1)\,ib}_{\alpha}(\psi^{(0)})$,
$\tilde{\eta}^{(2)ib}_{\alpha}(A^{(0)},\psi^{(0)})$ are defined by
Eqs.\,(\ref{eq:4e}), (\ref{eq:5s}) and (\ref{eq:6u}), (\ref{eq:6i}). 
The effective source $\tilde{\eta}^{(1)\,ij}_{\alpha}(A^{(0)})$ in the second 
line is given by Eqs.\,(\ref{eq:4u}), (\ref{eq:4y}) and higher order effective
sources $\tilde{\eta}^{(2)ij}_{\alpha}(\bar{\psi}^{(0)},\psi^{(0)})$,
$\tilde{\eta}^{(2)ij}_{\alpha}(A^{(0)},A^{(0)})$ are defined by
Eqs.\,(\ref{eq:6t}), (\ref{eq:6y}) and (\ref{eq:6o}), (B.1), respectively.
Substituting (\ref{eq:10r}) into (\ref{eq:10w}), we find second in importance
contribution after (\ref{eq:10e}) to expression for the energy
losses of the energetic parton. We write down this expression as a sum of 
two different in structure (and physical meaning) parts:
\[
\left(\!-\frac{dE^{(1)}}{dx}\right)_{\!{\cal F}}=
\left(\!-\frac{dE^{(1)}}{dx}\right)_{\!{\rm diag}}\!+\;\,
\left(\!-\frac{dE^{(1)}}{dx}\right)_{\!{\rm nondiag}},
\]
where `diagonal' part equals
\begin{equation}
\left(\!-\frac{dE^{(1)}}{dx}\right)_{\!{\rm diag}}=
-\frac{(2\pi)^3}{\vert{\bf v}\vert\,}\,\,
C_2^{(\zeta)}T_F
\biggl(\frac{\alpha_s}{2{\pi}^2}\biggr)^{\!2}\!\!
\sum\limits_{\lambda,\,\lambda_1=\pm\,}
\!\int\!q^0dq\,dq_1
\label{eq:10t}
\end{equation}
\[
\times
\biggl\{{\rm Im}(^{\ast}{\!\Delta}_{+}(q))\Bigl[\,
\tilde{\Upsilon}_{+}(q_1,x)
|\,\bar{u}_{\alpha}(\hat{\bf q},\lambda)
K_{\alpha\alpha_1}^{(Q)}(\chi,\bar{\chi}|\,q,-q_1)
u_{\alpha_1}(\hat{\bf q}_1,\lambda_1)|^{\,2}
\]
\[
\hspace{2.7cm}
+\,\tilde{\Upsilon}_{-}(q_1,x)
|\,\bar{u}_{\alpha}(\hat{\bf q},\lambda)
K_{\alpha\alpha_1}^{(Q)}(\chi,\bar{\chi}|\,q,-q_1)
v_{\alpha_1}(\hat{\bf q}_1,\lambda_1)|^{\,2}\Bigr]
\]
\[
\hspace{3.6cm}
+\,\Bigl(\,^{\ast}{\!\Delta}_{+}(q)\rightarrow
\,^{\ast}{\!\Delta}_{-}(q),\;
\bar{u}(\hat{\bf q},\lambda)\rightarrow
\bar{v}(\hat{\bf q},\lambda)\Bigr)\biggl\}
\,\delta(v\cdot(q-q_1))
\]
\[
\hspace{0.3cm}
+\,\frac{(2\pi)^3}
{\vert{\bf v}\vert\,}\,\,
C_{\theta}C_F
\Biggl(\frac{\alpha_s}{2{\pi}^2}\Biggr)^{\!2}\!
\sum\limits_{\lambda=\pm\,}
\!\int\!q^0dq\,dk
\]
\[
\times
\biggl\{{\rm Im}(^{\ast}{\!\Delta}_{+}(q))\Bigl[\,
I^t(k,x)\!\sum\limits_{\xi=1,\,2}
|\,\bar{u}_{\alpha}(\hat{\bf q},\lambda)
K_{\alpha}^{(Q)i}({\bf v},\chi|\,k,-q)
{\rm e}^i(\hat{\bf k},\xi)|^{\,2}
\]
\[
\hspace{2.7cm}
+\,I^l(k,x)\,\left(\frac{k^2}{k_0^2}\right)
|\,\bar{u}_{\alpha}(\hat{\bf q},\lambda)
K_{\alpha}^{(Q)i}({\bf v},\chi|\,k,-q)
\hat{k}^i|^{\,2}\Bigr]
\]
\[
\hspace{4.3cm}
+\,\Bigl(\,^{\ast}{\!\Delta}_{+}(q)\rightarrow
\,^{\ast}{\!\Delta}_{-}(q),\;
\bar{u}(\hat{\bf q},\lambda)\rightarrow
\bar{v}(\hat{\bf q},\lambda)\Bigr)\biggl\}
\,\delta(v\cdot(q-k)),
\]
and in turn `nondiagonal' part equals
\begin{equation}
\left(\!-\frac{dE^{(1)}}{dx}\right)_{\!{\rm nondiag}}=
-2\,\frac{(2\pi)^3}{\vert{\bf v}\vert\,\,}\,\,
(\theta_0^{\dagger j}\theta_0^{\,j^{\prime}})
\biggl(\frac{\alpha_s}{2{\pi}^2}\biggr)^{\!2}\!\!
\sum\limits_{\lambda,\,\lambda_1=\pm\,}
\!\int\!q^0dq\,dq_1\biggl[{\rm Im}(^{\ast}{\!\Delta}_{+}(q))
\label{eq:10y}
\end{equation}
\[
\times
\Bigl\{
\tilde{\Upsilon}_{+}(q_1,x)\,
{\rm Re}\Bigl[\,(\bar{\chi}\,u(\hat{\bf q},\lambda))\!
\left(\bar{u}_{\alpha}(\hat{\bf q},\lambda)\!
\,\bar{u}_{\alpha_1}(\hat{\bf q}_1,\lambda_1)
K_{\alpha\alpha_1\alpha_2}^{(Q)jiij^{\prime}}
(\chi,\chi,\bar{\chi}|\,q,q_1;-q_1)
u_{\alpha_2}(\hat{\bf q}_1,\lambda_1)\right)\Bigr]
\]
\[
\hspace{0.4cm}
+\,
\tilde{\Upsilon}_{\!-}(q_1,x)\,
{\rm Re}\Bigl[\,(\bar{\chi}\,u(\hat{\bf q},\lambda))\!
\left(\bar{u}_{\alpha}(\hat{\bf q},\lambda)\!
\,\bar{v}_{\alpha_1}(\hat{\bf q}_1,\lambda_1)
K_{\alpha\alpha_1\alpha_2}^{(Q)jiij^{\prime}}
(\chi,\chi,\bar{\chi}|\,q,q_1;-q_1)
v_{\alpha_2}(\hat{\bf q}_1,\lambda_1)\right)\Bigr]\Bigr\}
\]
\[
\hspace{2.5cm}
+\,\Bigl(\,^{\ast}{\!\Delta}_{+}(q)\rightarrow
\,^{\ast}{\!\Delta}_{-}(q),\quad
u(\hat{\bf q},\lambda)\rightarrow
v(\hat{\bf q},\lambda),\quad
\bar{u}(\hat{\bf q},\lambda)\rightarrow
\bar{v}(\hat{\bf q},\lambda)\Bigr)\biggl]
\,\delta(v\cdot q)
\]
\[
+\,\frac{(2\pi)^3}{\vert{\bf v}\vert\,\,}\,\,
(\theta_0^{\dagger j}\theta_0^{\,j^{\prime}})
\biggl(\frac{\alpha_s}{2{\pi}^2}\biggr)^{\!2}
\sum\limits_{\lambda=\pm}
\int\!q^0dq\,dk\biggl[{\rm Im}(^{\ast}{\!\Delta}_{+}(q))
\]
\[
\times
\Bigl\{
I^t(k,x)\!\sum\limits_{\xi=1,\,2}\!
{\rm Re}\Bigl[\,(\bar{\chi}\,u(\hat{\bf q},\lambda))\!
\left(\bar{u}_{\alpha}(\hat{\bf q},\lambda)
K_{ii^{\prime}\!,\,\alpha}^{(Q)aa,\,jj^{\prime}}
({\bf v},{\bf v},\chi|\,q\,;-k,k)\,{\rm e}^{\ast i}(\hat{\bf k},\xi)
{\rm e}^{i^{\prime}}(\hat{\bf k},\xi)
\right)\Bigr]
\]
\[
+\,I^l(k,x)\!\left(\frac{k^2}{k_0^2}\right)\!
{\rm Re}\Bigl[\,(\bar{\chi}\,u(\hat{\bf q},\lambda))\!
\left(\bar{u}_{\alpha}(\hat{\bf q},\lambda)
K_{ii^{\prime}\!,\,\alpha}^{(Q)aa,\,jj^{\prime}}
({\bf v},{\bf v},\chi|\,q\,;-k,k)\,\hat{k}^i\hat{k}^{i^{\prime}}
\right)\Bigr]\Bigr\}
\hspace{1.5cm}
\]
\[
\hspace{2.5cm}
+\,\Bigl(\,^{\ast}{\!\Delta}_{+}(q)\rightarrow
\,^{\ast}{\!\Delta}_{-}(q),\quad
u(\hat{\bf q},\lambda)\rightarrow
v(\hat{\bf q},\lambda),\quad
\bar{u}(\hat{\bf q},\lambda)\rightarrow
\bar{v}(\hat{\bf q},\lambda)\Bigr)\biggr]
\,\delta(v\cdot q).
\]
In deriving (\ref{eq:10t}), (\ref{eq:10y}) we have used the following 
decompositions of the spectral densities on the right-hand side of 
Eqs.\,(\ref{eq:8p}) and (\ref{eq:8k})
\[
\Upsilon(q)= h_{+}(\hat{\bf q}) \tilde{\Upsilon}_{+}(q,x) +
h_{-}(\hat{\bf q}) \tilde{\Upsilon}_{-}(q,x),
\]
\[
\hspace{0.2cm}
I_{\mu\nu}(k)= P_{\mu\nu}(k)I^{t}(k,x) + \tilde{Q}_{\mu\nu}(k)I^{l}(k,x),
\]
where longitudinal projector $\tilde{Q}_{\mu\nu}(k)$ in the last equation is 
determined in the temporal gauge. The dependence on $x$ takes into account 
weak non-homogeneity and slow evolution of the medium in time.

The diagonal part (\ref{eq:10t}) is not vanishing both for the
scattering of the high-energy parton off on-shell and off-shell
soft-quark and soft-gluon excitations. In the first case by using
the quasiparticle approximation for the spectral densities
$\tilde{\Upsilon}_{\!\pm}(q_1,x)$, $I^t(k,x)$ and $I^l(k,x)$ the 
integrand in (\ref{eq:10t}) can be expressed in terms of the
scattering probabilities ${\it w}_{\,q\rightarrow
q}^{(\zeta)(f;\,f_1)}({\bf v}|\,{\bf q},{\bf q}_1)$, ${\it
w}_{\,q\rightarrow {\rm g}}^{(f;\,b)}({\bf v}|\,{\bf q},{\bf k}),
$ etc. determined in Section 8. The nondiagonal part (\ref{eq:10y})
is different from zero only for the scattering processes by off-shell soft 
plasma excitations. In the following section we discuss this contribution
in full measure.

\section{\bf `Nondiagonal' contribution to energy losses}
\setcounter{equation}{0}

At the beginning we consider the latter contribution in
Eq.\,(\ref{eq:10y}) proportional to the gluon spectral densities
$I^t$ and $I^l$. For this purpose we note first of all that taking
into account color decomposition (I.5.19) for the HTL-induced
vertex $\delta\Gamma^{(Q)a_1a_2}_{\mu_1\mu_2}$, we can present the
coefficient function (B.1) in the following form
\begin{equation}
K^{(Q)a_1a_2,\,ij}_{\mu_1\mu_2,\,\alpha}
({\bf v},{\bf v},\chi|\,q;-k_1,-k_2)=
\frac{1}{2}\,\{t^{a_1}\!,t^{a_2}\}^{ij}\,
K^{({\cal S})}_{\mu_1\mu_2,\,\alpha}
({\bf v},{\bf v},\chi|\,q;-k_1,-k_2)
\label{eq:11q}
\end{equation}
\[
\hspace{4.9cm}
+\,
\frac{1}{2}\,[\,t^{a_1}\!,t^{a_2}]^{\,ij}\,
K^{({\cal A})}_{\mu_1\mu_2,\,\alpha}
({\bf v},{\bf v},\chi|\,q;-k_1,-k_2),
\]
where the functions $K^{({\cal S,A})}_{\mu_1\mu_2,\,\alpha}$ possess the
properties
\[
K^{({\cal S,\,A})}_{\mu_1\mu_2,\,\alpha}
({\bf v},{\bf v},\chi|\,q;-k_1,-k_2)=\pm\,
K^{({\cal S,\,A})}_{\mu_2\mu_1,\,\alpha}
({\bf v},{\bf v},\chi|\,q;-k_2,-k_1).
\]
The explicit form of these functions can be easily recovered by (B.1). The
coefficient function in the integrand in the last term of (\ref{eq:10y})
can be written, in view of (\ref{eq:11q}), as
\begin{equation}
K^{(Q)aa,\,jj^{{\prime}}}_{ii^{{\prime}}\!,\,\alpha}
({\bf v},{\bf v},\chi|\,q;-k,k)=
\delta^{jj^{{\prime}}} C_F\,
K^{({\cal S})}_{ii^{{\prime}}\!,\,\alpha}
({\bf v},{\bf v},\chi|\,q;-k_1,-k_2)\Bigl|_{\,k_1=-k_2=k},
\label{eq:11w}
\end{equation}
where
\begin{equation}
K^{({\cal S})}_{ii^{{\prime}}\!,\,\alpha}
({\bf v},{\bf v},\chi|\,q;-k_1,-k_2)\Bigl|_{\,k_1=-k_2=k\,}=
-\,{\cal T}^{({\cal S})}_{ii^{{\prime}}\!,\,\alpha\beta}(k,-k;q,-q)
\,^{\ast}\!S_{\beta\beta^{\prime}}(q)\chi_{\beta^{\prime}}
\label{eq:11e}
\end{equation}
\[
-\,\frac{1}{(v\cdot k)^2}\,v_{i}\,v_{i^{{\prime}}}\chi_{\alpha}\,
-\,\frac{1}{(v\cdot k)}\,\,v_i
\,^{\ast}\Gamma^{(Q)}_{i^{{\prime}}\!,\,\alpha\beta}(-k;q+k,-q)
\,^{\ast}\!S_{\beta\beta^{\prime}}(q+k)\chi_{\beta^{\prime}}
\]
\[
\hspace{3.2cm}
+\,\,\frac{1}{(v\cdot k)}\,\,v_{i^{\prime}}
\,^{\ast}\Gamma^{(Q)}_{i,\,\alpha\beta}(k;q-k,-q)
\,^{\ast}\!S_{\beta\beta^{\prime}}(q-k)\chi_{\beta^{\prime}}.
\]
The function ${\cal T}^{({\cal S})}_{ii^{{\prime}}}$ in (\ref{eq:11e})
is defined by Eq.\,(I.5.23). We recall that this function and also
${\cal T}^{({\cal A})}_{ii^{{\prime}}}$ enter into the matrix elements of the
elastic scattering process of soft-quark excitation off soft-gluon excitation
and the process of quark-antiquark annihilation into two soft-gluon excitations
with different parity of state of final two\,--\,soft-gluon system.
The symbols ${\cal S}$ and ${\cal A}$
belong to states of two gluons being in even and odd states correspondingly
(in the c.m.s. of these gluons). The decomposition (\ref{eq:11q}) suggests that the
scattering process of soft-quark excitation off hard parton with the subsequent
radiation of two soft-gluon excitations proceed through two physical
independent channels determined by a parity of final two\,--\,(soft) gluon
system.

Furthermore, the coefficient function
$K_{\alpha\alpha_1\alpha_2}^{(Q)jiij^{\prime}}$ in the former contribution on
the right-hand side of (\ref{eq:10y}) according to definition
(\ref{eq:6y}) has the following structure:
\begin{equation}
K_{\alpha\alpha_1\alpha_2}^{(Q)jiij^{\prime}}
(\chi,\chi,\bar{\chi}|\,q,q_1;-q_1) =
\delta^{jj^{{\prime}}} C_F\,
\biggl\{
-\,{\rm M}^{({\cal S})}_{\alpha\alpha_1\alpha_2\beta}(-q,-q;q_1,q_1)
\,^{\ast}\!S_{\beta\beta^{\prime}}(q)\chi_{\beta^{\prime}}
\label{eq:11r}
\end{equation}
\[
\hspace{2cm}
+\,\beta_1\,
\frac{1}{(v\cdot q_1)^2}\,\,
\chi_{\alpha}\chi_{\alpha_1}\bar{\chi}_{\alpha_2}\,
+\,
\frac{1}{(v\cdot q_1)}\,\,\chi_{\alpha_1}\!
\,^{\ast}\Gamma^{(Q)\mu}_{\alpha\alpha_2}(q-q_1;q_1,-q)
\,^{\ast}{\cal D}_{\mu\nu}(q-q_1) v^{\nu}\biggr\},
\]
where
\[
{\rm M}^{({\cal S})}_{\alpha\alpha_1\alpha_2\beta}(-q,-q;q_1,q_1)\equiv
\,^{\ast}\Gamma^{(Q)\mu}_{\alpha\alpha_2}(q-q_1;q_1,-q)
\,^{\ast}{\cal D}_{\mu\nu}(q-q_1)
\,^{\ast}\Gamma^{(G)\nu}_{\alpha_1\beta}(q-q_1;q_1,-q).
\]
Here we note also that the function ${\rm M}^{({\cal
S})}_{\alpha\alpha_1\alpha_2\beta}$ enters into the matrix element
of the elastic scattering process of soft-quark excitation off
soft-(anti)quark one (Section 6 in Paper I). The symbol
${\cal S}$ means that in nondiagonal contribution (\ref{eq:10y})
only `symmetric' part of the function ${\rm
M}_{\alpha\alpha_1\alpha_2\alpha_3}$ connected with even parity of
final state of soft-quark quasiparticle system survives. For
subsequent purposes we present the soft-quark propagator
$\,^{\ast}\!S_{\beta\beta^{\prime}}(q)$ in the first terms on the
right-hand side of Eqs.\,(\ref{eq:11e}) and (\ref{eq:11r}) in an
identical form
\begin{equation}
^{\ast}\!S_{\beta\beta^{\prime}}(q)=
\sum\limits_{\lambda^{\prime}=\pm}\Bigl\{
[\,u_{\beta}(\hat{\bf q},\lambda^{\prime})
\bar{u}_{\beta^{\prime}}(\hat{\bf q},\lambda^{\prime})]
\,^{\ast}\!\Delta_{+}(q)+
[\,v_{\beta}(\hat{\bf q},\lambda^{\prime})
\bar{v}_{\beta^{\prime}}(\hat{\bf q},\lambda^{\prime})]
\,^{\ast}\!\Delta_{-}(q)\Bigr\}.
\label{eq:11t}
\end{equation}

Now we substitute Eqs.\,(\ref{eq:11w})\,--\,(\ref{eq:11r}) into
(\ref{eq:10y}) and take into account the equation written just above. We add
polarization losses (\ref{eq:10e}) to the expression obtained.
After some regrouping of the terms we obtain the following final
expression:
\[
\left(\!-\frac{dE^{(0)}}{dx}\right)_{\!{\cal F}}\,+\,
\left(\!-\frac{dE^{(1)}}{dx}\right)_{\!{\rm nondiag}}\!=\,
\Lambda_1+\Lambda_2.
\]
Here the function $\Lambda_1$ is
\begin{equation}
\Lambda_1=\frac{1}{2E}\,
\frac{1}{\vert{\bf v}\vert}\,C_{\theta}
\Biggl(\frac{\alpha_s}{2{\pi}^2}\Biggr)\!
\int\!q^0dq
\,\Biggl\{{\rm Im}\,[^{\ast}{\!\Delta}_{+}(q)]
\Biggl[\,\sum\limits_{\lambda,\,\lambda^{\prime}=\pm\,}
\biggl([\,\bar{u}(\hat{\bf q},\lambda^{\prime})\varrho({\bf v})
u(\hat{\bf q},\lambda)]\,\delta^{\lambda\lambda^{\prime}}
\label{eq:11y}
\end{equation}
\[
+\,{\rm Re}\,\Bigl\{[\,\bar{u}(\hat{\bf q},\lambda^{\prime})\varrho({\bf v})
u(\hat{\bf q},\lambda)]\,\Sigma_{++}^{(1)}(q;\lambda,\lambda^{\prime})
\,^{\ast}\!\Delta_{+}(q)\Bigr\}
\]
\[
\hspace{0.5cm}
+\,{\rm Re}\,\Bigl\{[\,\bar{v}(\hat{\bf q},\lambda^{\prime})\varrho({\bf v})
u(\hat{\bf q},\lambda)]\,\Sigma_{+-}^{(1)}(q;\lambda,\lambda^{\prime})
\,^{\ast}\!\Delta_{-}(q)\Bigr\}\biggr)\Biggr]
\]
\[
+{\rm Im}[^{\ast}{\!\Delta}_{-}(q)]\!
\Biggl[\sum\limits_{\,\lambda,\,\lambda^{\prime}=\pm\,}\!\!\!
\biggl([\,\bar{v}(\hat{\bf q},\lambda^{\prime})\varrho({\bf v})
v(\hat{\bf q},\lambda)]\,\delta^{\lambda\lambda^{\prime}}
\!+\!\,{\rm Re}\Bigl\{[\bar{v}(\hat{\bf q},\lambda^{\prime})\varrho({\bf v})
v(\hat{\bf q},\lambda)]\Sigma_{--}^{(1)}(q;\lambda,\lambda^{\prime})
\!\,^{\ast}\!\Delta_{-}(q)\!\Bigr\}
\]
\[
\hspace{2.2cm}
+\,{\rm Re}\,\Bigl\{[\,\bar{u}(\hat{\bf q},\lambda^{\prime})\varrho({\bf v})
v(\hat{\bf q},\lambda)]\,\Sigma_{-+}^{(1)}(q;\lambda,\lambda^{\prime})
\,^{\ast}\!\Delta_{+}(q)\Bigr\}\biggr)\Biggr]
\Biggr\}\,\delta(v\cdot q),
\]
where in turn we have
\[
\Sigma_{++}^{(1)}(q;\lambda,\lambda^{\prime})\equiv
2g^2\,C_F\!\!\sum\limits_{\lambda_1=\pm\,}\!\int\!dq_1\biggl\{
\tilde{\Upsilon}_{+}(q_1,x)
\]
\[
\times
\Bigl[\,\bar{u}_{\alpha}(\hat{\bf q},\lambda)
\bar{u}_{\alpha_1}(\hat{\bf q}_1,\lambda_1)\,
{\rm M}_{\alpha\alpha_1\alpha_2\beta}(-q,-q;q_1,q_1)
u_{\alpha_2}(\hat{\bf q}_1,\lambda_1)
u_{\beta}(\hat{\bf q},\lambda^{\prime})\Bigr]
\]
\[
+\,\Bigl(\tilde{\Upsilon}_{+}(q_1,x)\rightarrow
\tilde{\Upsilon}_{-}(q_1,x),\,
\bar{u}_{\alpha_1}(\hat{\bf q}_1,\lambda_1)\rightarrow
\bar{v}_{\alpha_1}(\hat{\bf q}_1,\lambda_1),\,
u_{\alpha_2}(\hat{\bf q}_1,\lambda_1)
\rightarrow v_{\alpha_2}(\hat{\bf q}_1,\lambda_1)\Bigr)\biggr\}
\]
\[
-\,g^2C_F\!\int\!dk\,\biggl\{I^t(k,x)
\!\sum\limits_{\xi=1,\,2}
\Bigl[\,\bar{u}_{\alpha}(\hat{\bf q},\lambda)
{\cal T}^{({\cal S})ii^{{\prime}}}_{\alpha\beta}\!(k,-k;q,-q)
u_{\beta}(\hat{\bf q},\lambda^{\prime})
\,{\rm e}^{\ast i}(\hat{\bf k},\xi)
{\rm e}^{i^{\prime}}(\hat{\bf k},\xi)\Bigr]
\]
\[
+\,\Bigl(I^t(k,x)\rightarrow I^l(k,x),\,
{\rm e}^{i}(\hat{\bf k},\xi)\rightarrow
\sqrt{\frac{k^2}{k_0^2}}\,\hat{k}^i\Bigr)\biggr\}.
\]
The function $\Sigma_{+-}^{(1)}$ is obtained from $\Sigma_{++}^{(1)}$ by
the replacement $u_{\beta}(\hat{\bf q},\lambda^{\prime})\rightarrow
v_{\beta}(\hat{\bf q},\lambda^{\prime})$, and the functions
$\Sigma_{--}^{(1)}\,$, $\Sigma_{-+}^{(1)}$ are accordingly obtained by the
replacements
\[\bar{u}_{\alpha}(\hat{\bf q},\lambda)\rightarrow
\bar{v}_{\alpha}(\hat{\bf q},\lambda),\quad
u_{\beta}(\hat{\bf q},\lambda^{\prime})\rightarrow
v_{\beta}(\hat{\bf q},\lambda^{\prime})
\]
and $\bar{u}_{\alpha}(\hat{\bf q},\lambda)\rightarrow
\bar{v}_{\alpha}(\hat{\bf q},\lambda)$.

The function $\Lambda_2$ is conveniently represented in the form of a sum of
two parts:
\[
\Lambda_2\,=\,\Lambda_2[\,\tilde{\Upsilon}_{\pm}]
\,+\,\Lambda_2[I^{t,\,l}]\,,
\]
where
\begin{equation}
\Lambda_2[\,\tilde{\Upsilon}_{\pm}]\equiv
-\,2\,\frac{(2\pi)^3}{\vert{\bf v}\vert\,\,}\,\,
C_{\theta}\,C_F
\biggl(\frac{\alpha_s}{2{\pi}^2}\biggr)^{\!2}\!\!
\sum\limits_{\lambda,\,\lambda_1=\pm\,}
\!\int\!q^0dq\,dq_1\biggl\{{\rm Im}(^{\ast}{\!\Delta}_{+}(q))
\label{eq:11u}
\end{equation}
\[
\times
\biggl[\,
\tilde{\Upsilon}_{\!+}(q_1,x)\,
{\rm Re}\biggl(\beta_1\,
\frac{\vert\bar{\chi}\,u(\hat{\bf q},\lambda)\vert^{\,2}
\vert\,\bar{u}(\hat{\bf q}_1,\lambda_1)\chi\vert^{\,2}}
{(v\cdot q_1)^2}
\]
\[
+\,\frac{(\bar{\chi}\,u(\hat{\bf q},\lambda))
(\bar{u}(\hat{\bf q}_1,\lambda_1)\chi)}{(v\cdot q_1)}\,\,
[\,\bar{u}(\hat{\bf q},\lambda)
\,^{\ast}\Gamma^{(Q)\mu}(q-q_1;q_1,-q)u(\hat{\bf q}_1,\lambda_1)]
\,^{\ast}{\cal D}_{\mu\nu}(q-q_1) v^{\nu}\biggr)
\]
\[
\hspace{0.3cm}
+\,\tilde{\Upsilon}_{\!-}(q_1,x)\,
{\rm Re}\biggl(\beta_1\,
\frac{\vert\bar{\chi}\,u(\hat{\bf q},\lambda)\vert^{\,2}
\vert\,\bar{v}(\hat{\bf q}_1,\lambda_1)\chi\vert^{\,2}}
{(v\cdot q_1)^2}
\]
\[
\hspace{0.2cm}
+\,\frac{(\bar{\chi}\,u(\hat{\bf q},\lambda))
(\bar{v}(\hat{\bf q}_1,\lambda_1)\chi)}{(v\cdot q_1)}\,\,
[\,\bar{u}(\hat{\bf q},\lambda)
\,^{\ast}\Gamma^{(Q)\mu}(q-q_1;q_1,-q)v(\hat{\bf q}_1,\lambda_1)]
\,^{\ast}{\cal D}_{\mu\nu}(q-q_1) v^{\nu}\biggr)\biggr]
\]
\[
+\,\Bigr(\,^{\ast}{\!\Delta}_{+}(q)\rightarrow
\,^{\ast}{\!\Delta}_{-}(q),\,
u(\hat{\bf q},\lambda)\rightarrow v(\hat{\bf q},\lambda)\Bigr)\biggl\}\,
\delta(v\cdot q)
\]
and
\begin{equation}
\Lambda_2[I^{t,\,l}]\equiv
\frac{(2\pi)^3}{\vert{\bf v}\vert\,\,}\,\,
C_{\theta}\,C_F
\biggl(\frac{\alpha_s}{2{\pi}^2}\biggr)^{\!2}\!
\sum\limits_{\lambda=\pm}
\int\!q^0dq\,dk\biggl\{{\rm Im}(^{\ast}{\!\Delta}_{+}(q))
\label{eq:11i}
\end{equation}
\[
\times\biggl[\,I^t(k,x)
\!\sum\limits_{\xi=1,\,2}\!
\biggl(-\,\frac{|({\bf e}(\hat{\bf k},\xi)\cdot{\bf v})|^{\,2}
|\bar{\chi}\,u(\hat{\bf q},\lambda)|^{\,2}}
{(v\cdot k)^2}
\]
\[
+\,\frac{1}{(v\cdot k)}\,\,{\rm Re}\,\Bigl\{
(\bar{\chi}\,u(\hat{\bf q},\lambda))
({\bf e}(\hat{\bf k},\xi)\cdot {\bf v})
[\,\bar{u}(\hat{\bf q},\lambda)
\,^{\ast}\Gamma^{(Q)i}(k;q-k,-q)\,^{\ast}\!S(q-k)\chi]
\,{\rm e}^{\ast i}(\hat{\bf k},\xi)\Bigl\}
\]
\[
\hspace{0.45cm}
-\,\frac{1}{(v\cdot k)}\,\,{\rm Re}\,\Bigl\{
(\bar{\chi}\,u(\hat{\bf q},\lambda))
({\bf e}^{\ast}(\hat{\bf k},\xi)\cdot {\bf v})
[\,\bar{u}(\hat{\bf q},\lambda)
\,^{\ast}\Gamma^{(Q)i}(-k;q+k,-q)\,^{\ast}\!S(q+k)\chi]
\,{\rm e}^{i}(\hat{\bf k},\xi)\Bigl\}\biggr)
\]
\[
+\,\Bigl(I^t(k,x)\rightarrow I^l(k,x),\,
{\rm e}^{i}(\hat{\bf k},\xi)\rightarrow
\sqrt{\frac{k^2}{k_0^2}}\,\hat{k}^i\Bigr)\biggr]
\hspace{2.1cm}
\]
\[
\hspace{2.7cm}
+\,\Bigr(\,^{\ast}{\!\Delta}_{+}(q)\rightarrow
\,^{\ast}{\!\Delta}_{-}(q),\;
u(\hat{\bf q},\lambda)\rightarrow
v(\hat{\bf q},\lambda),\;
\bar{u}(\hat{\bf q},\lambda)\rightarrow
\bar{v}(\hat{\bf q},\lambda)\Bigr)\biggl\}\,
\delta(v\cdot q).
\]

First we discuss the function $\Lambda_1$. As it will be shown
just below the functions
$\Sigma_{++}(q;\lambda,\lambda^{\prime})$,
$\Sigma_{+-}(q;\lambda,\lambda^{\prime})$ etc. in the integrand of
Eq.\,(\ref{eq:11y}) represent nothing but the linear in
$\tilde{\Upsilon}_{\pm}$ and $I^{t,\,l}$ corrections to various
components of the soft-quark self-energy $\delta\Sigma(q)$ in the
HTL approximation. These corrections take into account a change of
dispersion properties of hot QCD plasma induced by the processes
of nonlinear interaction of soft-quark and soft-gluon excitations
among themselves (see Paper I). For low excited state of the
plasma corresponding to level of thermal fluctuations, the
corrections $\Sigma_{++}$, $\Sigma_{+-},\,\ldots$ are suppressed
by more power of $g$ in comparison with $\delta\Sigma(q)$. However,
we can expect that in the limiting case of strong soft fields as
far as possible these corrections (and also the corrections of
higher powers in $\tilde{\Upsilon}_{\pm}$ and $I^{t,\,l}$) is the
same order in $g$ as the soft-quark HTL-induced self-energy and
therefore consideration of an influence of the nonlinear
self-interaction of soft excitations on the energy losses of the
energetic parton becomes necessary.

We write out an {\it effective} inverse quark propagator
$\,^{\ast}\!\tilde{S}^{-1}(q)$ that takes into account additional
contributions considering the nonlinear effects of self-interaction
of soft plasma excitations
\[
\,^{\ast}\!\tilde{S}^{-1}(q)\,=
\,^{\ast}\!{S}^{-1}(q) -
\Sigma^{(1)}[\,\tilde{\Upsilon}_{\pm},\,I^{t,\,l}\,](q)
-\Sigma^{(2)}[\,\tilde{\Upsilon}_{\pm},\,I^{t,\,l}\,](q)-\,\ldots\,,
\]
where $\Sigma^{(1)},\,\Sigma^{(2)},\ldots$ are linear, quadratic, and so on
corrections in powers of the spectral densities
$\tilde{\Upsilon}_{\pm},\,I^{t,\,l}$.
For low excited state when $\,^{\ast}\!{S}^{-1}(q)\gg \Sigma^{(1)}\gg
\Sigma^{(2)}\gg\ldots$ from this equation to the first order in
$\tilde{\Upsilon}_{\pm}$ and $I^{t,\,l}$, we obtain
\[
\,^{\ast}\!\tilde{S}(q)\,\simeq\,^{\ast}\!{S}(q)\,+
\,^{\ast}\!{S}(q)\Sigma^{(1)}[\,\tilde{\Upsilon}_{\pm},\,I^{t,\,l}\,](q)
\,^{\ast}\!{S}(q)\,+\ldots\,.
\]
Taking into account (\ref{eq:11t}) the last equation can be identically
rewritten in the following form:
\[
\,^{\ast}\!\tilde{S}_{\beta\beta^{\prime}}(q)\simeq
\sum\limits_{\lambda,\,\lambda^{\prime}=\pm}
\Bigl[\,\delta^{\lambda\lambda^{\prime}}
(u_{\beta}(\hat{\bf q},\lambda)
\bar{u}_{\beta^{\prime}}(\hat{\bf q},\lambda^{\prime}))
\,^{\ast}\!\Delta_{+}(q)
\]
\[
\hspace{0.1cm}
+\,\,^{\ast}\!\Delta_{+}(q)
(u_{\beta}(\hat{\bf q},\lambda)
\bar{u}_{\beta^{\prime}}(\hat{\bf q},\lambda^{\prime}))\,
[\bar{u}(\hat{\bf q},\lambda)\Sigma^{(1)}(q)
u(\hat{\bf q},\lambda^{\prime})]
\,^{\ast}\!\Delta_{+}(q)
\]
\[
+\,\,^{\ast}\!\Delta_{+}(q)
(u_{\beta}(\hat{\bf q},\lambda)
\bar{v}_{\beta^{\prime}}(\hat{\bf q},\lambda^{\prime}))\,
[\bar{u}(\hat{\bf q},\lambda)\Sigma^{(1)}(q)
v(\hat{\bf q},\lambda^{\prime})]
\,^{\ast}\!\Delta_{-}(q)
\]
\[
\hspace{1.6cm}
+\,\Bigr(\,^{\ast}{\!\Delta}_{+}(q)\rightleftharpoons
\,^{\ast}{\!\Delta}_{-}(q),\,
u(\hat{\bf q},\lambda)\rightleftharpoons v(\hat{\bf q},\lambda),\,
u(\hat{\bf q},\lambda^{\prime})
\rightleftharpoons v(\hat{\bf q},\lambda^{\prime}),\,\ldots\Bigr)\Bigr].
\]
Hereinafter, we designate for brevity $\Sigma^{(1)}(q)\equiv
\Sigma^{(1)}[\,\tilde{\Upsilon}_{\pm},\,I^{t,\,l}\,](q)$.
We convolve this expression with $\chi_{\beta}\bar{\chi}_{\beta^{\prime}}$ and
take the imaginary part
\[
{\rm Im}\!\left(\bar{\chi}\,^{\ast}\!\tilde{S}(q)\chi\right)=
\frac{1}{2E}\,{\rm Im}\Bigl[\,{\rm Sp}
\left(\varrho({\bf v})\,^{\ast}\!\tilde{S}(q)\right)\Bigr]
\hspace{7.5cm}
\]
\begin{equation}
=\frac{1}{2E}\,{\rm Im}\left(\,^{\ast}{\!\Delta}_{+}(q)\right)\!
\sum\limits_{\lambda,\,\lambda^{\prime}=\pm}
\Bigl\{\,\delta^{\lambda\lambda^{\prime}}
[\bar{u}(\hat{\bf q},\lambda^{\prime})\varrho({\bf v})
u(\hat{\bf q},\lambda)]
\hspace{0.8cm}
\label{eq:11o}
\end{equation}
\[
+\,{\rm Re}\,\Bigl([\bar{u}(\hat{\bf q},\lambda^{\prime})\varrho({\bf v})
u(\hat{\bf q},\lambda)]\,
[\bar{u}(\hat{\bf q},\lambda)\Sigma^{(1)}(q)
u(\hat{\bf q},\lambda^{\prime})]
\,^{\ast}\!\Delta_{+}(q)\Bigr)
\]
\[
\hspace{0.1cm}
+\,{\rm Re}\,\Bigl([\bar{v}(\hat{\bf q},\lambda^{\prime})\varrho({\bf v})
u(\hat{\bf q},\lambda)]
[\bar{u}(\hat{\bf q},\lambda)\Sigma^{(1)}(q)
v(\hat{\bf q},\lambda^{\prime})]
\,^{\ast}\!\Delta_{-}(q)\Bigr)\Bigr\}
\]
\[
+\,\frac{1}{2E}\,{\rm Re}\left(\,^{\ast}{\!\Delta}_{+}(q)\right)\!
\sum\limits_{\lambda,\,\lambda^{\prime}=\pm}
\Bigl\{{\rm Im}\,\Bigl([\bar{u}(\hat{\bf q},\lambda^{\prime})\varrho({\bf v})
u(\hat{\bf q},\lambda)]\,
[\bar{u}(\hat{\bf q},\lambda)\Sigma^{(1)}(q)
u(\hat{\bf q},\lambda^{\prime})]
\,^{\ast}\!\Delta_{+}(q)\Bigr)
\]
\[
\hspace{4.4cm}
+\,{\rm Im}\,\Bigl([\bar{v}(\hat{\bf q},\lambda^{\prime})\varrho({\bf v})
u(\hat{\bf q},\lambda)]\,
[\bar{u}(\hat{\bf q},\lambda)\Sigma^{(1)}(q)
v(\hat{\bf q},\lambda^{\prime})]
\,^{\ast}\!\Delta_{-}(q)\Bigr)\Bigr\}
\]
\[
\hspace{1.4cm}
+\,\Bigr(\,^{\ast}{\!\Delta}_{\pm}(q)\rightleftharpoons
\,^{\ast}{\!\Delta}_{\mp}(q),\,
u(\hat{\bf q},\lambda)\rightleftharpoons v(\hat{\bf q},\lambda),\,
u(\hat{\bf q},\lambda^{\prime})
\rightleftharpoons v(\hat{\bf q},\lambda^{\prime}),\,\ldots\Bigr).
\]
Comparing (\ref{eq:11o}) with (\ref{eq:11y}), we see that the contributions
proportional to ${\rm Im}\left(\,^{\ast}{\!\Delta}_{\pm}(q)\right)$ in
equation (\ref{eq:11o}) exactly reproduces the integrand in
(\ref{eq:11y}) if we identify
\[
\Sigma_{++}^{(1)}(q;\lambda,\lambda^{\prime})\equiv
[\bar{u}(\hat{\bf q},\lambda)\Sigma^{(1)}(q)
u(\hat{\bf q},\lambda^{\prime})]\,,\quad
\Sigma_{+-}^{(1)}(q;\lambda,\lambda^{\prime})\equiv
[\bar{u}(\hat{\bf q},\lambda)\Sigma^{(1)}(q)
v(\hat{\bf q},\lambda^{\prime})]\,,
\]
etc. Thus if in equation (\ref{eq:11o}) there is no corrections
proportional to ${\rm
Re}\left(\,^{\ast}{\!\Delta}_{\pm}(q)\right)$, then the function
$\Lambda_1$ could be exclusively interpreted as the polarization
losses taking into account to the first approximation a change of
dispersion properties of the QCD medium induced by nonlinear
interaction of soft-quark and soft-gluon excitations among
themselves. It could be effectively presented as a replacement of
the HTL-resummed quark propagator $\,^{\ast}\!S(q)$ in expression
(\ref{eq:10e}) by the effective one
\[
\,^{\ast}\!S(q)\,\rightarrow\,^{\ast}\!\tilde{S}(q).
\]
Unfortunately, the existence of the terms proportional to
${\rm Re}\left(\,^{\ast}{\!\Delta}_{\pm}(q)\right)$ in (\ref{eq:11o})
doesn't allow us an opportunity to reduce everything to such a simple
replacement. The physical meaning of this circumstance is not clear.

Let us discuss the contribution $\Lambda_2$ defined by a sum of
Eqs.\,(\ref{eq:11u}) and (\ref{eq:11i}). Importance and necessity
of accounting this contribution will be completely brought to light from
its comparison with `diagonal' contribution (\ref{eq:10t}). Taking
into account the explicit form of the coefficient functions
$K_{\alpha\alpha_1}^{(Q)}(\chi,\bar{\chi}|\,q,-q_1)$ and
$K_{\alpha}^{(Q)i}({\bf v},\chi|\,k,-q)$, we write out
Eq.\,(\ref{eq:10t}) once more opening the modulus squared
$|\bar{u}K^{(Q)}u|^{\,2}$, $|\bar{u}K^{(Q)i}{\rm
e}^i|^{\,2}$ etc. By analogy with Eqs.\,(\ref{eq:11u}),
(\ref{eq:11i}) the `diagonal' contribution can be also presented as a
sum of two parts:
\[
\left(\!-\frac{dE^{(1)}}{dx}\right)_{\!{\rm diag}}=\,
\Phi_1[\,\tilde{\Upsilon}_{\pm}] + \Phi_2[I^{t,\,l}],
\]
where
\begin{equation}
\Phi_1[\,\tilde{\Upsilon}_{\pm}]\equiv
-\frac{(2\pi)^3}{\vert{\bf v}\vert\,}\,\,
C_2^{(\zeta)}T_F
\biggl(\frac{\alpha_s}{2{\pi}^2}\biggr)^{\!2}\!\!\!
\sum\limits_{\lambda,\,\lambda_1=\pm\,}
\int\!q^0dq\,dq_1\,\biggl\{{\rm Im}(^{\ast}{\!\Delta}_{+}(q))
\label{eq:11p}
\end{equation}
\[
\times\biggl[\,\tilde{\Upsilon}_{\!+}(q_1,x)\,\biggl(\,
\alpha^2\,\frac{\vert\bar{\chi}\,u(\hat{\bf q},\lambda)\vert^{\,2}
\vert\,\bar{u}(\hat{\bf q}_1,\lambda_1)\chi\vert^{\,2}}
{(v\cdot q_1)^2}
\]
\[
-\,2\alpha\,\frac{(\bar{\chi}\,u(\hat{\bf q},\lambda))
(\bar{u}(\hat{\bf q}_1,\lambda_1)\chi)}{(v\cdot q_1)}\,\,
[\,\bar{u}(\hat{\bf q},\lambda)
\,^{\ast}\Gamma^{(Q)\mu}(q-q_1;q_1,-q)u(\hat{\bf q}_1,\lambda_1)]
\,^{\ast}{\cal D}_{\mu\nu}(q-q_1) v^{\nu}
\]
\[
\hspace{5.3cm}
+\left|\,[\,\bar{u}(\hat{\bf q},\lambda)
\,^{\ast}\Gamma^{(Q)\mu}(q-q_1;q_1,-q)u(\hat{\bf q}_1,\lambda_1)]
\,^{\ast}{\cal D}_{\mu\nu}(q-q_1) v^{\nu}\right|^{\,2}
\biggr)
\]
\[
+\,\Bigl(\,\tilde{\Upsilon}_{\!+}(q_1,x)\rightarrow
\tilde{\Upsilon}_{\!-}(q_1,x),\;
u(\hat{\bf q}_1,\lambda_1)\rightarrow v(\hat{\bf q}_1,\lambda_1),\;
\bar{u}(\hat{\bf q}_1,\lambda_1)\rightarrow \bar{v}(\hat{\bf q}_1,\lambda_1)
\Bigr)\biggr]
\]
\[
\hspace{1.3cm}
+\,\Bigr(\,^{\ast}{\!\Delta}_{+}(q)\rightarrow
\,^{\ast}{\!\Delta}_{-}(q),\;
u(\hat{\bf q},\lambda)\rightarrow
v(\hat{\bf q},\lambda),\;
\bar{u}(\hat{\bf q},\lambda)\rightarrow
\bar{v}(\hat{\bf q},\lambda)\Bigr)\biggl\}\,
\delta(v\cdot (q-q_1))
\hspace{0.65cm}
\]
and
\begin{equation}
\Phi_2[I^{t,\,l}]\equiv
\frac{(2\pi)^3}{\vert{\bf v}\vert\,\,}\,\,
C_{\theta}\,C_F
\biggl(\frac{\alpha_s}{2{\pi}^2}\biggr)^{\!2}\!
\sum\limits_{\lambda=\pm}
\int\!q^0dq\,dk\,\biggl\{{\rm Im}(^{\ast}{\!\Delta}_{+}(q))
\label{eq:11a}
\end{equation}
\[
\times\biggl[\,I^t(k,x)
\!\sum\limits_{\xi=1,\,2}\!
\biggl(\,\frac{|\,({\bf e}(\hat{\bf k},\xi)\cdot{\bf v})|^{\,2}
|\,\bar{\chi}\,u(\hat{\bf q},\lambda)|^{\,2}}
{(v\cdot k)^2}
\]
\[
-\,2\,\frac{1}{(v\cdot k)}\,\,{\rm Re}\,\Bigl\{
(\bar{\chi}\,u(\hat{\bf q},\lambda))
({\bf e}(\hat{\bf k},\xi)\cdot {\bf v})\,
[\,\bar{u}(\hat{\bf q},\lambda)
\,^{\ast}\Gamma^{(Q)i}(k;q-k,-q)\,^{\ast}\!S(q-k)\chi]
\,{\rm e}^{\ast i}(\hat{\bf k},\xi)\Bigl\}
\]
\[
+\left|\,[\,\bar{u}(\hat{\bf q},\lambda)
\,^{\ast}\Gamma^{(Q)i}(k;q-k,-q)\,^{\ast}\!S(q-k)\chi]
\,{\rm e}^{\ast i}(\hat{\bf k},\xi)\right|^{\,2}
\,\biggr)
\]
\[
+\,\Bigl(I^t(k,x)\rightarrow I^l(k,x),\,
{\rm e}^{i}(\hat{\bf k},\xi)\rightarrow
\sqrt{\frac{k^2}{k_0^2}}\,\hat{k}^i\Bigr)\biggr]
\]
\[
\hspace{2.5cm}
+\,\Bigr(\,^{\ast}{\!\Delta}_{+}(q)\rightarrow
\,^{\ast}{\!\Delta}_{-}(q),\;
u(\hat{\bf q},\lambda)\rightarrow
v(\hat{\bf q},\lambda),\;
\bar{u}(\hat{\bf q},\lambda)\rightarrow
\bar{v}(\hat{\bf q},\lambda)\Bigr)\biggl\}\,
\delta(v\cdot (k-q)).
\]
The integrands of Eqs.\,(\ref{eq:11p}), (\ref{eq:11a}) contain the factors
\begin{equation}
\frac{1}{(v\cdot q_1)^2}\,,\quad \frac{1}{(v\cdot q_1)}\,,\quad
\frac{1}{(v\cdot k)^2}\,,\quad \frac{1}{(v\cdot k)}\,.
\label{eq:11s}
\end{equation}
If we define the functions $\Phi_1[\,\tilde{\Upsilon}_{\pm}]$ and
$\Phi_2[I^{t,\,l}]$ on mass-shell of soft plasma
excitations, i.e., we set
\[
{\rm Im}\,^{\ast}\!\Delta_{\pm}(q)\simeq
\mp\,\pi\,{\rm Z}_{\pm}({\bf q})\,
\delta (q^0 - \omega_{\bf q}^{\pm})
\pm\,\pi\,{\rm Z}_{\mp}({\bf q})\,
\delta (q^0 + \omega_{\bf q}^{\mp}),
\hspace{0.7cm}
\]
\[
\hspace{0.6cm}
{\rm Im}\,^{\ast}\!\Delta^{t,\,l}(k)\simeq -\pi\,{\rm sign}(k^0)\!
\left(\frac{{\rm Z}_{t,\,l}({\bf k})}{2\omega_{{\bf k}}^{t,\,l}}\right)
[\,\delta (k^0 - \omega_{{\bf k}}^{t,\,l})
+\delta (k^0 + \omega_{{\bf k}}^{t,\,l})],
\]
and choose the spectral densities in a form of the quasiparticle approximation,
Eqs.\,(\ref{eq:8a}) and (\ref{eq:8k}), then factors (\ref{eq:11s})
will not be singular. This takes place due to the fact that the linear Landau
damping process is absent in the QGP. However, for off mass-shell excitations
of the medium when a frequency and momentum of plasma excitations approach to
the ``Cherenkov cone''
\[
(v\cdot q_1)\rightarrow 0,\quad
(v\cdot k)\rightarrow 0,
\]
these factors become singular that results in divergence of the integrals on
the right-hand sides of Eqs.\,(\ref{eq:11p}), (\ref{eq:11a}).

There exist precisely the same singularities in the integrands of
the `nondiagonal' contributions (\ref{eq:11u}) and (\ref{eq:11i}).
In order that the expression for energy loss had a finite value it is
necessary that these singularities should exactly compensated with
those in Eqs.\,(\ref{eq:11p}), (\ref{eq:11a}). From a comparison between
(\ref{eq:11a}) and (\ref{eq:11i}) we see that singularities
$1/(v\cdot k)^2,\,1/(v\cdot k)$ are exactly compensated in the
limit $(v\cdot k)\rightarrow 0$. We will require that a similar
reduction should take place for expressions (\ref{eq:11p}) and
(\ref{eq:11u}). This requirement results in the following
conditions of cancellation of the singularities
\begin{equation}
-\,\alpha^2C_2^{(\zeta)}T_F=2\,C_{\theta}C_F\,{\rm Re}\,\beta_1,
\label{eq:11d}
\end{equation}
\[
\alpha\,C_2^{(\zeta)}T_F = C_{\theta}C_F.
\]
Hence in particular it immediately follows that
\[
{\rm Re}\,\beta_1=-\frac{1}{2}\,\,\alpha,\quad
\alpha\neq 0.
\]
The last relation suggests that introduction of additional
sources and currents (\ref{eq:5a}), (\ref{eq:5h}), and so on is
necessary ingredient of the theory for its self-consistency. Let
us analyze the second relation in (\ref{eq:11d}). It is
natural to require that the constant $\alpha$, which enters as a
multiplier into the first term of the coefficient function
(\ref{eq:5s}) would independent of a type of the energetic parton
just as the second term. In view of this circumstance it is easy
to see that there exist the only reasonable choice of pairs of the
constants $C_2^{(\zeta)}$ and $C_{\theta}$ for which the second
relation in (\ref{eq:11d}) will be fulfilled identically: for the
values $\zeta=Q,\,\bar{Q}$ it is necessary to take
$C_{\theta}=-\,C_F$ and for $\zeta=G$ it should be set
$C_{\theta}=-n_fT_F$, i.e.,
\[
\alpha\, C_FT_F=-\,C_F^{2},
\]
\[
\alpha\, C_AT_F=-n_fT_FC_F.
\]
The requirement for independence of the constant $\alpha$ of a type of the
energetic parton results in the relation
\begin{equation}
n_fT_F=C_A\,(\equiv N_c),
\label{eq:11f}
\end{equation}
then
\[
\alpha =-\,C_F/T_F.
\]
The relation (\ref{eq:11f}) is fulfilled for $n_f=6$, $T_F=1/2$ and $N_c=3$.
Unfortunately, in spite of the fact that relation (\ref{eq:11f}) seems quite
reasonable from the physical point of view, it is true for extremely high
temperatures when one can neglect by mass of the heaviest $t$ quark.
Only under these conditions the complete sum of all (`diagonal' and
`nondiagonal') contributions to energy losses of the energetic parton will
have a finite value for scattering on off-shell soft plasma excitations.

\section{\bf `Nondiagonal' contribution to energy losses (continuation)}
\setcounter{equation}{0}

Now we proceed to discussion of expression (\ref{eq:10q}). Let us take the
effective currents $\tilde{j}_{\mu}^{\,\Psi ab}$ and
$\tilde{j}_{\mu}^{\,\Psi aj}$ on the right-hand side of the first equation in
(\ref{eq:8w}) in approximation of the second order in powers of the soft free
fields
\begin{equation}
\tilde{j}_{\mu}^{\,\Psi ab}({\bf v},\chi|\,k) \simeq
\delta^{ab}\,\tilde{j}^{(0)}_{Q\mu}(k) +
\tilde{j}_{\mu}^{(2)ab}(\bar{\psi}^{(0)},\psi^{(0)})(k),\quad
\tilde{j}^{(0)}_{Q\mu}(k)\equiv
\frac{g}{(2\pi)^3}\,v_{\mu}\delta(v\cdot k),
\label{eq:12q}
\end{equation}
\[
\tilde{j}_{\mu}^{\,\Psi aj}({\bf v},\bar{\chi}|\,k)\simeq
\tilde{j}_{\mu}^{(1)aj}(\psi^{(0)})(k)+
\tilde{j}_{\mu}^{(2)aj}(A^{(0)},\psi^{(0)})(k),
\hspace{5.1cm}
\]
where the current $\tilde{j}_{\mu}^{(2)ab}(\bar{\psi}^{(0)},\psi^{(0)})$ is
defined by Eqs.\,(\ref{eq:6q}), (\ref{eq:6w}) and the currents
$\tilde{j}_{\mu}^{(1)aj}(\psi^{(0)})$,
$\tilde{j}_{\mu}^{(2)aj}(A^{(0)},\psi^{(0)})$ are defined
by Eqs.\,(\ref{eq:5tt}), (\ref{eq:5r}) and (\ref{eq:6ee}), (\ref{eq:6r}),
respectively. By the following step we consider the coefficient function 
$K^{(G)ab,\,ij}_{\mu,\,\alpha\beta}$
that enters into the definition of the effective current
$\tilde{j}_{\mu}^{(2)ab}(\bar{\psi}^{(0)},\psi^{(0)})$. By analogy with 
(\ref{eq:11q}) we present it in the following form of a color 
decomposition:
\begin{equation}
K^{(G)ab,\,ij}_{\mu,\,\alpha\beta}({\bf v},\chi,\bar{\chi}|\,k;q_1,-q_2)=
\frac{1}{2}\,\{t^{a}\!,t^{b}\}^{ij}\,
K^{({\cal S})}_{\mu,\,\alpha\beta}
({\bf v},\chi,\bar{\chi}|\,k;q_1,-q_2)
\label{eq:12w}
\end{equation}
\[
\hspace{5.1cm}
+\,
\frac{1}{2}\,[\,t^{a}\!,t^{b}\,]^{\,ij}\,
K^{({\cal A})}_{\mu,\,\alpha\beta}
({\bf v},\chi,\bar{\chi}|\,k;q_1,-q_2).
\]
Here `symmetric' part equals
\begin{equation}
K^{({\cal S})}_{\mu,\,\alpha\beta}({\bf v},\chi,\bar{\chi}|\,k;q_1,-q_2)=
-{\cal M}_{\mu\nu,\,\alpha\beta}^{({\cal S})}(-k,k+q_1-q_2;-q_1,q_2)
\,^{\ast}{\cal D}^{\nu\nu^{\prime}}(k+q_1-q_2)v_{\nu^{\prime}}
\label{eq:12e}
\end{equation}
\[
+\,2\sigma\,\frac{v_{\mu}\,\chi_{\alpha}\,\bar{\chi}_{\beta}}
{(v\cdot q_1)(v\cdot q_2)}\,+\,
\alpha\,\frac{1}{(v\cdot q_1)}\,\chi_{\alpha}\Bigl[\,
\bar{\chi}_{\gamma}\,^{\ast}\!S_{\gamma\gamma^{\prime}}(k-q_2)
\,^{\ast}\Gamma^{(G)}_{\mu,\,\gamma^{\prime}\beta}(k;-q_2,-k+q_2)
\Bigr]
\]
\[
\hspace{3.5cm}
-\,\alpha\,\frac{1}{(v\cdot q_2)}\,\bar{\chi}_{\beta}\Bigl[\,
\,^{\ast}\Gamma^{(G)}_{\mu,\,\alpha\gamma}(k;q_1,-k-q_1)
\,^{\ast}\!S_{\gamma\gamma^{\prime}}(k+q_1)\chi_{\gamma^{\prime}}\Bigr].
\]
The explicit form of the `antisymmetric' part
$K^{({\cal A})}_{\mu,\,\alpha\beta}$
is not needed for our subsequent consideration and therefore we do not give it
here. The function ${\cal M}_{\mu\nu,\,\alpha\beta}^{({\cal S})}$ (and also
${\cal M}_{\mu\nu,\,\alpha\beta}^{({\cal A})}$) was introduced in Paper I (the
expressions following after Eq.\,(I.7.15)). We recall that these functions
enter into the matrix elements of two processes: the elastic scattering process
of soft-gluon excitation off soft-quark excitation and the process of pair
production by fusion of two soft-gluon excitations.

Further, we substitute effective currents (\ref{eq:12q}) into
(\ref{eq:10q}) taking into account Eqs.\,(\ref{eq:12w}),
(\ref{eq:12e}). We present again the expression obtained as a sum of two parts
different in structure:
\[
\left(\!-\frac{dE^{(1)}}{dx}\right)_{\!{\cal B}}=
\left(\!-\frac{dE^{(1)}}{dx}\right)_{\!{\rm diag}}\!+\;\,
\left(\!-\frac{dE^{(1)}}{dx}\right)_{\!{\rm nondiag}}.
\]
The `diagonal' part is defined by the following expression:
\begin{equation}
\left(\!-\frac{dE^{(1)}}{dx}\right)_{\!{\rm diag}}=
\frac{(2\pi)^3}{\vert{\bf v}\vert\,}\,
C_{\theta}C_F
\biggl(\frac{\alpha_s}{2{\pi}^2}\biggr)^{\!2}\!
\sum\limits_{\lambda=\pm\,}
\!\int\!k^0dk\,dq
\,\biggl\{{\rm Im}(^{\ast}{\!\Delta}^{t}(k))
\label{eq:12r}
\end{equation}
\[
\times\sum\limits_{\xi=1,\,2}\biggl(
\Bigl[\,|\,\bar{K}^{(G)i}({\bf v},\bar{\chi}|\,k,-q)
u(\hat{\bf q},\lambda){\rm e}^i(\hat{\bf k},\xi)|^{\,2}
\,\tilde{\Upsilon}_{+}(q,x)
\hspace{1cm}
\]
\[
\hspace{2.8cm}
+\,|\,\bar{K}^{(G)i}({\bf v},\bar{\chi}|\,k,-q)
v(\hat{\bf q},\lambda){\rm e}^i(\hat{\bf k},\xi)|^{\,2}
\,\tilde{\Upsilon}_{-}(q,x)\Bigr]\,
\delta(v\cdot(k-q))
\]
\[
\hspace{1.1cm}
+\Bigl[|\,\bar{u}(\hat{\bf q},\lambda)K^{(G)i}({\bf v},\chi|\,k,q)
{\rm e}^i(\hat{\bf k},\xi)|^{\,2}
\,\tilde{\Upsilon}_{+}(q,x)
\hspace{1cm}
\]
\[
\hspace{2.8cm}
+\,|\,\bar{v}(\hat{\bf q},\lambda)K^{(G)i}({\bf v},\chi|\,k,q)
{\rm e}^i(\hat{\bf k},\xi)|^{\,2}
\,\tilde{\Upsilon}_{-}(q,x)\Bigr]\,
\delta(v\cdot(k+q))\biggr)
\]
\[
\hspace{0.3cm}
+\,\Bigl(\,^{\ast}{\!\Delta}^{t}(k)\rightarrow
\,^{\ast}{\!\Delta}^{l}(k),\quad
{\rm e}^{i}(\hat{\bf k},\xi)\rightarrow
\sqrt{\frac{k^2}{k_0^2}}\,\hat{k}^i\Bigr)\biggr\}.
\]
We add the expression for the polarization losses connected with the initial 
color current of hard parton $\tilde{j}^{(0)a}_{Q\mu}(k)\equiv
Q_0^a\tilde{j}^{(0)}_{Q\mu}(k)$
\begin{equation}
\left(\!-\frac{dE^{(0)}}{dx}\right)_{\!{\cal B}}=
-\,\frac{1}{\,\vert{\bf v}\vert\,}\,\,
C_2^{(\zeta)}\biggl(\frac{\alpha_s}{2{\pi}^2}\biggr)\!
\int\!k^0dk\,{\rm Im}\,\langle
\tilde{j}^{(0)\mu}_{Q}(k)
\,^{\ast}{\cal D}_{\mu\nu}(k)
\tilde{j}^{(0)\nu}_{Q}(k)\rangle
\label{eq:12t}
\end{equation}
\[
\equiv
-\,\frac{1}{\,\vert{\bf v}\vert\,}\,\,
C_2^{(\zeta)}\biggl(\frac{\alpha_s}{2{\pi}^2}\biggr)\!
\int\!k^0dk\,\Biggl\{
{\rm Im}(\,^{\ast}{\!\Delta}^{t}(k))\!\!\!
\sum\limits_{\,\,\xi,\,\xi^{\prime}=1,\,2}\!
({\bf e}^{\ast}(\hat{\bf k},\xi)\cdot {\bf v})
({\bf e}(\hat{\bf k},\xi^{\prime})\cdot {\bf v})\,
\delta^{\xi\xi^{\prime}}
\hspace{1cm}
\]
\[
\hspace{9.6cm}
+\,
\biggl(\frac{k^2}{k_0^2}\biggr)
{\rm Im}(\,^{\ast}{\!\Delta}^{l}(k))
({\bf v}\cdot\hat{\bf k})^2
\Biggr\}\,\delta(v\cdot k)
\]
to the `nondiagonal' contribution $(-dE^{(1)}/dx)_{\rm nondiag}$.
On the rightmost side of (\ref{eq:12t}) we have taken into account that in the 
temporal gauge the following replacement holds:
\begin{equation}
\,^{\ast}{\cal D}_{\mu\nu}(k)\rightarrow
\,^{\ast}{\!\Delta}^{t}(k)\!\sum\limits_{\xi=1,\,2}\!
{\rm e}^{\ast i}(\hat{\bf k},\xi){\rm e}^{i^{\prime}}
(\hat{\bf k},\xi)\,+
\,^{\ast}{\!\Delta}^{l}(k)\biggl(\frac{k^2}{k_0^2}\biggr)\,
\hat{k}^i\hat{k}^{i^{\prime}}.
\label{eq:12y}
\end{equation}
As in previous section we present expression for a sum of
(\ref{eq:12t}) and `nondiagonal' contribution in the form of a sum of two 
terms:
\[
\left(\!-\frac{dE^{(0)}}{dx}\right)_{\!{\cal B}}\,+\,
\left(\!-\frac{dE^{(1)}}{dx}\right)_{\!{\rm nondiag}}\!=\,
\breve{\Lambda}_1+\breve{\Lambda}_2.
\]
Here the function $\breve{\Lambda}_1$ is
\[
\breve{\Lambda}_1\equiv
-\,\frac{1}{\,\vert{\bf v}\vert\,}\,\,
C_2^{(\zeta)}\biggl(\frac{\alpha_s}{2{\pi}^2}\biggr)\!
\int\!k^0dk\,\Biggl\{{\rm Im}(\,^{\ast}{\!\Delta}^{t}(k))
\Biggl(\,\sum\limits_{\,\xi,\,\xi^{\prime}=1,\,2}\biggl[
({\bf e}^{\ast}(\hat{\bf k},\xi)\cdot {\bf v})
({\bf e}(\hat{\bf k},\xi^{\prime})\cdot {\bf v})\,
\delta^{\xi\xi^{\prime}}
\]
\begin{equation}
+\,{\rm Re}\Bigl[\,
({\bf e}^{\ast}(\hat{\bf k},\xi)\cdot {\bf v})
({\bf e}(\hat{\bf k},\xi^{\prime})\cdot {\bf v})
\Pi_{tt}^{(1)}(k;\xi,\xi^{\prime})\,^{\ast}{\!\Delta}^{t}(k)\Bigr]\biggr]
\label{eq:12u}
\end{equation}
\[
+\,\sqrt{\frac{k^2}{k_0^2}}\sum\limits_{\xi=1,\,2}
{\rm Re}\Bigl[\,
({\bf e}^{\ast}(\hat{\bf k},\xi)\cdot {\bf v})
(\hat{{\bf k}}\cdot{\bf v})
\Pi_{tl}^{(1)}(k;\xi)\,^{\ast}{\!\Delta}^{l}(k)\Bigr]\Biggr)
\]
\[
\hspace{0.3cm}
+\,\Bigl(\,^{\ast}{\!\Delta}^{t}(k)\rightleftharpoons
\,^{\ast}{\!\Delta}^{l}(k),\;
{\rm e}^{i}(\hat{\bf k},\xi)\rightleftharpoons
\sqrt{\frac{k^2}{k_0^2}}\,\hat{k}^i,\;
\Pi_{tt}^{(1)}\rightarrow \Pi_{ll}^{(1)},\;
\Pi_{tl}^{(1)}\rightarrow \Pi_{lt}^{(1)}\Bigr)\Biggr\}\,
\delta(v\cdot k),
\]
where in turn
\[
\Pi_{tt}^{(1)}(k;\xi,\xi^{\prime})\!\equiv\!
-2g^2T_F\!\!\sum\limits_{\lambda=\pm\,}\!\!\int\!\!dq
\!\left\{\tilde{\Upsilon}_{+}(q,x)\!
\Bigl[\,{\rm e}^{\ast i}(\hat{\bf k},\xi)\bar{u}(\hat{\bf q},\lambda)
{\cal M}^{({\cal S})ii^{\prime}}(-k,k;-q,q)u(\hat{\bf q},\lambda)
{\rm e}^{i^{\prime}}(\hat{\bf k},\xi^{\prime})\Bigr]\right.
\]
\[
\hspace{4.8cm}
\left.
+\,\tilde{\Upsilon}_{-}(q,x)\!
\Bigl[\,{\rm e}^{\ast i}(\hat{\bf k},\xi)\bar{v}(\hat{\bf q},\lambda)
{\cal M}^{({\cal S})ii^{\prime}}(-k,k;-q,q)v(\hat{\bf q},\lambda)
{\rm e}^{i^{\prime}}(\hat{\bf k},\xi^{\prime})\Bigr]\right\}\!,
\]
\vspace{0.1cm}
\[
\Pi_{tl}^{(1)}(k;\xi)\!\equiv\!
-2g^2T_F\!\!\sum\limits_{\lambda=\pm\,}\!\int\!\!dq
\!\left\{\tilde{\Upsilon}_{+}(q,x)
\sqrt{\frac{k^2}{k_0^2}}
\Bigl[\,{\rm e}^{\ast i}(\hat{\bf k},\xi)\bar{u}(\hat{\bf q},\lambda)
{\cal M}^{({\cal S})ii^{\prime}}(-k,k;-q,q)u(\hat{\bf q},\lambda)
\hat{k}^{i^{\prime}})\Bigr]\right.
\]
\[
\hspace{4.7cm}
\left.
+\,\tilde{\Upsilon}_{-}(q,x)
\sqrt{\frac{k^2}{k_0^2}}
\Bigl[\,{\rm e}^{\ast i}(\hat{\bf k},\xi)\bar{v}(\hat{\bf q},\lambda)
{\cal M}^{({\cal S})ii^{\prime}}(-k,k;-q,q)v(\hat{\bf q},\lambda)
\hat{k}^{i^{\prime}})\Bigr]\!\right\}
\]
and so on. The functions $\Pi_{tt}^{(1)}$, $\Pi_{tl}^{(1)}$
represent linear in the soft-quark spectral densities
$\tilde{\Upsilon}_{\pm}$ corrections to various components of
the soft-gluon self-energy $\delta\Pi_{\mu\nu}(k)$ in the HTL
approximation. These corrections take into account a change of dispersion
properties of medium induced by the processes of nonlinear interaction of
soft-gluon and soft-quark excitations among themselves.
It is possible to make sure in this if we define an {\it effective} gluon
propagator $\,^{\ast}\tilde{\cal D}_{\mu\nu}(k)$ that takes into account
additional contributions considering nonlinear effects of the
self-interaction of soft excitations
\[
^{\ast}\tilde{\cal D}_{\mu\nu}^{-1}(k)\equiv
\,^{\ast}{\cal D}_{\mu\nu}^{-1}(k)-
\Bigl\{\Pi^{(1)}_{\mu\nu}[\tilde{\Upsilon}_{\pm}](k) +
\Pi^{(1)}_{\mu\nu}[I^{t,\,l}](k)\Bigr\} -
\Pi^{(2)}_{\mu\nu}[\tilde{\Upsilon}_{\pm},I^{t,\,l}](k)-\,\ldots\,.
\]
For low excited state when $\,^{\ast}{\cal D}_{\mu\nu}^{-1}(k)
\gg\Pi^{(1)}_{\mu\nu}\gg\Pi^{(2)}_{\mu\nu}\gg\ldots$ from equation above to
the first order in $\tilde{\Upsilon}_{\pm}$ and $I^{t,\,l}$, we obtain
\[
^{\ast}\tilde{\cal D}_{\mu\nu}(k)\simeq
\,^{\ast}{\cal D}_{\mu\nu}(k)\,+
\,^{\ast}{\cal D}_{\mu\mu^{\prime}}(k)
\Bigl\{\Pi^{(1)\mu^{\prime}\nu^{\prime}}[\tilde{\Upsilon}_{\pm}](k) +
\Pi^{(1)\mu^{\prime}\nu^{\prime}}[I^{t,\,l}](k)\Bigr\}
\!\,^{\ast}{\cal D}_{\nu^{\prime}\nu}(k) +\ldots\,.
\]
The term with $\Pi^{(1)\mu^{\prime}\nu^{\prime}}[I^{t,\,l}](k)$ was considered
in the paper \cite{markov_AOP_04}. It connected with self-interaction of
soft-gluon excitations. Convolving this expression with $v^{\mu}v^{\nu}$,
considering (\ref{eq:12y}) and taking the imaginary part, we obtain an
expression similar in the form to Eq.\,(\ref{eq:11o}). The terms proportional
to ${\rm Im}(\!\,^{\ast}{\!\Delta}^{t,\,l}(k))$ exactly reproduce the integrand
in (\ref{eq:12u}) if we identify
\[
\Pi_{tt}^{(1)}(k;\xi,\xi^{\prime})\equiv
{\rm e}^{\ast i}(\hat{\bf k},\xi)
\Pi^{(1)ii^{\prime}}[\tilde{\Upsilon}_{\pm}](k)
{\rm e}^{i^{\prime}}(\hat{\bf k},\xi^{\prime}),
\]
\[
\Pi_{tl}^{(1)}(k;\xi)\equiv \sqrt{\frac{k^2}{k_0^2}}\,
\Bigl({\rm e}^{\ast i}(\hat{\bf k},\xi)
\Pi^{(1)ii^{\prime}}[\tilde{\Upsilon}_{\pm}](k)
\hat{k}^{i^{\prime}}\Bigr),
\]
etc. If the contributions proportional to 
${\rm Re}(\!\,^{\ast}{\!\Delta}^{t,\,l}(k))$ are absent, then function
(\ref{eq:12u}) can be exceptionally interpreted as the
polarization losses taking into account to the first approximation
a change of the dispersion properties of the QCD medium induced by 
self-interaction of soft excitations. This could
be effectively presented as a replacement of the HTL-resummed
soft-gluon propagator $\,^{\ast}{\cal D}_{\mu\nu}(k)$ in the first
line of Eq.\,(\ref{eq:12t}) by the effective one
\[
\,^{\ast}{\cal D}_{\mu\nu}(k)\rightarrow
\,^{\ast}\tilde{\cal D}_{\mu\nu}(k).
\]
Unfortunately, the existence of the terms proportional to
${\rm Re}(\,^{\ast}{\!\Delta}^{t,\,l}(k))$ gives no way of
reducing everything to such a simple replacement.

Furthermore, the function $\breve{\Lambda}_2$ is
\begin{equation}
\breve{\Lambda}_2\equiv-2\,
\frac{(2\pi)^3}{\vert{\bf v}\vert\,}\,\,
C_2^{(\zeta)}T_F
\biggl(\frac{\alpha_s}{2{\pi}^2}\biggr)
\sum\limits_{\lambda=\pm\,}
\!\int\!k^0dk\,dq
\,\Biggl\{\biggl[\,{\rm Im}(^{\ast}{\!\Delta}^{t}(k))
\label{eq:12i}
\end{equation}
\[
\times\tilde{\Upsilon}_{+}(q,x)
\sum\limits_{\xi=1,\,2}
\biggl(
2\!\,{\rm Re}\,\sigma\,\frac{|({\bf e}(\hat{\bf k},\xi)\cdot {\bf v})|^{\,2}
|\bar{u}(\hat{\bf q},\lambda)\chi|^{\,2}}
{(v\cdot q)^2}
\]
\[
+\,\alpha\,\frac{1}{(v\cdot q)}\,{\rm Re}\,\Bigl[
({\bf e}^{\ast}(\hat{\bf k},\xi)\cdot {\bf v})\,
(\bar{u}(\hat{\bf q},\lambda)\chi)
\Bigl(\bar{\chi}\,^{\ast}\!S(k-q)\,^{\ast}\Gamma^{(G)i}(k;-q,-k+q)
u(\hat{\bf q},\lambda){\rm e}^i(\hat{\bf k},\xi)\Bigr)\Bigr]
\]
\[
-\,\alpha\,\frac{1}{(v\cdot q)}\,{\rm Re}\,\Bigl[
({\bf e}(\hat{\bf k},\xi)\cdot {\bf v})\,
(\bar{\chi}\,u(\hat{\bf q},\lambda))
\Bigl(\bar{u}(\hat{\bf q},\lambda)
\,^{\ast}\Gamma^{(G)i}(k;q,-k-q)
\,^{\ast}\!S(k+q)\chi\,{\rm e}^{\ast i}(\hat{\bf k},\xi)\Bigr)\Bigr]
\biggr)
\hspace{0.2cm}
\]
\[
\hspace{1.5cm}
+\,\Bigl(\,\tilde{\Upsilon}_{\!+}(q,x)\rightarrow
\tilde{\Upsilon}_{\!-}(q,x),\;
u(\hat{\bf q},\lambda)\rightarrow v(\hat{\bf q},\lambda),\;
\bar{u}(\hat{\bf q},\lambda)\rightarrow \bar{v}(\hat{\bf q},\lambda)
\Bigr)\biggr]
\]
\[
+\,\Bigl(\,^{\ast}{\!\Delta}^{t}(k)\rightarrow
\,^{\ast}{\!\Delta}^{l}(k),\quad
{\rm e}^{i}(\hat{\bf k},\xi)\rightarrow
\sqrt{\frac{k^2}{k_0^2}}\,\hat{k}^i\Bigr)\Biggr\}\,
\delta(v\cdot k).
\hspace{0.65cm}
\]

By virtue of the definition of the coefficient functions
$K_{\alpha}^{(G)i}$, $\bar{K}_{\alpha}^{(G)i}$ (Eqs.\,(\ref{eq:5e}),
(\ref{eq:5r})) the integrand of `diagonal' contribution (\ref{eq:12r})
has singularities of a type
\[
\frac{1}{(v\cdot q)^2},\quad \frac{1}{(v\cdot q)}
\]
for off-shell excitations of medium when frequency and momentum approach
to the ``Cherenkov cone''
\[
(v\cdot q)\rightarrow 0.
\]
We require that these singularities in accuracy should be compensated by
similar ones in the integrand of Eq.\,(\ref{eq:12i}). This results in
the following conditions of cancellation of the singularities
\begin{equation}
2C_2^{(\zeta)}T_F\,{\rm Re}\,\sigma=
C_{\theta}\,C_F,
\label{eq:12o}
\end{equation}
\[
\alpha\,C_2^{(\zeta)}T_F=C_{\theta}\,C_F.
\]
Hence it follows
\[
{\rm Re}\,\sigma=\frac{1}{2}\,\alpha\,.
\]
The second condition in (\ref{eq:12o}) exactly coincides with the
second one in (\ref{eq:11d}) and thus here, with the help of the
same reasoning we obtain the relation (\ref{eq:11f}) and value for the
constant $\alpha$: $\alpha=-C_F/T_F$. The conditions of
cancellation of singularities obtained in present and previous
sections do not contradict each other. We can expect that similar
cancellation takes place for all contributions to energy losses 
(\ref{eq:10q}), (\ref{eq:10w}) of higher orders in powers of the soft 
free fields. This in particular leads to (partial) fixing
undetermined constants of the additional currents and sources introduced in
Section 5.

\section{\bf Conclusion}
\setcounter{equation}{0} 

In this paper within the framework of the
semiclassical approximation we have presented successive scheme of
construction of the effective theory for the processes of interaction
of soft and hard quark-gluon plasma excitations for both the Fermi and Bose
statistics. In the third part \cite{paper_III}, we completed an
analysis of dynamics of soft fermion excitations taking into
account also radiative processes. Unfortunately, in view of great amount 
of this work we did not give the concrete analysis of the
expressions obtained for energy losses, as it has been made in
our early papers \cite{markov_AOP_04}, \cite{markov_AOP_05}. This is
expected to be made in separate publication.

As has been shown during all this work, introduction into
consideration of the Grassmann color charges for hard particle
and color sources with them associated turns out to be rather powerful
method in the analysis of dynamics of the soft (anti)quark modes. This
allows obtaining almost completely self-consistent and
self-sufficient calculation scheme of the effective color currents
and sources and the matrix elements for the scattering processes
we are interested in.
Note that the effective currents and sources derived within the framework 
of this calculation scheme possess striking symmetry with respect to
free soft-gluon and soft-quark fields, the usual and Grassmann
charges. This circumstance suggests that there should be
(super?)transformations touching by nontrivial way both boson and fermion 
degrees of freedom of hard and soft excitations of system under consideration
transforming effective currents and sources into each
other\footnote{It is curious to note that the relation of
(\ref{eq:11f}) type (without the multiplier $T_F$) was mentioned
for the first time in the paper \cite{dokshitzer} in the context of
analysis of various symmetry relations among quark and gluon
decay probabilities.}. In view of observable high symmetry with
respect to fermion and boson degrees of freedom it is possible to
raise the question about a possibility of supersymmetric
generalization of the approach suggested in this work.
Supersymmetric formulation of the effective theory would allow to
look from the more general point of view on the dynamics of
interaction processes in QGP and possibly to predict
existence of qualitatively new phenomena. Partly, a tool necessary for 
such generalization was considered in literature. So, for example, the
problem of supersymmetrization of classical point particle with
spin and isospin has been investigated in detail both within the
framework of the local supersymmetric formulation and in the
super-space one (see, e.g., Ref.\,\cite{salomonson} and references
therein). Unfortunately, at present there is no a supersymmetric
generalization of the most important ingredient of the effective theory, 
namely, the concept of the hard thermal loops. This in itself can be a 
subject for separate research.

\section*{\bf Acknowledgments}
This work was supported in part by the Russian Foundation for Basic Research
(project no 03-02-16797).

\newpage
\section*{\bf Appendix A}
\setcounter{equation}{0}

The dynamics of the classical color particle in the external gauge and quark
fields (when we neglect by the change in trajectory and spin state of particle)
can be described by the following action
$$
{\cal S}=\int\limits_{t_0}^t\!\!{\cal L}(t)\,dt,\quad
{\cal L}(t)= {\cal L}_I(t)+{\cal L}_{II}(t),
$$
where
$$
{\cal L}_I(t)=i\,\vartheta^{\dagger\, i}\dot{\vartheta^i}-
gv^{\mu}A^a_{\mu}\vartheta^{\dagger\, i}(t^a)^{ij}\vartheta^j-
g\Bigl\{\vartheta^{\dagger\, i}
(\bar{\chi}_{\alpha}\psi_{\alpha}^i)+
(\bar{\psi}_{\alpha}^i{\chi}_{\alpha})\vartheta^i\Bigr\}
\eqno{({\rm A}.1)}
$$
and
$$
{\cal L}_{II}(t)=-g\,[\vartheta^{\dagger\, i}(t^a)^{ij}\vartheta^j]
\,\Bigl\{\hat{\alpha}\,(\bar{\psi}_{\alpha}^k{\chi}_{\alpha})(t^a)^{ks}
\Omega^s +
\hat{\alpha}^{\ast}\,\Omega^{\dagger\, s}(t^a)^{sk}
(\bar{\chi}_{\alpha}\psi_{\alpha}^k)\Bigr\}
\eqno{({\rm A}.2)}
$$
$$
-\,g\,\Bigl\{\hat{\beta}\,[\vartheta^{\dagger\, i}(t^a)^{ij}\Omega^j]
[\,(\bar{\psi}_{\alpha}^k{\chi}_{\alpha})(t^a)^{ks}\vartheta^s]+
\hat{\beta}^{\ast}\,[\,\Omega^{\dagger\, j}(t^a)^{ji}\vartheta^i]
[\vartheta^{\dagger\, s}(t^a)^{sk}
(\bar{\chi}_{\alpha}\psi_{\alpha}^k)]\Bigr\}.
$$
In Eq.\,(A.2) $\hat{\alpha}$ and $\hat{\beta}$ are (complex)
parameters\footnote{We introduce the hat above to distinguish the parameters
from those in the additional sources in Section 5.}. The action ${\cal S}$ is
real and it is invariant under gauge transformation
$$
A_{\mu}^at^a\rightarrow S A_{\mu}^a t^a S^{-1}-
(i/g) S\partial_{\mu}S^{-1},\quad
\psi_{\alpha}\rightarrow S\psi_{\alpha},\quad
\vartheta\rightarrow S\vartheta.
\eqno{({\rm A}.3)}
$$
The soft stochastic gauge and quark fields in the Lagrangian ${\cal L}$
are determined on the {\it linear} trajectory
$$
A_{\mu}^a\equiv A_{\mu}^a(t,{\bf v}t),\quad
\psi^i_{\alpha}\equiv\psi^i_{\alpha}(t,{\bf v}t),\quad
\bar{\psi}^i_{\alpha}\equiv\bar{\psi}^i_{\alpha}(t,{\bf v}t)
$$
and the function $\Omega^i=\Omega^i(t)$ is defined by equation (\ref{eq:5p}).
Here $t$ is the coordinate time. The action with Lagrangian (A.1) results in
equations of motion (\ref{eq:5y}) and Lagrangian (A.2) defines additional
interaction terms in these equations. Lagrangian (A.2) (at least the first
term) describes current-current interaction in a system, where the first
current represents color that of hard component of system, and the second
current represents color that of soft component. In view of (A.2) the
following terms should be added to the left-hand side of the first equation
in (\ref{eq:5y})
$$
ig\,(t^a)^{ij}\vartheta^j
\,\Bigl\{\hat{\alpha}\,(\bar{\psi}_{\alpha}^k{\chi}_{\alpha})(t^a)^{ks}
\Omega^s +
\hat{\alpha}^{\ast}\,\Omega^{\dagger\, s}(t^a)^{sk}
(\bar{\chi}_{\alpha}\psi_{\alpha}^k)\Bigr\}
\eqno{({\rm A}.4)}
$$
$$
+\,ig\,\Bigl\{\hat{\beta}\,(t^a)^{ij}\Omega^j\,
[\,(\bar{\psi}_{\alpha}^k{\chi}_{\alpha})(t^a)^{ks}\vartheta^s]+
\hat{\beta}^{\ast}\,(t^a)^{ij}
(\bar{\chi}_{\alpha}\psi_{\alpha}^j)
[\,\Omega^{\dagger\, k}(t^a)^{ks}\vartheta^s]\Bigr\}.
$$

The equation of motion (\ref{eq:5y}) with additional terms (A.4) has the
following general solution
$$
\vartheta^{i}(t) = {\cal U}^{\,ij}(t,t_0)\theta_0^{j}
-ig\!\int\limits_{t_0}^{t}\!
{\cal U}^{\,ij}(t,\tau)
\Bigl(\bar{\chi}_{\alpha}\psi^{j}_{\alpha}(\tau,{\bf v}\tau)\Bigr)d\tau,
\eqno{({\rm A}.5)}
$$
where we have introduced extended evolution operator
$$
{\cal U}(t,\tau) =
{\rm T}\exp\Biggl\{-ig\!\int\limits_{\tau}^{t}v^{\mu}
A_{\mu}^a(\tau^{\,\prime},{\bf v}\tau^{\,\prime\,})t^a d\tau^{\,\prime}
\eqno{({\rm A}.6)}
$$
$$
-ig\!\int\limits_{\tau}^{t}\Bigl[\,
\hat{\alpha}\,(\bar{\psi}_{\alpha}^k(\tau^{\,\prime},{\bf v}\tau^{\,\prime\,})
{\chi}_{\alpha})(t^a)^{ks}
\Omega^s(\tau^{\,\prime\,}) +
\hat{\alpha}^{\ast}\,\Omega^{\dagger\, s}(\tau^{\,\prime\,})(t^a)^{sk}
(\bar{\chi}_{\alpha}
\psi_{\alpha}^k(\tau^{\,\prime},{\bf v}\tau^{\,\prime\,}))\Bigr]t^a
d\tau^{\,\prime}
$$
$$
-ig\!\int\limits_{\tau}^{t}\Bigl[\,
\hat{\beta}\,t^a\Omega(\tau^{\,\prime\,})\otimes
(\bar{\psi}_{\alpha}(\tau^{\,\prime},{\bf v}\tau^{\,\prime\,})
{\chi}_{\alpha})t^a+
\hat{\beta}^{\ast}\,t^a(\bar{\chi}_{\alpha}
\psi_{\alpha}(\tau^{\,\prime},{\bf v}\tau^{\,\prime\,}))\otimes
\Omega^{\dagger}(\tau^{\,\prime\,})t^a\Bigr] d\tau^{\,\prime}\Biggr\}.
$$
Here $\otimes$ is a sign of the direct production. The evolution operator (A.6)
takes into account the effect of rotation of color charge in color space
induced by both soft gauge and soft quark stochastic fields. The influence of
soft quark field on the rotation should be already taken into account in the
scattering processes of a third order in the coupling constant. Under gauge
transformation (A.3) this evolution operator transforms by covariant fashion
$$
{\cal U}(t,\tau)\rightarrow S(t)\,{\cal U}(t,\tau)S^{-1}(\tau).
$$
Instead of identity (\ref{eq:5uu}) now we have
$$
{\cal U}(\tau,t)t^a{\cal U}(t,\tau)=\tilde{\cal U}^{\,ab}(t,\tau)t^b,
\eqno{({\rm A}.7)}
$$
where the extended evolution operator in the adjoint representation
$\tilde{{\cal U}}(t,\tau)$ is
$$
\tilde{\cal U}(t,\tau) =
{\rm T}\exp\Biggl\{-ig\!\int\limits_{\tau}^{t}v^{\mu}
A_{\mu}^a(\tau^{\,\prime},{\bf v}\tau^{\,\prime\,})T^a d\tau^{\,\prime}
$$
$$
-\,ig(\hat{\alpha}+\hat{\beta}/N_c)\!\int\limits_{\tau}^{t}
\Bigl[\,(\bar{\psi}_{\alpha}^k(\tau^{\,\prime},{\bf v}\tau^{\,\prime\,})
{\chi}_{\alpha})(t^a)^{ks}
\Omega^s(\tau^{\,\prime\,})\Bigr]T^ad\tau^{\,\prime}
$$
$$
-\,ig(\hat{\alpha}^{\ast}+\hat{\beta}^{\ast}/N_c)\!\int\limits_{\tau}^{t}
\Bigl[\,\Omega^{\dagger\, s}(\tau^{\,\prime\,})(t^a)^{sk}
(\bar{\chi}_{\alpha}
\psi_{\alpha}^k(\tau^{\,\prime},{\bf v}\tau^{\,\prime\,}))\Bigr]T^a
d\tau^{\,\prime}\Biggr\}.
$$
The relation (A.7) is easily proved with the Fierz identity for the matrices
$t^a$
$$
(t^a)^{ij}(t^a)^{ks}=-\frac{1}{N_c}\,(t^a)^{is}(t^a)^{kj}
+\frac{C_F}{N_c}\,\delta^{is}\delta^{kj}.
$$
We add the action for the soft gluon and quark fields to the action determining
dynamics of hard test particle
$$
-\frac{1}{4}\int\!d^4x\, F^a_{\mu\nu}(x)F^{a\,\mu\nu}(x)\,+\,
i\!\int\!d^4x\,\bar{\psi}(x)\gamma_{\mu}D^{\mu}(x)\psi(x)\,+\,...\,,
$$
where $D^{\mu}(x)\equiv\partial/\partial x_{\mu}+igA^{a\mu}(x)t^a$. Hereafter
the dots mean contributions caused by HTL effects. The equation of motion for
soft quark field $\psi$ now has the following form
$$
i(\gamma_{\mu}D^{\mu}(x)\psi(x))^i_{\alpha}=g\chi_{\alpha}\vartheta^i(t)
\,\delta({\bf x}-{\bf v}t)
+g\,\hat{\alpha}\,\chi_{\alpha}(t^a)^{ij}\Omega^j(t)
[\,\vartheta^{\dagger\, k}(t)(t^a)^{ks}\vartheta^s(t)]
\,\delta({\bf x}-{\bf v}t)
\hspace{1cm}
$$
$$
\hspace{7.1cm}
+\,g\hat{\beta}\,\chi_{\alpha}(t^a)^{ij}\vartheta^j(t)
[\,\vartheta^{\dagger\, k}(t)(t^a)^{ks}\Omega^s(t)]
\,\delta({\bf x}-{\bf v}t)\,+\,...\,.
$$
Let us substitute solution (\ref{eq:5u}) (here we neglect
temporarily by contribution of $\psi$-fields to the evolution operator
${\cal U}(t,t_0))$ into the second and third terms on the right-hand side
of the last equation. If we require fulfilment of condition
$$
\hat{\alpha}+\hat{\beta}=0,
\eqno{({\rm A}.8)}
$$
then these terms result in the form\footnote{The condition (A.8) enables one to
cancel out the terms not containing color charges at all.}
$$
g\,\Bigl\{
\hat{\alpha}\,\chi_{\alpha}(t^a)^{ij}\Omega^j(t)Q^a(t)+
\hat{\alpha}\,\chi_{\alpha}(t^a)^{ij}\Omega^j(t)
[\,\Omega^{\dagger\,k}(t)(t^a)^{ks}\theta^s(t)]
\hspace{1cm}
\eqno{({\rm A}.9)}
$$
$$
\hspace{1.2cm}
+\,\hat{\beta}\,\chi_{\alpha}(t^a)^{ij}\theta^j(t)
[\,\theta^{\dagger\,k}(t)(t^a)^{ks}\Omega^s(t)]
+\hat{\beta}\,\chi_{\alpha}(t^a)^{ij}\theta^j(t)\Xi^a(t)\Bigr\}
\,\delta({\bf x}-{\bf v}t).
$$
The first, second and last terms here reproduce entered by hands additional
sources (\ref{eq:5a}), (\ref{eq:5h}) and (\ref{eq:5g}) correspondingly.
The next to last term represents new additional rather remarkable source.
Its existence suggests necessity of introduction of one more color structure
for description of color charge in the semiclassical approximation, namely,
$$
\Theta^{ij}\equiv[(t^a)^{ii^{\,\prime}}\theta^{i^{\,\prime}}\!(t)]
\,[\theta^{\dagger\, j^{\,\prime}}\!(t)(t^a)^{j^{\,\prime}\!j}],\quad
\Theta^{\dagger\,ij}=\Theta^{ji}.
$$
When such a combination in final expressions occurs, it is
necessary to identify it as independent function, by analogy to the
structure $Q^a(t)\equiv \theta^{\dagger\, i}(t^a)^{ij}\theta^j(t)$.
In particular, to lowest order in the coupling constant the third term in (A.9)
defines the eikonal contribution to the effective source
$$
\left.\frac{\delta^2\eta_{\alpha}^i(q)}
{\delta \Theta^{js}_0
\delta\psi^{(0)k}_{\beta}(q_1)}\,
\right|_{\,0}=\hat{\beta}\,
\frac{\,g^2}{(2\pi)^3}\,
\frac{\chi_{\alpha}\bar{\chi}_{\beta}}{(v\cdot q_1)}
\,\delta^{ij}\delta^{sk}
\,\delta(v\cdot (q-q_1)).
$$

The equation of motion for the soft gluon field $A_{\mu}^a(x)$ have former form
$$
[D^{\mu}(x),F_{\mu\nu}(x)]=
gv_{\nu}[\vartheta^{\dagger\, i}(t)(t^a)^{ij}\vartheta^j(t)]\,t^a\,
\delta({\bf x}-{\bf v}t)\,+\,...\,,
\eqno{({\rm A}.10)}
$$
where as the Grassmann charge $\vartheta^i(t)$ it is necessary to mean solution
(A.5). The contribution of additional interaction terms (A.2) here
`is hid' in extended evolution operator (A.6). Some new current structures do
not appear. Unfortunately, generalized action suggested in this Appendix not
enables one to generate additional current (\ref{eq:5k}) containing
anticommutator $\{t^a,t^b\}$. The current on the right-hand side of (A.10)
enables us to determine the same expression (\ref{eq:5k}), but with commutator
$[t^a,t^b]$ only.

\section*{\bf Appendix B}
\setcounter{equation}{0}

\begin{figure}[hbtp]
\begin{center}
\includegraphics[width=1\textwidth]{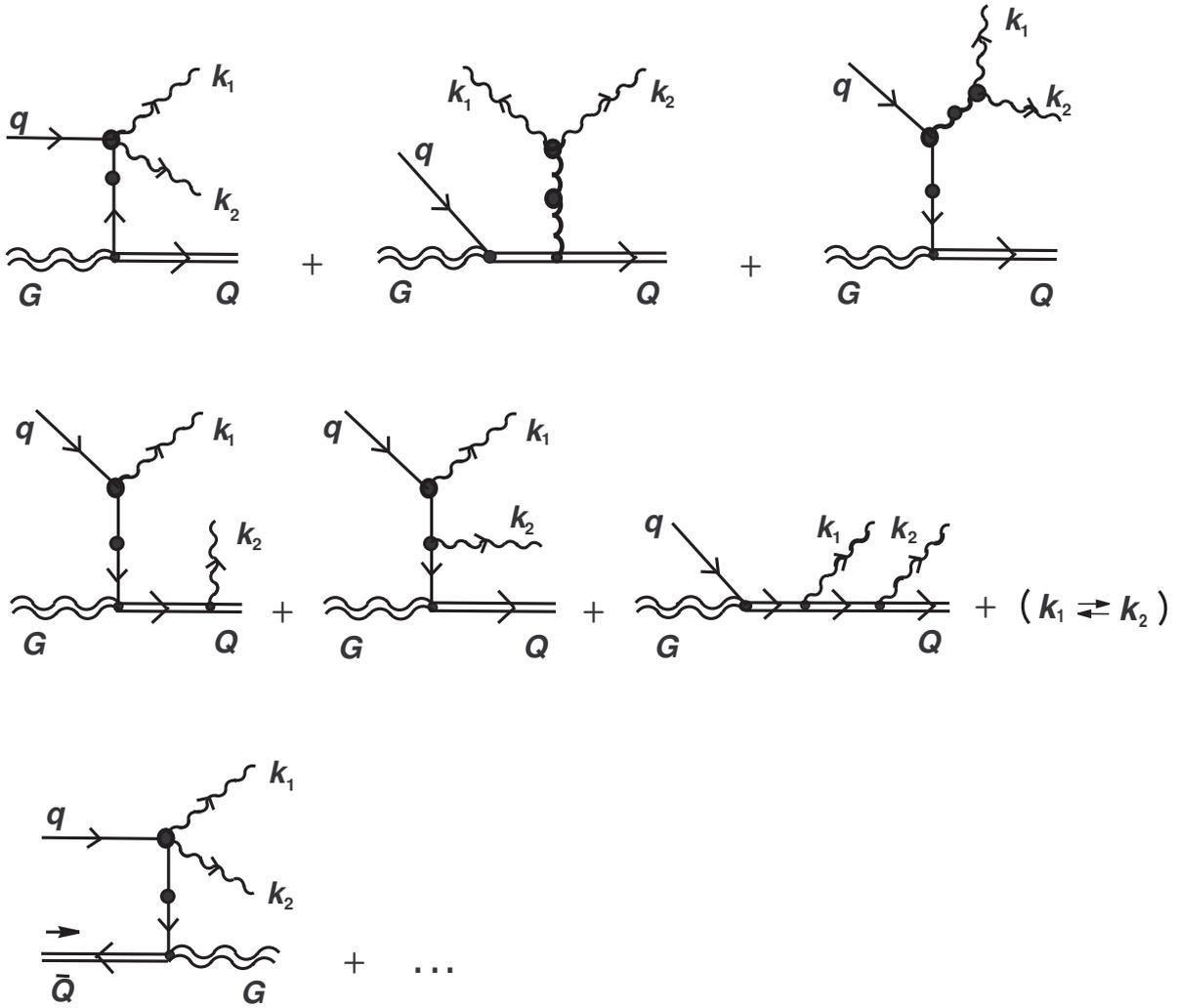}
\end{center}
\caption{\small The scattering processes of soft-quark elementary
excitation off the hard test particle followed by emission of two
soft-gluon excitations.}
\label{fig13}
\end{figure}

Here we give complete expression for the coefficient function
$K_{\mu_1\mu_2,\,\alpha}^{(Q)a_1a_2,\,ij}
({\bf v},{\bf v},\chi|\,q;-k_1,-k_2)$ defining the scattering process of
soft-quark excitation off hard parton followed by emission of two
soft-gluon excitations
$$
K^{(Q)a_1a_2,\,ij}_{\mu_1\mu_2,\,\alpha}
({\bf v},{\bf v},\chi|\,q;-k_1,-k_2)\equiv
-\,\delta{\Gamma}^{(Q)a_1a_2,\,ij}_{\mu_1\mu_2,\,\alpha\beta}
(k_1,k_2;q-k_1-k_2,-q)
\,^{\ast}\!S_{\beta\beta^{\,\prime}}(q-k_1-k_2)
\chi_{\beta^{\,\prime}}
$$
$$
+\,[t^{a_1},t^{a_2}]^{ij}
K_{\mu^{\,\prime}\!,\,\alpha}^{(Q)}({\bf v},\chi|\,k_1+k_2,-q)
\,^{\ast}{\cal D}^{\mu^{\,\prime}\mu}(k_1+k_2)
\!\,^\ast\Gamma_{\mu\mu_1\mu_2}(k_1+k_2,-k_1,-k_2)\!
$$
$$
-\,(t^{a_1}t^{a_2})^{ij}
\,^{\ast}\Gamma^{(Q)}_{\mu_1,\,\alpha\beta}(k_1;q-k_1,-q)
\,^{\ast}\!S_{\beta\beta^{\,\prime}}(q-k_1)
K_{\mu_2,\,\beta^{\,\prime}}^{(Q)}({\bf v},\chi|\,k_2,-q+k_1)\!\!
\eqno{({\rm B}.1)}
$$
$$
\hspace{0.3cm}
-\,(t^{a_2}t^{a_1})^{ij}
\,^{\ast}\Gamma^{(Q)}_{\mu_2,\,\alpha\beta}(k_2;q-k_2,-q)
\,^{\ast}\!S_{\beta\beta^{\,\prime}}(q-k_2)
K_{\mu_1,\,\beta^{\,\prime}}^{(Q)}({\bf v},\chi|\,k_1,-q+k_2)
\hspace{1.1cm}
$$
$$
+\,(t^{a_1}t^{a_2})^{ij}\,
\frac{1}{(v\cdot q)(v\cdot k_2)}\,\,
v_{\mu_1}v_{\mu_2}\chi_{\alpha}
+\,(t^{a_2}t^{a_1})^{ij}\,
\frac{1}{(v\cdot q)(v\cdot k_1)}\,\,
v_{\mu_1}v_{\mu_2}\chi_{\alpha}\,.
\hspace{0.9cm}
$$
The partial coefficient functions
$K_{\mu^{\,\prime}\!,\,\alpha}^{(Q)}({\bf v},\chi|\,k_1+k_2,-q)$ and so on
are defined by expression (\ref{eq:4y}). The graphic interpretation
of various terms on the right-hand side of Eq.\,(B.1) is presented in
Fig.\,\ref{fig13}. The dots means graphs of the scattering processes with
participation of hard thermal antiquark.


\section*{\bf Appendix C}
\setcounter{equation}{0}

In this Appendix we consider in more detail a structure of the
terms in Eq.\,(\ref{eq:7q}) containing the singularity
$\delta(v\cdot q^{\,\prime\,})/(v\cdot q^{\,\prime\,})$. In the
fist stage of our analysis it is more convenient to use the
HTL-resummed quark propagator in the representation suggested by
Weldon \cite{weldon}
$$
\,^{\ast}\!S(q)=\frac{1}{(1+a)\!\!\not\!q+b\!\!\not\!u}
=\frac{(1+a)\!\!\not\!q+b\!\!\not\!u}{D}\,,
\eqno{({\rm C}.1)}
$$
where $\not\!q\equiv q^{\mu}\gamma_{\mu}$, $u$ is a global four-velocity of
plasma, $a,\,b$ are Lorentz-invariant functions, and
$D=(1+a)^2q^2+2(1+a)b(q\cdot u)+b^2$. Let us define a polarization matrix of
hard parton (quark or antiquark) $\varrho=(\varrho_{\alpha\beta})$ such that
in pure state it is reduced to a product
$$
\varrho_{\alpha\beta}=\chi_{\alpha}\bar{\chi}_{\beta}.
$$
We consider a case of fully unpolarized state. By virtue of the fact
that this matrix can depend only on the energy $E$ and the velocity ${\bf v}$
of hard test particle, we can write out at once its explicit form
$$
\varrho=\varrho(E,{\bf v})=\frac{1}{2E}\,\varrho({\bf v}),
\eqno{({\rm C}.2)}
$$
where
$$
\varrho({\bf v})=\frac{1}{2}\,\,(v\cdot\gamma),\quad
v=(1,{\bf v}).
$$
The multiplier $1/2E$ is chosen for reasons of dimension.
In this case only both terms in amplitude (\ref{eq:5s}) have the same
dimension. Thus in view of the structure of soft-quark propagator (C.1)
and polarization matrix (C.2) the following replacement is correct:
$$
\bar{\chi}_{\alpha}
\,^{\ast}\!S_{\alpha\alpha^{\,\prime}}(q^{\,\prime\,})\chi_{\alpha^{\,\prime}}\,
\Rightarrow\,\frac{1}{2E}\,
{\rm Sp}\Bigl[\,\varrho({\bf v})\,^{\ast}\!S(q^{\,\prime\,})\Bigr]=
\frac{1}{2E}\,\left[\frac{2(1+a)}{D}\,(v\cdot q^{\,\prime\,}) +
\frac{2b}{D}\,(v\cdot u)\right].
$$
Furthermore, by using the property $\,^{\ast}\!S(-q^{\,\prime\,})=
-\gamma^0\,^{\ast}\!S^{\dagger}(q^{\,\prime\,})\,\gamma^0$ the first
two terms in integrand of (\ref{eq:7q}) can be resulted in the
following form:
$$
\frac{2}{E}\!\int\!dq^{\,\prime}\,{\rm Re}
\biggl[\frac{(1+a)}{D}\biggr]\,\delta(v\cdot q^{\,\prime\,})+
\frac{2}{E}\!\int\!dq^{\,\prime}\,
\frac{\delta(v\cdot q^{\,\prime\,})}{(v\cdot q^{\,\prime\,})}
\,\,{\rm Re}\biggl[\frac{b}{D}\biggr]\,.
\eqno{({\rm C}.3)}
$$
The expression obtained in particular shows that a contribution of the term
proportional to $\not\!q$ in soft-quark propagator (C.1) (containing
zero-temperature part) has no a singularity. Now we are coming from the
functions $a$ and $b$ to the scalar propagators
$\,^{\ast}\!\Delta_{\pm}(q^{\,\prime\,})$ according to formulae
$$
\frac{1+a}{D}=\frac{1}{2|{\bf q}^{\,\prime}|}\,\,
\Bigl\{\,^{\ast}\!\Delta_{+}(q^{\,\prime\,})\,-
\,^{\ast}\!\Delta_{-}(q^{\,\prime\,})\Bigr\}\,,\;\;
\frac{b}{D}=
\frac{1}{2}\left(1-\frac{q^{\,\prime\,0}}{|{\bf q}^{\,\prime}|}\right)
\!\,^{\ast}\!\Delta_{+}(q^{\,\prime\,})+
\frac{1}{2}\left(1+\frac{q^{\,\prime\,0}}{|{\bf q}^{\,\prime}|}\right)
\!\,^{\ast}\!\Delta_{-}(q^{\,\prime\,}).
$$
Taking into account the property of the quark scalar propagators
$\,^{\ast}\!\Delta_{-}(q^{\,\prime\,})=
-\Bigl(\!\,^{\ast}\!\Delta_{+}(-q^{\,\prime\,})\Bigr)^{\!\ast}\!,$
we result (C.3) in final form
$$
\frac{2}{E}\int\!dq^{\,\prime}\,\delta(v\cdot q^{\,\prime\,})\,\frac{1}{|{\bf q}^{\,\prime}|}
\,{\rm Re}\Bigl[\,^{\ast}\!\Delta_{+}(q^{\,\prime\,})\Bigr]
+
\frac{2}{E}\int\!dq^{\,\prime}\,\frac{\delta(v\cdot q^{\,\prime\,})}{(v\cdot q^{\,\prime\,})}
\,\left(1-\frac{q^{\,\prime\,0}}{|{\bf q}^{\,\prime}|}\right)
{\rm Re}\,\Bigl[\,^{\ast}\!\Delta_{+}(q^{\,\prime\,})\Bigr].
$$
The close analysis of initial presuppositions shows that the
assumption of straightness of a hard parton trajectory is origin
of the singularity $\delta(v\cdot q^{\,\prime\,})/(v\cdot q^{\,\prime\,})$.
So the function $\delta(v\cdot q^{\,\prime\,})$ arises under the Fourier
transformation of the function $\delta({\bf x}-{\bf v}t)$ entering
into the initial source of hard particle
$\eta_{\theta\alpha}^{(0)}(x)$ and the factor $1/(v\cdot q^{\,\prime\,})$
appears in calculation of the integral
$$
\int\limits_{t_0}^t\!\Bigl(\bar{\chi}_{\alpha}\psi_{\alpha}^i
(\tau,{\bf v}\tau)\Bigr)\,d\tau
=\int\!dq^{\,\prime}\!
\!\int\limits_{t_0}^t\!
{\rm e}^{-i(v\cdot q^{\,\prime\,})\tau}
\Bigl(\bar{\chi}_{\alpha}\psi_{\alpha}^i(q^{\,\prime\,})\Bigr)
\,d\tau
\eqno{({\rm C}.4)}
$$
$$
=\int\!\frac{i\,dq^{\,\prime}}{(v\cdot q^{\,\prime\,})}
\Bigl[\,{\rm e}^{-i(v\cdot q^{\,\prime\,})t} -
{\rm e}^{-i(v\cdot q^{\,\prime\,})t_0}\Bigr]
\Bigl(\bar{\chi}_{\alpha}\psi_{\alpha}^i(q^{\,\prime\,})\Bigr).
$$
Here the soft-quark field $\psi_{\alpha}^i(x)$ is defined on the linear
trajectory ${\bf x}={\bf v}t$. Thus for regularization of the
singularity it is necessary to take into account a change of hard
parton trajectory\footnote{In the general case along with a change of the
trajectory it is necessary to take into account a change of
a polarization state of hard particle defined by the spinor $\chi$. The
equation describing a change of $\chi=\chi(t)$ in semiclassical
approximation may be in principle obtained from the following
reasonings. Let us define the spin 4-vector $S^{\mu}$ (or the spin tensor
$S^{\mu\nu}$) setting
$S^{\mu}(t)\equiv\bar{\chi}(t)\gamma_5\gamma^{\mu}\chi(t)$
($S^{\mu\nu}(t)\equiv\bar{\chi}(t)\sigma^{\mu\nu}\chi(t)$). By using
known the semiclassical equation for spin motion in external field, one
can attempt to restore the equation of motion for the spinor
$\chi(t)$. This is in a certain sense similar to restoring
equation of motion for the Grassmann charge $\theta(t)$ entering
into the definition of usual charge
$Q^a(t)=\theta^{\dagger}(t)t^a\theta(t)$ when we know
the equation of motion for the charge $Q^a(t)$ (Eq.\,(\ref{eq:3w})).} in
`collisions' of hard parton
with soft fluctuation field of system (although this is beyond the frameworks
of the HTL approximation accepted in this work). If we present a weakly
perturbative trajectory in the form
${\bf x}(t)={\bf v}t+\triangle{\bf x}(t),\,
\vert\triangle{\bf x}(t)\vert\ll\vert{\bf v}\vert\,t$, then instead of
(C.4), e.g., we have
$$
\int\!\!dq^{\,\prime}
\Bigl(\bar{\chi}_{\alpha}\psi_{\alpha}^i(q^{\,\prime\,})\Bigr)
\!\int\limits_{t_0}^t\!d\tau\,
{\rm e}^{-i(v\cdot q^{\,\prime\,})\tau\,+\,
i{\bf q}^{\,\prime}\cdot\triangle{\bf x}(\tau)}.
$$
The integral in $d\tau$ here should be exactly calculated, since
expansion in a series with respect to $\triangle{\bf x}(\tau)$
generates immediately the singularity. An explicit form of the
function $\triangle{\bf x}(t)$ was obtained in Appendix of
Ref.\,\cite{JOP_G_2000} in the linear approximation in the
soft-gluon field $A^{(0)}$. This function is suppressed by the
factor $1/E$, where $E$ is energy of hard parton with respect to a
change of its color charge $\triangle Q^a(t)$ obtained in
the same approximation.

In the remainder of this Appendix we consider a singular contribution
connected with the soft-gluon loop (the second term in Eq.\,(\ref{eq:7e})). We
use an explicit expression for gluon propagator (in covariant gauge)
$$
\,^{\ast}{\cal D}_{\mu \nu}(k)= -
P_{\mu \nu}(k) \,^{\ast}\!\Delta^t(k) -
Q_{\mu \nu}(k) \,^{\ast}\!\Delta^l(k) +
\xi D_{\mu \nu}(k)\Delta^0(k),
\eqno{({\rm C}.5)}
$$
where Lorentz matrices are defined by
$$
P_{\mu\nu}(k) = g_{\mu\nu}-D_{\mu\nu}(k)-Q_{\mu\nu}(k),\quad
Q_{\mu \nu}(k) =
\frac{\bar{u}_{\mu}(k)\bar{u}_{\nu}(k)}{\bar{u}^2(k)}\,,\quad
D_{\mu \nu}(k)=\frac{k_{\mu}k_{\nu}}{k^2}\,,\quad
\eqno{({\rm C}.6)}
$$
$$
\Delta^0(k)=1/k^2,\quad
\bar{u}_{\mu}=k^2u_{\mu}-k_{\mu}(k\cdot u),
$$
and $\xi$ is a gauge fixing parameter. With the help of (C.5), (C.6) the
second term in Eq.\,(\ref{eq:7e}) can be presented in the following form:
$$
\int\!dk^{\,\prime}\,\delta(v\cdot k^{\,\prime\,})\,
\Bigl[\,^{\ast}\!\Delta^{t}(k^{\,\prime\,})-
\,^{\ast}\!\Delta^{l}(k^{\,\prime\,})\Bigr]
\frac{1}{{\bf k}^{\,\prime\,2}}
\left\{2k^{\,\prime\,0}-\frac{k^{\,\prime\,2}}{(v\cdot k^{\,\prime\,})}
\right\}\,-\,
v^2\!\!
\int\!dk^{\,\prime}\,\frac{\delta(v\cdot k^{\,\prime\,})}{(v\cdot k^{\,\prime\,})}
\,\,^{\ast}\!\Delta^{t}(k^{\,\prime\,}).
$$
This expression also contains both finite and singular contributions. The
last term vanishes in the case of massless hard particle $(v^2=0)$.
In opposite case for heavy particle with the mass $M$ this term is proportional
to $M^2/E^2$. The term with the gauge parameter exactly equals zero. The factor
$\delta(v\cdot k^{\,\prime\,})/(v\cdot k^{\,\prime\,})$ arises from
expression similar to (C.4), where instead of the function
$\Bigl(\chi_{\alpha}\psi_{\alpha}^i(\tau,{\bf v}\tau)\Bigr)$ it is necessary
to mean the function $\Bigl(v^{\mu}A_{\mu}^a(\tau,{\bf v}\tau)\Bigr)$, i.e.,
the soft-gluon field defined on the parton linear trajectory.


\section*{\bf Appendix D}
\setcounter{equation}{0}

\begin{figure}[hbtp]
\begin{center}
\includegraphics[width=0.95\textwidth]{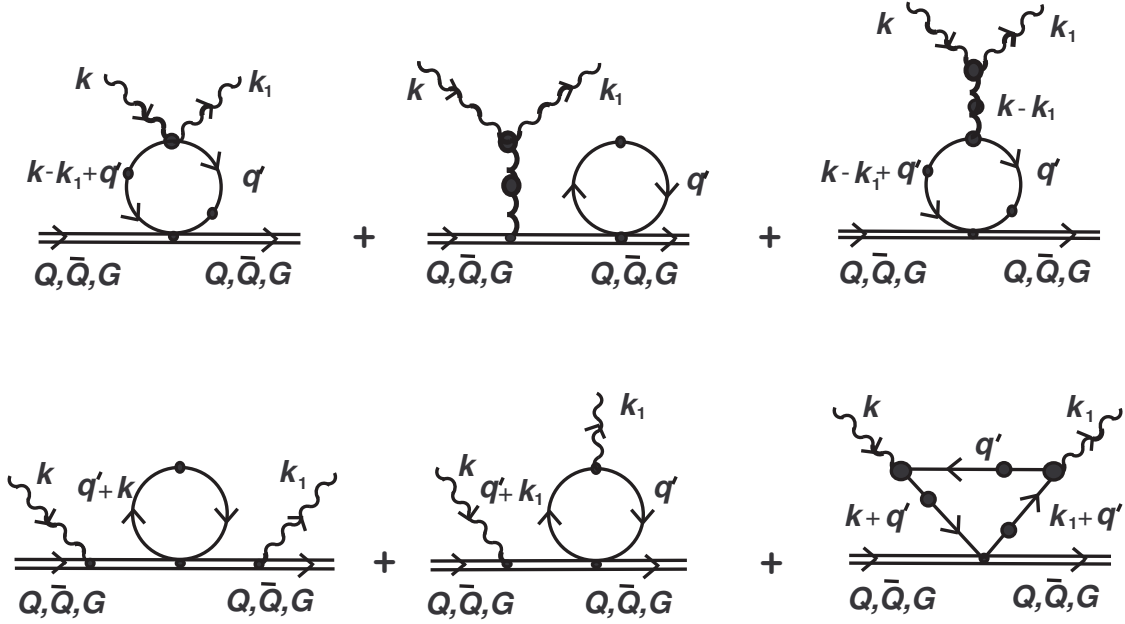}
\end{center}
\caption{\small The soft-quark loop corrections to the elastic scattering
of soft gluon elementary excitation
off the hard test parton drawn on Figs.\ref{fig1} and \ref{fig2} in
Ref.\,\cite{markov_AOP_04}.}
\label{fig14}
\end{figure}

The third order derivative of relation (\ref{eq:3f}) with respect to the
Grassmann charges $\theta_0^{\dagger}$, $\theta_0$ and free soft-gluon
field $A^{(0)}$ has the following form:
$$
\left.\frac{\delta^{3}\!j^{a}_{\mu}[A,\bar{\psi},\psi,Q_0,\theta^{\dagger}_0,
\theta_0](k)}
{\delta \theta_0^{\dagger\, i}\, \delta \theta_0^{j}\,
\,\delta A^{(0)a_1\mu_1}(k_1)}
\right|_{\,0}
=
\left.\frac{\delta^3\tilde{j}^{a}_{\mu}
[A^{(0)},\bar{\psi}^{(0)},\psi^{(0)},Q_0,\theta_0^{\dagger},\theta_0,\,](k)}
{\delta \theta_0^{\dagger\, i}\, \delta \theta_0^{j}
\,\delta A^{(0)a_1\mu_1}(k_1)}\,
\right|_{\,0}
\eqno{({\rm D}.1)}
$$
$$
=\frac{\,g^4}{(2\pi)^6}\,
\Biggl\{\int\biggl[\,\Bigl(\bar{\chi}\,^{\ast}\!S(-q^{\,\prime\,})
\delta{\Gamma}^{(G)aa_1,\,ij}_{\mu\mu_1}
(k,-k_1;q^{\,\prime},-k+k_1-q^{\,\prime\,})
S(k-k_1+q^{\,\prime\,})\chi\Bigr)
$$
$$
-\,[t^{a},t^{a_1}]^{ij}
\,^{\ast}\Gamma_{\mu\nu\mu_1}(k,-k+k_1,-k_1)
\,^\ast{\cal D}^{\nu\nu^{\,\prime}}\!(k-k_1)
\Biggl(
\frac{v_{\nu^{\,\prime}}}{(v\cdot q^{\,\prime\,})}\,
\Bigl[\,\Bigl(\bar{\chi}\,^{\ast}\!S(q^{\,\prime\,})\chi\Bigr)
-\Bigl(\bar{\chi}\,^{\ast}\!S(-q^{\,\prime\,})\chi\Bigr)\Bigr]
\hspace{1cm}
$$
$$
\hspace{4.5cm}
+\,\Bigl[\,
\bar{\chi}\,^{\ast}\!S(k-k_1-q^{\,\prime\,})
\,^{\ast}\Gamma^{(G)}_{\mu}(k-k_1;-k+k_1+q^{\,\prime},-q^{\,\prime\,})
\,^{\ast}\!S(q^{\,\prime\,})\chi\,\Bigr]\Bigg)
$$
$$
+\,(t^{a}t^{a_1})^{ij}
\Bigl[\bar{K}^{(G)}_{\mu}
({\bf v},\bar{\chi}|\,k,-k-q^{\,\prime\,})
\,^{\ast}\!S(k+q^{\,\prime\,})
K_{\mu_1}^{(Q)}({\bf v},\chi|\,k_1,-k-q^{\,\prime\,})\Bigr]
\hspace{4.8cm}
$$
$$
-\,(t^{a_1}t^{a})^{ij}
\Bigl[\bar{K}^{(Q)}_{\mu_1}
({\bf v},\bar{\chi}|\,-k_1,k-q^{\,\prime\,})
\,^{\ast}\!S(k-q^{\,\prime\,})
K_{\mu}^{(G)}({\bf v},\chi|\,k,-k+q^{\,\prime\,})\Bigr]
\,\biggr]
\,\delta(v\cdot q^{\,\prime\,})dq^{\,\prime}
\Biggr\}\,
\delta(v\cdot (k-k_1)).
$$
The graphic interpretation of various terms in this expression is presented
in Fig.\,\ref{fig14}.


\section*{\bf Appendix E}
\setcounter{equation}{0}

\begin{figure}[hbtp]
\begin{center}
\includegraphics*[scale=0.6]{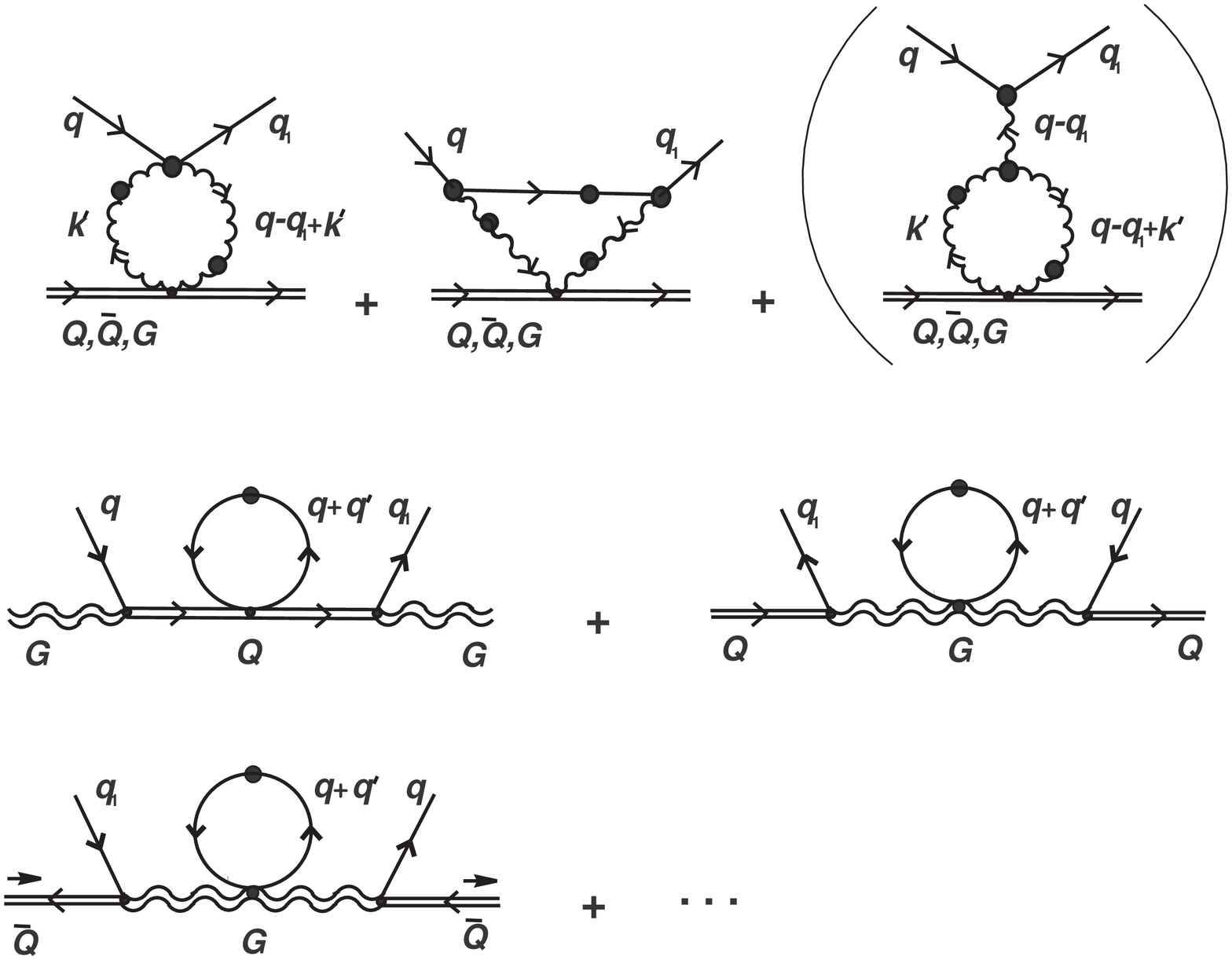}
\end{center}
\caption{\small The soft-quark loop corrections to the elastic scattering
processes of soft quark elementary excitation off the hard test particle
(Figs.\,\ref{fig1} and \ref{fig3}). These diagrams are additional to ones
depicted in Fig.\,\ref{fig10}.}
\label{fig15}
\end{figure}

In this Appendix we give an explicit form of third order derivative
of relation (\ref{eq:3g}) with respect to the color charges $Q_0^a$,
$Q_0^b$ and free soft-quark field $\psi^{(0)}$:
$$
\left.\frac{\delta^3\eta^i_{\alpha}[A,\bar{\psi},\psi,Q_0,\theta_0](q)}
{\delta Q_0^a\,\delta Q_0^b
\,\delta\psi_{\alpha_1}^{(0)i_1}(q_1)}\,
\right|_{\,0}=
\left.\frac{\delta^2\tilde{\eta}^i_{\alpha}
[A^{(0)},\bar{\psi}^{(0)},\psi^{(0)},Q_0,\theta_0](q)}
{\delta Q_0^a\,\delta Q_0^b
\,\delta\psi_{\alpha_1}^{(0)i_1}(q_1)}\,
\right|_{\,0}
\eqno{({\rm E}.1)}
$$
$$
=-\frac{\,g^4}{(2\pi)^6}\,\{t^{a},t^{b}\}^{ii_1}\Biggl\{\int\biggl[\,
\alpha\,\chi_{\alpha}\bar{\chi}_{\alpha_1}\,\frac{1}{(v\cdot q)(v\cdot q_1)}\,
(v_{\mu}\,^{\ast}{\cal D}^{\mu\mu^{\prime}}(k^{\,\prime\,})v_{\mu^{\prime}})
$$
$$
-\,\frac{1}{2}\,\delta{\Gamma}^{(Q\!;\,{\cal S})}_{\mu\nu,\,\alpha\alpha_1}
(k^{\,\prime},q-q_1-k^{\,\prime};q_1,-q)
\,^\ast{\cal D}^{\mu\mu^{\prime}}\!(k^{\,\prime\,})
v_{\mu^{\prime}}
\,^\ast{\cal D}^{\nu\nu^{\prime}}\!(q-q_1-k^{\,\prime\,})
v_{\nu^{\prime}}\biggr]\,
\delta(v\cdot k^{\,\prime\,})dk^{\,\prime}
$$
$$
+\!\int\Bigl[\,
K_{\alpha\beta}^{(Q)}(\chi,\bar{\chi}|\,q,-q+q^{\,\prime\,})
\,^{\ast}\!S_{\beta\beta^{\prime}}(q-q^{\,\prime\,})
K^{(Q)}_{\beta^{\prime}\alpha_1}(\chi,\bar{\chi}|\,q-q^{\,\prime},-q_1)
\Bigl]\,\delta(v\,\cdot\,q^{\,\prime\,})dq^{\,\prime}\Biggr\}
\,\delta(v\cdot (q-q_1)).
\hspace{4.7cm}
$$
The diagrammatic interpretation of different terms on the
right-hand side is presented in Fig.\,\ref{fig15}.
The vertex function
$\delta{\Gamma}^{(Q\!;\,{\cal S})}_{\mu\nu,\,\alpha\alpha_1}$ is the
`symmetric' part of the HTL-induced vertex between quark pair and two gluons
$\delta{\Gamma}^{(Q)ab,\,ii_1}_{\mu\nu,\,\alpha\alpha_1}$. It is defined by
Eq.\,(I.5.19). To the diagram in parentheses in Fig.\,\ref{fig15} there
corresponds a term
\[
\,^{\ast}\Gamma^{(Q)}_{\mu,\,\alpha\alpha_1}(q-q_1;q_1,-q)
\,^{\ast}{\cal D}^{\mu\mu^{\prime}}(q-q_1)\int\!\!
\,^{\ast}\Gamma_{\mu^{\prime}\nu^{\prime}\lambda^{\prime}}
(q-q_1,-k^{\,\prime},-q+q_1+k^{\,\prime\,})
\,^{\ast}{\cal D}^{\lambda^{\prime}\lambda}(q-q_1-k^{\,\prime\,})v_{\lambda}
\]
\[
\times
\,^{\ast}{\cal D}^{\nu^{\prime}\nu}(k^{\,\prime\,})
v_{\nu}\,\delta(v\cdot k^{\,\prime\,})dk^{\,\prime}\,\delta(v\cdot (q-q_1)).
\]
By virtue of a property of three-gluon HTL-vertex:
$\!\,^{\ast}\Gamma_{\mu\mu_1\mu_2}(k,k_1,k_2)=-
\,^{\ast}\Gamma_{\mu\mu_2\mu_1}(k,k_2,k_1)$, the integrand is odd function
under the replacement of integration variable
$k^{\,\prime}\rightarrow q-q_1-k^{\,\prime}$ and therefore in integrating
it is equal to zero.


\section*{\bf Appendix F}
\setcounter{equation}{0}

In Section 9, considering a structure of the scattering probability
${\it w}_{\,q\rightarrow {\rm g}}^{(f;\,b)}$
we have faced with contractions of the HTL-resummed vertex\footnote{In
high-temperature QCD, under conditions when we can neglect by mass of medium
constituents, there is no (linear) Landau damping of on-shell soft excitations
Fermi and Bose statistics. Consequence of this fact is equality of the
HTL-resummed two-quark\,--\,one-gluon vertices
$\,^{\ast}\Gamma^{(Q)}_{\mu}=\,^{\ast}\Gamma^{(G)}_{\mu}\equiv
\,^{\ast}\Gamma_{\mu}$, as it is supposed in this Appendix.} between quark
pair and gluon $\,^{\ast}\Gamma^{i}(k;l,-q)$ ($\equiv \,^{\ast}\Gamma^{i}$)
with soft momentum $k^i$ of plasmon mode and the polarization vectors
${\rm e}^i(\hat{\bf k},\xi)$ of transverse soft-gluon mode. In this Appendix
we give various forms of representations of the vertex function
$\,^{\ast}\Gamma^{i}$, which have been actively used in Section 9.

From analysis of the explicit expression for the vertex
$\,^{\ast}\Gamma^{i}$ derived by Frenkel and Taylor (Eq.\,(3.38) in
Ref.\,\cite{frenkel}), it is easy to see that this function can be presented
in the form of the expansion
$$
\,^{\ast}\Gamma^{\,i} = \gamma^0
\delta\!{\it \Gamma}_0^{\,i} +
({\bf l}\cdot \vec{\gamma})
\,^{\ast}\!{\it \Gamma}_{\parallel}^{\,i} +
(({\bf n}\times {\bf l})\cdot \vec{\gamma})
\,^{\ast}\!{\it \Gamma}_{\perp}^{\,i} +
({\bf n}\cdot\vec{\gamma})
\,^{\ast}\!{\it \Gamma}_{\!1\perp}^{\,i},
\eqno{({\rm F}.1)}
$$
where ${\bf l}\equiv {\bf q}-{\bf k}$,
${\bf n} \equiv {\bf q}\times {\bf k}$ and the `scalar' coefficient
functions are defined as
$$
\delta\!{\it \Gamma}_0^{\,i}  = \omega_0^2\!\int\!
\frac{{\rm d}\Omega_{\bf v}}{4\pi}
\,\frac{v^{i} }{(v\cdot l + i\epsilon )(v\cdot q)},
\quad \epsilon\rightarrow +0\,,
\hspace{1cm}
$$
$$
\,^{\ast}\!{\it \Gamma}_{\parallel}^{\,i}  =
\frac{l^{i} }{{\bf l}^2}\, +\, \delta\!{\it \Gamma}_{\parallel}^{\,i}
\equiv\frac{l^{i} }{{\bf l}^2} \,-
\frac{\omega_0^2}{{\bf l}^2}\!\int\!\frac{{\rm d}\Omega_{\bf v}}{4\pi}
\,\frac{v^{i} \,({\bf v}\cdot {\bf l})}
{(v\cdot l + i\epsilon)(v\cdot q)}\,,
$$
$$
\hspace{2.4cm}
\,^{\ast}\!{\it \Gamma}_{\perp}^{\,i}  =
\frac{({\bf n}\times {\bf l})^{i} }{{\bf n}^2{\bf l}^2}\,+\,
\delta\!{\it \Gamma}_{\perp}^{\,i}  \equiv
\frac{({\bf n}\times {\bf l})^{i} }{{\bf n}^2 {\bf l}^2} \,-
\frac{\omega_0^2}{{\bf n}^2\,{\bf l}^2}\!\int\!\frac{{\rm d}\Omega_{\bf v}}
{4\pi}\,\frac{v^{i} \,({\bf v}\cdot ({\bf n}
\times {\bf l}))}{(v\cdot l + i\epsilon )(v\cdot q)}\,,
$$
$$
\hspace{0.3cm}
\,^{\ast}\!{\it \Gamma}_{\!1\perp}^{\,i}  =
\frac{{\bf n}^{i}}{{\bf n}^2}\,+\,
\delta\!{\it \Gamma}_{\!1\perp}^{\,i}  \equiv
\frac{{\bf n}^{i}}{{\bf n}^2} \,-
\frac{\omega_0^2}{{\bf n}^2}\!\int\!\frac{{\rm d}\Omega_{\bf v}}
{4\pi}\,\frac{v^{i} \,({\bf v}\cdot {\bf n})}
{(v\cdot l + i\epsilon )(v\cdot q)}\,.
$$
The matrix basis in expansion (F.1) is convenient by virtue of its
`orthogonality' in computing traces. The `transverse' vertex functions
possess obvious properties:
$$
l^i\,^{\ast}\!{\it \Gamma}_{\perp}^{\,i}=
n^i\,^{\ast}\!{\it \Gamma}_{\perp}^{\,i}=0,
$$
$$
k^i\,^{\ast}\!{\it \Gamma}_{\!1\perp}^{\,i}=
q^i\,^{\ast}\!{\it \Gamma}_{\!1\perp}^{\,i}=0.
\eqno{({\rm F}.2)}
$$
However, in concrete applications it is considerably more convenient to
use another representation of the decomposition (F.1) \cite{markov_PRD_01}
$$
\,^{\ast}\Gamma^i =
-\,h_{-}(\hat{\bf l})
\,^{\ast}\!{\it \Gamma}_{+}^{\,i}
-h_{+}(\hat{\bf l})
\,^{\ast}\!{\it \Gamma}_{-}^{\,i}
+ 2h_{-}(\hat{\bf q})\,{\bf l}^2 \vert {\bf q}\vert
\,^{\ast}\!{\it \Gamma}_{\perp}^{\,i}
+({\bf n}\cdot\vec{\gamma})
\,^{\ast}\!{\it \Gamma}_{\!1\perp}^{\,i},
\eqno{({\rm F}.3)}
$$
where the `scalar' vertex functions $\!\,^{\ast}\!{\it \Gamma}_{\pm}^{\,i}$
are connected with the previous ones by relations
$$
\,^{\ast}\!{\it \Gamma}_{\pm}^{\,i} \equiv -\,\delta\!{\it \Gamma}_0^{\,i} \mp
\vert {\bf l}\vert \,^{\ast}\!{\it \Gamma}_{\parallel}^{\,i}
+ \frac{{\bf n}^2}{\vert {\bf q} \vert}\,\frac{1}{1\mp \hat{\bf q}\cdot
\hat{\bf l}} \,^{\ast}\!{\it\Gamma}_{\perp}^{\,i}\,.
\eqno{({\rm F}.4)}
$$
The expansion (F.3) with the matrix $h_{-}(\hat{\bf q})$ in the last but one
term is specially adapted to studying of plasmino branch of fermion excitations.
In the case of a branch describing normal-particle excitations, instead of
(F.3), (F.4) it is necessary to use the following decomposition:
$$
\,^{\ast}\Gamma^i =
-\,h_{-}(\hat{\bf l})
\,^{\ast}\!\acute{{\it \Gamma}}_{+}^{\,i}
-h_{+}(\hat{\bf l})
\,^{\ast}\!\acute{{\it \Gamma}}_{-}^{\,i}
- 2h_{+}(\hat{\bf q})\,{\bf l}^2 \vert {\bf q} \vert
\,^{\ast}\!{\it \Gamma}_{\perp}^{\,i}
+({\bf n}\cdot\vec{\gamma})
\,^{\ast}\!{\it \Gamma}_{\!1\perp}^{\,i},
\eqno{({\rm F}.5)}
$$
where the `scalar' vertex functions
$\,^{\ast}\!\acute{{\it \Gamma}}_{\pm}^{\,i}$ are defined as
$$
\,^{\ast}\!\acute{{\it \Gamma}}_{\pm}^{\,i}
\equiv -\,\delta\!{\it \Gamma}_0^{\,i} \mp
\vert {\bf l}\vert \,^{\ast}\!{\it \Gamma}_{\parallel}^{\,i}
- \frac{{\bf n}^2}{\vert {\bf q} \vert}\,
\frac{1}{1\pm \hat{\bf q}\cdot
\hat{\bf l}} \,^{\ast}\!{\it \Gamma}_{\perp}^{\,i}\,.
\eqno{({\rm F}.6)}
$$

In the paper \cite{markov_PRD_01} the system of kinetic equations describing a
change of the plasmino and plasmon number densities generated by induced
scattering of plasmino (plasmon) off the hard test particle with transition in
plasmon (plasmino) has been also obtained:
$$
\frac{\partial n_{\bf q}^{-}}{\partial t} +
{\bf v}_{\bf q}^{-}\cdot \frac{\partial n_{\bf q}^{-}}
{\partial {\bf x}} =
-\,g^2C_F\,n_{\bf q}^{-}\!\int\!{\rm d}{\bf k}\,
{\cal Q}({\bf q},{\bf k}) N_{\bf k}^l\,,
\eqno{({\rm F}.7)}
$$
$$
\frac{\partial N_{\bf k}^{l}}{\partial t} +
{\bf v}_{\bf k}^{l}\cdot \frac{\partial N_{\bf k}^{l}}
{\partial {\bf x}} =
g^2n_fT_FN_{\bf k}^{l}\!\int\!{\rm d}{\bf q}\,
{\cal Q}({\bf q},{\bf k})\,n_{\bf q}^{-}.
\eqno{({\rm F}.8)}
$$
Here in the latter equation the factor $n_f$ accounts all
kinematically accessible quark flavors. The scattering kernel
${\cal Q}({\bf q},{\bf k})$ has the following structure:
$$
{\cal Q}({\bf q},{\bf k})\!=\!
\omega_0^2\!
\left(\frac{{\rm Z}_{l}({\bf k})}{2\omega_{\bf k}^l}\right)\!
\left(\frac{{\rm Z}_{-}({\bf q})}{2}\right)\!
\frac{k^2}{(\omega_{\bf k}^l)^2
{\bf k}^2}\!\int\!\frac{{\rm d}\Omega_{\bf v}}{4\pi}\,
\Bigl\{\rho_{+}({\bf v};\hat{\bf q},\hat{\bf l})
{\it w}_{\bf v}^{+}({\bf q},{\bf k})
+ \rho_{-}({\bf v};\hat{\bf q},\hat{\bf l})
{\it w}_{\bf v}^{-}({\bf q},{\bf k})\Bigr\}
$$
$$
\times\,2\pi\delta(\omega_{\mathbf q}^{-} -
\omega_{\bf k}^l - {\bf v}\cdot ({\bf q} - {\bf k})),
\eqno{({\rm F}.9)}
$$
where the coefficient functions $\rho_{\pm}$ are
$$
\rho_{\pm}({\bf v};\hat{\bf q},\hat{\bf l}) =
1+{\bf v}\cdot\hat{\bf q}\mp(\hat{\bf q}\cdot\hat{\bf l}
+{\bf v}\cdot\hat{\bf l}\,)\,,
\eqno{({\rm F}.10)}
$$
the `scalar' scattering probabilities ${\it w}_{\bf v}^{\pm}$ are defined
by expressions
$$
{\it w}_{\bf v}^{\pm}({\bf q},{\bf k}) =
\biggl\vert\,\frac{{\bf v}\cdot {\bf k}}{v\cdot q}
\,\,+\,^{\ast}\!\Delta_{\pm}(l)\,^{\ast}\!{\it \Gamma}_{\pm}(k;l-q)
\,\biggr\vert^{\,2}_{\,\,{\rm on-shell}},
\eqno{({\rm F}.11)}
$$
and $\omega_0^2=g^2C_FT^2/8$ is the plasma frequency of the quark
sector of plasma excitations.

\end{document}